\documentclass[runningheads,envcountsame]{llncs}

\makeatletter
\RequirePackage[bookmarks,unicode,colorlinks=true]{hyperref}%
   \def\@citecolor{blue}%
   \def\@urlcolor{blue}%
   \def\@linkcolor{blue}%

\def\orcidID#1{\smash{\href{http://orcid.org/#1}{\protect\raisebox{-1.25pt}{\protect\includegraphics{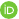}}}}}
\makeatother

\usepackage{xcolor,colortbl}
\usepackage{mathtools}
\usepackage{multirow}
\usepackage{stmaryrd}
\usepackage{cancel}
\usepackage{amsmath,amssymb}
\usepackage{thmtools, thm-restate}
\usepackage[shortlabels]{enumitem}
\usepackage{microtype}
\usepackage{wrapfig}
\usepackage{pbox}
\usepackage{marvosym}
\usepackage{listings}
\usepackage[ruled, vlined, linesnumbered]{algorithm2e}
\hypersetup{
	pdftitle={Proving Almost-Sure Innermost Termination of Probabilistic Term
          Rewriting Using Dependency Pairs}, colorlinks=true, linkcolor=blue, citecolor=olive, filecolor=magenta, urlcolor=cyan
}
\usepackage{todonotes}
\usepackage{placeins}
\usepackage{float}
\usepackage{tikz}
\usepackage{caption}
\usepackage{subcaption}
\usepackage{mymatrix}
\usepackage{nicefrac,xfrac}
\usetikzlibrary{shapes,calc,arrows,automata,decorations.pathmorphing,backgrounds}

\RequirePackage{makecell}

\usepackage[capitalize,nameinlink]{cleveref}
\usepackage{IEEEtrantools}

\newenvironment{myproof}{
	\noindent{\it Proof.}
}{\qed
	\medskip
}
\makeatletter \newcommand*\bigcdot{\mathpalette\bigcdot@{.4}}
\newcommand*\bigcdot@[2]{\mathbin{\vcenter{\hbox{\scalebox{#2}{$\m@th#1\bullet$}}}}}
\makeatother

\renewcommand{\emptyset}{\varnothing}

\newcommand{\disabledcomment}[1]{}
\newcommand{\oldcomment}[1]{}

\newcommand{\dontprint}[1]{}

\renewcommand{\epsilon}{\varepsilon}

\newcommand{\IN}{\mathbb{N}}
\newcommand{\IR}{\mathbb{R}}

\newcommand{\R}{\mathcal{R}}

\newcommand{\PP}{\mathcal{P}}
\newcommand{\SSS}{\mathcal{S}}

\newcommand{\aprove}{\textsf{AProVE}}

\crefname{definition}{Def.}{Def.}
\crefname{example}{Ex.}{Ex.}
\crefname{counterexample}{Counterex.}{Counterex.}
\crefname{appendix}{App.}{App.}
\crefname{ex}{Ex.}{Ex.}
\crefname{theorem}{Thm.}{Thm.}
\crefname{lemma}{Lemma}{Lemmas}
\crefname{remark}{Rem.}{Rem.}
\crefname{section}{Sect.}{Sect.}
\crefname{subsection}{Sect.}{Sect.}
\crefname{subsubsection}{Sect.}{Sect.}
\crefname{line}{Line}{Lines}
\crefname{corollary}{Cor.}{Cor.}
\crefname{figure}{Fig.}{Fig.}
\crefname{enumi}{}{}
\crefname{algorithm}{Alg.}{Alg.}

\renewcommand{\ts}{\mathsf{s}}
\renewcommand{\O}{\mathcal{O}}
\newcommand{\tf}{\mathsf{f}}
\newcommand{\tg}{\mathsf{g}}
\renewcommand{\th}{\mathsf{h}}
\newcommand{\ta}{\mathsf{a}}
\newcommand{\tb}{\mathsf{b}}

\newcommand{\tminus}{\mathsf{minus}}
\newcommand{\tdiv}{\mathsf{div}}
\newcommand{\tpdiv}{\mathsf{pdiv}}

\newcommand{\trw}{\mathsf{rw}}
\newcommand{\tcons}{\mathsf{cons}}
\newcommand{\tnil}{\mathsf{nil}}

\newcommand{\ttrue}{\mathsf{true}}
\newcommand{\tfalse}{\mathsf{false}}
\newcommand{\tM}{\mathsf{M}}
\newcommand{\tD}{\mathsf{D}}
\newcommand{\tF}{\mathsf{F}}
\newcommand{\tG}{\mathsf{G}}
\newcommand{\tH}{\mathsf{H}}
\newcommand{\tA}{\mathsf{A}}
\newcommand{\tB}{\mathsf{B}}

\newcommand{\trotate}{\mathsf{rotate}}
\newcommand{\tapp}{\mathsf{app}}
\newcommand{\tqs}{\mathsf{qs}}
\newcommand{\tqshelp}{\mathsf{qsHelp}}
\newcommand{\tifhigh}{\mathsf{ifHigh}}
\newcommand{\thigh}{\mathsf{high}}
\newcommand{\tiflow}{\mathsf{ifLow}}
\newcommand{\tlow}{\mathsf{low}}
\newcommand{\tleq}{\mathsf{leq}}
\newcommand{\tstop}{\mathsf{stop}}
\newcommand{\xs}{\mathit{xs}}
\newcommand{\ys}{\mathit{ys}}
\newcommand{\zs}{\mathit{zs}}
\newcommand{\tbogo}{\mathsf{bogo}}
\newcommand{\tbogohelp}{\mathsf{bogoHelp}}
\newcommand{\tifsorted}{\mathsf{ifSorted}}
\newcommand{\tifleq}{\mathsf{ifLeq}}
\newcommand{\tsortr}{\mathsf{sortR}}
\newcommand{\tdecreaseX}{\mathsf{decreaseX}}
\newcommand{\tloopGuard}{\mathsf{loopGuard}}
\newcommand{\tloop}{\mathsf{loop}}
\newcommand{\tdecreaseY}{\mathsf{decreaseY}}

\newcommand{\C}[1]{\mathcal{#1}}
\newcommand{\F}[1]{\mathfrak{#1}}

\newcommand{\ctroot}{\F{r}}
\newcommand{\ctleaf}{\operatorname{Leaf}}

\newcommand{\ctdepth}{\operatorname{d}}

\newcommand{\ctlevelTwo}{\C{L}_{2}}
\newcommand{\ctlevelTwowithborder}{\F{L}_2}

\newcommand{\Pol}{\operatorname{Pol}}
\newcommand{\Proc}{\operatorname{Proc}}

\newcommand{\Com}[1]{\mathsf{c}_{#1}}
\newcommand{\SigmaDP}{\Sigma \uplus \Sigma^{\#}}

\newcommand{\TSet}[2]{\mathcal{T}\left(#1,#2\right)}

\newcommand{\VSet}{\mathcal{V}}

\newcommand{\DTuple}[1]{\mathcal{DT}(#1)}
\newcommand{\DPair}[1]{\mathcal{DP}(#1)}

\newcommand{\FDist}{\operatorname{FDist}}
\newcommand{\Supp}{\operatorname{Supp}}

\newcommand{\rootsym}{\operatorname{root}}

\newcommand{\rules}{\operatorname{Rules}}
\newcommand{\urules}{\mathcal{U}}

\newcommand{\projOne}{\mathrm{proj}_1}
\newcommand{\projTwo}{\mathrm{proj}_2}
\newcommand{\Subd}{\mathrm{Sub}_D}
\newcommand{\MSubd}{\mathrm{MSub}_D}

\newcommand{\Junk}{\mathrm{Junk}}

\newcommand\nameft\textrm

\makeatletter
\newcommand{\oset}[2]{%
  {\mathop{#2}\limits^{\vbox to 1\ex@{\kern-\tw@\ex@
   \hbox{\scriptsize #1}\vss}}}}
\makeatother

\makeatletter
\newcommand{\osetthree}[2]{%
  {\mathop{#2}\limits^{\vbox to 3\ex@{\kern-\tw@\ex@
   \hbox{\scriptsize #1}\vss}}}}
\makeatother

\makeatletter
\newcommand{\osetfive}[2]{%
  {\mathop{#2}\limits^{\vbox to 5\ex@{\kern-\tw@\ex@
   \hbox{\scriptsize #1}\vss}}}}
\makeatother

\makeatletter
\newcommand{\osetminus}[2]{%
  {\mathop{#2}\limits^{\vbox to -2\ex@{\kern-\tw@\ex@
   \hbox{\scriptsize #1}\vss}}}}
\makeatother

\newcommand{\itor}{\mathrel{\smash{\stackrel{\raisebox{3.4pt}{\scriptsize $\mathsf{i}\:$}}%
{\smash{\rightarrow}}}_{\R}}}

\newcommand{\irightrightarrowsR}{\mathrel{\smash{\stackrel{\raisebox{4.6pt}{\scriptsize $\mathsf{i}\:$}}%
{\smash{\rightrightarrows}}}_{\R}}}
\newcommand{\irightrightarrowsRprobexample}{\mathrel{\smash{\stackrel{\raisebox{4.6pt}{\scriptsize $\mathsf{i}\:$}}%
{\smash{\rightrightarrows}}}_{\R_{3}}}}
\newcommand{\irightrightarrowsRandomWalk}{\mathrel{\smash{\stackrel{\raisebox{4.6pt}{\scriptsize $\mathsf{i}\:$}}%
{\smash{\rightrightarrows}}}_{\R_{\trw}}}}

\newcommand{\irightrightarrowsRstar}{\mathrel{\smash{\stackrel{\raisebox{4.6pt}{\scriptsize $\mathsf{i}\:$}}%
{\smash{\rightrightarrows}}}_{\R}^*}}

\newcommand{\irightrightarrowsDPRprobexample}{\mathrel{\smash{\stackrel{\raisebox{4.6pt}{\scriptsize
                 $\mathsf{i}\:$}}%
             {\smash{\rightrightarrows}}}_{\C{D}_3}}}

\newcommand{\cont}{cont}

\newcommand{\nonprob}{\mathrm{np}}

\newcommand{\setitops}{\mathrel{\oset{\scriptsize $\mathsf{i}$\qquad}{{\rightarrowtriangle}_{\PP,\SSS}}}}
\newcommand{\setitopsgt}{\mathrel{\oset{\scriptsize $\mathsf{i}$\qquad\;\;\,}{{\rightarrowtriangle}_{\PP_>,\SSS}}}}
\newcommand{\setitoUsablePairsps}{\mathrel{\oset{\scriptsize $\mathsf{i}$\qquad\qquad\qquad}{{\rightarrowtriangle}_{\C{T}_\mathtt{UT}(\PP, \SSS),\SSS}}}}
\newcommand{\setitozs}{\mathrel{\oset{\scriptsize $\mathsf{i}$\qquad}{{\rightarrowtriangle}_{Z,\SSS}}}}

\newcommand{\itononprobPS}{\mathrel{\oset{\scriptsize $\mathsf{i}$\qquad \qquad \;\;\;\;\;\;\,}{{\to}_{\nonprob(\PP),\nonprob(\SSS)}}}}
\newcommand{\itodr}{\mathrel{\oset{\scriptsize $\mathsf{i}$\qquad\;}{{\to}_{\C{D},\R}}}}
\newcommand{\itodrDIV}{\mathrel{\oset{\scriptsize $\mathsf{i}$\qquad\qquad\qquad\,}{{\to}_{\DPair{\R_{\tdiv}},\R_{\tdiv}}}}}

\newcommand{\setlifting}{\mathrel{\ooalign{\raisebox{2pt}{$\rightarrowtriangle$}\cr\hfil\raisebox{-2pt}{$\rightarrowtriangle$}\hfil}}}

\newcommand{\setiliftingS}{\mathrel{\oset{$\mathsf{i}$\;}{\setlifting}_{\SSS}}}
\newcommand{\setiliftingPS}{\mathrel{\oset{$\mathsf{i}$\;}{\setlifting}_{\PP, \SSS}}}

\newcommand{\setiliftingDRRprobexample}{\mathrel{\oset{$\mathsf{i}$\;}{\setlifting}_{\DTuple{\R_{3}}, \R_{3}}}}
\newcommand{\setiliftingDRRincompl}{\mathrel{\oset{$\mathsf{i}$\;}{\setlifting}_{\DTuple{\R_{\mathsf{incompl}}}, \R_{\mathsf{incompl}}}}}

\newcommand{\setiliftingRincompl}{\mathrel{\oset{$\mathsf{i}$\;}{\setlifting}_{\R_{\mathsf{incompl}}}}}
\newcommand{\setiliftingSstar}{\mathrel{\oset{$\mathsf{i}$\;}{\setlifting}_{\SSS}^{\raisebox{-4pt}{\tiny *}}}}

\newcommand{\setitodrr}{\mathrel{\oset{$\mathsf{i}$\;}{\rightarrowtriangle}_{\DTuple{\R},\R}}}

\newcommand{\itos}{\mathrel{\oset{\scriptsize $\mathsf{i}$ \;\;\,}{{\to}_{\SSS}}}}

\newcommand{\setitos}{\mathrel{\oset{\scriptsize $\mathsf{i}$\;\;\;\,}{{\rightarrowtriangle}_{\SSS}}}}

\newcommand{\itorstar}{\mathrel{\smash{\stackrel{\raisebox{2pt}{\scriptsize $\mathsf{i}$\,}}%
{\smash{\rightarrow}}}_{\R}^*}}
\newcommand{\itorstarDIV}{\mathrel{\smash{\stackrel{\raisebox{2pt}{\scriptsize $\mathsf{i}$\,}}%
{\smash{\rightarrow}}}_{\R_{\tdiv}}^*}}
\newcommand{\itononprobs}{\mathrel{\smash{\stackrel{\raisebox{3pt}{\scriptsize $\mathsf{i}$}}%
{\smash{\rightarrow}}}_{\nonprob(\SSS)}}}
\newcommand{\itononprobsstar}{\mathrel{\smash{\stackrel{\raisebox{3pt}{\scriptsize $\mathsf{i}$}}%
{\smash{\rightarrow}}}_{\normalfont{\nonprob}(\SSS)}^*}}

\newcommand{\upairs}{\mathcal{U}\mathcal{T}}

\newcommand{\val}{\mathit{\C{V}al}}
\newcommand{\adval}{\overline{\mathit{\C{V}al}}}
\newcommand{\advaltwo}{\widehat{\mathit{\C{V}al}}}

\newcommand\leftleadsto{\mathrel{\reflectbox{$\leadsto$}}}

\newcommand{\SubmultPoss}{\mathrm{MSub}_{\mathit{com}}}
\newcommand{\SubDPoss}{\mathrm{MSub}_{\mathit{term}}}
\newcommand{\SubmultMain}{\mathrm{MSub}_{\mathit{com}}^{\mathsf{i}}}
\newcommand{\SubDMain}{\mathrm{MSub}_{\mathit{term}}^{\mathsf{i}}}

\SetKwInOut{Input}{input}
\SetKwInOut{Output}{output}

\definecolor{Gray}{gray}{0.85}
\definecolor{LightCyan}{rgb}{0.88,1,1}

\newcolumntype{a}{>{\columncolor{Gray}}c}
\newcolumntype{b}{>{\columncolor{white}}c}

\newcommand{\makeproof}[2]{}
\newcommand{\paper}[1]{}
\newcommand{\report}[1]{#1} 

\report{
  \setlength{\textwidth}{125mm}
  \setlength{\textheight}{198mm}
}

\newcommand{\notnew}[1]{}
 
\usepackage{cite}
\AtBeginDocument{\renewcommand{\doi}[1]{\url{https://doi.org/#1}}}

\title{\hspace*{-.45cm}\mbox{Proving Almost-Sure Innermost Termination of}
  \\\mbox{Probabilistic Term Rewriting} \\Using Dependency Pairs\thanks{funded by the
    Deutsche Forschungsgemeinschaft (DFG, German Research Foundation) - 235950644 (Project
    GI 274/6-2) and DFG Research Training Group 2236 UnRAVeL}}
\titlerunning{Proving Almost-Sure Innermost Termination of Probabilistic Term Rewriting}
\author{Jan-Christoph Kassing\paper{$^{(\mbox{\Letter})}$}\orcidID{0009-0001-9972-2470} \and Jürgen Giesl\paper{$^{(\mbox{\Letter})}$}\orcidID{0000-0003-0283-8520}}
\institute{LuFG Informatik 2, RWTH Aachen University, Aachen, Germany}
\authorrunning{J.-C.\ Kassing, J.\ Giesl}

\begin{document}
\allowdisplaybreaks

\maketitle \begin{abstract}
	Dependency pairs are one of the most powerful techniques to analyze
        termination of term rewrite systems (TRSs) automatically.
	We adapt the dependency pair framework to the probabilistic setting in order\linebreak to
        prove almost-sure innermost termination of probabilistic TRSs.
	To eval\-uate its power, we implemented the new 
        framework in our tool
        \aprove.
\end{abstract}

\section{Introduction}\label{sec-introduction}

Techniques
      and tools to analyze innermost termination of term rewrite systems
(TRSs) automatically are successfully used for termination analysis
of programs in many 
languages (e.g., \textsf{Java} \cite{MoserIC18,otto2010JavaToTRS,CAV12}, \textsf{Haskell}
\cite{TOPLAS11},
and \textsf{Prolog}
\cite{PPDP12}).
While there exist several classical orderings for proving termination of TRSs (e.g., based on polynomial interpretations
\cite{lankford1979proving}), a \emph{direct} application of these orderings
is usually too weak for TRSs that
result from actual programs. However, these orderings can be used successfully within the
\emph{dependency pair} (DP) framework \cite{arts2000termination,giesl2006mechanizing,gieslLPAR04dpframework}. This
framework allows for modular termination proofs (e.g., which apply different orderings in different
sub-proofs) and  is
one of the most powerful techniques for 
termination analysis of TRSs that is used in essentially all current  termination tools for
TRSs, e.g., \textsf{AProVE}~\cite{JAR-AProVE2017},
\textsf{MU-TERM}~\cite{gutierrez_mu-term_2020},
\textsf{NaTT}~\cite{natt_sys_2014}, \textsf{TTT2}~\cite{ttt2_sys},
etc.

On the other hand, \emph{probabilistic} programs are used to describe randomized
algorithms and probability distributions, with applications in many areas. To use TRSs
also for such programs,  \emph{probabilistic term rewrite systems}
(PTRSs) were introduced in \cite{BournezRTA02,bournez2005proving}. 
In the probabilistic setting, there are several notions of ``termination''.
A program is \emph{almost-surely terminating} (AST) if
the probability for termination is $1$. As remarked in \cite{dblp:journals/pacmpl/huang0cg19}: ``AST is the classical and most widely-studied problem that extends
termination of non-probabilistic programs, and is considered as a core problem in the
programming languages community''.
A strictly stronger notion is \emph{positive almost-sure termination}
(PAST), which requires that the expected runtime is finite.
While there exist many automatic approaches to prove (P)AST of imperative programs
on numbers (e.g.,
\cite{kaminski2018weakest,mciver2017new,TACAS21,lexrsm,FoundationsTerminationMartingale2020,FoundationsExpectedRuntime2020,rsm,cade19,dblp:journals/pacmpl/huang0cg19,amber,ecoimp,absynth}),
there are only  few automatic approaches for programs with complex non-tail recursive
structure \cite{beutner2021probabilistic,Dallago2017ProbSizedTyping}, and even less
approaches which are also suitable for algorithms on recursive
data structures
\cite{wang2020autoexpcost,LeutgebCAV2022amor,KatoenPOPL23,avanzini2020probabilistic}.
The approach of~\cite{wang2020autoexpcost} focuses on algorithms on lists
and~\cite{LeutgebCAV2022amor} mainly targets algorithms on trees, but they cannot
easily be adjusted to other (possibly user-defined) data structures. 
The calculus of \cite{KatoenPOPL23} considers imperative programs with  stack, heap, and
pointers, but
it is not   yet automated. 
Moreover, the approaches of
\cite{wang2020autoexpcost,LeutgebCAV2022amor,KatoenPOPL23,avanzini2020probabilistic}
analyze expected runtime, while we focus on AST\@.

PTRSs can be used to model algorithms (possibly with complex recursive struc\-ture)
operating on algebraic data types.
While PTRSs were introduced in~\cite{BournezRTA02,bournez2005proving},\linebreak
the first (and up to now only) tool to analyze their termination automatically was presented in~\cite{avanzini2020probabilistic}, where orderings
based on interpretations were adapted to prove PAST.
Moreover, \cite{Faggian2019ProbabilisticRN} extended general concepts of abstract rewrite
systems (e.g., confluence and uniqueness of normal forms) to the probabilistic setting.

 As mentioned, already for non-probabilistic TRSs a \emph{direct} 
 application of order\-ings (as in~\cite{avanzini2020probabilistic})
 is limited in power.
To obtain a powerful approach, one should\linebreak combine such orderings in a modular way, as in
the DP framework. In this paper, we show for the first time that an adaption
of dependency pairs to the probabilistic setting is possible and present the first
DP framework for probabilistic term rewriting. Since the crucial idea of
dependency pairs is the modularization of the termination proof, we analyze AST instead of
PAST, because it is 
well known that AST is compositional, while PAST is not (see, e.g., \cite{kaminski2018weakest}).
We also present
a novel
adaption of the technique from~\cite{avanzini2020probabilistic} for the direct
application of 
polynomial
interpretations in order to prove
AST (instead of PAST) of PTRSs.

We start by briefly recapitulating the DP framework for non-probabilistic
TRSs in \cref{DP Framework}.
Then we recall the definition of PTRSs based on
\cite{bournez2005proving,avanzini2020probabilistic,Faggian2019ProbabilisticRN} in
\cref{Probabilistic Term Rewriting} and introduce a novel way to prove AST using polynomial interpretations automatically.
In \cref{Probabilistic Dependency Pairs} we present our new probabilistic DP framework.
The implementation of our approach in the tool \aprove{} is evaluated in \cref{Evaluation}.
We refer to\report{ App.\ \ref{appendix}}\paper{ \cite{report}} for all proofs (which
are much more involved than the original proofs for the non-probabilistic DP framework from \cite{arts2000termination,giesl2006mechanizing,gieslLPAR04dpframework}).

\section{The DP Framework}\label{DP Framework}

We assume familiarity with term rewriting
\cite{baader_nipkow_1999} and
regard TRSs over a finite signature
$\Sigma$ and a set of variables $\VSet$.
A \emph{polynomial interpretation} $\Pol$ is a $\Sigma$-algebra with carrier set $\IN$ which maps every function symbol $f \in \Sigma$ to a polynomial $f_{\Pol} \in \IN[\VSet]$. 
For a term $t \in \TSet{\Sigma}{ \VSet}$, $\Pol(t)$ denotes
the interpretation of $t$ by the $\Sigma$-algebra $\Pol$.
An arithmetic
inequation like $\Pol(t_1) > \Pol(t_2)$ \emph{holds}
if it is true for all instantiations of its variables by natural numbers.

\begin{restatable}[Termination With Polynomial Interpretations \cite{lankford1979proving}]{theorem}{polyinterpretations}Let\label{direct-application-poly-interpretations}
   $\R$ be a TRS and let
 $\Pol:\TSet{\Sigma}{\VSet} \to \IN[\VSet]$ be a monotonic polynomial interpretation
 (i.e., $x >
y$ implies $f_{\Pol}(\ldots, x, \ldots) > f_{\Pol}(\ldots, y, \ldots)$ for all $f
\in \Sigma$).
  If for every $\ell \to r \in \R$, we have $\Pol(\ell) >
  \Pol(r)$,
  then $\R$ is terminating.
\end{restatable}

  The search for polynomial interpretations is usually automated by SMT solving.
Instead of  polynomials over the naturals,
\Cref{direct-application-poly-interpretations} (and the other termination criteria in the paper) can also be extended to 
polynomials over the non-negative reals, by requiring that whenever a term is ``strictly
decreasing'', then its interpretation decreases at least by a certain fixed amount $\delta >
0$.

\begin{example}
  \label{example:TRS-int-div}
	Consider the TRS $\R_{\tdiv} = \{\eqref{R-div-1}, \ldots,  \eqref{R-div-4} \}$ for 
        division from \cite{arts2000termination}.

\vspace*{-.6cm}
        
    \hspace*{-.8cm}
    \begin{minipage}[t]{5.3cm}
      \begin{align}
		\label{R-div-1} \tminus(x,\O) &\!\to\! x \\
		\label{R-div-2} \tminus(\ts(x),\ts(y)) &\!\to\! \tminus(x,y)\!
       \end{align}
\end{minipage}
\hspace*{.1cm}
\begin{minipage}[t]{6.8cm}
     \begin{align}
		\label{R-div-3} \tdiv(\O,\ts(y)) & \!\to\! \O \\
		\label{R-div-4} \tdiv(\ts(x),\ts(y)) & \!\to\! \ts\left(\tdiv(\tminus(x,y),\ts(y))\right)\!
	\end{align}
\end{minipage} 

\vspace*{.2cm}

\noindent
Termination of $\R_{\tminus} = \{\eqref{R-div-1}, 
\eqref{R-div-2} \}$ can be proved 
    by the polynomial interpretation
     that maps $\tminus(x,y)$ to $x+y+1$, $\ts(x)$ to $x+1$, and $\O$ to $0$.
     However,  a direct application of classical techniques
     like polynomial interpretations fails for
     $\R_{\tdiv}$. These techniques correspond to so-called
     (quasi-)simplification orderings \cite{Dershowitz82} which cannot handle rules like 
 \eqref{R-div-4} where the right-hand side is  embedded in the left-hand side if $y$ is
 instantiated with $\ts(x)$.
    In contrast, the dependency pair framework is able to prove termination of $\R_{\tdiv}$ automatically.
\end{example}

We now recapitulate the  DP framework and its
core processors, and refer to, e.g.,
\cite{arts2000termination,gieslLPAR04dpframework,giesl2006mechanizing,hirokawa2005automating,gieslLPAR04dpframework}
for more details.
In this paper, we restrict ourselves to  the
DP framework for \emph{innermost} rewriting (denoted ``$\itor$''), because  our adaption to
the probabilistic setting relies on this evaluation strategy (see \cref{Dependency Tuples and Chains for Probabilistic Term Rewriting}).

\begin{definition}[Dependency Pair]\label{dependency-pair}
  Let $\R$ be a (finite) TRS\@.
	We decompose its signature $\Sigma = \Sigma_{C} \uplus \Sigma_{D}$ such that $f \in \Sigma_D$ if $f = \rootsym(\ell)$ for some rule $\ell \to r \in \R$.
	The symbols in $\Sigma_{C}$
and  $\Sigma_{D}$
        are called \emph{constructors} and \emph{defined symbols}, respectively.
	For every $f \in \Sigma_{D}$, we introduce a fresh \emph{tuple symbol} $f^{\#}$ of the same arity.
  Let $\Sigma^{\#}$ denote the set of all tuple symbols.
  To ease readability, we often write $\tF$ instead of $\tf^\#$.
  For any term $t = f(t_1,\ldots,t_n) \in \TSet{\Sigma}{\VSet}$ with $f \in \Sigma_{D}$, let 
    $t^{\#} = f^{\#}(t_1,\ldots,t_n)$.
  Moreover, for 
  any $r \in \TSet{\Sigma}{\VSet}$, let $\Subd(r)$ be the set of all subterms of $r$ with defined root symbol.
  For a rule $\ell \to r$ with $\Subd(r) = \{ t_1,\ldots,t_n\}$, one obtains the $n$
  dependency pairs (DPs) $\ell^\# \to t_i^\#$ with $1 \leq i \leq n$.
  $\DPair{\R}$ denotes the set of all dependency pairs of $\R$.
\end{definition}

\begin{example}
  \label{example:dependency-pair}
    For the TRS $\R_{\tdiv}$ from \cref{example:TRS-int-div}, we get the following
    dependency pairs.

    \vspace*{-.6cm}
    
    \hspace*{-.8cm}
\begin{minipage}[t]{4.5cm}
	\begin{align}
	  \label{R-div-deppair-1} \tM(\ts(x),\ts(y)) &\to \tM(x,y)
        \end{align}
\end{minipage}
\hspace*{.5cm}
\begin{minipage}[t]{7cm}
	\begin{align}
	\label{R-div-deppair-2} \tD(\ts(x),\ts(y)) &\to \tM(x,y) \\
		\label{R-div-deppair-3} \tD(\ts(x),\ts(y)) &\to \tD(\tminus(x,y),\ts(y))
	\end{align}
\end{minipage}
\end{example}

\vspace*{-.1cm}

The DP framework uses \emph{DP problems} $(\C{D}, \R)$ where
$\C{D}$ is a (finite) set of 
DPs\linebreak and  $\R$ is a (finite) TRS. 
A (possibly infinite) sequence $t_0^\#, t_1^\#, t_2^\#,
\ldots$ with $t_i^\# \itodr\linebreak \circ \itorstar t_{i+1}^\#$ for all $i$ is an (innermost)
$(\C{D}, \R)$-\emph{chain}.
Here, $\itodr$ is the restriction of $\to_{\C{D}}$ to rewrite steps
where
the used redex  is in normal form w.r.t.\ $\R$.
A chain repre\-sents subsequent ``function calls''  in evaluations. Between two function calls (corres\-ponding to
steps with $\C{D}$) one can evaluate the arguments
with $\R$.
For example, $\tD(\ts^2(\O), \ts(\O)), \;
\tD(\ts(\O), \ts(\O))$ is a $(\DPair{\R_{\tdiv}}, \R_{\tdiv})$-chain, as
$\tD(\ts^2(\O), \ts(\O))$\linebreak {\small $\itodrDIV
  \tD(\tminus(\ts(\O),\O),\ts(\O)) 
  \itorstarDIV  \tD(\ts(\O), \ts(\O))$}, where
  $\ts^2(\O)$ is $\ts(\ts(\O))$.

A DP problem $(\C{D}, \R)$ is called \emph{innermost terminating} (iTerm) if there is no infinite innermost $(\C{D}, \R)$-chain.
The main result on dependency pairs is the \emph{chain criterion} which states
that a TRS $\R$ is iTerm iff $(\DPair{\R},\R)$ is iTerm.
The key idea of the DP framework is  a \emph{divide-and-conquer}
approach which
applies \emph{DP processors} to
transform DP problems into simpler  sub-problems.
A \emph{DP processor} $\Proc$ has the form $\Proc(\C{D}, \R) = \{(\C{D}_1,\R_1), \ldots,
(\C{D}_n,\R_n)\}$, where
$\C{D}, \C{D}_1, \ldots, \C{D}_n$ are sets of dependency pairs and  $\R, \R_1, \ldots,
\R_n$ are TRSs. 
A processor $\Proc$ is \emph{sound} if $(\C{D}, \R)$  is iTerm whenever 
$(\C{D}_i,\R_i)$ is iTerm for all 
$1 \leq i \leq n$. 
It is \emph{complete} if $(\C{D}_i,\R_i)$ is iTerm for all 
$1 \leq i \leq n$ whenever  $(\C{D}, \R)$  is  iTerm.

So given a TRS $\R$, one starts with the initial
DP problem $(\DPair{\R}, \R)$ and applies sound 
(and preferably complete) DP processors repeatedly until all sub-problems are ``solved'' 
(i.e., sound processors transform them to the empty set). This allows for modular termination
proofs, since different techniques can be applied on each resulting ``sub-problem'' $(\C{D}_i,\R_i)$.
The following three theorems recapitulate the three most important processors of the DP framework.

The (innermost) \emph{$(\C{D}, \R)$-dependency graph} is a control flow graph that indicates
which dependency pairs can be used after each other in a chain.
Its node set is $\C{D}$
and there is an edge from $\ell_1^\# \to t_1^\#$ to $\ell_2^\# \to t_2^\#$ if there exist
\begin{wrapfigure}[8]{r}{0.08\textwidth}
  \begin{center}
  \scriptsize
    \vspace*{-1.3cm}
    \begin{tikzpicture}
        \node[shape=rectangle,draw=black!100] (A) at (0,0) {\eqref{R-div-deppair-1}};
        \node[shape=rectangle,draw=black!100] (B) at (0,.7) {\eqref{R-div-deppair-2}};
        \node[shape=rectangle,draw=black!100] (C) at (0,1.4) {\eqref{R-div-deppair-3}};
   
        \path [->,in=290,out=250,looseness=5] (A) edge (A);
        \path [->] (B) edge (A);
        \path [->] (C) edge (B);
        \path [->,in=110,out=70,looseness=5] (C) edge (C);
    \end{tikzpicture}
    \caption*{}
  \end{center}
\end{wrapfigure}
substitutions $\sigma_1, \sigma_2$ such that $t_1^\# \sigma_1 \itorstar \ell_2^\# \sigma_2$, and both
$\ell_1^\#  \sigma_1$ and $\ell_2^\# \sigma_2$ are in normal form w.r.t.\ $\R$.
Any infinite $(\C{D}, \R)$-chain corresponds to
  an infinite path in the dependency graph, and since the graph is finite, this infinite
  path must end in some strongly connected component (SCC).\footnote{Here, a
  set $\C{D}'$ of dependency pairs is  an \emph{SCC} if it is a maximal cycle,
  i.e., it is a maximal set such that for any $\ell_1^\# \to t_1^\#$ and $\ell_2^\# \to
  t_2^\#$ in $\C{D}'$ there is
  a non-empty path from $\ell_1^\# \to t_1^\#$ to $\ell_2^\# \to
  t_2^\#$ which only traverses nodes from $\C{D}'$.}
Hence, it suffices to consider the SCCs of this graph independently.
The $(\DPair{\R_{\tdiv}}, \R_{\tdiv})$-dependency graph can be seen on the right.

\begin{restatable}[Dep.\ Graph Processor]{theorem}{depgraph}\label{DGP}
For the SCCs 
$\C{D}_1, ..., \C{D}_n$
  of the $(\C{D}, \R)$-dependency graph,  
 $\Proc_{\mathtt{DG}}(\C{D},\R) = \{(\C{D}_1,\R), ..., (\C{D}_n,\R)\}$ is sound and
  complete. 
\end{restatable}
While the exact dependency graph is not computable in general, there are sev\-eral
techniques to over-approximate it automatically, see, e.g.,
\cite{arts2000termination,giesl2006mechanizing,hirokawa2005automating}.
In our example, 
 applying $\Proc_{\mathtt{DG}}$  to the initial problem
$(\DPair{\R_{\tdiv}}, \R_{\tdiv})$
results
in the smaller problems $\bigl(\{\eqref{R-div-deppair-1}\}, \R_{\tdiv}\bigr)$ and
$\bigl(\{\eqref{R-div-deppair-3}\}, \R_{\tdiv}\bigr)$ that can be treated separately.

The next processor removes rules that cannot be used to evaluate
right-hand sides of dependency pairs when their variables are instantiated with normal
forms.\linebreak

\vspace*{-.4cm}

\begin{restatable}[Usable Rules Processor]{theorem}{usablerules}\label{URP}
  Let $\R$ be a TRS\@.
  For every $f \in \SigmaDP$ let $\rules_\R(f) = \{\ell \to r \in \R \mid \rootsym(\ell) = f\}$.
  For any  $t \in  \TSet{\SigmaDP}{ \VSet}$,  
  its \emph{usable rules} $\urules_\R(t)$ are
    the smallest set
    such that $\urules_\R(x) = \emptyset$ for all $x \in \VSet$ and $\urules_\R(f(t_1, \ldots, t_n)) = \rules_\R(f) \cup \bigcup_{i = 1}^n \urules_\R(t_i) \; \cup \; \bigcup_{\ell \to r \in \rules_\R(f)} \urules_\R(r)$.
    The  \emph{usable rules} for the DP problem $(\C{D}, \R)$ are $\urules(\C{D},\R) =
    \bigcup_{\ell^\# \to t^\# \in \C{D}} \urules_\R(t^\#)$.
    Then $\Proc_{\mathtt{UR}}(\C{D},\R) = \{(\C{D},\urules(\C{D},\R))\}$ is sound but not
    complete.\footnote{For a complete version of the usable rules processor, one has
    to use a more involved notion of DP problems with more components that we omit here
    for readability
    \cite{gieslLPAR04dpframework}.}
\end{restatable}
\begingroup\emergencystretch=1.5em
For the DP problem $\bigl(\{\eqref{R-div-deppair-3}\}, \R_{\tdiv}\bigr)$ only the $\tminus$-rules are usable and thus $\Proc_{\mathtt{UR}}\bigl(\{\eqref{R-div-deppair-3}\}, \R_{\tdiv}\bigr) = \{\bigl(\{\eqref{R-div-deppair-3}\}, \{\eqref{R-div-1}, \eqref{R-div-2}\}\bigr)\}$.
For $\bigl(\{\eqref{R-div-deppair-1}\}, \R_{\tdiv}\bigr)$ there are no usable rules at
all, and thus $\Proc_{\mathtt{UR}}\bigl(\{\eqref{R-div-deppair-1}\}, \R_{\tdiv}\bigr) = \{\bigl(\{\eqref{R-div-deppair-1}\}, \emptyset\bigr)\}$.\par\endgroup

The last processor adapts classical orderings like polynomial interpretations to\linebreak DP
problems.\footnote{In this paper, we only regard the reduction pair processor with
polynomial interpretations, because for most other classical orderings it
is not clear how to extend them to probabilistic TRSs, where one has to
consider ``expected values of terms''.}
In contrast to their direct application in \cref{direct-application-poly-interpretations},
we may now use weakly  monotonic polynomials
$f_{\Pol}$ that do not have to depend on all of their arguments. The reduction pair processor 
requires that  all rules and dependency pairs  are weakly decreasing
and it removes those
DPs that are strictly decreasing.

\begin{restatable}[Reduction Pair Processor with Polynomial Interpretations]{theorem}{rpp}\label{RPP}
  Let $\Pol:\TSet{\SigmaDP}{\VSet} \to \IN[\VSet]$ be a weakly monotonic
  polynomial interpretation (i.e., $x \geq y$ implies $f_{\Pol}(\ldots, x, \ldots) \geq
  f_{\Pol}(\ldots, y, \ldots)$ for all $f \in \Sigma \uplus \Sigma^\#$).
Let $\C{D} = \C{D}_{\geq} \uplus \C{D}_{>}$  with $\C{D}_{>} \neq
\emptyset$ such that:
	\begin{itemize}
		\item[(1)] For every $\ell \to r \in \R$, we have $\Pol(\ell) \geq \Pol(r)$.
		\item[(2)] For every $\ell^\# \to t^\# \in \C{D}$, we have
                  $\Pol(\ell^\#) \geq \Pol(t^\#)$.
	  \item[(3)] For every $\ell^\# \to t^\# \in \C{D}_{>}$, we have $\Pol(\ell^\#) > \Pol(t^\#)$.
		\end{itemize}
    Then $\Proc_{\mathtt{RP}}(\C{D},\R) = \{(\C{D}_{\geq},\R)\}$ is sound and complete.
\end{restatable}
The constraints of the reduction pair processor for the remaining
DP problems $(\{\eqref{R-div-deppair-3}\}, \{\eqref{R-div-1}, \eqref{R-div-2}\})$ and
$(\{\eqref{R-div-deppair-1}\}, \emptyset)$ are satisfied by the polynomial interpretation
which maps $\O$ to $0$, $\ts(x)$ to $x + 1$, and all other non-constant function symbols
to the projection on their first arguments. Since $\eqref{R-div-deppair-3}$ and
$\eqref{R-div-deppair-1}$ are strictly decreasing, $\Proc_{\mathtt{RP}}$ transforms both $(\{\eqref{R-div-deppair-3}\}, \{\eqref{R-div-1}, \eqref{R-div-2}\})$ and
$(\{\eqref{R-div-deppair-1}\}, \emptyset)$ into DP problems of the form $(\emptyset,
\ldots)$. As $\Proc_{\mathtt{DG}}(\emptyset,
\ldots) = \emptyset$ and all processors used are sound, this means that there is no
infinite innermost chain for the initial DP problem 
$(\DPair{\R_{\tdiv}}, \R_{\tdiv})$ and thus, $\R_{\tdiv}$ is innermost
terminating.

\section{Probabilistic Term Rewriting}\label{Probabilistic Term Rewriting}

Now we recapitulate  \emph{probabilistic TRSs}
\cite{avanzini2020probabilistic,Faggian2019ProbabilisticRN,bournez2005proving}
and present a novel criterion to prove almost-sure termination automatically 
by adapting the direct application of polynomial interpretations from
\Cref{direct-application-poly-interpretations} to PTRSs.
In contrast to TRSs, a PTRS has
finite\footnote{Since our goal is the automation of termination analysis, in this paper we restrict ourselves to finite PTRSs with finite
multi-distributions.}
multi-distributions on the right-hand side of rewrite rules.

\begin{definition}[Multi-Distribution] \label{def:multi-distribution}
A finite \emph{multi-distribution} $\mu$ on a set $A\linebreak \neq \emptyset$ is a  finite multiset
of pairs $(p:a)$, where $0 < p \leq 1$ is a probability and $a \in A$,\linebreak such that  $\sum _{(p:a) \in \mu}p = 1$.
$\FDist(A)$ is the set of all finite multi-distributions on $A$.
For  $\mu\in\FDist(A)$, its \emph{support}  is the multiset $\Supp(\mu)\!=\!\{a \mid (p\!:\!a)\!\in\!\mu$ for some $p\}$.\linebreak
\end{definition}

\vspace*{-.6cm}

\begin{definition}[PTRS] \label{def:PTRS}
	A \emph{probabilistic rewrite rule} is a pair $\ell \to \mu \in
        \TSet{\Sigma}{ \VSet}\linebreak \times \FDist(\TSet{\Sigma}{\VSet})$ such that
 $\ell \not\in \VSet$ and 
        $\VSet(r) \subseteq \VSet(\ell)$ for every $r \in \Supp(\mu)$.
	A \emph{probabilistic TRS} (PTRS) is a finite set $\R$ of probabilistic rewrite rules.
	Similar to TRSs, the PTRS $\R$ induces a \emph{rewrite relation}
        ${\to_{\R}} \subseteq \TSet{\Sigma}{\VSet} \times \FDist(\TSet{\Sigma}{\VSet})$
        where $s \to_{\R} \{p_1:t_1, \ldots, p_k:t_k\}$  if there is
         a position $\pi$, a rule $\ell \to \{p_1:r_1,\linebreak \ldots, p_k:r_k\} \in \R$,
          and a substitution $\sigma$
    such that $s|_{\pi}=\ell\sigma$ and $t_j = s[r_j\sigma]_{\pi}$ for all $1 \leq j \leq k$.
	We call $s \to_{\R} \mu$ an \emph{innermost} rewrite step (denoted $s \itor \mu$) if every proper subterm of the used redex $\ell\sigma$ is in normal form w.r.t.\ $\R$.
\end{definition}

\begin{example} 
  \label{example:PTRS-random-walk}
  As an example, 
  consider the PTRS $\R_{\trw}$  with the only rule $\tg(x) \to
\{\nicefrac{1}{2}:x, \; \nicefrac{1}{2}:\tg(\tg(x))\}$, which 
corresponds to a  symmetric  
random walk.
\end{example}

As 
proposed in \cite{avanzini2020probabilistic}, we \emph{lift} $\to_{\R}$ to a
rewrite relation between multi-distributions in order to
track all probabilistic rewrite sequences (up to non-determinism) at once.
 For any $0 < p \leq 1$ and any $\mu \in \FDist(A)$, let $p \cdot \mu = \{ (p\cdot q:a) \mid (q:a) \in \mu \}$.

\begin{definition}[Lifting] \label{def:lifting}
$\!\!\!$	The \emph{lifting} ${\rightrightarrows} \subseteq \FDist(\TSet{\Sigma}{\VSet}) \times \FDist(\TSet{\Sigma}{\VSet})$ of a  relation ${\to} \subseteq \TSet{\Sigma}{\VSet} \times \FDist(\TSet{\Sigma}{\VSet})$ is the smallest relation with:
	\begin{itemize}
	\item[$\bullet$] If $t \in \TSet{\Sigma}{\VSet}$
          is in normal form w.r.t.~$\rightarrow$, then $\{1: t\} \rightrightarrows \{1:t\}$.
		\item[$\bullet$] If $t \to \mu$, then $\{1: t\} \rightrightarrows \mu$.
		\item[$\bullet$]
                  If  for all $1 \leq j \leq k$ there are
$\mu_j, \nu_j \in \FDist(\TSet{\Sigma}{\VSet})$    with  $\mu_j \rightrightarrows \nu_j$
and  $0 < p_j \leq 1$ with $\sum_{1 \leq j \leq k} p_j = 1$,
then $\bigcup_{1 \leq j \leq k} p_j \cdot \mu_j \rightrightarrows \bigcup_{1 \leq j \leq k} p_j \cdot \nu_j$.
	\end{itemize}
\end{definition}
For a PTRS $\R$, we write $\rightrightarrows_\R$ and $\irightrightarrowsR$
for the liftings of $\to_{\R}$ and  $\itor$, respectively.

\begin{example} \label{example:PTRS-random-walk-lifting-sequence}
  For instance, we obtain the following $\irightrightarrowsRandomWalk$-rewrite sequence:
	\[\begin{array}{lll}
	     \{1:\tg(\O)\} &\irightrightarrowsRandomWalk&
             \{\nicefrac{1}{2}:\O, \nicefrac{1}{2}:\tg^2(\O)\}
            \; \irightrightarrowsRandomWalk \;
            \{\nicefrac{1}{2}:\O,
            \nicefrac{1}{4}:\tg(\O),
            \nicefrac{1}{4}:\tg^3(\O)
             \}\\
             &\irightrightarrowsRandomWalk&
             \{
             \nicefrac{1}{2}:\O,
             \nicefrac{1}{8}:\O,
             \nicefrac{1}{8}:\tg^2(\O),
             \nicefrac{1}{8}:\tg^2(\O),
             \nicefrac{1}{8}:\tg^4(\O)
\}
\end{array}
	\]
        Note that the two occurrences of
    $\O$ and    $\tg^2(\O)$
 in the multi-distribution above 
could be rewritten differently if the PTRS had rules resulting in different terms.
    So it should be distinguished from 
    $\{\nicefrac{5}{8}:\O,
\nicefrac{1}{4}:\tg^2(\O),
    \nicefrac{1}{8}:\tg^4(\O)
\}$. 
\end{example}

To express the concept of almost-sure termination,
one has
to determine the probability for normal forms in a multi-distribution.

\begin{definition}[$|\mu|_{\R}$] \label{def:prob-abs-value}
    For a PTRS $\R$,  $\mathtt{NF}_{\R} \subseteq \TSet{\Sigma}{\VSet}$ denotes
    the set of all normal forms w.r.t.\ $\R$.
    For any $\mu \in \FDist(\TSet{\Sigma}{\VSet})$, let
  $|\mu|_{\R} = \sum_{(p:t) \in \mu, t \in \mathtt{NF}_{\R}} p$.
\end{definition}

\begin{example}
	Consider the multi-distribution
       $\{   \nicefrac{1}{2}:\O,
             \nicefrac{1}{8}:\O,
             \nicefrac{1}{8}:\tg^2(\O),
             \nicefrac{1}{8}:\tg^2(\O),
             \nicefrac{1}{8}:\tg^4(\O)
             \}$
from \Cref{example:PTRS-random-walk-lifting-sequence} and $\R_{\trw}$ from \Cref{example:PTRS-random-walk}.
	Then $|\mu|_{\R_{\trw}} = \nicefrac{1}{2} + \nicefrac{1}{8}  = \nicefrac{5}{8}$.
\end{example}

\begin{definition}[(Innermost) AST]\label{def:ptrs-innermost-term-innermost-AST}
	Let $\R$ be a PTRS and $(\mu_n)_{n \in \IN}$ be an infinite $\rightrightarrows_{\R}$-rewrite sequence, i.e., $\mu_n \rightrightarrows_{\R} \mu_{n+1}$ for all $n \in \IN$. 
	Note that
$\lim\limits_{n \to \infty}|\mu_n|_{\R}$ exists, since
        $|\mu_n|_{\R} \leq |\mu_{n+1}|_{\R}\leq 1$ for all $n \in \IN$.
	$\R$ is \emph{almost-surely terminating (AST)} (\emph{innermost almost-surely terminating (iAST)})
        if   $\lim\limits_{n \to \infty}
  |\mu_n|_{\R} = 1$ holds
        for every infinite
        $\rightrightarrows_{\R}$-rewrite
        sequence
($\irightrightarrowsR$-rewrite
        sequence) $(\mu_n)_{n \in \IN}$.
 \end{definition}

\begin{example}
For the (unique) infinite extension of the  $\irightrightarrowsRandomWalk$-rewrite
  sequence
$(\mu_n)_{n \in \IN}$ in
  \cref{example:PTRS-random-walk-lifting-sequence},
we have   $\lim\limits_{n \to \infty}
|\mu_n|_{\R} = 1$. Indeed,  $\R_{\trw}$ is  AST (but not PAST, i.e., the expected number
of rewrite steps is infinite for every term containing $\tg$). 
\end{example}

\Cref{theorem:ptrs-direct-application-poly-interpretations}
introduces a novel technique
to prove AST automatically using a direct application of polynomial \pagebreak[2]
interpretations.

\begin{restatable}[Proving AST with Polynomial Interpretations]{theorem}{ASTPolInt}\label{theorem:ptrs-direct-application-poly-interpretations}
	Let $\R$ be a PTRS,
        let $\Pol:\TSet{\Sigma}{\VSet} \to \IN[\VSet]$ be a monotonic, multilinear\footnote{As in
\cite{avanzini2020probabilistic}, 
 multilinearity ensures ``monotonicity'' w.r.t.\ expected values, since multilinearity implies 
$f_{\Pol}(\ldots, \sum_{1 \leq j \leq k}p_j \cdot \Pol(r_j), \ldots) = 
\sum_{1 \leq j \leq k}p_j \cdot \Pol(f(\ldots, r_j, \ldots))$.}
       polynomial interpretation
(i.e., for all $f \in \Sigma$, all monomials of $f_{\Pol}(x_1,\ldots,x_n)$ have the form
        $c \cdot x_1^{e_1} \cdot \ldots \cdot x_n^{e_n}$ with $c \in \IN$ and
        $e_1,\ldots,e_n \in \{0,1\}$). If for every rule
         $\ell \to \{p_1:r_1, \ldots, p_k:r_k\} \in \R$,
        \begin{itemize}
		\item[(1)] there exists a $1 \leq j \leq k$ with $\Pol(\ell) >
                  \Pol(r_j)$ and
		\item[(2)] $\Pol(\ell) \geq \sum_{1 \leq j \leq k} \; p_j \cdot \Pol(r_j)$,
	\end{itemize}
	then $\R$ is AST.
\end{restatable}

In \cite{avanzini2020probabilistic}, it was shown that PAST can be proved by using
multilinear polynomials and requiring a
strict decrease in the expected value of each rule.
In contrast, we only require a weak decrease of the expected value in (2) and in addition, 
at least one term in the support of the right-hand side must become strictly smaller (1).
As mentioned, 
the proof for \Cref{theorem:ptrs-direct-application-poly-interpretations}
(and for all our other new results and observations) can be found in\report{
  App.\ \ref{appendix}}\paper{ \cite{report}}. The proof idea is
based on \cite{mciver2017new},  but it extends  their approach from while-programs on
integers to terms. 
However, in contrast to \cite{mciver2017new}, PTRSs can only deal 
with constant probabilities, since all variables stand for
terms, not for numbers.
Note that the constraints (1) and (2) of our new criterion in
\Cref{theorem:ptrs-direct-application-poly-interpretations}  are equivalent to
the constraint of the classical \Cref{direct-application-poly-interpretations} in the
special case where
the PTRS is in fact a TRS (i.e., 
all rules 
have the form $\ell \to \{ 1: r \}$).

\begin{example}\label{example:direct-application-AST}
	To prove that $\R_{\trw}$ is AST with \Cref{theorem:ptrs-direct-application-poly-interpretations}, we can use the polynomial interpretation that maps $\tg(x)$ to $x+1$ and $\O$ to $0$. 
\end{example}

\section{Probabilistic Dependency Pairs}\label{Probabilistic Dependency Pairs}

We introduce our new adaption of DPs to the probabilistic setting
in \Cref{Dependency Tuples and Chains for Probabilistic Term Rewriting}. Then we present the processors for the probabilistic DP framework in
\Cref{The Probabilistic DP Framework}.

\subsection{Dependency Tuples and Chains for Probabilistic Term Rewriting}\label{Dependency Tuples and Chains for Probabilistic Term Rewriting}

We first show why straightforward adaptions
are  unsound.
A natural idea to define\linebreak DPs for probabilistic rules $\ell \to
\{p_1:r_1,\dots, p_k:r_k\} \in \R$ would be 
\eqref{dp A} or \eqref{dp B}: 
\begin{align}
    &\{ \, \ell^{\#} \to \{p_1:r_1,\dots, p_j:t_j^\#, \dots, p_k:r_k\} \; \mid \; t_j \in
    \Subd(r_j) \text{ with } 1 \leq j \leq k \, \} \label{dp A}\\
    &\{ \, \ell^{\#} \to \{p_1:t_1^\#,\dots, p_k:t_k^\#\} \; \mid \; t_j \in \Subd(r_j)
    \text{ for all } 1 \leq j \leq k \, \} \label{dp B}
\end{align}
For \eqref{dp B}, 
if $\Subd(r_j) = \emptyset$, then we insert a fresh constructor $\bot$ into $\Subd(r_j)$ that does not occur in $\R$.
So in both \eqref{dp A} and \eqref{dp B}, we replace $r_j$ by
a single term $t_j^\#$ in the right-hand side.
The following example shows that this notion of probabilistic DPs does not
yield a sound chain criterion.
Consider the PTRSs $\R_1$ and $\R_2$:
\begin{equation}
\label{R1R2}
	\R_1 = \{ \tg \rightarrow \{\nicefrac{1}{2}: \O, \nicefrac{1}{2}: \tf(\tg,\tg)
        \} \} \qquad
	\R_2  =  \{ \tg \rightarrow \{\nicefrac{1}{2}: \O, \nicefrac{1}{2}:
        \tf(\tg,\tg,\tg)\} \} \pagebreak[2]
\end{equation}
$\R_1$ is AST  since it corresponds to a symmetric random walk stopping at $0$, where the
number of $\tg$s denotes the current position.
In contrast, $\R_2$ is not AST as it corresponds to a random walk 
where
there is an equal chance of reducing the number of $\tg$s
by $1$ or increasing it by $2$.
For both $\R_1$ and $\R_2$,  \eqref{dp A} and \eqref{dp B}
would result in the only dependency pair
$\tG \to \{\nicefrac{1}{2}: \O, \nicefrac{1}{2}: \tG  \}$ and
$\tG \to \{\nicefrac{1}{2}: \bot, \nicefrac{1}{2}: \tG  \}$, resp.
Rewriting with this DP is clearly AST,
since it corresponds to a program that flips a coin 
until one gets head  and then terminates.
So the definitions \eqref{dp A} and \eqref{dp B}
would not yield a sound approach for proving AST\@.

$\R_1$ and $\R_2$  show that the number of occurrences of the same subterm in the
right-hand side $r$ of a rule matters for AST\@.
Thus, we now regard the \emph{multiset} $\MSubd(r)$ of all subterms of $r$ with
 defined root symbol to ensure that 
 multiple occurrences of the same subterm in $r$ are taken
into account. 
Moreover,
instead of pairs
we regard \emph{dependency tuples} which 
consider all subterms with defined root in $r$ at once.
 Dependency tuples were
already used when adapting DPs for complexity analysis of (non-probabilistic)
TRSs \cite{noschinski2013analyzing}.
We now adapt them to the probabilistic setting and present a novel rewrite relation for
dependency tuples.

\begin{definition}[Transformation $dp$]\label{Transformation dp revised}
    If $\MSubd(r) =\{t_{1}, \dots, t_{n}\}$, then we define $dp(r) = \Com{n}(t^\#_1, \ldots, t^\#_n)$. 
    To make $dp(r)$ unique, we use the lexicographic
    ordering $<$ on positions where $t_{i} = r|_{\pi_{i}}$ and $\pi_{1} < \ldots < \pi_{n}$.
    Here, we extend $\Sigma_{C}$ by fresh \emph{compound} 
 constructor symbols $\Com{n}$  of arity $n$ for $n \in \mathbb{N}$.
\end{definition}

When rewriting a subterm $t^\#_i$ of $\Com{n}(t^\#_1, \ldots, t^\#_n)$ with a dependency
tuple, one obtains terms with nested compound symbols.
To abstract from nested compound symbols and from the order of their arguments, we introduce the following normalization.
\begin{definition}[Normalizing Compound Terms]\label{normalization}
  For any term $t$, its \emph{content} $\cont(t)$ is the multiset
defined by
  $\cont(\Com{n}(t_1,\ldots,t_n)) = \cont(t_1) \cup \ldots \cup  \cont(t_n)$ and $\cont(t) = \{t\}$ otherwise.
    For any term $t$ with $\cont(t) = \{ t_1, \ldots, t_n \}$, the term $\Com{n}(t_1,\ldots,t_n)$ is a \emph{normalization} of $t$.
    For two terms $t, t'$, we define $t \approx t'$ if $\cont(t) = \cont(t')$. 
   We define $\approx$ on multi-distributions in a similar way:
whenever $t_j \approx t_j'$ for all $1 \leq j \leq k$, then  $\{p_1:t_1, \ldots, p_k: t_k
\}
\approx  \{p_1:t_1', \ldots, p_k: t_k'\}$.
\end{definition}

So for example, $\Com{3}(x, x, y)$ is a normalization of $\Com{2}(\Com{1}(x), \Com{2}(x,y))$.
We do not distinguish between terms and multi-distributions that are
equal w.r.t.\ $\approx$ and we write $\Com{n}(t_1,\ldots,t_n)$ for any term $t$ with a compound root symbol where
$\cont(t) = \{t_1,\ldots,t_n\}$, i.e., we consider all such $t$ to be normalized.

For any rule $\ell \to \{p_1:r_1,\dots, p_k:r_k\} \in \R$, the natural idea would be to
define its \emph{dependency tuple (DT)} as
$\ell^{\#} \to \{p_1:dp(r_1), \dots, p_k:dp(r_k)\}$.
Then
innermost \emph{chains} in the
probabilistic setting would result from alternating a DT-step
with  an arbitrary number of $\R$-steps
(using $\irightrightarrowsRstar$). 
However, such chains would not necessarily correspond to the
original rewrite sequence
and thus, the resulting chain criterion would not be sound.

\begin{example}\label{example:problem-with-dependency-tuples}
  Consider the PTRS $\R_3 = \{\tf(\O) \to \{1: \tf(\ta) \},
\ta \to \{\nicefrac{1}{2}:\tb_1, \nicefrac{1}{2}:\tb_2\},
 \tb_1 \to \{1: \O \},
  \tb_2 \to \{1: \tf(\ta)
        \}   \}$.
	Its DTs would be
        $\C{D}_3 = \{\tF(\O) \to \{1:\Com{2}(\tF(\ta), \tA)\},
 \tA \to \{\nicefrac{1}{2}:\Com{1}(\tB_1), \nicefrac{1}{2}:\Com{1}(\tB_2) \},
         \tB_1 \to \{1:\Com{0}\},
        \tB_2 \to \{1:\Com{2}(\tF(\ta), \tA)\}       
        \}$.
        $\R_3$ is not iAST, 
        because one can extend the rewrite \pagebreak[2] sequence 
        \begin{equation}
          \label{R3 rewrite sequence}
  \mbox{\small $\{ 1\!:\!\tf(\O)\} \irightrightarrowsRprobexample \{ 1\!:\!\tf(\ta)\} \irightrightarrowsRprobexample \{
      \nicefrac{1}{2}\!:\!\tf(\tb_1), \nicefrac{1}{2}\!:\!\tf(\tb_2)\} \irightrightarrowsRprobexample \{
      \nicefrac{1}{2}\!:\!\tf(\O), \nicefrac{1}{2}\!:\!\tf(\tf(\ta))\}$}
  \end{equation}
  to an infinite sequence without normal forms.
	The resulting chain starts  with
	\[\mbox{\small
		$\begin{array}{lc@{}r@{}l}
    &\{&1:\Com{1}(\tF(\O))&\}\\
	  \irightrightarrowsDPRprobexample &\{&1:\Com{2}(\tF(\ta), \tA)&\} \\
		\irightrightarrowsDPRprobexample &\{&\nicefrac{1}{2}: \Com{2}(\tF(\ta), \tB_1)&,\nicefrac{1}{2}: \Com{2}(\tF(\ta), \tB_2) \}\\
		\irightrightarrowsRprobexample &\{&\nicefrac{1}{4}: \Com{2}(\tF(\tb_1), \tB_1)&, \nicefrac{1}{4}: \underline{\Com{2}(\tF(\tb_2), \tB_1)}, \nicefrac{1}{4}: \Com{2}(\tF(\tb_1), \tB_2), \nicefrac{1}{4}: \Com{2}(\tF(\tb_2), \tB_2) \}.
	  \end{array}$}	
	\]
        The second and third term in the last distribution do
        not correspond
              to terms in\linebreak the original rewrite sequence \eqref{R3 rewrite sequence}. 
        After the next $\C{D}_3$-step which removes $\tB_1$, 
no further $\C{D}_3$-step can be applied to 
        the underlined term anymore, because $\tb_2$ 
cannot be rewritten to $\O$.
  Thus, the resulting chain criterion would be unsound, as every
   chain $(\mu_n)_{n \in \IN}$ in this example
contains such $\C{D}_3$-normal forms and therefore, it
   is AST (i.e.,
  $\lim\limits_{n \to \infty}|\mu_n|_{\C{D}_3} =1$ where 
$|\mu_n|_{\C{D}_3}$ is
        the probability for $\C{D}_3$-normal forms in $\mu_n$).
	So we have to ensure that when $\tA$ is rewritten to $\tB_1$ via a DT from $\C{D}_3$,
        then the ``copy'' $\ta$ of the redex $\tA$ is rewritten via $\R_3$\linebreak to the corresponding term
        $\tb_1$ instead of $\tb_2$. Thus, after the step with $\irightrightarrowsRprobexample$
        we should have $\Com{2}(\tF(\tb_1), \tB_1)$ and $\Com{2}(\tF(\tb_2),
        \tB_2)$, but not $\Com{2}(\tF(\tb_2), \tB_1)$ or $\Com{2}(\tF(\tb_1),
        \tB_2)$.
\end{example}

Therefore, for our new adaption of DPs to the probabilistic setting, we
operate\linebreak on \emph{pairs}.
Instead of having a rule $\ell \to \{ p_1:r_1, \ldots, p_k:r_k \}$ from $\R$ and its
corres\-ponding dependency tuple $\ell^\# \to \{ p_1:dp(r_1), \ldots, p_k:dp(r_k)\}$
separately, we cou\-ple them together to
$\langle \ell^\#,\ell \rangle \to \{ p_1:\langle dp(r_1),r_1 \rangle, \ldots, p_k:\langle dp(r_k),r_k \rangle\}$.
This type\linebreak of rewrite system is called a \emph{probabilistic pair term rewrite system (PPTRS)}, and its rules are called \emph{coupled dependency tuples}.
Our new DP framework works on
\emph{(probabilistic) DP problems}
$(\PP,\SSS)$, where $\PP$ is a PPTRS and
$\SSS$ is a PTRS. 

\begin{definition}[Coupled Dependency Tuple] \label{def:coupled-dependency-pairs}
	Let $\R$ be a PTRS.
	For every $\ell \to \mu = \{p_1:r_1, \ldots, p_k:r_k\} \in \R$, its 
     \emph{coupled dependency tuple} (or simply \emph{dependency tuple},
        \emph{DT})
        is $\DTuple{\ell \to \mu} = \langle \ell^\#,\ell \rangle \to \{ p_1 : \langle dp(r_1),r_1 \rangle, \ldots, p_k : \langle dp(r_k),r_k \rangle\}$.
	The set of all coupled dependency tuples of $\R$ is denoted by $\DTuple{\R}$.
\end{definition}
\begin{example}\label{example:ptrs-int-div-coupled-positional-dependency-pairs}
  The following PTRS $\R_{\tpdiv}$
 adapts $\R_{\tdiv}$ 
  to the probabilistic setting.

  \vspace*{-.6cm}
  \hspace*{-.57cm}
  \begin{minipage}[t]{4.5cm}
    \begin{align}
            \label{eq:ptrs-div-1} \tminus(x,\O) & \to \{ 1:x\}               
    \end{align}
  \end{minipage}
  \hspace*{.2cm}
  \begin{minipage}[t]{7.2cm}
    \begin{align}
            \label{eq:ptrs-div-2} \tminus(\ts(x),\ts(y)) & \to \{ 1:\tminus(x,y)\} 
    \end{align}
  \end{minipage}

  \vspace*{-.4cm}

  \hspace*{-2.13cm}\begin{minipage}[t]{13.8cm}
    \begin{align}
    \label{eq:ptrs-div-3} \tdiv(\O,\ts(y)) & \to \{ 1:\O\}\\
    \label{eq:ptrs-div-4} \hspace*{-1.1cm} \tdiv(\ts(x),\ts(y)) & \to \{
    \nicefrac{1}{2}:\tdiv(\ts(x),\ts(y)),
    \nicefrac{1}{2}:\ts(\tdiv(\tminus(x,y),\ts(y))) \}
  \end{align} 
   \end{minipage}

\vspace*{.2cm}

   In  \eqref{eq:ptrs-div-4}, we now do the actual rewrite step with
  a chance of $\nicefrac{1}{2}$ or the terms stay the same.
Our new probabilistic DP framework  can prove automatically that  $\R_{\tpdiv}$
is iAST, while (as in
the non-probabilistic setting) a direct application of polynomial
  interpretations via \cref{theorem:ptrs-direct-application-poly-interpretations}
  fails.
  We get $\DTuple{\R_{\tpdiv}} = \{ \eqref{R-div-deptup-1}, \ldots,
  \eqref{R-div-deptup-4}\}$:

\vspace*{-.4cm}
  
  {\small
 \begin{align}
          \label{R-div-deptup-1}  \langle\tM(x,\O), \tminus(x,\O)\rangle & \to \{1:\langle\Com{0},\,x\rangle\}\\
          \label{R-div-deptup-2}  \langle\tM(\ts(x),\ts(y)), \tminus(\ts(x),\ts(y))\rangle &\to \{1:\langle\Com{1}(\tM(x,y)),\,\tminus(x,y)\rangle\}\\ 
          \label{R-div-deptup-3}  \langle\tD(\O,\ts(y)), \tdiv(\O,\ts(y))\rangle &\to \{1:\langle\Com{0},\,\O\rangle\}\\
          \langle\tD(\ts(x),\ts(y)), \tdiv(\ts(x),\ts(y))\rangle
& \to \{\nicefrac{1}{2}: \langle\Com{1}(\tD(\ts(x),\ts(y))),\,
          \tdiv(\ts(x),\ts(y))\rangle, \hspace*{2cm} \nonumber
          \\
 \label{R-div-deptup-4} 
 \rlap{\hspace*{-2.1cm}$\nicefrac{1}{2}:\langle\Com{2}(\tD(\tminus(x,y),\ts(y)), \tM(x,y)),\,\ts(\tdiv(\tminus(x,y),\ts(y)))\rangle\}$}
  \end{align}
  }
\end{example}

\begin{definition}[PPTRS,$\setitops$] \label{def:PPTRS-rewriting-on-sets}
Let $\PP$ be a finite set of rules of the form
$\langle \ell^\#,\ell \rangle \to  \{ p_1:\langle d_1,r_1 \rangle, \ldots, p_k:\langle
d_k,r_k \rangle\}$. 
        For every such rule, let
        $\projOne(\PP)$ contain $\ell^\# \to \{ p_1:d_1, \ldots, p_k:d_k \}$ and let
        $\projTwo(\PP)$ contain $\ell \to \{ p_1:r_1, \ldots, p_k:r_k \}$.
        If  $\projTwo(\PP)$ is a PTRS
        and $\cont(d_j) \subseteq \cont(dp(r_j))$ holds\footnote{The reason for
        $\cont(d_j) \subseteq \cont(dp(r_j))$
instead of   $\cont(d_j) = \cont(dp(r_j))$ is that in this way processors can
remove terms from the right-hand sides of DTs, see \Cref{theorem:prob-UPP}.
        }
        for all $1 \leq j \leq k$,
        then $\PP$ 
        is a  \emph{probabilistic pair term rewrite system (PPTRS)}.
    
  Let $\SSS$ be a PTRS\@.
	Then a normalized term $\Com{n}(s_1,\ldots,s_n)$ \emph{rewrites} with the \mbox{PPTRS} $\PP$ to
        $\{p_1:b_1, \ldots, p_k:b_k\}$ w.r.t.\ $\SSS$ (denoted $\setitops$) if there are an
        $1 \leq i \leq n$, an $\langle \ell^\#,\ell \rangle \to \{ p_1:\langle d_1,r_1
        \rangle, \ldots, p_k:\langle d_k,r_k \rangle\} \in \PP$, 
         a substitution
        $\sigma$ with $s_i = \ell^\# \sigma \in \mathtt{NF}_{\SSS}$,
	and for all $1 \leq j \leq k$ we have $b_j = \Com{n}(t_1^j,\ldots,t_n^j)$ where
	\begin{itemize}
	\item[$\bullet$] $t_i^j = d_j \sigma$ for all $1 \leq j \leq k$, i.e., we rewrite the term $s_i$ using
  $\projOne(\PP)$.
		\item[$\bullet$] For every $1 \leq i' \leq n$ with $i \neq i'$ we have
                  \begin{itemize}
                  \item[\normalfont{(i)}]  $t_{i'}^j = s_{i'}$ for all $1 \leq j \leq k$ \quad or
  \item[\normalfont{(ii)}]       $t_{i'}^j =   s_{i'}[r_j \sigma]_{\tau}$    for all $1 \leq j \leq k$,\\             if
      $s_{i'}|_\tau = \ell \sigma$ for some position $\tau$ and if $\ell \to  \{ p_1:r_1, \ldots, p_k:r_k\} \in \SSS$. 
                    \end{itemize}
                  So  
                  $s_{i'}$ stays the same in all $b_j$ or we can apply the
                  rule from  $\projTwo(\PP)$ to rewrite
                  $s_{i'}$ in all $b_j$, provided that
      this rule is also contained in $\SSS$. Note
      that even if the rule is applicable, the term $s_{i'}$ can still stay the same in all $b_j$. 
	\end{itemize}
\end{definition}
\begin{example}
  \label{example:solving-problem-with-dep-pair}
  For $\R_3$ from \cref{example:problem-with-dependency-tuples},
	the (coupled) dependency tuple for the $\tf$-rule\linebreak is $\langle \tF(\O),\tf(\O)\rangle
        \to \{1:\langle\Com{2}(\tF(\ta), \tA), \tf(\ta)\rangle\}$ and the DT
        for the $\ta$-rule is $\langle\tA,\ta\rangle \to
        \{\nicefrac{1}{2}:\langle\Com{1}(\tB_1),\tb_1\rangle,
        \nicefrac{1}{2}:\langle\Com{1}(\tB_2),\tb_2\rangle\}$. With the lifting
        $\setiliftingPS$  of $\setitops$,
        we get the following sequence  which corresponds to
        the rewrite sequence \eqref{R3 rewrite sequence}
        from \cref{example:problem-with-dependency-tuples}.
	\begin{equation}
          \label{R3 chain}
		\begin{array}{rcl}
								\{1: \Com{1}(\tF(\O))\}
		&	\setiliftingDRRprobexample &\{1:\Com{2}(\tF(\ta), \tA)\} \\
			&\setiliftingDRRprobexample &\{\nicefrac{1}{2}: \Com{2}(\tF(\tb_1), \tB_1), \nicefrac{1}{2}: \Com{2}(\tF(\tb_2), \tB_2) \}
		\end{array}	
	\end{equation}
        So with the PPTRS, when rewriting $\tA$ to $\tB_1$ in the second step,
        we can simultaneously\ rewrite
        the in\-ner subterm $\ta$ of $\tF(\ta)$
to $\tb_1$ or keep $\ta$ unchanged, but we cannot rewrite $\ta$ to $\tb_2$. This is
ensured by $\tb_1$ in the second component of $\langle\tA,\ta\rangle \to
\{\nicefrac{1}{2}:\langle\Com{1}(\tB_1),\tb_1\rangle, \ldots \}$, since by
\Cref{def:PPTRS-rewriting-on-sets}, if $s_{i'}$ contains $\ell\sigma$ at some arbitrary position $\tau$, then one can
(only) use the rule in the second component of the DT to rewrite $\ell\sigma$ (i.e.,
here we have $s_{i'}=\tF(\ta)$, $s_i=\tA$, and $s_{i'}|_\tau=\ta$). A similar observation
holds when rewriting $\tA$ to $\tB_2$.
Recall that with the notion of chains in \Cref{example:problem-with-dependency-tuples},
one \emph{cannot simulate} every possible rewrite sequence, which leads to unsoundness. In
contrast, with the notion of coupled DTs and PPTRSs, every possible rewrite sequence
\emph{can be simulated} which ensures soundness of the chain criterion. 
Of course, due to the ambiguity in (i) and (ii) of \Cref{def:PPTRS-rewriting-on-sets},
one could also create other ``unsuitable''
        $\setiliftingDRRprobexample$-sequences where $\ta$ is
        not reduced to $\tb_1$ and $\tb_2$ in the second step, but is kept unchanged. This does not affect the soundness of the chain
        criterion, since every rewrite sequence of the original PTRS can be simulated
        by a ``suitable'' chain. To obtain completeness of the chain criterion, one would
        have
        to avoid such ``unsuitable''  sequences.
\end{example}

We also introduce an analogous rewrite relation for PTRSs, where we can apply the same
rule simultaneously to the same subterms in a \pagebreak[2] single rewrite step.

\begin{definition}[$\setitos$] \label{def:PTRS-rewriting-on-sets}
  For a PTRS $\SSS$ and a normalized term $\Com{n}(s_1,\ldots,s_n)$, we\linebreak  define $\Com{n}(s_1,...,s_n) \setitos  \{p_1\!:\!b_1,..., p_k\!:\!b_k\}$ if there are an $1\!\leq\!i\!\leq n$,  an $\ell\!\to\!\{
  p_1\!:\!r_1,\linebreak \ldots, p_k\!:\!r_k\} \in \SSS$, a position $\pi$, a substitution
  $\sigma$ with $s_i|_{\pi}\!=\!\ell \sigma$ such that every proper\linebreak subterm of
  $\ell \sigma$ is in $\mathtt{NF}_{\SSS}$, and for all $1\!\leq j\!\leq k$ we have $b_j = \Com{n}(t_1^j,\ldots,t_n^j)$ where
	\begin{itemize}
		\item[$\bullet$]
		$t_i^j = s_i[r_j \sigma]_{\pi}$ for all $1 \leq j \leq k$, i.e., we rewrite the term $s_i$ using
                  $\SSS$.
                \item[$\bullet$] For every $1 \leq i' \leq n$ with $i \neq i'$ we have
\begin{itemize}
\item[\normalfont{(i)}]  $t_{i'}^j = s_{i'}$  for all $1 \leq j \leq k$ \quad or
\item[\normalfont{(ii)}]   $t_{i'}^j = s_{i'}[r_j \sigma]_{\tau}$ for all $1 \leq j \leq k$, if
  $s_{i'}|_\tau = \ell \sigma$ for some position $\tau$.
   \end{itemize}
 \end{itemize}
\end{definition}
So for example, the lifting $\setiliftingS$ of $\setitos$ for $\SSS = \R_3$ rewrites
$\{1:\Com{2}(\tf(\ta), \ta)\}$  to both
$\{\nicefrac{1}{2}: \Com{2}(\tf(\tb_1), \tb_1), \nicefrac{1}{2}: \Com{2}(\tf(\tb_2),
\tb_2) \}$ and 
$\{\nicefrac{1}{2}: \Com{2}(\tf(\ta), \tb_1), \nicefrac{1}{2}: \Com{2}(\tf(\ta), \tb_2)
\}$.

A straightforward adaption of ``chains'' to the probabilistic setting
 using $\setiliftingPS\linebreak \circ \setiliftingSstar$
 would  force us 
 to use
steps with DTs from $\PP$ at the same time for all  terms in\linebreak a
multi-distribution.
Therefore,  instead we view a rewrite sequence on multi-dis\-tri\-bu\-tions as a tree (e.g.,
the tree 
representation of the rewrite sequence
\eqref{R3 chain}
from
\begin{wrapfigure}[5]{r}{0.36\textwidth}
  \vspace*{-.7cm}
  \scriptsize
	\hspace*{-.25cm}\begin{tikzpicture}
			\tikzstyle{adam}=[rectangle,thick,draw=black!100,fill=white!100,minimum size=4mm]
			\tikzstyle{empty}=[rectangle,thick,minimum size=4mm]
			
			\node[adam,pin={[pin distance=0.1cm, pin edge={,-}] 180:\scriptsize \textcolor{blue}{$P$}}] at (0, 0)  (a) {$1:\Com{1}(\tF(\O))$};
			\node[adam,pin={[pin distance=0.1cm, pin edge={,-}] 180:\scriptsize \textcolor{blue}{$P$}}] at (0, -.7)  (b) {$1:\Com{2}(\tF(\ta), \tA)$};
			\node[adam] at (-1.19, -1.4)  (c) {$\nicefrac{1}{2}\!:\Com{2}(\tF(\tb_1), \tB_1)\!$};
			\node[adam] at (1.19, -1.4)  (d) {$\nicefrac{1}{2}\!:\Com{2}(\tF(\tb_2), \tB_2)\!$};
		
			\draw (a) edge[->] (b);
			\draw (b) edge[->] (c);
			\draw (b) edge[->] (d);
		\end{tikzpicture}
\end{wrapfigure}
 \cref{example:solving-problem-with-dep-pair}  is on the right).
Regarding the paths in this tree (which represent  rewrite sequences of terms with certain
probabilities)
allows us to adapt the\linebreak idea of chains, i.e.,
that one uses only finitely many  $\SSS$-steps
before the next step with a DT from $\PP$.

{\small \begin{definition}[Chain Tree] \label{def:chain-tree}
   $\!\!\F{T}\!=\!(V,E,L,P)$ {\normalsize is an (innermost)}
    $(\PP\!,\SSS)${\normalsize\emph{-chain tree}} if\linebreak

\vspace*{-.6cm}

    {\normalsize	\begin{enumerate}
		\item $V \neq \emptyset$ is a possibly infinite set of nodes and
	          $E \subseteq V \times V$ is a set of directed edges, such that
                  $(V, E)$ is a (possibly infinite) directed tree where $vE
= \{ w \mid (v,w) \in E \}$
                   is
                   finite for every $v \in V$.
            \item  $L:V\rightarrow(0,1]\times\TSet{\SigmaDP}{\VSet}$ labels every node
              $v$
              by a probability $p_v$ and a term $t_v$.
               For the root $v \in V$ of the tree, we have $p_v = 1$.
	     \item $P \subseteq V \setminus \ctleaf$ (where $\ctleaf$ are all leaves)
               is a subset of the
                  inner nodes to\linebreak indicate whether we use the PPTRS $\PP$
or the PTRS $\SSS$ for the rewrite step.
		 $S = V \setminus (\ctleaf \cup P)$ are all inner nodes that are not in $P$.
		Thus, $V = P \uplus S \uplus \ctleaf$.
		\item For all  $v \in P$: If
                  $vE = \{w_1, \ldots, w_k\}$, then 
                  $t_v \setitops \{\tfrac{p_{w_1}}{p_v}:t_{w_1}, \ldots,
                  \tfrac{p_{w_k}}{p_v}:t_{w_k}\}$.
		\item For all  $v \in S$: If
                  $vE = \{w_1, \ldots, w_k\}$, then
 $t_v \setitos \{\tfrac{p_{w_1}}{p_v}:t_{w_1}, \ldots, \tfrac{p_{w_k}}{p_v}:t_{w_k}\}$.
		\item Every infinite path in $\F{T}$ contains infinitely many
                  nodes from $P$. 
	\end{enumerate}}
\end{definition}}

Conditions 1--5 ensure that the tree represents a valid rewrite sequence and the
last condition is the main property for chains.

\begin{definition}[$|\F{T}|_{\ctleaf}$, iAST] \label{def:chain-tree-convergence-notation}
	For any innermost $(\PP,\SSS)$-chain tree $\F{T}$
        we define $|\F{T}|_{\ctleaf} = \sum_{v \in \ctleaf} \,p_v$.        
We say that $(\PP,\SSS)$ is \emph{iAST} if we have
$|\F{T}|_{\ctleaf} = 1$ for
every innermost
$(\PP,\SSS)$-chain tree $\F{T}$.
\end{definition}
While we have $|\F{T}|_{\ctleaf} = 1$ for every finite chain tree $\F{T}$, 
for infinite chain trees $\F{T}$ we may have  $|\F{T}|_{\ctleaf} <1$ or
even $|\F{T}|_{\ctleaf} = 0$ if  $\F{T}$ has no leaf at all.

With this new type of DTs and chain trees, we now obtain
an analogous chain criterion to the non-probabilistic setting.

\begin{restatable}[Chain Criterion]{theorem}{ProbChainCriterion}\label{theorem:prob-chain-criterion}
    A PTRS $\R$ is iAST if $(\DTuple{\R},\R)$ \pagebreak[2] is iAST.
\end{restatable}

In contrast to the non-probabilistic case, our chain criterion
as presented in the paper is \emph{sound} but not \emph{complete}
(i.e., we do not have ``iff'' in \Cref{theorem:prob-chain-criterion}).
However, we also developed a refinement where
our chain criterion is made complete by also 
storing the positions of the defined symbols in $dp(r)$
\cite{Kassing:Thesis:2022}.
In this way, one can avoid ``unsuitable'' chain trees, as discussed at the end of
\cref{example:solving-problem-with-dep-pair}.

Our notion of DTs and chain trees
is only suitable for  \emph{innermost}
evaluation. To see this, consider the PTRSs $\R_1'$ and $\R_2'$ which both contain
$\tg \to \{ \nicefrac{1}{2}:\O, \nicefrac{1}{2}:\th(\tg) \}$, but in addition
$\R_1'$ has the rule $\th(x) \to \{ 1: \tf(x,x) \}$ and $\R_2'$ has the rule $\th(x) \to
\{1:\tf(x,x,x)\}$. Similar to $\R_1$ and $\R_2$ in \eqref{R1R2},
$\R_1'$ is AST while  $\R_2'$ is not. In contrast,  both  $\R_1'$ and $\R_2'$ are iAST, since the
innermost evaluation strategy prevents  the application of the $\th$-rule to terms
containing $\tg$.
Our DP framework handles   $\R_1'$ and $\R_2'$ in the same way,
as both 
have the same DT $\langle \tG, \tg \rangle \to
\{ \nicefrac{1}{2}:\langle \Com{0}, \O \rangle,
\nicefrac{1}{2}:\langle \Com{2}(\tH(\tg),\tG), \th(\tg) \rangle \}$ and
a DT $\langle \tH(x), \th(x) \rangle \to
\{ 1: \langle \Com{0}, \tf(\ldots) \rangle\}$.
Even if we allowed the application of the second DT to terms of the form $\tH(\tg)$, we
would still obtain $|\F{T}|_{\ctleaf} = 1$
for every chain tree $\F{T}$.
So a DP framework to analyze ``full'' instead of innermost AST would be considerably more
involved.

\subsection{The Probabilistic DP Framework}\label{The Probabilistic DP Framework}

Now we introduce the probabilistic dependency pair framework which keeps the
core ideas of the non-probabilistic framework.
So instead of applying one ordering for a PTRS directly as in
\Cref{theorem:ptrs-direct-application-poly-interpretations}, we want to 
benefit from modularity. 
Now a \emph{DP processor} $\Proc$  is of the form $\Proc(\PP, \SSS) = \{(\PP_1,\SSS_1),
\ldots, (\PP_n,\SSS_n)\}$, where
 $\PP, \PP_1, \ldots, \PP_n$ are PPTRSs and  $\SSS, \SSS_1, \ldots, \SSS_n$ are PTRSs. 
A processor $\Proc$ is \emph{sound} if
$(\PP, \SSS)$ is iAST whenever $(\PP_i, \SSS_i)$ is iAST for all $1 \leq i \leq n$. 
It is \emph{complete} if $(\PP_i, \SSS_i)$ is iAST for all $1 \leq i \leq n$ whenever 
$(\PP, \SSS)$ is iAST. 
In the following, we adapt the three  main processors from \Cref{DGP,URP,RPP} to the
probabilistic setting
and present two additional processors.

The (innermost)  $(\PP,\SSS)$-\emph{dependency graph} indicates which DTs from $\PP$ can rewrite to
each other using the PTRS $\SSS$.
The possibility of rewriting with $\SSS$ is not related to the probabilities.
Thus, for the dependency graph, we can use the \emph{non-probabilistic variant}
$\nonprob(\SSS) = \{\ell \to r_j \mid \ell \to \{p_1:r_1, \ldots, p_k:r_k\} \in \SSS, 1
\leq j \leq k\}$.\linebreak

\vspace*{-.3cm}

\begin{definition}[Dep.\ Graph]
  The node set of the
  \emph{$(\PP,\SSS)$-dependency graph} is $\PP$ and 
	there is an edge from $\langle \ell^\#_1,\ell_1 \rangle \to \{ p_1:\langle d_1,r_1 \rangle, \ldots,
  p_k:\langle d_k,r_k \rangle\}$ to $\langle \ell^\#_2, \ell_2 \rangle \to \ldots$ if there are substitutions
  $\sigma_1, \sigma_2$ and
  $t^\# \in \cont(d_j)$ for some $1 \leq j \leq k$ such that $t^\# \sigma_1 \itononprobsstar
  \ell^\#_2 \sigma_2$ and both  $\ell_1^\# \sigma_1$ and $\ell_2^\#
  \sigma_2$ are in $\mathtt{NF}_{\SSS}$.
\end{definition}

\begin{wrapfigure}[3]{r}{0.17\textwidth}
  \scriptsize
  \vspace*{-1cm}
    \hspace*{-.15cm}\begin{tikzpicture}
    		\node[shape=rectangle,draw=black!100, minimum size=3mm] (A) at (0,0) {$\eqref{R-div-deptup-1}$};
    		\node[shape=rectangle,draw=black!100, minimum size=3mm] (B) at (1,0) {$\eqref{R-div-deptup-2}$};
    		\node[shape=rectangle,draw=black!100, minimum size=3mm] (C) at (0,.7) {$\eqref{R-div-deptup-3}$};
    		\node[shape=rectangle,draw=black!100, minimum size=3mm] (D) at (1,.7) {$\eqref{R-div-deptup-4}$};
    	
    		\path [->,loop below,in=340,out=20,looseness=6] (B) edge (B);
    		\path [->,loop above,in=340,out=20,looseness=6] (D) edge (D);
    		\path [->] (B) edge (A);
    		\path [->] (D) edge (C);
    		\path [->] (D) edge (B);
            \path [->] (D) edge (A);                
    	\end{tikzpicture}      
  \end{wrapfigure}
For $\R_{\tpdiv}$ from \Cref{example:ptrs-int-div-coupled-positional-dependency-pairs}, the
$(\DTuple{\R_{\tpdiv}}, \R_{\tpdiv})$-dependency graph is on the side.
In
the non-probabilistic DP framework, every\linebreak step with $\itodr$ corresponds to an
edge in the $(\C{D}, \R)$-dependency graph. Similarly, in the probabilistic setting,  every
path from one node of $P$ to the next node of $P$ in a $(\PP,\SSS)$-chain tree corresponds to an edge in the
$(\PP,\SSS)$-\linebreak dependency graph. Since every infinite path in a chain tree contains infinitely
many nodes from $P$, when tracking the arguments of the compound symbols,
every such path traverses a
cycle of the dependency graph infinitely often. Thus, it again suffices \pagebreak[2] 
to consider the SCCs of the dependency graph separately.
So for our example, we obtain  $\Proc_{\mathtt{DG}}(\DTuple{\R_{\tpdiv}},\R_{\tpdiv}) = \{(\{\eqref{R-div-deptup-2}\},\R_{\tpdiv}), (\{\eqref{R-div-deptup-4}\},\R_{\tpdiv})\}$.
To automate the following two processors, the same over-approximation techniques as for the
non-probabilistic dependency graph can be used.

\begin{restatable}[Prob.\ Dep.\ Graph Processor]{theorem}{ProbDepGraphProc}\label{theorem:prob-DGP}
For the SCCs $\PP_1, ..., \PP_n$ of the\linebreak
  $(\PP\!,\SSS)$-dependency graph,
  \mbox{\small $\Proc_{\mathtt{DG}}(\PP\!,\SSS)\!=\!\{(\PP_1,\SSS), ...,
  (\PP_n,\SSS)\}$} is sound and complete.
\end{restatable}

Next, we introduce a new \emph{usable terms processor} (a similar processor was also
proposed for the DTs in \cite{noschinski2013analyzing}).
Since we regard dependency \emph{tuples} instead of pairs, after applying
 $\Proc_{\mathtt{DG}}$, the right-hand sides of DTs $\langle \ell^\#_1, \ell_1 \rangle \to
\ldots$ might still contain terms $t^\#$ where no instance
$t^\#\sigma_1$ rewrites to an instance $\ell_2^\#\sigma_2$ of a left-hand side of a DT
(where we only consider instantiations such that $\ell^\#_1\sigma_1$ and $\ell^\#_2\sigma_2$ are in
$\mathtt{NF}_{\SSS}$,
because  only such instantiations
are regarded in chain trees).
Then
$t^\#$ can be removed from the right-hand side of the DT. For example, in the DP 
problem $(\{\eqref{R-div-deptup-4}\},\R_{\tpdiv})$,
the only DT \eqref{R-div-deptup-4} has the left-hand side $\tD(\ts(x),\ts(y))$. As the term 
$\tM(x,y)$ in \eqref{R-div-deptup-4}'s right-hand side cannot ``reach''
$\tD(\ldots)$, the following processor removes it, i.e.,
 $\Proc_{\mathtt{UT}}(\{\eqref{R-div-deptup-4}\},\R_{\tpdiv}) =
\{(\{\eqref{R-div-deptup-5}\},\R_{\tpdiv})\}$, where \eqref{R-div-deptup-5}
is

\vspace*{-.3cm}

{\small
  \begin{align}
    \langle\tD(\ts(x),\ts(y)), \tdiv(\ts(x),\ts(y))\rangle
& \to \{\nicefrac{1}{2}: \langle\Com{1}(\tD(\ts(x),\ts(y))),
          \tdiv(\ts(x),\ts(y))\rangle, \hspace*{2cm} \nonumber
          \\
  \label{R-div-deptup-5}
 \rlap{\hspace*{-.5cm}$\nicefrac{1}{2}:\langle\Com{1}(\tD(\tminus(x,y),\ts(y))),\ts(\tdiv(\tminus(x,y),\ts(y)))\rangle\}$.}
  \end{align}
}

So both \Cref{theorem:prob-DGP,theorem:prob-UPP}  are needed to fully simulate the
dependency graph processor in the probabilistic 
setting, i.e., they are both necessary to guarantee that the probabilistic DP
processors work analogously to the non-probabilistic ones (which in turn 
ensures that the probabilistic DP framework is similar in power to its non-probabilistic
counterpart). This is also confirmed by our experiments in \Cref{Evaluation} which
show that disabling the processor of \Cref{theorem:prob-UPP} affects the power of our
approach.
For example, without \Cref{theorem:prob-UPP}, the 
proof that $\R_{\tpdiv}$ is iAST in the probabilistic DP framework
would require a more complicated polynomial interpretation. In contrast, when using both
processors of \Cref{theorem:prob-DGP,theorem:prob-UPP}, then one can prove iAST of
$\R_{\tpdiv}$
with the same polynomial
interpretation that was used to prove iTerm of $\R_{\tdiv}$ (see \Cref{example:RPP}).

\vspace*{-.1cm}

\begin{restatable}[Usable Terms Processor]{theorem}{UsableTermsProc}\label{theorem:prob-UPP}
Let $\ell^\#_1$ be a term and
  $(\PP, \SSS)$ be a DP problem.
  We call a term $t^\#$ \emph{usable} w.r.t.\ $\ell^\#_1$ and
  $(\PP, \SSS)$ if there is a  $\langle\ell^\#_2,\ell_2 \rangle \to \ldots \in \PP$
  and substitutions $\sigma_1, \sigma_2$ such that $t^\# \sigma_1 \itononprobsstar
  \ell^\#_2 \sigma_2$ and both $\ell^\#_1 \sigma_1$ and $\ell^\#_2
  \sigma_2$ are  in $\mathtt{NF}_{\SSS}$.
  If $d= \Com{n}(t_1^\#, \ldots, t_n^\#)$, then 
$\upairs(d)_{\ell^\#_1\!,\PP,\SSS}$ denotes the term $\Com{m}(t_{i_1}^\#, \ldots,
    t_{i_m}^\#)$, where
    $1 \leq i_1 < \ldots < i_m \leq n$ are the indices 
    of all terms
$t^\#_i$ that are usable w.r.t.\  $\ell^\#_1$ and
  $(\PP, \SSS)$. 
	The transformation that removes all non-usable terms in the right-hand sides of dependency tuples is denoted by:
	\[
		\begin{array}{l}
		  \C{T}_\mathtt{UT}(\PP, \SSS) = \{ \langle \ell^\#,\ell \rangle \to
                  \{ p_1:\langle\upairs(d_1)_{\ell^\#\!,\PP,\SSS},r_1\rangle, \ldots,
                  p_k:\langle\upairs(d_k)_{\ell^\#\!,\PP,\SSS},r_k\rangle\} \\
		  \hspace{5cm} \mid \langle\ell^\#,\ell\rangle
                  \to  \{ p_1:\langle d_1,r_1\rangle, \ldots, p_k:\langle d_k,r_k\rangle\} \in \PP \}
		\end{array}
	\]
  Then $\Proc_{\mathtt{UT}}(\PP,\SSS) = \{(\C{T}_\mathtt{UT}(\PP, \SSS),\SSS)\}$ is sound and complete.
\end{restatable}

To adapt the \emph{usable rules processor}, we
adjust the definition of usable rules such that it
regards every term in the support of the
distribution on the right-hand side of a rule. \pagebreak[2] 
The usable rules processor only deletes non-usable rules from $\SSS$, but not from
$\projTwo(\PP)$. This is sufficient, because according to \Cref{def:PPTRS-rewriting-on-sets}, rules from 
$\projTwo(\PP)$ can only be applied if they also occur in $\SSS$.

\begin{restatable}[Probabilistic Usable Rules Processor]{theorem}{ProbUsRulesProc}\label{def:prob-usable-rules}
    Let  $(\PP, \SSS)$ be a DP problem.
    For every $f \in \SigmaDP$ let $\rules_\SSS(f) = \{\ell \to \mu \in \SSS \mid \rootsym(\ell) = f\}$.
    For any term  $t \in  \TSet{\SigmaDP}{ \VSet}$,
    its \emph{usable rules} $\urules_\SSS(t)$ are the smallest set
    such that
    $\urules_\SSS(x) = \emptyset$ for all $x \in \VSet$
    and $\urules_\SSS(f(t_1, \ldots, t_n)) = \rules_\SSS(f) \cup \bigcup_{i = 1}^n
    \urules_\SSS(t_i)\linebreak 
     \cup \; \bigcup_{\ell \to \mu \in \rules_\SSS(f), r \in \Supp(\mu)} \urules_\SSS(r)$.
    The \emph{usable rules} for  $(\PP, \SSS)$ are
    $\urules(\PP,\SSS) =\linebreak \bigcup_{\ell^\# \to \mu \in \projOne(\PP), d \in \Supp(\mu)} \urules_\SSS(d)$.
    Then $\Proc_{\mathtt{UR}}(\PP,\SSS) = \{(\PP,\urules(\PP,\SSS))\}$ is sound.
\end{restatable}
\begin{example}\label{example:prob-usable-rules}
	For the DP problem $(\{\eqref{R-div-deptup-5}\},\R_{\tpdiv})$ only the $\tminus$-rules
  are usable and
thus $\Proc_{\mathtt{UR}}(\{\eqref{R-div-deptup-5}\},\R_{\tpdiv}) =
        \{ (\{\eqref{R-div-deptup-5}\},\{\eqref{eq:ptrs-div-1},\eqref{eq:ptrs-div-2}\}) \}$.
	For $(\{\eqref{R-div-deptup-2}\},\R_{\tpdiv})$
  there are no usable rules at all, hence
        $\Proc_{\mathtt{UR}}(\{\eqref{R-div-deptup-2}\},\R_{\tpdiv}) =
        \{ (\{\eqref{R-div-deptup-2}\},\emptyset)\}$. 
\end{example}

For the \emph{reduction pair processor}, we again restrict ourselves to multilinear polynomials and use analogous
  constraints as in our new  criterion for the direct application of polynomial
  interpretations to PTRSs (\cref{theorem:ptrs-direct-application-poly-interpretations}),
  but adapted to DP problems $(\PP, \SSS)$. Moreover, as in the original reduction pair
  processor of \cref{RPP}, the polynomials only have to be weakly monotonic.
For every rule in $\SSS$ or $\projOne(\PP)$, we require that the expected value is weakly
decreasing.
The reduction pair processor then removes those
DTs $\langle \ell^\#,\ell \rangle \to \{ p_1:\langle d_1,r_1 \rangle, \ldots,p_k:\langle d_k,r_k \rangle \}$
 from $\PP$ where in addition there is
at least one term $d_j$ that is strictly decreasing.
Recall that we can also rewrite with
the original rule 
$\ell \to \{ p_1:r_1, \ldots,p_k:r_k \}$
from $\projTwo(\PP)$, provided
that it is also contained in $\SSS$. Therefore, to remove the dependency tuple, we also have to require that the rule
$\ell \to r_j$ is weakly decreasing.
Finally,  we have to use \emph{$\Com{}$-additive}
interpretations (with ${\Com{n}}_{\Pol}(x_1,
\ldots, x_n) = x_1 + \ldots + x_n$) to handle compound symbols and their normalization
correctly.

\begin{restatable}[Probabilistic Reduction Pair Processor]{theorem}{ProbRPP}\label{theorem:prob-RPP}
  Let $\Pol:
\mathcal{T}(\Sigma \,\uplus$\linebreak $\Sigma^\#, \VSet)
\to \IN[\VSet]$ be a
weakly monotonic, multilinear, and $\Com{}$-additive
polynomial interpretation. Let $\PP = \PP_{\geq} \uplus \PP_{>}$
with $\PP_> \neq \emptyset$ such that:
	\begin{itemize}
		\item[(1)] For every $\ell \to \{ p_1:r_1, ...,p_k:r_k \} \in \SSS$, we have 
		$\Pol(\ell) \geq \sum_{1 \leq j \leq k} p_j \cdot \Pol(r_j)$.
		\item[(2)] For every $\langle \ell^\#,\ell \rangle \to \{ p_1:\langle
                  d_1,r_1 \rangle, \ldots,p_k:\langle d_k,r_k \rangle \} \in \PP$,
                we have $\Pol(\ell^\#) \geq \sum_{1 \leq j \leq k} p_j \cdot \Pol(d_j)$.   
		\item[(3)]
		For every $\langle \ell^\#,\ell \rangle \to \{ p_1:\langle d_1,r_1
                \rangle, \ldots,p_k:\langle d_k,r_k \rangle \} \in \PP_{>}$, 
                there exists a $1 \leq j \leq k$ with $\Pol(\ell^\#) > \Pol(d_j)$.\\
		If $\ell \to \{ p_1:r_1, \ldots,p_k:r_k \} \in \SSS$, 
                then we additionally
                have   $\Pol(\ell) \geq \Pol(r_j)$. 
	\end{itemize}
	Then $\Proc_{\mathtt{RP}}(\PP,\SSS) = \{(\PP_{\geq},\SSS)\}$ is sound and complete.
\end{restatable}
\begin{example}\label{example:RPP}
  The constraints of the reduction pair processor for the two
  DP problems from \cref{example:prob-usable-rules} are satisfied by the
  $\Com{}$-additive polynomial
  interpretation which again maps $\O$ to $0$, $\ts(x)$ to $x+1$, and all other non-constant function symbols to the projection on their first arguments.
  As in the non-probabilistic case, this results in DP problems
 of the form $(\emptyset,
\ldots)$ and subsequently, $\Proc_{\mathtt{DG}}(\emptyset,
\ldots)$ yields $\emptyset$. By the soundness of all processors, this proves that
$\R_{\tpdiv}$ is iAST.
\end{example}

So with the new probabilistic DP framework, the proof that  $\R_{\tpdiv}$ is iAST is
analogous \pagebreak[2] to the proof  that  $\R_{\tdiv}$ is iTerm in the original DP framework
(the proofs even use the same polynomial
interpretation in the respective reduction pair processors). This indicates that 
our novel framework for PTRSs
has the same essential concepts and advantages 
as the original DP
framework for TRSs.
This is different from our previous adaption of dependency pairs for complexity analysis of
TRSs, which also relies on dependency tuples~\cite{noschinski2013analyzing}. 
  There, the power is considerably restricted, because one does not have full modularity
  as one cannot decompose the proof according to the SCCs of the dependency
  graph.

 In proofs with the probabilistic DP framework, one may obtain
DP problems $(\PP,\SSS)$ that have a non-probabilistic structure 
  (i.e., every DT in $\PP$ has the form $\langle\ell^\#, \ell \rangle \to \{1:\langle d,r\rangle\}$
and every  rule in $\SSS$ has the form $\ell' \to \{1:r'\}$). We now introduce a processor
that allows us
to switch to the original non-probabilistic DP framework
for such (sub-)problems.
This is advantageous,  because
due to the use of dependency \emph{tuples} instead of pairs
in $\PP$, in general the constraints of the 
probabilistic reduction pair processor of \cref{theorem:prob-RPP} are 
harder than the ones of the reduction pair processor of \cref{RPP}.
Moreover, 
\cref{RPP}
is not restricted to  multilinear polynomial interpretations and the original DP framework
has many additional processors that have not yet been adapted to the probabilistic
setting.

\begin{restatable}[Probability Removal Processor]{theorem}{NPP}\label{theorem:prob-NPP}
	Let
        $(\PP, \SSS)$ be a probabilistic DP problem where every DT in $\PP$ has the
        form $\langle \ell^\#, \ell \rangle \to \{1:\langle d,r\rangle\}$ and every
        rule in $\SSS$ has the form $\ell' \to \{1\!:r'\}$.
        Let $\nonprob(\PP) = \{\ell^\# \to t^\# \mid \ell^\# \to \{1\!:d\} \in \projOne(\PP),\linebreak
        t^\# \in \cont(d) \}$. Then
        $(\PP,\SSS)$ is iAST iff the
 non-probabilistic DP problem $(\nonprob(\PP),\linebreak
        \nonprob(\SSS))$ is
        iTerm.  So
 if
$(\nonprob(\PP),
        \nonprob(\SSS))$ is
        iTerm, then
        the processor 
        $\Proc_{\mathtt{PR}}(\PP,\SSS) = \emptyset$
        is
        sound and complete.
\end{restatable}

\section{Conclusion and Evaluation}\label{Evaluation}

Starting
with a new ``direct'' technique to prove almost-sure termination of probabilistic
TRSs (\Cref{theorem:ptrs-direct-application-poly-interpretations}),
we presented the first adaption of the dependency pair framework to the probabilistic
setting in order to prove  innermost AST automatically.
This is not at all obvious, since most straightforward ideas for such an adaption are 
unsound (as discussed in \Cref{Dependency Tuples and Chains for Probabilistic Term
  Rewriting}).
So the challenge was to find  a suitable
  definition of dependency pairs (resp.\ tuples) and chains (resp.\ chain trees)
 such that
  one can define \pagebreak[2] DP processors which are sound and work analogously   to the
  non-probabilistic setting  (in order to obtain a framework which is similar in power to the
  non-probabilistic one).
 While the soundness proofs for our new processors
 are much more
 involved than in the non-probabilistic case,
the new processors themselves are quite analogous to
  their non-probabilistic counterparts and thus,
  adapting an existing implementation of the
non-probabilistic DP framework to the probabilistic one does not require
much effort.

We implemented our contributions in our termination prover
\textsf{AProVE},
which yields the first tool to prove almost-sure innermost termination of
PTRSs on arbitrary data structures (including 
PTRSs
that are not PAST). In our experiments, 
we compared the direct application of polynomials for proving AST (via
our new \cref{theorem:ptrs-direct-application-poly-interpretations}) with the
probabilistic DP framework. 
We evaluated \textsf{AProVE}
on a collection \pagebreak[2] of 67 PTRSs
which includes many typical probabilistic algorithms.
For example,
it contains the following  PTRS $\R_{\tqs}$ for
probabilistic quicksort.

\vspace*{-.4cm}

{\small
  \begin{align*}
  \trotate(\tcons(x,\xs)) & \!\to\! \{ \nicefrac{1}{2} : \tcons(x,\xs), \nicefrac{1}{2} : \trotate(\tapp(\xs, \tcons(x,\tnil)))\}\\
  \tqs(\tnil) & \!\to\! \{ 1 : \tnil\}\\
  \tqs(\tcons(x,\xs)) & \!\to\! \{ 1 : \tqshelp(\trotate(\tcons(x,\xs)))\}\\
    \tqshelp(\tcons(x,\xs)) & \!\to\! \{ 1 :
    \tapp(\tqs(\tlow(x,\xs)),\tcons(x,\tqs(\thigh(x,\xs))))\}
\end{align*}
    }

\vspace*{-.1cm}

The
 $\trotate$-rules rotate
a list  randomly often (they are 
AST, but not terminating). Thus, by choosing the first element
of the resulting list, one obtains a random pivot element for the recursive call of quicksort. 
In addition to the  rules above, $\R_{\tqs}$
contains
rules for
list concatenation ($\tapp$), and rules such that
$\tlow(x,\xs)$ (resp.\ $\thigh(x,\xs)$)
returns all elements of the list $\xs$ that are smaller (resp.\ greater or equal) than
$x$, see\report{ App.\ \ref{Examples}}\paper{ \cite{report}}. Using the probabilistic DP framework,
\textsf{AProVE} can prove iAST of $\R_{\tqs}$ and many other typical programs.

61 of the 67 examples in our collection are iAST and \textsf{AProVE}
can prove iAST for 53 (87\%) of
them.
Here, the DP framework proves iAST for 51 examples and
the direct application of polynomial interpretations via
 \cref{theorem:ptrs-direct-application-poly-interpretations}
 succeeds for 27 examples. (In contrast, 
 proving PAST via the direct application of
 polynomial interpretations as in \cite{avanzini2020probabilistic} only works for 22 examples.)
 The average
runtime of \textsf{AProVE} per example was 2.88~s (where no example took longer than 8~s).
So our experiments indicate
that the power of the DP framework
can now also be used for probabilistic TRSs.

We also performed experiments where we disabled individual processors
of the probabilistic DP
framework. More precisely,
we disabled either the usable terms\linebreak processor (\Cref{theorem:prob-UPP}), both the dependency graph and the
usable terms processor (\Cref{theorem:prob-DGP,theorem:prob-UPP}), or all processors except the reduction pair processor of \Cref{theorem:prob-RPP}.
Our experiments show that disabling  processors
indeed affects the power of the approach, in particular for larger examples with several
defined symbols (e.g., 
  then \textsf{AProVE}
 cannot prove iAST of $\R_{\tqs}$ anymore). So all of our 
processors are needed to obtain a powerful technique for termination analysis of PTRSs.

Due to the  use of dependency \emph{tuples} instead of pairs, 
 the probabilistic
DP framework does not (yet) subsume the direct application of polynomials completely  (two examples in our
collection can only be proved by the latter, see\report{ App.\ \ref{Examples}}\paper{ \cite{report}}).
Therefore, currently \textsf{AProVE} uses the direct approach of  \cref{theorem:ptrs-direct-application-poly-interpretations} in addition to
the probabilistic DP framework.
  In future work, we will adapt further
  processors of the original DP framework to the probabilistic setting, which will also
  allow us to 
   integrate the direct approach of
  \cref{theorem:ptrs-direct-application-poly-interpretations}
  into the probabilistic DP framework in a modular way.
Moreover, we will
  develop processors  to prove AST of full (instead of innermost) rewriting. 
Further work may also include processors to disprove (i)AST and possible extensions
to analyze PAST and expected runtimes as well. Finally, one could also modify the formalism
of PTRSs in order to allow non-constant probabilities which depend on the sizes of terms.

For details on our experiments and for instructions on how to run our implementation
in \textsf{AProVE} via its \emph{web interface} or locally, we refer to \url{https://aprove-developers.github.io/ProbabilisticTermRewriting/}.

\medskip

\noindent
\textbf{Acknowledgements.} We are grateful to Marcel Hark, Dominik Meier, and Florian
Frohn for help and\vspace*{-.4cm}\pagebreak[2] advice.

\bibliographystyle{splncs04}
\newcommand{\noopsort}[1]{}
\bibliography{biblio}

\report{
  \clearpage

\appendix
\renewcommand{\theHsection}{appendix.\Alph{section}}

\vspace*{-.4cm}

\section*{Appendix}

\vspace*{-.1cm}

In \Cref{Examples},  we present several examples to demonstrate certain aspects of the DP
framework. Afterwards, \Cref{appendix} contains all proofs for our new contributions and
observations.

\section{Examples}\label{Examples}

In this section, we present several examples to
illustrate specific strengths and weaknesses of the new probabilistic DP framework. All of
these examples are contained in the collection that we used in our experiments. The
collection also contains the examples from our paper (\Cref{example:PTRS-random-walk},
\ref{example:problem-with-dependency-tuples}, and \ref{example:ptrs-int-div-coupled-positional-dependency-pairs},
and $\R_1, \R_2, \R_1', \R_2'$ from \Cref{Dependency Tuples and Chains for Probabilistic Term Rewriting}).

\subsection{Probabilistic Quicksort}

As discussed in \Cref{Evaluation}, a classic algorithm which can benefit from
randomization is quicksort. The following PTRS 
implements a probabilistic quicksort algorithm.

\vspace*{-.3cm}

{\small
\begin{align*}
  \trotate(\tnil) & \!\to\! \{ 1 : \tnil \}\\
  \trotate(\tcons(x,\xs)) & \!\to\! \{ \nicefrac{1}{2} : \tcons(x,\xs), \nicefrac{1}{2} : \trotate(\tapp(\xs, \tcons(x,\tnil)))\}\\
  \tqs(\tnil) & \!\to\! \{ 1 : \tnil\}\\
  \tqs(\tcons(x,\xs)) & \!\to\! \{ 1 : \tqshelp(\trotate(\tcons(x,\xs)))\}\\
  \tqshelp(\tcons(x,\xs)) & \!\to\! \{ 1 : \tapp(\tqs(\tlow(x,\xs)),\tcons(x,\tqs(\thigh(x,\xs))))\}\\
  \tlow(x,\tnil)  & \!\to\! \{ 1 : \tnil\}\\
  \tlow(x,\tcons(y,\ys)) & \!\to\! \{ 1 : \tiflow(\tleq(x,y),x,\tcons(y,\ys))\}\\
  \tiflow(\ttrue,x,\tcons(y,\ys)) & \!\to\! \{ 1 : \tlow(x,\ys)\}\\
  \tiflow(\tfalse,x,\tcons(y,\ys)) & \!\to\! \{ 1 : \tcons(y,\tlow(x,\ys))\}\\
  \thigh(x,\tnil) & \!\to\! \{ 1 : \tnil\}\\
  \thigh(x,\tcons(y,\ys)) & \!\to\! \{ 1 : \tifhigh(\tleq(x,y),x,\tcons(y,\ys))\}\\
  \tifhigh(\ttrue,x,\tcons(y,\ys)) & \!\to\! \{ 1 : \tcons(y,\thigh(x,\ys))\}\\
  \tifhigh(\tfalse,x,\tcons(y,\ys)) & \!\to\! \{ 1 : \thigh(x,\ys)\}\\
  \tleq(0,x) & \!\to\! \{ 1 : \ttrue\}\\
  \tleq(\ts(x),0) & \!\to\! \{ 1 : \tfalse\}\\
  \tleq(\ts(x),\ts(y)) & \!\to\! \{ 1 : \tleq(x,y)\}\\
  \tapp(\tnil,\ys) & \!\to\! \{ 1 : \ys\}\\
  \tapp(\tcons(x,\xs),\ys) & \!\to\! \{ 1 : \tcons(x,\tapp(\xs,\ys))\}\!
\end{align*}
}

The first two $\trotate$ rules are used to determine a random pivot element.
This is achieved by rotating the list and always moving the head element to the end of the list.
With a chance of $\nicefrac{1}{2}$ we stop this iteration and use the current head element as the next pivot element.
The rest of the rules represent the classical quicksort algorithm without any probabilities.
Here, $\tapp$ computes list concatenation,
$\tlow(x,\xs)$ returns all elements of the list $\xs$ that are smaller than $x$, and
$\thigh$ works analogously.

Using our new probabilistic DP framework,
\textsf{AProVE} can
automatically prove that this PTRS $\R_{\tqs}$ is iAST.
Using the dependency graph, the initial DP problem is split
into six sub-problems which are then analyzed on their own.
For four of them, the probabilistic $\trotate$-rules can be removed via the usable rules processor.
Hence, here one can use the non-probabilistic DP framework instead via the probability removal processor.
For the sub-problem with the dependency tuple resulting from
the $\trotate$-rules and for the sub-problem with the dependency tuples from the $\tqs$- and $\tqshelp$-rules,
the probabilistic reduction pair processor is used to prove that they are iAST.

\textsf{AProVE} can also 
prove non-termination of $\nonprob(\R_{\tqs})$.
So one can conclude that the reason for  non-termination
must be the probabilistic $\trotate$-rules. Indeed,
with a chance of $\nicefrac{1}{2}$, $\trotate$
calls itself on the rotated list and this can be repeated infinitely often.

\subsection{Disadvantage of Dependency Tuples Compared to Non-Probabilistic Reduction Pair
  Processor}

The probability removal processor (\Cref{theorem:prob-NPP}) was motivated by the observation 
that in general, the constraints of the 
probabilistic reduction pair processor of \cref{theorem:prob-RPP} are 
harder than the ones of the original reduction pair processor of \cref{RPP}. To illustrate that,
consider the following PTRS.

\vspace*{-.5cm}

{\small
\begin{align*}
  \tg(\ts(x)) & \!\to\!  \{ 1 : \tf(\tg(x),\tg(x),\tg(x)) \}
  \end{align*}
}
If one wants to prove iAST with the DP framework using
linear polynomial interpretations, then the probabilistic 
reduction pair processor has to map $\ts(x)$ to a polynomial of the form $s_0 + s_1 \cdot
x$ with $s_1 \geq 3$, because $\tG(\ts(x))$ must be greater than $\tG(x) + \tG(x) +
\tG(x)$.
The reason is that the dependency tuple contains $\Com{3}(\tG(x), \tG(x),\tG(x))$ on the
right-hand side. 
Similarly, if one considers a variant of the above
TRS where the right-hand side has the form $\tf(\tg(x), \ldots, \tg(x))$ with $k$
occurrences of $\tg(x)$, then the probabilistic DP framework only succeeds if one uses a
polynomial with $s_1 \geq k$. In contrast, the non-probabilistic reduction pair processor
only needs a linear polynomial interpretation with coefficients from $\{0,1\}$, because
here one only has the dependency \emph{pair} $\tG(\ts(x)) \to \tG(x)$ (instead of
dependency \emph{tuples}), irrespective of the number
of occurrences of $\tg(x)$ on the right-hand side.  So it suffices to map $\tG(x)$ to $x$
and $\ts(x)$ to $x + 1$.

\subsection{Disadvantage of Dependency Tuples Compared to Direct Application of Polynomials}

The following PTRS describes a random walk that increases or decreases the
number of $\tg$s in a term by $2$ with a chance of $\nicefrac{1}{2}$. 
Hence, this PTRS is AST as it corresponds to a fair random walk.

\vspace*{-.5cm}

{\small
\begin{align*}
  \tg(\tg(x)) & \!\to\!  \{ \nicefrac{1}{2} : \tg(\tg(\tg(\tg(x)))), \nicefrac{1}{2} : x \}
\end{align*}
}

The direct application of polynomial interpretations  in \Cref{theorem:ptrs-direct-application-poly-interpretations}
can prove its almost-sure termination using the polynomial interpretation that maps $\tg(x)$ to the polynomial $x+1$.

However, currently our implementation of the DP framework fails in proving iAST.
We would first create the following DP tuple for the only rule:

\vspace*{-.3cm}

{\small
  \begin{align*}
    \langle \tG(\tg(x)),\tg^2(x) \rangle
    & \!\to\!  \{ \nicefrac{1}{2} : \langle \Com{4}(\tG(\tg^3(x)), \tG(
    \tg^2(x)),   \tG(\tg(x)), \tG(x)), \tg^4(x) \rangle, 
    \nicefrac{1}{2} : \langle \Com{0}, x \rangle \}
    \end{align*}
}

Now the terms $\tG(\tg^2(x))$ and $\tG(x)$ are not usable and can be removed by the usable terms processor.
Thus, we result in the following DP tuple:

\vspace*{-.3cm}

{\small
  \begin{align*}
    \langle \tG(\tg(x)),\tg^2(x) \rangle
    & \!\to\!  \{ \nicefrac{1}{2} : \langle \Com{2}(\tG(\tg^3(x)),   \tG(\tg(x))),  \tg^4(x)  \rangle, 
    \nicefrac{1}{2} : \langle \Com{0}, x \rangle \}
    \end{align*}
}

At this point we cannot apply any processor except the reduction pair processor.
It can remove this dependency tuple with the interpretation that maps $\tG(x)$ to $1$ and $\tg(x)$ to $x+1$.
But if $\tG(\tg^2(x))$ and $\tG(x)$ are kept (as with the approximation of usable terms used in our implementation), the reduction pair processor is unable to find an interpretation for the first
dependency tuple above.
The reason is that we have to take the sum of the polynomial values of the terms
$\tG(\tg^3(x))$, $\tG(\tg^2(x))$, $\tG(\tg(x))$, and $\tG(x)$, since they all occur as arguments of
the same compound symbol.
In contrast, with the direct application of polynomials, we only have to regard the
polynomial value of $\tg^4(x)$.

Note that the alternative idea of interpreting compound symbols by the maximum instead of
the sum would not be sound in the probabilistic setting, since in this way one cannot
distinguish anymore between examples like $\R_1$ and $\R_2$ in \Cref{Dependency Tuples and Chains for Probabilistic Term Rewriting}, where $\R_1$ is
AST and $\R_2$ is not AST. 

For that reason, our implementation uses the DP framework first and if it cannot prove
iAST, we apply polynomial interpretations directly via \Cref{theorem:ptrs-direct-application-poly-interpretations}.
In future work, we plan to integrate the direct approach into the DP framework, e.g.,  by adapting
the rule removal processor of the non-probabilistic DP framework \cite[Thm.\ 7]{gieslLPAR04dpframework}.

\subsection{Incompleteness of the Probabilistic Chain Criterion}\label{Incompleteness of the Probabilistic Chain Criterion}

The following PTRS $\R_{\mathsf{incompl}}$
shows that the chain criterion of \Cref{theorem:prob-chain-criterion} is not complete.

\vspace*{-.7cm}

{\small
\begin{align*}
  \tg & \!\to\!  \{ \nicefrac{5}{8} : \tf(\tg), \nicefrac{3}{8} : \tstop \}\\
  \tg & \!\to\! \{ 1 : \tb\}\\
  \tf(\tb) & \!\to\! \{ 1 : \tg\}
\end{align*}
}

\textsf{AProVE} can prove AST via \Cref{theorem:ptrs-direct-application-poly-interpretations} with
the polynomial interpretation that maps $\tf(x)$ to $x + 2$, $\tg$ to $4$, $\tb$ to $3$, and $\tstop$ to $0$.
On the other hand, the dependency tuples for this PTRS are:

\vspace*{-.3cm}

{\small
\begin{align*}
  \langle \tG, \tg \rangle & \!\to\!  \{ \nicefrac{5}{8} : \langle \Com{2}(\tF(\tg), \tG),
  \tf(\tg) \rangle, \nicefrac{3}{8} : \langle \Com{0}, \tstop \rangle \}\\
   \langle \tG, \tg \rangle & \!\to\! \{ 1 : \langle \Com{0}, \tb \rangle\}\\
    \langle \tF(\tb),  \tf(\tb) \rangle  & \!\to\! \{ 1 : \langle \Com{1}(\tG), \tg
    \rangle \}
\end{align*}
}

The DP problem $(\DTuple{\R_{\mathsf{incompl}}},\R_{\mathsf{incompl}})$ is not iAST.
To see this, consider the following rewrite sequence.
	\[
	\begin{array}{rll}
\{ 1 :  \Com{1}(\tG) \} 
&	\setiliftingDRRincompl &  \{ \nicefrac{5}{8} :  \Com{2}(\tF(\tg), \tG),
\nicefrac{3}{8} :  \Com{0} \} \\
&	\setiliftingRincompl &  \{ \nicefrac{5}{8} :  \Com{2}(\tF(\tb), \tG), \nicefrac{3}{8} :  \Com{0} \}\\
&	\setiliftingDRRincompl &  \{ \nicefrac{5}{8} :  \Com{2}(\tG, \tG), \nicefrac{3}{8} :  \Com{0} \}
		\end{array}	
	\]

        Using these combinations of steps with the first and the third
        dependency tuple  and the  rewrite rule   $\tg \to \{ 1 :
        \tb\}$, 
we essentially end up with a biased random walk where the number of $\tg$s is increased
with probability $\nicefrac{5}{8}$ and decreased with probability $\nicefrac{3}{8}$.
Hence,  $(\DTuple{\R_{\mathsf{incompl}}},\R_{\mathsf{incompl}})$ is not iAST.
The problem here is that the terms $\tG$ and $\tF(\tg)$ can both be rewritten to $\tG$. 
When creating the dependency tuples, we lose the information that the $\tg$ inside the
term $\tF(\tg)$
corresponds to the second argument $\tG$ of the compound symbol $\Com{2}$.
In order to obtain a complete chain criterion, one has to extend dependency tuples by positions. Then
the second argument $\tG$ of the compound symbol $\Com{2}$ would be augmented 
by the position of $\tg$ in the right-hand side of the first rule, because this rule
was used to generate the dependency tuple.

As mentioned at the end of \Cref{Dependency Tuples and Chains for Probabilistic Term
  Rewriting}, we also developed a refinement where our chain 
criterion is made complete. However, the current DP processors do not yet exploit this
refinement. Therefore, 
this is an example that can only be handled by the direct technique of
\Cref{theorem:ptrs-direct-application-poly-interpretations}, but not by the current
DP framework.

\subsection{The Limits of the DP Framework}\label{The Limits of the DP Framework}

The following PTRS $\R_{\mathsf{bogo}}$ describes an implementation of bogosort.

\vspace*{-.3cm}

{\small
\begin{align*}
  \tbogo(\xs) & \!\to\! \{ 1 : \tbogohelp(\tsortr(\xs)) \}\\
  \tbogohelp(\xs) & \!\to\! \{ 1 : \tifsorted(\xs,\xs)\}\\
  \tsortr(\tnil) & \!\to\! \{ 1 : \tnil\}\\
  \tsortr(\tcons(x,\tnil)) & \!\to\! \{ 1 : \tcons(x,\tnil)\}\\
  \tsortr(\tcons(x,\tcons(y,\xs))) & \!\to\! \{ \nicefrac{1}{2} :
  \tcons(x,\tsortr(\tcons(y,\xs))),\\
  & \phantom{\!\to\! \{} \;\nicefrac{1}{2} : \tcons(y,\tsortr(\tcons(x,\xs)))\}\\
  \tifsorted(\tnil,\zs)  & \!\to\! \{ 1 : \zs\}\\
  \tifsorted(\tcons(x,\tnil),\zs)
  & \!\to\! \{ 1 : \zs\}\\
  \tifsorted(\tcons(x,\tcons(y,\ys)),\zs) & \!\to\! \{ 1 : \tifleq(\tleq(x,y),\tcons(y,\ys),\zs)\}\\
  \tifleq(\ttrue,\tcons(y,\xs),\zs) & \!\to\! \{ 1 : \tifsorted(\tcons(y,\xs),\zs)\}\\
  \tifleq(\tfalse,\tcons(y,\xs),\zs) & \!\to\! \{ 1 : \tbogo(\zs)\}\\
  \tleq(\O,x) & \!\to\! \{ 1 : \ttrue\}\\
  \tleq(\ts(x),\O)  & \!\to\! \{ 1 : \tfalse\}\\
  \tleq(\ts(x),\ts(y))  & \!\to\! \{ 1 : \tleq(x,y)\}\!
\end{align*}
}

Here, we randomly shuffle the list until we eventually reach a sorted list and stop.
This PTRS is AST as in every recursive call of the main function $\tbogo$, we use the function $\tsortr$ that randomly switches some neighboring elements of the list.
Hence, there is a certain chance to move towards a ``more sorted'' list so that we eventually stop the algorithm.
However, in order to prove this via the reduction pair processor or via
\Cref{theorem:ptrs-direct-application-poly-interpretations}, we would have to find an ordering that distinguishes between lists that are ``more sorted'' than others.
Polynomial interpretations are not able to express this comparison by ``more sortedness'' and this also holds for most other typical orderings on terms that are amenable to automation.

\subsection{Comparing \Cref{theorem:ptrs-direct-application-poly-interpretations} with the Criterion for PAST \cite{avanzini2020probabilistic}}

The following PTRS shows that there are examples where the direct application of
polynomial interpretations in \cite{avanzini2020probabilistic} can prove PAST, whereas our
direct application of polynomial interpretations in \Cref{theorem:ptrs-direct-application-poly-interpretations} to prove AST fails. 

\begin{align}
\tdecreaseX(\O, y)     &\to  \{ 1: \tloopGuard(\O, y) \} \nonumber\\
 \tdecreaseX(\ts(x), y)  &\to \{ 1: \tloopGuard(x, y) \}\nonumber\\
 \tloop(x, y)          &\to \{\nicefrac{1}{2} : \tdecreaseX(x, y), \nicefrac{1}{2} :
 \tdecreaseY(x, y)  \}\label{loopRule}\\
 \tloopGuard(\ts(x), y)  &\to \{1 : \tloop(\ts(x), y)\} \nonumber\\
 \tloopGuard(\O, \O)     &\to \{ 1 : \tstop \} \nonumber\\
 \tdecreaseY(x, \O)     &\to \{ 1 : \tloopGuard(x, \O) \} \nonumber\\
 \tdecreaseY(x, \ts(y))  &\to \{ 1 : \tloopGuard(x, y)\} \nonumber\\
 \tloopGuard(x, \ts(y))  &\to \{ 1 : \tloop(x, \ts(y)) \}\nonumber
\end{align}

The PAST criterion of
\cite{avanzini2020probabilistic} finds the following polynomial interpretation that proves
PAST:
\begin{align*}
\Pol(\O) &= 0\\
    \Pol(\tdecreaseX(x_1, x_2)) &= 2 + x_1 + 2\cdot x_1\cdot x_2 + 3\cdot x_2\\
    \Pol(\tdecreaseY(x_1, x_2)) &= 2 + 3\cdot x_1 + x_1\cdot x_2 + x_2\\
    \Pol(\tloop(x_1, x_2)) &= 4 + 2\cdot x_1 + 2\cdot x_1\cdot x_2 + 2\cdot x_2\\
    \Pol(\tloopGuard(x_1, x_2)) &= 1 + 3\cdot x_1 + 3\cdot x_1\cdot x_2 + 3\cdot x_2\\
    \Pol(\ts(x_1)) &= 4 + 3\cdot x_1\\
    \Pol(\tstop) &= 0
    \end{align*}

\noindent
In Rule \eqref{loopRule},  the expected value is indeed strictly decreasing:
\[ \begin{array}{cl}
& 4 + 2\cdot x + 2\cdot x\cdot y + 2\cdot y \\
 > &
\nicefrac{1}{2} \cdot (2 + x + 2\cdot x\cdot y + 3\cdot y) + \nicefrac{1}{2} \cdot (2 +
3\cdot x + x\cdot y + y)\\
=& 2 + 2\cdot x + \nicefrac{3}{2} \cdot x\cdot y + 2\cdot y
\end{array}
\]

However, our criterion of \Cref{theorem:ptrs-direct-application-poly-interpretations}
cannot be applied with this polynomial interpretation. We have neither a strict decrease
from
$\tloop(x, y)$ to $\tdecreaseX(x,y)$ (since $4 + 2 \cdot x + 2 \cdot x \cdot y + 2 \cdot y$ is not greater than
$2 + x + 2 \cdot x \cdot y + 3 \cdot y$) nor from $\tloop(x, y)$ to $\tdecreaseY(x,y)$
(since $4 + 2 \cdot x + 2 \cdot x \cdot y + 2 \cdot y$ is not greater than
$2 + 3 \cdot x + x \cdot y + y$). Indeed, the implementation of our AST criterion from
\Cref{theorem:ptrs-direct-application-poly-interpretations}
fails for this example, as it does not find any polynomial interpretation satisfying its
constraints. However, our implementation can prove iAST using the probabilistic DP framework.

\section{Proofs}\label{appendix}

In this section, we prove all of our contributions and observations in the paper.

\begin{lemma}[Difficulty of $\R_{\tdiv}$]\label{Difficulty of Rdiv}
    Termination of $\R_{\tdiv}$ from \Cref{example:TRS-int-div} cannot be proved using \Cref{direct-application-poly-interpretations}.
\end{lemma}

\begin{myproof}
 Every  monotonic polynomial interpretation over the naturals yields a quasi-simplification
ordering that satisfies the weak subterm property, i.e., if $t'$ is a subterm of $t$ then we have $\Pol(t) \geq \Pol(t')$. This
implies that we cannot have $\Pol(\tdiv(\ts(x),\ts(y))) >
\Pol(\ts(\tdiv(\tminus(x,y),\ts(y))))$ for the fourth rule \eqref{eq:ptrs-div-4}. The
reason is that otherwise, when instantiating $y$ with $\ts(x)$, we would get
 $\Pol(\tdiv(\ts(x),\ts^2(x))) >
\Pol(\ts(\tdiv(\tminus(x,\ts(x)),\ts^2(x))))$. But the weak subterm property implies
$\Pol(\tminus(x,\ts(x))) \geq \Pol(\ts(x))$ and by monotonicity, we obtain
$\Pol(\ts(\tdiv(\tminus(x,\ts(x)),\ts^2(x)))) \geq
\Pol(\tdiv(\ts(x),\ts^2(x)))$, which is a contradiction to the irreflexivity of $>$.
\end{myproof}

\ASTPolInt*

\begin{myproof}
 Let $\mu = (\mu_n)_{n \in \IN}$ be an arbitrary infinite rewrite sequence.
	The main idea for the proof comes from~\cite{mciver2017new}, but we extend their
        approach from while-programs on integers to terms. The core steps of the proof are the following:
	\begin{enumerate}
		\item[1.] We extend the conditions (1) and (2) to rewrite steps instead of just rules.
		\item[2.] We create a rewrite sequence $(\mu_n^{\leq N})_{n \in \IN}$ for any $N \in \IN$.
		\item[3.] We prove that $\lim_{n \to \infty} |\mu_n^{\leq N}|_{\R} \geq p_{min}^{N}$ for any $N \in \IN$.
		\item[4.] We prove that $\lim_{n \to \infty} |\mu_n^{\leq N}|_{\R}=1$ for any $N \in \IN$.
		\item[5.] Finally, we prove that $\lim_{n \to \infty} |\mu_n|_{\R}=1$
	\end{enumerate}
	Here, let $p_{min}>0$ be the minimal probability that occurs in the rules of $\R$.
	As $\R$ has only finitely many rules and all occurring multi-distributions are finite, this minimum is well defined.
    Note that the $N$ in $p_{min}^{N}$ denotes the exponent and is not an index, i.e., we have $p_{min}^{N} = \underbrace{p_{min} \cdot \ldots \cdot p_{min}}_\text{$N$ times}$.

	\medskip
        
  \noindent
	\textbf{\underline{1. We extend (1) and (2) to rewrite steps instead of just rules}}

  \noindent
	We first show that the conditions (1) and (2) of the theorem also hold for rewrite steps instead of just rules. 
	Let $s \in \TSet{\Sigma}{\VSet}$ and $\{p_1:t_1, \ldots, p_k:t_k\} \in \FDist(\TSet{\Sigma}{\VSet})$ with $s \to_{\R} \{p_1:t_1, \ldots, p_k:t_k\}$.
	We want to show that:
        
	\begin{enumerate}
		\item[(a)] There exists a $1 \leq j \leq k$ with $Pol(s) > Pol(t_j)$.
		\item[(b)] We have $Pol(s) \geq \sum_{1 \leq j \leq k}p_j \cdot \Pol(t_j)$.
	\end{enumerate}

        By the definition of rewriting, there exists a rule $\ell \to \{p_1:r_1, \ldots, p_k:r_k\} \in \R$, a substitution $\sigma$, and a position $\pi$ of $s$ such that $s|_\pi =\ell\sigma$ and $t_j = s[r_j \sigma]_\pi$ for all $1 \leq j \leq k$.

	\begin{enumerate}
		\item[(a)] We perform structural induction on $\pi$.
		      So in the induction base, let $\pi = \epsilon$.
		      Hence, we have $s = \ell\sigma$ and $t_j = r_j \sigma$ for all $1 \leq j \leq k$.
		      By Requirement (1), there is a $1 \leq j \leq k$ with
                      $\Pol(\ell)> \Pol(r_j)$.
		      As these inequations hold for all instantiations of the occurring variables, for $t_j = r_j\sigma$ we have
		      \[ 
		        \Pol(s) = \Pol(\ell\sigma) > \Pol(r_j\sigma) = \Pol(t_j).
		      \]

		      In the induction step, we have $\pi = i.\pi'$, $s = f(s_1,\ldots,s_i,\ldots,s_n)$, $s_i \to_{\R} \{p_1:t_{i,1}, \ldots, p_k:t_{i,k}\}$, and $t_j = f(s_1,\ldots,t_{i,j},\ldots,s_n)$ with $t_{i,j} = s_i[r_j \sigma]_{\pi'}$ for all $1 \leq j \leq k$.
		      Then by the induction hypothesis there is a $1 \leq j \leq k$ with
                      $\Pol(s_i) > \Pol(t_{i,j})$.
		      For $t_j = f(s_1,\ldots,t_{i,j},\ldots,s_n)$ we obtain
		      \[
			      \begin{array}{lcl}
				      \Pol(s) & = & \Pol(f(s_1,\ldots,s_i,\ldots,s_n)) \\
				              & = & f_{\Pol}(\Pol(s_1),\ldots,\Pol(s_i),\ldots,\Pol(s_n)) \\
				              & > & f_{\Pol}(\Pol(s_1),\ldots,\Pol(t_{i,j}),\ldots,\Pol(s_n))\\
							  &  & \hspace*{1cm} \text{(by monotonicity of $f_{\Pol}$ and $Pol(s_i) > Pol(t_{i,j})$)} \\
				              & = & \Pol(f(s_1,\ldots,t_{i,j},\ldots,s_n)) \\
				              & = & \Pol(t_j).
			      \end{array}
		      \]

		\item[(b)] We again perform structural induction on $\pi$.
		      So in the induction base $\pi = \epsilon$ we again have $s = \ell\sigma$ and $t_j = r_j \sigma$ for all $1 \leq j \leq k$.
		     By (2), we know that $\Pol(\ell) \geq \sum_{1 \leq j \leq k}p_j \cdot \Pol(r_j)$.
		      This inequation must hold for all instantiations of the occurring variables.
		      Thus, we obtain
		      \[
			      \Pol(s) \;=\; \Pol(\ell\sigma) \;\geq\;\sum_{1 \leq j \leq k}p_j \cdot \Pol(r_j \sigma) \;=\; \sum_{1 \leq j \leq k}p_j \cdot \Pol(t_j). 
			  \]

		      In the induction step, we have $\pi = i.\pi'$, $s = f(s_1,\ldots,s_i,\ldots,s_n)$, $s_i \to_{\R} \{p_1:t_{i,1}, \ldots, p_k:t_{i,k}\}$, and $t_j = f(s_1,\ldots,t_{i,j},\ldots,s_n)$ with $t_{i,j} = s_i[r_j \sigma]_{\pi'}$ for all $1 \leq j \leq k$.
		      Then by the induction hypothesis we have $\Pol(s_i) \geq \sum_{1 \leq j \leq k} p_j \cdot \Pol(t_{i,j})$.
			  Thus, we obtain
		      \[
			      \begin{array}{lcl}
				      \Pol(s) & = & \Pol(f(s_1,\ldots,s_i,\ldots,s_n)) \\
				              & = & f_{\Pol}(\Pol(s_1),\ldots,\Pol(s_i),\ldots,\Pol(s_n)) \\
				              & \geq & f_{\Pol}(\Pol(s_1),\ldots,\sum_{1 \leq j \leq k} p_j \cdot \Pol(t_{i,j}),\ldots,\Pol(s_n)) \\
							  &  & \hspace*{1cm}
                                      \text{(by weak monotonicity of $f_{\Pol}$}\\
                                      	  &  & \hspace*{1cm} \text{and $\Pol(s_i) \geq \sum_{1 \leq j \leq k} p_j \cdot \Pol(t_{i,j})$)} \\
				              & = & \sum_{1 \leq j \leq k} p_j \cdot f_{\Pol}(\Pol(s_1),\ldots,\Pol(t_{i,j}),\ldots,\Pol(s_n)),\\
							  &  & \hspace*{1cm} \text{(as $f_{\Pol}$ is multilinear)}\\
				              & = & \sum_{1 \leq j \leq k} p_j \cdot \Pol(f(s_1,\ldots,t_{i,j},\ldots,s_n)) \\
				              & = & \sum_{1 \leq j \leq k} p_j \cdot \Pol(t_{j}).
			      \end{array}
		      \]
		    	\end{enumerate}

	\medskip
\pagebreak[2]
        
  \noindent 
	\textbf{\underline{2. We create a rewrite sequence $(\mu_n^{\leq N})_{n \in \IN}$ for any $N \in \IN$}}

  \noindent
  We define the \emph{value} $V(t)$ of any $t \in \TSet{\Sigma}{\VSet}$ by 
    \[
        V: \TSet{\Sigma}{\VSet} \to \IN, \quad t \mapsto
        \begin{cases}
        \Pol_{0}(t) + 1, & \text{ if } t\in \TSet{\Sigma}{\VSet} \setminus \mathtt{NF}_{\R} \\
        0,             & \text{ if } t \in \mathtt{NF}_{\R}
        \end{cases}
    \]
    Here, $\Pol_{0}$ denotes the function $\TSet{\Sigma}{\VSet} \to \IN$, $t \mapsto \Pol(t)(0,\ldots,0)$, i.e., we use the polynomial interpretation of $t$ and instantiate all variables with $0$.
    Furthermore, we define the value of any $\mu \in \FDist(\TSet{\Sigma}{\VSet})$ by
    \[
        V: \FDist(\TSet{\Sigma}{\VSet}) \to \IR_{\geq 0}, \quad \mu \mapsto \sum_{(p:t) \in \mu}p \cdot V(t)
    \]
    Note that we indeed have $V(t) \in \IN$ for any $t \in \TSet{\Sigma}{\VSet}$ and that we have $V(t) = 0$ if and only if $t$ is a normal form w.r.t.\ $\R$.
  For any rewrite sequence $(\mu_n)_{n \in \mathbb{N}}$ and any $N \in \IN$, we now consider a modified sequence $(\mu^{\leq N}_n)_{n \in \mathbb{N}}$, where we inductively replace each $\mu_n$ by $\mu^{\leq N}_n$ starting from $\mu_0$. 
  To obtain $\mu^{\leq N}_n$, we replace $(p:t) \in \mu_n$ by $(p:\top)$ if $V(t) \geq N + 1$. 
  Otherwise, we keep $(p:t)$.
  Here, $\top$ is a newly introduced function symbol with $V(\top)=N+1$.
  If at least one term is replaced in a $\mu_n$, we replace $(\mu_m)_{m > n}$ by a new sequence before proceeding; starting from $\mu^{\leq N}_n$, we use the same rewrite steps as the original rewrite sequence, except for the new normal forms $\top$, which stay the same. 
  Of course, the result is not a valid $\rightrightarrows_{\R}$-rewrite sequence anymore.
    
  Our goal is to prove that in
  every valid $\rightrightarrows_{\R}$-rewrite sequence $(\mu_n)_{n \in \mathbb{N}}$, the probability for normal forms converges towards one, i.e., that $\lim_{n \to \infty} |\mu_n|_{\R} = 1$.
	The proof now proceeds similarly to the approach in~\cite{mciver2017new}.
	First, we prove that $\lim_{n \to \infty} |\mu^{\leq N}_n|_{\R}=1$ for any $N \in \IN$, i.e., we prove that this new kind of sequence converges with probability one.
	Note that we now also consider $\top \in \mathtt{NF}_{\R}$. 
	However, this does not yet prove that $\lim_{n \to \infty} |\mu_n|_{\R}=1$, which we will show afterwards.

	\medskip
       
  \noindent 
	\textbf{\underline{3. We prove that $\lim_{n \to \infty} |\mu_n^{\leq N}|_{\R} \geq p_{min}^{N}$ for any $N \in \IN$}}

  \noindent
	Due to (a), for every term $s$ that is not in normal form, there
	is a rewrite step $s \to_{\R} \mu$ such that there is a $(p' : t) \in \mu$ with $p' \geq p_{min}$
     and $\Pol(s) > \Pol(t)$ (and thus, also $\Pol_0(s) > \Pol_0(t)$, and hence $V(s) \geq V(t) + 1$) for some $t \in \TSet{\Sigma}{\VSet}$.
	So for any $N \in \IN$ and any $\mu_0$, we have $|\mu^{\leq N}_{N}|_{\R} \ge
        p_{min}^{N}$. The reason is that for every term in $\Supp(\mu^{\leq N}_n)$ that is not in normal form, there is always a term in $\Supp(\mu^{\leq N}_{n+1})$ which decreases the value of $V$ by at least $1$. 
	So for all terms in $\Supp(\mu_0^{\leq N})$ that are not in normal form, we reach a normal form in at most $N$ steps with at least probability $p_{min}^{N}$.
        
	\medskip
        
  \noindent 
	\textbf{\underline{4. We prove that $\lim_{n \to \infty} |\mu_n^{\leq N}|_{\R} = 1$ for any $N \in \IN$}}

  \noindent 
	Note that $|\mu^{\leq N}_n|_{\R}$ is bounded and weakly monotonically increasing, so that $\lim_{n \to \infty}|\mu^{\leq N}_n|_{\R}$ must exist and we have $\lim_{n \to \infty}|\mu^{\leq N}_n|_{\R} \ge |\mu^{\leq N}_{N}|_{\R}$.
	Hence, for any $N \in \IN$, we have
	\begin{equation}
        \label{pNstar}
	    p^{\star}_{N}:=\inf_{(\mu_n)_{n \in \IN} \text{ is a rewrite sequence}} (\lim_{n \to \infty}|\mu^{\leq N}_{n}|_{\R}) \geq p_{min}^{N} >0
	\end{equation}
	We now prove by contradiction that this is enough to ensure $p^{\star}_N = 1$.
	So assume that $p^{\star}_N < 1$. 
	Then we define $\epsilon := \frac{p^{\star}_N \cdot (1-p^{\star}_N)}{2}>0$. 
	By definition of the infimum, $p^{\star}_N + \epsilon$ is not a lower bound of $\lim_{n \to \infty}|\mu_n^{\leq N}|_{\R}$ for all rewrite sequences.
	Hence, there must exist a rewrite sequence $(\mu_n)_{n \in \IN}$ such that
	\begin{equation} \label{pstarEpsilon}
        p^{\star}_N\leq \lim_{n \to \infty}|\mu^{\leq N}_{n}|_{\R} < p^{\star}_N + \epsilon.
    \end{equation}
	By the monotonicity of $|\cdot|$ w.r.t.\ $\rightrightarrows_{\R}$ steps, there must exist a natural number $m^{\star} \in \IN$ such that
	\begin{equation} \label{p star half}
	    |\mu^{\leq N}_{m^*}|_{\R}> \tfrac{p^{\star}_N}{2}.
	\end{equation}
  Then we have 
	\begin{align} \label{lim1t}
		\lim_{n \to \infty}|\mu^{\leq N}_n|_{\R} = |\mu_{m^*}^{\leq N}|_{\R} + \sum_{(p:t) \in \mu_{m^*}^{\leq N},  t \notin \mathtt{NF}_{\R}} p \cdot \lim_{n \to \infty} | \{1:t\}^{\leq N}_n |_{\R},
	\end{align}
	where $(\{1:t\}^{\leq N}_n)_{n \in \IN}$ denotes the sequence starting with $\{1:t\}$,
        where the same rules are applied as for $t$ in $(\mu^{\leq N}_{n + m^*})_{n \in \IN}$.
	Furthermore, we have
	\begin{equation} \label{1 - observation}
	    \sum_{(p:t) \in \mu_{m^*}^{\leq N}, t \notin \mathtt{NF}_{\R}} p = 1 - |\mu_{m^*}^{\leq N}|_{\R}
	\end{equation}
	So we obtain
	\begin{align*}
		& p^{\star}_N+\epsilon 
		\\ 
		{}>{}    & \lim_{n \to \infty} |\mu_n^{\leq N}|_{\R} \qquad \text{(by \eqref{pstarEpsilon})}
		\\
		{}={}    & |\mu_{m^*}^{\leq N}|_{\R} + \sum_{(p:t) \in \mu_{m^*}^{\leq N}, t
                  \notin \mathtt{NF}_{\R}} p \cdot                
		\underbrace{\lim_{n \to \infty} | \{1:t\}^{\leq N}_n |_{\R}}_{\geq \,
                  p^{\star}_N} \qquad \text{(by \eqref{lim1t} and \eqref{pNstar})}
		\\
		{}\geq{}    & |\mu_{m^*}^{\leq N}|_{\R} + \sum_{(p:t) \in \mu_{m^*}^{\leq N}, t
                  \notin \mathtt{NF}_{\R}} p \cdot p^{\star}_N
		\\
		{}={} &|\mu_{m^*}^{\leq N}|_{\R} + p^{\star}_N \cdot \sum_{(p:t) \in \mu_{m^*}^{\le
                    N}, t \notin \mathtt{NF}_{\R}} p
		\\ 
		{}={}    & |\mu_{m^*}^{\leq N}|_{\R} + p^{\star}_N \cdot (1 - |\mu_{m^*}^{\leq N}|_{\R})  \qquad	\text{(by \eqref{1 - observation})}
		\\   
		{}={}    & p^{\star}_N + |\mu_{m^*}^{\leq N}|_{\R} - p^{\star}_N \cdot |\mu_{m^*}^{\leq N}|_{\R}
		\\       
		{}={}    & p^{\star}_N + |\mu_{m^*}^{\leq N}|_{\R} \cdot (1-p^{\star}_N)
		\\       
		{}>{}    & p^{\star}_N + \left(1-p^{\star}_N\right)\cdot
		\tfrac{p^{\star}_N}{2} \qquad
		\text{(by \eqref{p star half})}\\
		{}={}    & p^{\star}_N + \epsilon, \quad \textcolor{red}{\lightning}
	\end{align*}
	a contradiction.
	So $p^{\star}_N=1$. 
	In particular, this means that for every $N \in \IN$ and every rewrite sequence $(\mu_n)_{n \in \IN}$, we have
    \begin{equation} \label{limitMuLeqNIs1}
        \lim_{n \to \infty}  |\mu^{\leq N}_n|_{\R} = 1.
    \end{equation}
	
	\medskip

\pagebreak[2]
        
  \noindent  
	\textbf{\underline{5. We prove that $\lim_{n \to \infty} |\mu_n|_{\R} = 1$}}

  \noindent 
  Finally, we have to prove that $\lim_{n \to \infty} |\mu_n|_{\R} = 1$ holds as well,
  i.e., that in
  every rewrite sequence, the probability for normal forms converges towards one.
  By (b), we have that $\mu \rightrightarrows_{\R} \mu'$ implies $\Pol(\mu) \ge \Pol(\mu')$ and therefore, $V(\mu) \geq V(\mu')$. 
  While our modified sequence $(\mu^{\leq N}_{n})_{n \in \IN}$ is not a rewrite sequence anymore, we still have $V(\mu^{\leq N}_n) \geq V(\mu^{\leq N}_{n+1})$ because either this is a valid $\rightrightarrows_{\R}$-step or a term with a larger value is replaced by a smaller or equal one, i.e., $\top$. 
	Additionally, $V(\mu^{\leq N}_n)$ is bounded from below by $0$, so that $\lim_{n \to \infty}(V(\mu_n^{\leq N}))$ exists and we have 
  \begin{equation} \label{VmuLeqN}
    V(\mu^{\leq N}_0) \geq \lim_{n \to \infty}(V(\mu_n^{\leq N})).
  \end{equation}
    
  Now we fix $N \in \IN$ and a rewrite sequence $(\mu_n)_{n \in \IN}$, and obtain the corresponding transformed sequence $(\mu^{\leq N}_n)_{n \in \IN}$.
  Note that by \eqref{limitMuLeqNIs1} we have $\lim_{n \to \infty} |\mu_n^{\leq N}|_{\R} = 1
  = q_N + q_N'$, where
  \[q_N = \lim_{n \to \infty} \sum_{(p:t) \in \mu_n^{\leq N}, t \in
    \mathtt{NF}_{\R}, t \ne \top} p \qquad
  \text{ and } \qquad q_N' = \lim_{n \to \infty} \sum_{(p:\top) \in \mu_n^{\leq N}} p.\]
  Hence, $q_N' = 1- q_N$.  
	Now we can determine $\lim_{n \to \infty}(V(\mu_n^{\leq N}))$.
	The probabilities of zero entries (i.e., terms $t \in
        \mathtt{NF}_{\R}\setminus\{\top\}$ where $V(t) = 0$) add up to $q_N$, while the
        probabilities for entries $\top$
        with value $V(\top) = N+1$ add up to probability $q_N' = 1-q_N$. 
	So $\lim_{n \to \infty}(V(\mu_n^{\leq N})) = q_N \cdot 0 + (1-q_N)\cdot (N+1)= (1-q_N)\cdot (N+1)$.
  Thus by \eqref{VmuLeqN},
  \[
    V(\mu_0) \geq V(\mu_0^{\leq N}) \geq \lim_{n \to \infty}(V(\mu_n^{\leq N})) = (1-q_N) \cdot (N+1),
  \]
  which implies $q_N \geq 1-\tfrac{V(\mu_0)}{N+1}$.
  
  Note that $q_N$ is weakly monotonically increasing and bounded from above by 1 for $N \to
  \infty$. 
  Hence, $q := \lim_{N \to \infty} q_N$ exists and $1 \geq q \geq \lim_{N \to \infty} (1-\tfrac{V(\mu_0)}{N+1}) = 1$, i.e., $q = 1$. 
  Hence, we obtain $\lim_{n \to \infty}  |\mu_n|_{\R} = \lim_{N \to \infty} q_N = q = 1$.
  So the PTRS $\R$ is AST.
\end{myproof}

\begin{lemma}[Difficulty of $\R_{\tpdiv}$]\label{Difficulty of Rpdiv}
  Proving that $\R_{\tpdiv}$ from
  \Cref{example:ptrs-int-div-coupled-positional-dependency-pairs} is iAST
    cannot be done using \Cref{theorem:ptrs-direct-application-poly-interpretations}.
\end{lemma}

\begin{myproof}
As shown in \Cref{Difficulty of Rdiv}, we cannot have \[\Pol(\tdiv(\ts(x),\ts(y))) >
\Pol(\ts(\tdiv(\tminus(x,y),\ts(y))))\] for the fourth rule \eqref{eq:ptrs-div-4}. 
Moreover, by irreflexivity of $>$, we also
cannot have \[\Pol(\tdiv(\ts(x),\ts(y))) > \Pol(\tdiv(\ts(x),\ts(y))).\]
  Hence, there exists no monotonic polynomial interpretation that
  satisfies the conditions of \Cref{theorem:ptrs-direct-application-poly-interpretations}
  for the direct application of polynomial interpretations to prove AST (and therefore,
  also iAST).
\end{myproof}

\begin{lemma}[Counterexample for Completeness of the Chain Criterion]
	There exists a PTRS $\R$ such that $\R$ is iAST (and even AST), but $(\DTuple{\R},\R)$ is not iAST\@.
\end{lemma}

\begin{myproof}
  Such an example is contained in our benchmark collection, see \Cref{Incompleteness of the
    Probabilistic Chain Criterion}.
\end{myproof}

For the proof of the chain criterion, we want to regard $\irightrightarrowsR$-rewrite sequences of a PTRS $\R$ as a tree, similar to the chain trees for a DP problem $(\PP, \SSS)$.

\begin{definition}[Rewrite Sequence Tree] \label{def:rewrite-sequence-tree}
  Let $\R$ be a PTRS\@.
  $\F{T}\!=\!(V,E,L)$ {\normalsize is an (innermost)} $\R${\normalsize\emph{-rewrite sequence tree}} if

 	\begin{enumerate}
  \item $V \neq \emptyset$ is a possibly infinite set of nodes and $E \subseteq V \times V$ is a set of directed edges, such that $(V, E)$ is a (possibly infinite) directed tree where $vE = \{ w \mid (v,w) \in E \}$ is finite for every $v \in V$.
  \item  $L : V \rightarrow (0,1] \times \TSet{\Sigma}{\VSet}$ labels every node $v$ by a probability $p_v$ and a term $t_v$.
  For the root $v \in V$ of the tree, we have $p_v = 1$.
  \item For all  $v \in V$: If $vE = \{w_1, \ldots, w_k\}$, then $t_v \itor \{\tfrac{p_{w_1}}{p_v}:t_{w_1}, \ldots, \tfrac{p_{w_k}}{p_v}:t_{w_k}\}$.
  \end{enumerate}
\end{definition}

When it is not clear about which rewrite sequence tree (or chain tree) we are talking, we will always explicitly indicate the tree. 
For instance, for the probability $p_v$ of the node $v \in V$ of some rewrite sequence tree $\F{T} = (V,E,L)$, we may also write $p_v^{\F{T}}$.

\begin{definition}[$|\F{T}|_{\ctleaf}$] \label{def:rewrite-sequence-tree-convergence-notation}
  Let $\R$ be a PTRS.
	For any innermost $\R$-rewrite sequence tree $\F{T}$ we again define $|\F{T}|_{\ctleaf} = \sum_{v \in \ctleaf} p_v$.
\end{definition}

We say that a rewrite sequence tree (or chain tree) $\F{T}$ \emph{converges with
probability} $p \in \IR$ if we have $|\F{T}|_{\ctleaf} = p$. Similarly, for a PTRS $\R$ we
say that an $\R$-rewrite sequence $(\mu_n)_{n \in \IN}$  \emph{converges with
probability} $p$ if $\lim_{n \to \infty} |\mu_n|_\R = p$.

It is now easy to see that a PTRS is iAST (i.e., for all $\irightrightarrowsR$-rewrite
sequences $(\mu_n)_{n \in \IN}$ we have $\lim_{n \to \infty} |\mu_n|_\R = 1$) iff for all
innermost $\R$-rewrite sequence trees $\F{T}$ we have $|\F{T}|_{\ctleaf} = 1$. 
To see this, note that every infinite $\irightrightarrowsR$-rewrite sequence $(\mu_n)_{n
  \in \IN}$ that begins
with a single start term (i.e., $\mu_0 = \{1:t\}$) can be represented by an infinite
innermost $\R$-rewrite sequence tree $\F{T}$ that is fully evaluated (i.e., for every leaf
$v$, $t_v$ is an  $\R$-normal form) such that $\lim_{n \to \infty} |\mu_n|_\R =
|\F{T}|_{\ctleaf}$ and vice versa. So $\R$ is iAST iff all fully evaluated innermost
$\R$-rewrite sequence trees with a single start term converge with probability 1.

\begin{example}\label{example:random-walk-rst}
  Let $\R_{\trw}$ be the PTRS from \Cref{example:PTRS-random-walk}.
  Consider the infinite rewrite sequence:
  \[\begin{array}{ll}
    &\{1:\tg(\O)\}\\
    \irightrightarrowsRandomWalk&\{\nicefrac{1}{2}:\tg^2(\O),\nicefrac{1}{2}:\O\}\\
    \irightrightarrowsRandomWalk&\{\nicefrac{1}{4}:\tg^3(\O), \nicefrac{1}{4}:\tg(\O), \nicefrac{1}{2}:\O\}\\
    \irightrightarrowsRandomWalk&\{\nicefrac{1}{8}:\tg^4(\O),\nicefrac{1}{8}:\tg^2(\O), \nicefrac{1}{8}:\tg^2(\O), \nicefrac{1}{8}:\O, \nicefrac{1}{2}:\O\}\\
    \irightrightarrowsRandomWalk& \ldots \!
  \end{array}\]
  This sequence can be represented by the following innermost $\R$-rewrite sequence tree:
  \begin{center}
    \small
    \begin{tikzpicture}
      \tikzstyle{adam}=[rectangle,thick,draw=black!100,fill=white!100,minimum size=4mm]
      \tikzstyle{empty}=[rectangle,thick,minimum size=4mm]
      
      \node[adam] at (-4, 0)  (a) {$1:\tg(\O)$};
      \node[adam] at (-6, -1)  (b) {$\nicefrac{1}{2}:\tg^2(\O)$};
      \node[adam] at (-2, -1)  (c) {$\nicefrac{1}{2}:\O$};
      \node[adam] at (-8, -2)  (d) {$\nicefrac{1}{4}:\tg^3(\O)$};
      \node[adam] at (-4, -2)  (e) {$\nicefrac{1}{4}:\tg(\O)$};
      \node[adam] at (-9, -3)  (f) {$\nicefrac{1}{8}:\tg^4(\O)$};
      \node[adam] at (-7, -3)  (g) {$\nicefrac{1}{8}:\tg^2(\O)$};
      \node[adam] at (-5, -3)  (h) {$\nicefrac{1}{8}:\tg^2(\O)$};
      \node[adam] at (-3, -3)  (i) {$\nicefrac{1}{8}:\O$};
      \node[empty] at (-9, -4)  (j) {$\ldots$};
      \node[empty] at (-7, -4)  (k) {$\ldots$};
      \node[empty] at (-5, -4)  (l) {$\ldots$};
    
      \draw (a) edge[->] (b);
      \draw (a) edge[->] (c);
      \draw (b) edge[->] (d);
      \draw (b) edge[->] (e);
      \draw (d) edge[->] (f);
      \draw (d) edge[->] (g);
      \draw (e) edge[->] (h);
      \draw (e) edge[->] (i);
      \draw (f) edge[->] (j);
      \draw (g) edge[->] (k);
      \draw (h) edge[->] (l);
    \end{tikzpicture}
  \end{center}
\end{example}

Furthermore, for every $\R$-rewrite sequence tree $\F{T}$ that is not fully evaluated, there exists a $\R$-rewrite sequence tree $\F{T}'$ that is fully evaluated such that $|\F{T}|_{\ctleaf} \geq |\F{T}'|_{\ctleaf}$.
To get from $\F{T}$ to $\F{T}'$ we can simply perform arbitrary (possibly infinitely many)
rewrite steps at the leaves that are not in normal form to fully evaluate the tree.
So $\R$ is iAST iff all innermost
$\R$-rewrite sequence trees with a single start term converge with probability 1.

It remains to prove that it suffices to only regard rewrite sequences that start with a single start term $\{1:t\}$.

\begin{lemma}[Single Start Terms Suffice for iAST] \label{lemma:PTRS-AST-single-start-term}
	Suppose that there exists an infinite $\irightrightarrowsR$-rewrite sequence $(\mu_n)_{n \in \mathbb{N}}$ that converges with probability $<1$.
	Then there is also an infinite $\irightrightarrowsR$-rewrite sequence $(\mu'_n)_{n \in \mathbb{N}}$ with a single start term 
	(i.e., $\mu'_0 = \{1:t\}$ for some term $t$) that converges with probability $<1$.
\end{lemma}

\begin{myproof}
	Let $(\mu_n)_{n \in \mathbb{N}}$ be an $\irightrightarrowsR$-rewrite sequence that converges with probability $<1$.
	Suppose that we have $\mu_0 = \{p_1:t_1, \ldots, p_k:t_k\}$.
	Let $(\mu^j_n)_{n \in \mathbb{N}}$ with $\mu^j_0 =
        \{1 : t_j\}$ denote the infinite $\irightrightarrowsR$-rewrite sequence that uses
        the same rules
        as $(\mu_n)_{n \in \mathbb{N}}$ does on the term $t_j$ for every $1 \leq j \leq k$.
	Assume for a contradiction that for every $1 \leq j \leq k$ the
        $\irightrightarrowsR$-rewrite sequence  $(\mu^j_n)_{n \in \mathbb{N}}$
      converges with probability $1$.
	Then we would have
	\[
	  \begin{array}{rl}
     &\lim_{n \to \infty}|\mu_{n}|_{\R}\\
		=&\lim_{n \to \infty} \sum_{(p:t) \in \mu_{n}, t \in \mathtt{NF}_{\R}} p\\
		=&\lim_{n \to \infty} \sum_{1 \leq j \leq k} p_j \cdot \sum_{(p:t) \in \mu^j_n, t \in \mathtt{NF}_{\R}} p\\
		=&\sum_{1 \leq j \leq k} p_j \cdot \lim_{n \to \infty} \sum_{(p:t) \in \mu^j_n, t \in \mathtt{NF}_{\R}} p\\
		=&\sum_{1 \leq j \leq k} p_j \cdot \lim_{n \to \infty}|\mu^j_n|_{\R}\\
		=&\sum_{1 \leq j \leq k} p_j \cdot 1 \\
		=&\sum_{1 \leq j \leq k} p_j \\
		=&1
		\end{array}
	\]
	which is a contradiction to our assumption that we have $\lim_{n \to \infty}|\mu_{n}|_{\R} < 1$.
	Therefore, we have at least one $1 \leq j \leq k$ such that $(\mu^j_n)_{n \in
          \mathbb{N}}$
        converges with probability $<1$.
\end{myproof}

By \Cref{lemma:PTRS-AST-single-start-term} and the considerations before, we now obtain the following corollary.

\begin{corollary}[Characterizing iAST with Rewrite Sequence Trees]\label{cor:Characterizing iAST with Rewrite Sequence Trees}
  Let $\R$ be a PTRS. Then $\R$ is iAST iff for all innermost $\R$-rewrite sequence trees
  $\F{T}$ we have $|\F{T}|_{\ctleaf} = 1$.
\end{corollary}

We need some more notions for the proof of the chain criterion.

\begin{definition}[$\SubDPoss,\SubDMain,\SubmultPoss,\SubmultMain$]\label{def:prop-important-sets}
	Let $\R$ be a PTRS\@.
	For any term $r$ we define the multiset
        $\SubDPoss(r,\R) := \{ t^{\#} \mid t \in \MSubd(r)$, $t \notin \mathtt{NF}_\R\, \}$
        and the multiset
        $\SubDMain(r,\R) := \{ t^{\#} \mid t\in \MSubd(r)$, 
 $t \notin \mathtt{NF}_\R$, and every proper subterm of $t$ is in
        $\mathtt{NF}_\R\}$.
	For any normalized term $\Com{n}(s_1^\#, \ldots, s_n^\#)$ we define the multiset
$\SubmultPoss(\Com{n}(s_1^\#, \ldots, s_n^\#),\R) := \{s_i^\# \mid 1 \leq i \leq n$, $s_i
        \notin \mathtt{NF}_\R\,\}$
        and the multiset
        $\SubmultMain(\Com{n}(s_1^\#, \ldots, s_n^\#),\R) := \{s_i^\# \mid 1 \leq i \leq
        n$,
$s_i  \notin \mathtt{NF}_\R,$ and every proper subterm of $s_i$ is in
        $\mathtt{NF}_\R\}$.
\end{definition}

$\SubDMain(t,\R)$ and $\SubmultMain(\Com{n}(s_1^\#, \ldots, s_n^\#),\R)$ contain the
redexes that may be used for the next innermost rewrite step. 
The multisets $\SubDPoss(t,\R)$ and
$\SubmultPoss(\Com{n}(s_1^\#, \ldots, s_n^\#),\R)$ contain all terms that may become
redexes now or in the future (after evaluating their subterms), because they have a defined root symbol and are not in normal form w.r.t.\ $\R$.

\ProbChainCriterion*

\begin{myproof}
  Assume that $\R$ is not iAST\@.
  Then by \Cref{cor:Characterizing iAST with Rewrite Sequence Trees} there exists an
  innermost $\R$-rewrite sequence tree $\F{T} = (V,E,L)$ such that $|\F{T}|_{\ctleaf} <
  1$. 
  We will construct a $(\DTuple{\R},\R)$-chain tree $\F{T}' = (V,E,L',V \setminus
  \ctleaf^{\F{T}})$ with the same underlying tree structure and adjusted labeling such
  that all nodes get the same probabilities as in $\F{T}$.
  Since the tree structure and the probabilities are the same, we then obtain $|\F{T}|_{\ctleaf} = |\F{T}'|_{\ctleaf}$.
  To be precise, the set of leaves in $\F{T}$ is equal to the set of leaves in $\F{T}'$,
  and they have the same probabilities.
  Since $|\F{T}|_{\ctleaf} < 1$, we thus have $|\F{T}'|_{\ctleaf} < 1$, and this means
  that $(\DTuple{\R},\R)$ is not iAST either.
	\begin{center}
    \small
    \begin{tikzpicture}
      \tikzstyle{adam}=[rectangle,thick,draw=black!100,fill=white!100,minimum size=4mm]
      \tikzstyle{empty}=[rectangle,thick,minimum size=4mm]
      
      \node[adam] at (-3.5, 0)  (a) {$1:t$};
      \node[adam] at (-5, -1)  (b) {$p_1:t_{1}$};
      \node[adam] at (-2, -1)  (c) {$p_2:t_{2}$};
      \node[adam] at (-6, -2)  (d) {$p_3:t_3$};
      \node[adam] at (-4, -2)  (e) {$p_4:t_4$};
      \node[adam] at (-2, -2)  (f) {$p_5:t_5$};
      \node[empty] at (-6, -3)  (g) {$\ldots$};
      \node[empty] at (-4, -3)  (h) {$\ldots$};
      \node[empty] at (-2, -3)  (i) {$\ldots$};

      \node[empty] at (-0.5, -1)  (arrow) {\Huge $\leadsto$};
      
      \node[adam,pin={[pin distance=0.1cm, pin edge={,-}] 135:\tiny \textcolor{blue}{$P$}}] at (3.5, 0)  (a2) {$1:dp(t)$};
      \node[adam,pin={[pin distance=0.1cm, pin edge={,-}] 135:\tiny \textcolor{blue}{$P$}}] at (2, -1)  (b2) {$p_1:u_{1}$};
      \node[adam,pin={[pin distance=0.1cm, pin edge={,-}] 45:\tiny \textcolor{blue}{$P$}}] at (5, -1)  (c2) {$p_2:u_{2}$};
      \node[adam,pin={[pin distance=0.1cm, pin edge={,-}] 135:\tiny \textcolor{blue}{$P$}}] at (1, -2)  (d2) {$p_3:u_3$};
      \node[adam,pin={[pin distance=0.1cm, pin edge={,-}] 45:\tiny \textcolor{blue}{$P$}}] at (3, -2)  (e2) {$p_4:u_4$};
      \node[adam,pin={[pin distance=0.1cm, pin edge={,-}] 45:\tiny \textcolor{blue}{$P$}}] at (5, -2)  (f2) {$p_5:u_5$};
      \node[empty] at (1, -3)  (g2) {$\ldots$};
      \node[empty] at (3, -3)  (h2) {$\ldots$};
      \node[empty] at (5, -3)  (i2) {$\ldots$};
    
      \draw (a) edge[->] (b);
      \draw (a) edge[->] (c);
      \draw (b) edge[->] (d);
      \draw (b) edge[->] (e);
      \draw (c) edge[->] (f);
      \draw (d) edge[->] (g);
      \draw (e) edge[->] (h);
      \draw (f) edge[->] (i);

      \draw (a2) edge[->] (b2);
      \draw (a2) edge[->] (c2);
      \draw (b2) edge[->] (d2);
      \draw (b2) edge[->] (e2);
      \draw (c2) edge[->] (f2);
      \draw (d2) edge[->] (g2);
      \draw (e2) edge[->] (h2);
      \draw (f2) edge[->] (i2);
    \end{tikzpicture}
  \end{center}
        We construct the new labeling $L'$ for the $(\DTuple{\R},\R)$-chain tree
with $L(x) = (p_x,t_x)$ and $L'(x) = (p_x, u_x)$ for all nodes $x \in V$ inductively
such that for all inner nodes $x \in V \setminus \ctleaf$ with children nodes $xE = \{y_1,\ldots,y_k\}$ we have $u_x \setitodrr \{\tfrac{p_{y_1}}{p_x}:u_{y_1}, \ldots, \tfrac{p_{y_k}}{p_x}:u_{y_k}\}$.
 All terms $u_x$ occurring in our chain tree have the form $\Com{n}(s_1^\#, \ldots, s_n^\#)$ and are normalized.
  Then every property of \cref{def:chain-tree} is satisfied so that $\F{T}'$ is indeed a $(\DTuple{\R},\R)$-chain tree.
  Let $X \subseteq V$ be the set of nodes where we have already defined the labeling $L'(x)$.
  During our construction, we ensure that the following property holds:
  \begin{equation}\label{chain-crit-1-soundness-induction-hypothesis}
    \parbox{.85\textwidth}{For every node $x \in X$ we have $\SubDPoss(t_x,\R) \subseteq \SubmultPoss(u_x,\R)$ and hence $\SubDMain(t_x,\R) \subseteq \SubmultMain(u_x,\R)$.}
  \end{equation}
  This means that the term $t_x$ for the node $x$ in $\F{T}$ has at most the same possible redexes as we have possible terms that can be used for the next $\setitodrr$-step for the term $u_x$ in $\F{T}'$.

If the root of $\F{T}$ is labeled with the term $t$, then we label the root of $\F{T'}$
with the term $dp(t)$. 
  Here, we have $\SubDPoss(t,\R) = \SubmultPoss(dp(t),\R)$, since $dp$ contains all subterms of $t$ with defined root symbol, hence also the ones that can be used for the rewrite steps.

  As long as there is still an inner node $x \in X$ such that its successors are not
  contained in $X$, we proceed as follows.
  Let $xE = \{y_1, \ldots, y_k\}$ be the set of its successors.
  We need to define the corresponding normalized compound terms $u_{y_1}, \ldots, u_{y_k}$ for the nodes $y_1, \ldots, y_k$.
  Since $x$ is not a leaf, we have $t_x \itor \{\tfrac{p_{y_1}}{p_x}:t_{y_1}, \ldots, \tfrac{p_{y_k}}{p_x}:t_{y_k}\}$.
  This means that there is a rule $\ell \to \{p_1:r_1, \ldots, p_k:r_k\} \in \R$, a position $\pi$, and a substitution $\sigma$ such that ${t_x}|_\pi = \ell\sigma$ and every proper subterm of $\ell\sigma$ is in normal form w.r.t.\ $\R$.
  Furthermore, we have $t_{y_j} = t_x[r_j \sigma]_{\pi}$ for all $1 \leq j \leq k$.

  Let $\mu = \{ p_1 : r_1, \ldots, p_k : r_k \}$.
  The corresponding dependency tuple for the rule $\ell \to \mu$ is $\DTuple{\ell \to \mu}
  = \langle \ell^{\#}, \ell \rangle \to \{ p_1 : \langle dp(r_1), r_1 \rangle, \ldots, p_k
  : \langle dp(r_k), r_k \rangle \}$.
  Furthermore, we have $\ell^{\#}\sigma \in \SubDMain(t_x,\R) \subseteq_{(IH)} \SubmultMain(u_x,\R)$.
  Let $u_x = \Com{n}(s_1^{\#}, \ldots, s_n^{\#})$ and let $1 \leq j \leq k$. 
  Then for all $1 \leq i \leq n$ we now define a term $v_i^\#$, and then $u_{y_j}$
is defined to be the term that results from the normalization of $\Com{n}(v_1^{\#}, \ldots, v_n^{\#})$.

We have $s_i^{\#} = \ell^{\#}\sigma$ for some $1 \leq i \leq n$, and hence, we can rewrite
$u_x$ with
$\langle \ell^{\#}, \ell \rangle \to \{ p_1 : \langle dp(r_1), r_1 \rangle, \ldots, p_k :
\langle dp(r_k), r_k \rangle \} \in \DTuple{\R}$ and the substitution $\sigma$, since
every proper subterm of $s_i^{\#} = \ell^{\#}\sigma$ is in normal form w.r.t.\ $\R$.
Thus, we define $v_i^\# = dp(r_j)\sigma$. 
  Furthermore, let $\{\tau_1, \ldots, \tau_m\}$ be all positions in the term $t_x$ that
  are strictly above $\pi$ (i.e., for every $\tau \in \{\tau_1, \ldots, \tau_m\}$ there
  exists a position $\chi \neq \epsilon$ such that $\tau.\chi = \pi$).
  For all $\tau \in \{\tau_1, \ldots, \tau_m\}$ such that $t_x|_{\tau} \in
  \SubDPoss(t_x,\R) \subseteq_{(IH)} \SubmultPoss(u_x,\R)$ we can find a (unique) $1 \leq
  i' \leq n$ with $i \neq i'$ such that $s_{i'}^{\#} = (t_x|_{\tau})^{\#}$.
  Then we define $v_{i'}^\# = s_{i'}^{\#}[r_j \sigma]_{\chi}$.
  All other arguments of the compound symbol remain the same, i.e., here we have
   $v_{i'}^\# =s_{i'}^{\#}$.

  It remains to show that our property~\eqref{chain-crit-1-soundness-induction-hypothesis}
  is still satisfied for this new labeling, i.e., that we have $\SubDPoss(t_{y_j}, \R)
  \subseteq \SubmultPoss(u_{y_j}, \R)$,
  i.e., that 
  \[\SubDPoss(t_x[r_j\sigma]_\pi, \R) \subseteq \SubmultPoss(u_{y_j}, \R) \quad \text{for all $1 \leq j \leq k$.}\]

  Let $1 \leq j \leq k$ and $(t_x[r_j\sigma]_\pi|_{\zeta})^{\#} \in \SubDPoss(t_x[r_j\sigma]_\pi, \R)$ be a subterm at position $\zeta$.
  We have the following possibilities:
  \begin{itemize}
    \item[$\bullet$] If $\pi \leq \zeta$, then there is a $\chi \in \mathbb{N}^*$ such that $\pi.\chi = \zeta$ and we have $(t_x[r_j\sigma]_\pi|_{\zeta})^{\#} = (r_j\sigma|_{\chi})^{\#}$.
    Due to the innermost strategy, $\chi$ is a position of $r_j$ and $r_j|_\chi \not\in \VSet$.
    This means that at least the defined root symbol of $r_j\sigma|_{\chi}$ must be inside
    of $r_j$ and thus we have
    $(r_j|_{\chi})^{\#} \in \cont(dp(r_j))$.
    Thus, $(t_x[r_j\sigma]_\pi|_{\zeta})^{\#} = (r_j\sigma|_{\chi})^{\#} \in \SubmultPoss(dp(r_j)\sigma, \R) \subseteq \SubmultPoss(u_{y_j}, \R)$.
  
    \item[$\bullet$] If $\zeta < \pi$, then there is a $\chi \in \mathbb{N}^+$ such that $\zeta.\chi = \pi$, hence $\zeta \in \{\tau_1, \ldots, \tau_m\}$, and we have $t_x[r_j\sigma]_\pi|_{\zeta} = t_x|_{\zeta}[r_j\sigma]_\chi$.
    Hence,  $(t_x|_{\zeta})^{\#} \in \SubDPoss(t_x, \R) \subseteq_{(IH)} \SubmultPoss(u_x, \R)$.
    By construction, we replaced $(t_x|_{\zeta})^{\#}$ by $(t_x|_{\zeta})^{\#}[r_j\sigma]_\chi$ so that $(t_x|_{\zeta})^{\#}[r_j\sigma]_\chi \in \SubmultPoss(u_{y_j}, \R)$.

    \item[$\bullet$] If $\zeta \bot \pi$, then $t_x[r_j\sigma]_\pi|_{\zeta} = t_x|_{\zeta}$.
    Hence, we have $(t_x|_{\zeta})^{\#} \in \SubDPoss(t_x, \R) \subseteq_{(IH)} \SubmultPoss(u_x, \R)$.
    Since $\zeta \bot \pi$, we kept $(t_x|_{\zeta})^{\#}$ as an argument of
    the compound symbol in our construction and thus $(t_x|_{\zeta})^{\#} \in \SubmultPoss(u_{y_j}, \R)$.
  \end{itemize}
\end{myproof}

For the rest of the appendix, let $(\PP, \SSS)$ be an arbitrary DP problem.
Before we can prove the soundness of our processors, we have to prove three auxiliary lemmas.
The first one (\cref{lemma:p-partition}) shows the modularity of iAST in chain trees, which is used in the proofs of the dependency graph and the reduction pair processor.
The second one (\cref{lemma:splitting}) shows that we can allow steps in a chain
tree which do not change the term but split it into multiple copies with certain probabilities.
This is used for the constructions in the proof of the dependency graph processor and the usable terms processor.
Finally, the third one (\cref{lemma:starting}) shows that we can restrict ourselves to chain trees that start with a term $\Com{1}(s_1^\#)$ such that every proper subterm of $s_1^\#$ is in normal form w.r.t.\ $\SSS$, which is especially needed for the usable rules processor but also used in the dependency graph processor and the usable terms processor.

Again, we need some more definitions for the following proofs.
We first define the notion of a \emph{sub chain tree}.

\begin{definition}[Sub Chain Tree] \label{def:chain-tree-induced-sub}
	Let $\F{T} = (V,E,L,P)$ be a $(\PP, \SSS)$-chain tree.
	Let $W \subseteq V$ be non-empty, weakly connected, and for all $x \in W$ we have $xE \cap W = \emptyset$ or $xE \cap W = xE$.
	The property of being non-empty and weakly connected ensures that the resulting graph $G^{\F{T}[W]} = (W, E \cap (W \times W))$ is a tree again.
	The last property regarding the successors of a node ensures that the sum of probabilities for the successors of a node $x$ is equal to the probability for the node $x$ itself.
	Then, we define the \emph{sub chain tree} $\F{T}[W]$ by
	\[
		\F{T}[W] \coloneqq (W,E \cap (W \times W),L^W,P \cap (W \setminus W_{\ctleaf}))
	\]
	Here, $W_{\ctleaf}$ denotes the leaves of the tree $G^{\F{T}[W]}$ so that the new set $P \cap (W \setminus W_{\ctleaf})$ only contains inner nodes.
	Let $w \in W$ be the root of $G^{\F{T}[W]}$.
	To ensure that the root of our  sub chain tree has the probability $1$ again,
	we use the labeling $L^W(x) = (\frac{p_{x}^{\F{T}}}{p_w^{\F{T}}}: t_{x}^{\F{T}})$ for all nodes $x \in W$.
  If $W$ contains the root of $(V,E)$, then we call the sub chain tree \emph{grounded}.
\end{definition}

\begin{example}
  Analogous to sub chain trees, we can also define sub rewrite sequence trees.
  Let $\F{T}$ be the rewrite sequence tree from \Cref{example:random-walk-rst}.
 A sub rewrite sequence tree that starts at the left successor of the root is:
  \begin{center}
    \small
    \begin{tikzpicture}
      \tikzstyle{adam}=[rectangle,thick,draw=black!100,fill=white!100,minimum size=4mm]
      \tikzstyle{empty}=[rectangle,thick,minimum size=4mm]
      
      \node[adam] at (-6, -1)  (b) {$1:\tg^2(\O)$};
      \node[adam] at (-8, -2)  (d) {$\nicefrac{1}{2}:\tg^3(\O)$};
      \node[adam] at (-4, -2)  (e) {$\nicefrac{1}{2}:\tg(\O)$};
      \node[adam] at (-9, -3)  (f) {$\nicefrac{1}{4}:\tg^4(\O)$};
      \node[adam] at (-7, -3)  (g) {$\nicefrac{1}{4}:\tg^2(\O)$};
      \node[adam] at (-5, -3)  (h) {$\nicefrac{1}{4}:\tg^2(\O)$};
      \node[adam] at (-3, -3)  (i) {$\nicefrac{1}{4}:\O$};
      \node[empty] at (-9, -4)  (j) {$\ldots$};
      \node[empty] at (-7, -4)  (k) {$\ldots$};
      \node[empty] at (-5, -4)  (l) {$\ldots$};
    
      \draw (b) edge[->] (d);
      \draw (b) edge[->] (e);
      \draw (d) edge[->] (f);
      \draw (d) edge[->] (g);
      \draw (e) edge[->] (h);
      \draw (e) edge[->] (i);
      \draw (f) edge[->] (j);
      \draw (g) edge[->] (k);
      \draw (h) edge[->] (l);
    \end{tikzpicture}
  \end{center}
  Note that we have to adjust the probabilities so that the root node has the probability $1$ again.
\end{example}

\begin{lemma}[Sub Chain Trees are Chain Trees] \label{lemma:chain-tree-induced-sub-well-defined}
	Let $\F{T} = (V,E,L,P)$ be a $(\PP, \SSS)$-chain tree and let $W \subseteq V$ satisfy
	the conditions of \cref{def:chain-tree-induced-sub}.
	Then $\F{T}[W] = (V', E', L', P')$ is a chain tree again.
\end{lemma}

\begin{myproof}
	We have to show that $\F{T}[W] = (V', E', L', P') = (W, E \cap (W \times W), L^W, (P \cap (W \setminus W_{\ctleaf})))$ satisfies the conditions of \cref{def:chain-tree}.
	Since $W$ is non-empty and weakly connected, we know that the graph $G^{\F{T}[W]} = (W, E \cap (W \times W))$ is a finitely branching, directed tree again.
	Let $w \in W$ be the root of $G^{\F{T}[W]}$.
	For all $x \in W$ with $xE' \neq \emptyset$ we have $xE' = xE$ and thus
	\[
		\sum_{y \in xE'} p_y^{\F{T}[W]} = \sum_{y \in xE'} \tfrac{p_{y}^{\F{T}}}{p_w^{\F{T}}} = \sum_{y \in xE} \tfrac{p_{y}^{\F{T}}}{p_w^{\F{T}}} = \tfrac{1}{p_w^{\F{T}}} \cdot \sum_{y \in xE} p_{y}^{\F{T}} = \tfrac{1}{p_w^{\F{T}}} \cdot p_{x}^{\F{T}} = \tfrac{p_{x}^{\F{T}}}{p_w^{\F{T}}} = p_x^{\F{T}[W]}
	\]
 	For the root $w \in W$, we have 
	$p_{w}^{\F{T}[W]} = \tfrac{p_{w}^{\F{T}}}{p_{w}^{\F{T}}} = 1$.
	The rest of the requirements follows from the fact that $\F{T}$ is a chain tree
        and that we remove the leaves of $G^{\F{T}[W]}$ from the set $P$ for our sub chain tree.
\end{myproof}

\begin{lemma}[P-Partition Lemma]\label{lemma:p-partition}
  Let $\F{T} = (V,E,L,P)$ be a $(\PP, \SSS)$-chain tree that converges with probability $<1$.
	Assume that we can partition $P = P_1 \uplus P_2$ such that every sub chain tree that only contains $P$ nodes from $P_1$ converges with probability $1$.
	Then there is a grounded sub chain tree $\F{T}'$ that converges with probability $<1$ such that every infinite path has an infinite number of nodes from $P_2$.
\end{lemma}

\begin{myproof}
  Let $\F{T} = (V,E,L,P)$ be a chain tree with $|\F{T}|_{\ctleaf} = c < 1$ for some $c \in \IR$.
	Since we have $0 \leq c < 1$, there is an $\varepsilon > 0$ such that $c + \varepsilon < 1$.
	Remember that the formula for the geometric series is:
	\[
		\sum_{n = 1}^{\infty} \left(\frac{1}{d}\right)^n = \frac{1}{d-1}, \text{ for all } \frac{1}{|d|} < 1
	\]
	We set $d = \frac{1}{\varepsilon} + 2$ and get
	\[
		\frac{1}{\varepsilon} + 1 < \frac{1}{\varepsilon} + 2 \Leftrightarrow \frac{1}{\varepsilon} + 1 < d \Leftrightarrow \frac{1}{\varepsilon} < d-1 \Leftrightarrow \frac{1}{d-1} < \varepsilon
	\]
	We will now construct a grounded sub chain tree $\F{T}' = (V',E',L',P')$ such that
        every infinite path
        has an infinite number of $P_2$ nodes and such that 
	\begin{equation}\label{eq:sum-after-all-cuts}
		|\F{T}'|_{\ctleaf} \leq |\F{T}|_{\ctleaf} + \sum_{n = 1}^{\infty} \left(\frac{1}{d}\right)^n.
	\end{equation}
	Then, we finally have
	\[
		|\F{T}'|_{\ctleaf} \leq |\F{T}|_{\ctleaf} + \sum_{n = 1}^{\infty} \left(\frac{1}{d}\right)^n = |\F{T}|_{\ctleaf} + \frac{1}{d-1} = c + \frac{1}{d-1} < c + \varepsilon < 1
	\]
	
	The idea of this construction is that we cut infinite sub-trees of pure $P_1$ and $S$ nodes as soon as the probability for normal forms is high enough.
	In this way, one then obtains paths where after finitely many $P_1$ nodes, there is always a $P_2$ node, or we reach a leaf.

	The construction works as follows.
	For any node $x \in V$, let $\ctlevelTwo(x)$ be the number of $P_2$ nodes in the path from the root to $x$.
	Furthermore, for any set $W \subseteq V$ and $k \in \IN$, let
	$\ctlevelTwowithborder(W,k) = \{x \in W \mid \ctlevelTwo(x) \leq k \lor (x
	\in P_2 \land \ctlevelTwo(x) \leq k+1)\}$ be the set of all nodes in $W$ that have at most $k$ nodes from $P_2$ in the path from the root to its predecessor.
	So if $x \in W$ is not in $P_2$, then we have at most $k$ nodes from $P_2$ in the
        path from the root to $x$
        and if $x \in W$ is in $P_2$, then we have at most $k+1$ nodes from $P_2$ in the path from the root to $x$.
	We will inductively define a set $U_k \subseteq V$ such that $U_k \subseteq
	\ctlevelTwowithborder(V,k)$ and then define the grounded sub chain tree as
	$\F{T}' := \F{T}[\bigcup_{k \in \IN} U_k]$.

	We start by considering the  sub chain tree $T_{0} = \F{T}[\ctlevelTwowithborder(V,0)]$.
	This tree only contains $P$ nodes from $P_1$.
	While the node set $\ctlevelTwowithborder(V,0)$ itself may contain nodes from $P_2$, they can only occur at the leaves of $T_{0}$, and by definition of a sub chain tree, we remove every leaf from $P$ in the creation of $T_{0}$.
	Using the prerequisite of the lemma, we get $|T_{0}|_{\ctleaf}=1$.
	In \cref{Possibilities for Te} one can see the different possibilities for $T_0$.
	Either $T_0$ is finite or $T_0$ is infinite.
	In the first case, we can add all the nodes to $U_0$ since there is no infinite path of pure $P_1$ and $S$ nodes.
	Hence, we define $U_0 = \ctlevelTwowithborder(V,0)$.
	In the second case, we have to cut the tree at a specific depth once the probability of leaves is high enough.
	Let $\ctdepth_{0}(y)$ be the depth of the node $y$ in the tree $T_{0}$.
	Moreover, let $D_{0}(k) = \{x \in \ctlevelTwowithborder(V,0) \mid \ctdepth_{0}(x) \leq k\}$ be the set of nodes in $T_0$ that have a depth of at most $k$.
	Since $|T_{0}|_{\ctleaf}=1$ and $|\cdot|_{\ctleaf}$ is monotonic w.r.t.\ the depth of
        the tree $T_{0}$, we can find an $N_{0} \in \IN$ such that
	\[
		\sum_{x \in \ctleaf^{T_{0}}, d_{0}(x) \leq N_{0}} p_x^{T_{0}} \geq  1 - \frac{1}{d}
	\]
	We include all nodes from $D_{0}(N_{0})$ in $U_0$ and delete every other node of $T_{0}$.
	In other words, we cut the tree after depth $N_{0}$.
	This cut can be seen in \cref{Possibilities for Te}, indicated by the red line.
  We now know that this cut may increase the probability of leaves by at most $\frac{1}{d}$.
	Therefore, we define $U_0 = D_{0}(N_{0})$ in this case.

	\begin{figure}
        \centering
        \begin{subfigure}[b]{0.45\textwidth}
            \centering
            \begin{tikzpicture}
                \tikzstyle{adam}=[circle,thick,draw=black!100,fill=white!100,minimum size=3mm]
                \tikzstyle{empty}=[circle,thick,minimum size=3mm]
                
                \node[adam] at (0, 0)  (a) {};
                \node[adam, label=center:{\tiny $P_1$}] at (2, -3)  (b) {};
                \node[adam, label=center:{\tiny $P_2$}] at (-2, -3)  (c) {};
                \node[adam, label=center:{\tiny $P_2$}] at (0.75, -3)  (d) {};
                \node[adam, label=center:{\tiny $P_1$}] at (-0.75, -3)  (e) {};
                \node[empty, label=center:{\small $S$}] at (0, -1.5)  (middleA) {};
				
                \node[adam, label=center:{\tiny $\mathtt{NF}$}] at (-1.5, -5)  (nf1) {};
                \node[adam, label=center:{\tiny $\mathtt{NF}$}] at (0, -5)  (nf2) {};
                
                \node[adam, label=center:{\tiny $\mathtt{NF}$}] at (2.75, -5)  (bb) {};
                \node[adam, label=center:{\tiny $P_2$}] at (1.25, -5)  (bc) {};
                \node[adam, label=center:{\tiny $P_2$}] at (2, -5)  (bd) {};
                \node[empty, label=center:{\small $S$}] at (2, -4)  (middleA) {};
    
                \node[empty, label=center:{\small $S$}] at (-0.75, -4)  (middleA) {};
                
                \node[empty] at (0, -6)  (stretch) {};
            
                \draw (a) edge[-] (b);
                \draw (b) edge[-] (d);
                \draw (d) edge[-] (e);
                \draw (e) edge[-] (c);
                \draw (a) edge[-] (c);
                
                \draw (b) edge[-] (bb);
                \draw (bb) edge[-] (bd);
                \draw (bd) edge[-] (bc);
                \draw (b) edge[-] (bc);
                
                \draw (e) -- (nf1) -- (nf2) -- (e);
                
                \begin{scope}[on background layer]
                  \fill[green!20!white,on background layer] (0, 0) -- (-2, -3) -- (2, -3);
                  \fill[green!20!white,on background layer] (2, -3) -- (1.25, -5) -- (2.75, -5);
                  \fill[green!20!white,on background layer] (-0.75, -3) -- (-1.5, -5) -- (0, -5);
               \end{scope}
            \end{tikzpicture}
            \caption{$T_{x}$ finite}
        \end{subfigure}
        \hspace{30px}
        \begin{subfigure}[b]{0.45\textwidth}
            \centering
            \begin{tikzpicture}
                \tikzstyle{adam}=[circle,thick,draw=black!100,fill=white!100,minimum size=3mm]
                \tikzstyle{empty}=[circle,thick,minimum size=3mm]
                
                \node[adam] at (0, 0)  (a) {};
                \node[adam, label=center:{\tiny $P_1$}] at (2, -3)  (b) {};
                \node[adam, label=center:{\tiny $P_2$}] at (-2, -3)  (c) {};
                \node[adam, label=center:{\tiny $P_2$}] at (0.75, -3)  (d) {};
                \node[adam, label=center:{\tiny $P_1$}] at (-0.75, -3)  (e) {};
                \node[empty, label=center:{\small $S$}] at (0, -1.5)  (middleA) {};
				
                \node[adam, label=center:{\tiny $\mathtt{NF}$}] at (-1.5, -5)  (nf1) {};
                \node[adam, label=center:{\tiny $\mathtt{NF}$}] at (0, -5)  (nf2) {};
                
                \node[adam, label=center:{\tiny $P_1$}] at (2.75, -5)  (bb) {};
                \node[empty] (bbi) at (2.75,-6)  {};
                \node[adam, label=center:{\tiny $P_2$}] at (1.25, -5)  (bc) {};
                \node[adam, label=center:{\tiny $P_1$}] at (2, -5)  (bd) {};
                \node[empty] (bdi) at (2,-6)  {};
                \node[empty, label=center:{\small $S$}] at (2, -4)  (middleA) {};
    
                \node[empty, label=center:{\small $S$}] at (-0.75, -4)  (middleA) {};

                \node[empty, label=center:{\small \textcolor{red}{$N_{x}$}}] at (3, -3.7)  (cut) {};
            
                \draw (a) edge[-] (b);
                \draw (b) edge[-] (d);
                \draw (d) edge[-] (e);
                \draw (e) edge[-] (c);
                \draw (a) edge[-] (c);
                
                \draw (b) edge[-] (bb);
                \draw (bb) edge[-] (bd);
                \draw (bd) edge[-] (bc);
                \draw (b) edge[-] (bc);
                
                \draw (e) -- (nf1) -- (nf2) -- (e);
                
                \draw[] (bb) edge ($(bb)!0.3cm!(bbi)$) edge [dotted] ($(bb)!0.7cm!(bbi)$);
                \draw[] (bd) edge ($(bd)!0.3cm!(bdi)$) edge [dotted] ($(bd)!0.7cm!(bdi)$);

                \draw[] (-2, -3.7) edge [red, dotted] (cut);
                
                \begin{scope}[on background layer]
                  \fill[green!20!white,on background layer] (0, 0) -- (-2, -3) -- (2, -3);
                  \fill[green!20!white,on background layer] (2, -3) -- (1.25, -5) -- (2.75, -5);
                  \fill[green!20!white,on background layer] (-0.75, -3) -- (-1.5, -5) -- (0, -5);
               \end{scope}
            \end{tikzpicture}
            \caption{$T_{x}$ infinite}
        \end{subfigure}
        \caption{Possibilities for $T_x$}\label{Possibilities for Te}
    \end{figure}
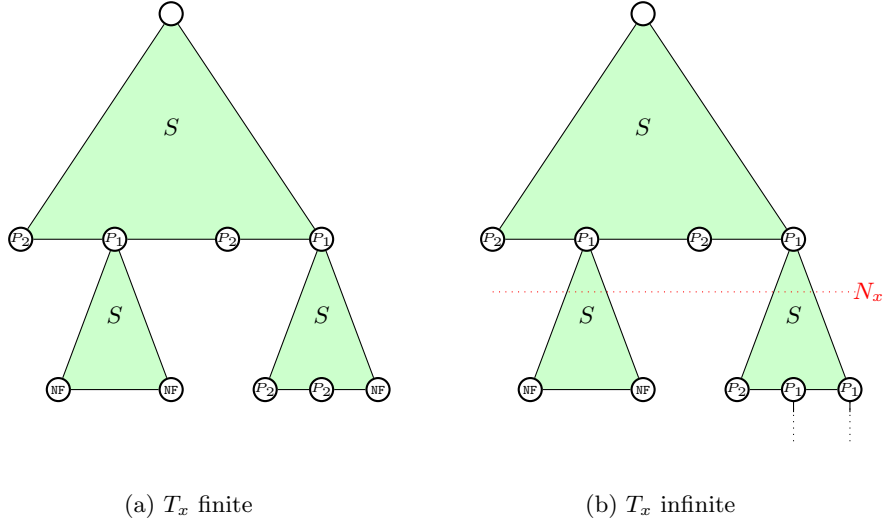

	For the induction step, assume that we have already defined a subset $U_i \subseteq \ctlevelTwowithborder(V,i)$.
	Let $H_i = \{x \in U_i \mid x \in P, \ctlevelTwo(x) = i+1\}$ be the set of leaves in $\F{T}[U_i]$ that are in $P_2$.
	For each $x \in H_i$, we consider the sub chain tree that starts at $x$ until we
        reach the next node from $P_2$, including the node itself.
	Everything below such a node will be cut.
	To be precise, we regard the tree $T_{x} = (V_x,E_x,L_x,P_x) =
        \F{T}[\ctlevelTwowithborder(xE^*,i+1)]$. Here, $xE^*$ is the set of all nodes that
        are reachable from $x$ by arbitrary many steps via the edges $E$.
	
	First, we show that $|T_{x}|_{\ctleaf} = 1$.
	For every direct successor $y$ of $x$, the sub chain tree $T_{y} = T_{x}[y E_x^*]$ of $T_{x}$ that starts at $y$ does not contain any nodes from $P_2$.
	Hence, we have $|T_{y}|_{\ctleaf} = 1$ by the prerequisite of the lemma, and hence
  \[
    |T_x|_{\ctleaf} = \sum_{y \in xE} p_y \cdot |T_{y}|_{\ctleaf} = \sum_{y \in xE} p_y \cdot 1 = \sum_{y \in xE} p_y = 1.
	\]
For the construction of $U_{i+1}$, we have the same cases as before, see \Cref{Possibilities for Te}.
	Either $T_x$ is finite or $T_x$ is infinite.
  Let $Z_{x}$ be the set of nodes that we want to add to our node set $U_{i+1}$ from the tree $T_x$.
	In the first case we can add all the nodes again and set $Z_{x} = V_x$.
	In the second case, we once again cut the tree at a specific depth once the probability for leaves is high enough.
	Let $\ctdepth_{x}(z)$ be the depth of the node $z$ in the tree $T_{x}$.
	Moreover, let $D_{x}(k) = \{z \in V_x \mid \ctdepth_{x}(z) \leq k\}$ be the set of nodes in $T_x$ that have a depth of at most $k$.
	Since $|T_{x}|_{\ctleaf}=1$ and $|\cdot|_{\ctleaf}$ is monotonic w.r.t.\ the depth of the tree $T_{x}$, we can find an $N_x \in \IN$ such that
	\[
		\sum_{y \in \ctleaf^{T_{x}}, d_{x}(y) \leq N_{x}} p_y^{T_{x}} \geq 1 - \left(\frac{1}{d}\right)^{i+1} \cdot \frac{1}{|H_i|}
	\]
	We will include all nodes from $D_{x}(N_{x})$ in $U_{i+1}$ and delete every other node of $T_{x}$.
	In other words, we cut the tree after depth $N_{x}$.
  We now know that this cut may increase the probability of leaves by at most $\left(\frac{1}{d}\right)^{i+1} \cdot \frac{1}{|H_i|}$.
	Therefore, we set $Z_{x} = D_{x}(N_{x})$ in this case.
	
	We do this for each $x \in H_i$ and in the end, we set
        $U_{i+1} = U_i \cup \bigcup_{x \in H_i} Z_{x}$.

	It is straightforward to see that $\bigcup_{k \in \IN} U_k$ satisfies the
        conditions of \Cref{def:chain-tree-induced-sub}, as we only cut after certain nodes in our construction.
	Hence, $\bigcup_{k \in \IN} U_k$ is non-empty and weakly connected, and for each
        of its nodes, it either contains no or all successors.
	Furthermore, $\F{T}' = \F{T}[\bigcup_{k \in \IN} U_k]$ is a grounded sub chain
        tree and it
        does not contain an infinite path of pure $P_1$ and $S$ nodes as we cut every such
        path after a finite depth.
        
	It remains to prove that $|\F{T}'|_{\ctleaf} \leq |\F{T}|_{\ctleaf} + \sum_{n = 1}^{\infty} \left(\frac{1}{d}\right)^n$ holds.
	During the $i$-th iteration of the construction, we may increase the value of $|\F{T}|_{\ctleaf}$ by the sum of all probabilities corresponding to the new leaves resulting from the cuts.
	As we cut at most $|H_i|$ trees in the $i$-th iteration and for each such tree, we added at most a total probability of $\left(\frac{1}{d}\right)^{i+1} \cdot \frac{1}{|H_i|}$, the value of $|\F{T}|_{\ctleaf}$ might increase by 
	\[
		|H_i| \cdot \left(\frac{1}{d}\right)^{i+1} \cdot \frac{1}{|H_i|} = \left(\frac{1}{d}\right)^{i+1}
	\]
	in the $i$-th iteration, and hence in total, we then get
	\[
		|\F{T}'|_{\ctleaf} \leq |\F{T}|_{\ctleaf} + \sum_{n = 1}^{\infty} \left(\frac{1}{d}\right)^n,
	        \]
                as desired (see \eqref{eq:sum-after-all-cuts}).
\end{myproof}

For the splitting lemma, we introduce a new definition of a chain tree \emph{with split-nodes}.

 \begin{definition}[Chain Tree with Split-Nodes] \label{def:chain-tree-with-splits}
  $\!\!\F{T}\!=\!(V,E,L,P)$ {\normalsize is an (innermost)}
   $(\PP\!,\SSS)${\normalsize\emph{-chain tree with split-nodes}} (or ``splits'', for
   short)
   if
 	\begin{enumerate}
   \item $V \neq \emptyset$ is a possibly infinite set of nodes and
           $E \subseteq V \times V$ is a set of directed edges, such that
                 $(V, E)$ is a (possibly infinite) directed tree where $vE
= \{ w \mid (v,w) \in E \}$
                  is
                  finite for every $v \in V$.
           \item  $L:V\rightarrow(0,1]\times\TSet{\SigmaDP}{\VSet}$ labels every node
             $v$
             by a probability $p_v$ and a term $t_v$.
              For the root $v \in V$ of the tree, we have $p_v = 1$.
   \item $P \subseteq V \setminus \ctleaf$ is a subset of the
                 inner nodes to indicate whether we use the\linebreak PPTRS $\PP$
or the PTRS $\SSS$ for the rewrite step.
  $\ctleaf$ are all leaves and
    $S = V \setminus (\ctleaf \cup P)$ are all inner nodes that are not in $P$.
   Thus, $V = P \uplus S \uplus \ctleaf$.
   \item For all  $v \in P$: If
                 $vE = \{w_1, \ldots, w_k\}$, then 
                 $t_v \setitops \{\tfrac{p_{w_1}}{p_v}:t_{w_1}, \ldots,
                 \tfrac{p_{w_k}}{p_v}:t_{w_k}\}$.
   \item For all $v \in S$: If
                 $vE = \{w_1, \ldots, w_k\}$, then
$t_v \setitos \{\tfrac{p_{w_1}}{p_v}:t_{w_1}, \ldots, \tfrac{p_{w_k}}{p_v}:t_{w_k}\}$, \underline{or $t_v = t_{w_j}$ for all $1 \leq j \leq k$}.
   \item Every infinite path in $\F{T}$ contains infinitely many
                 nodes from $P$.
 \end{enumerate}
\end{definition}

So the only difference between a chain tree and a chain tree with split-nodes is the
underlined part, i.e.,
that we are now able to split a node into multiple ones with a certain probability but the same term.
Hence, at an $S$ node, we do not have to use a rewrite rule anymore.
This does not interfere with the property of being iAST for DP problems, as shown by the following lemma.

\begin{lemma}[Splitting Lemma]\label{lemma:splitting}
  A DP problem $(\PP, \SSS)$ is iAST iff for all $(\PP, \SSS)$-chain trees $\F{T}$ \emph{with split-nodes} we have $|\F{T}|_{\ctleaf} = 1$.
\end{lemma}

\begin{myproof}
  We only have to show that if there exists a $(\PP, \SSS)$-chain tree $\F{T}$ with splits such that $|\F{T}|_{\ctleaf}<1$, then there is also a $(\PP, \SSS)$-chain tree $\F{T}'$ without splits such that $|\F{T}'|_{\ctleaf}<1$.
	Assume that there exists a $(\PP, \SSS)$-chain tree $\F{T} = (V,E,L,P)$ with splits such that $|\F{T}|_{\ctleaf}=c<1$.
	From $\F{T}$, we will now create an infinite, finitely branching, labeled tree
        $\F{F}$ whose nodes are also labeled with trees. More precisely,
        every node $X$ of
$\F{F}$ is labeled by a finite $(\PP, \SSS)$-chain tree $\F{T}_X = (V_X,E_X,L_X,P_X)$
        without splits, such that $V_X \subseteq V$, and such that the sum of all
        probabilities for leaves in $\F{T}_X$ that are also leaves in $\F{T}$ is at most $c$ (i.e., $\sum_{x \in \ctleaf^{\F{T}_X}, x \in \ctleaf^{\F{T}}} p_x^{\F{T}_X} \leq c$).
	Since this tree $\F{F}$ is infinite and finitely branching, it must have
        an infinite path by König's Lemma.
	By taking the ``limit'' of the trees on this infinite path, one obtains
        an infinite $(\PP, \SSS)$-chain tree $\F{T}_{\lim}$ without splits such that $\ctleaf^{\F{T}_{\lim}} \subseteq \ctleaf^{\F{T}}$ and thus, we have 
	\[|\F{T}_{\lim}|_{\ctleaf} = \sum_{x \in \ctleaf^{\F{T}_{\lim}}} p_x^{\F{T}_{\lim}} = \sum_{x \in \ctleaf^{\F{T}_{\lim}}, x \in \ctleaf^{\F{T}}} p_x^{\F{T}_{\lim}} \leq c < 1\] 
	so that $\F{T}_{\lim}$ converges with probability $<1$, and this ends the proof. 

  \medskip
			
  \noindent
  \underline{\textbf{Construction of $\F{F}$}}

  \noindent
	Now we explain the precise construction of the tree $\F{F}$.
	The root of $\F{F}$ (i.e., the node of $\F{F}$ at depth 0)
        is labeled with the  sub chain tree $\F{T}[\{v\}]$ of $\F{T}$ that only consists of the root $v$ of $\F{T}$.
	Thus, $\F{T}[\{v\}]$ is a finite $(\PP, \SSS)$-chain tree and $\sum_{x \in \ctleaf^{\F{T}[\{v\}]}, x \in \ctleaf^{\F{T}}} p_x^{\F{T}[\{v\}]} = 0 \leq c$, since the root of $\F{T}$ cannot be a leaf in $\F{T}$, otherwise, we would have $|\F{T}|_{\ctleaf} = 1$.

	Let $x_0, x_1, \ldots$ be an enumeration of $V$ such that there exists no $i < j$
        with $\ctdepth(x_i) > \ctdepth(x_j)$, where $\ctdepth(x_i)$ denotes the depth of
        $x_i$ in the tree $\F{T}$.
	This means that our enumeration starts with the root $x_0$, then it enumerates all nodes of depth $1$, then all nodes of depth $2$, and so on.
	
	In the $i$-th iteration of our construction of the tree $\F{F}$ (where $i \geq
        0$), we construct the nodes of depth $i+1$ in the tree $\F{F}$. To this end, we consider node
        $x_i$ of $\F{T}$.
	Let $V_i \coloneqq \{x_0, \ldots, x_i\} \cup \bigcup_{0 \leq j \leq i} x_jE$ and let $\F{T}_i \coloneqq \F{T}[V_i]$ be the  sub $(\PP, \SSS)$-chain tree consisting of the nodes $x_0, \ldots, x_i$ together with their successors.
	The set $V_i$ satisfies the conditions of \cref{def:chain-tree-induced-sub}
        because it is weakly connected by definition of our enumeration and for each of
        its nodes, it either contains all successors or none.
	We now create a set $\text{Split}(\F{T}_i)$ that contains all possible $(\PP,
        \SSS)$-chain trees without splits that are contained in our $(\PP, \SSS)$-chain tree $\F{T}_i$ with splits.
  On every split-node we split the chain tree into multiple ones with a certain probability.
	Then for the tree $\F{F}$, we have a node in the $(i+1)$-th depth for all those
        trees $T$ of $\text{Split}(\F{T}_i)$ where $\sum_{x \in \ctleaf^{T}, x \in \ctleaf^{\F{T}}} p_x^{T} \leq c$.

  \medskip
			
  \noindent
  \underline{\textbf{Intuition for $\text{Split}(\F{T}_i)$}}

  \noindent
 First, we give the intuition for the following construction of $\text{Split}(\F{T}_i)$.
	By this construction, we obtain a set $\text{Split}(\F{T}_i)$ that consists of pairs $(p_T,T)$, where $0 < p_T \leq 1$ is a probability and $T$ is a $(\PP, \SSS)$-chain tree. 
  Furthermore, it satisfies the following:
	\begin{itemize}
		\item[(A)] For all $(p_T,T) \in \text{Split}(\F{T}_i)$, the tree $T$ is a
                  finite $(\PP, \SSS)$-chain tree without splits.
		\item[(B)] $\sum_{(p_T,T) \in \text{Split}(\F{T}_i)} p_T = 1$. This means
                  that the sum of the probabilities for all possible $(\PP,
                  \SSS)$-computation trees in $\text{Split}(\F{T}_i)$ is one.
		\item[(C)] For every $x \in V_i$ that is not a split-node in $\F{T}_i$, we
                  have
                  \[ \begin{array}{rcl}
                    p_x^{\F{T}_i} &=& \sum_{(p_T,T) \in \text{Split}(\F{T}_i)} p_T \cdot
                  \overline{p_{(T,x)}}, \quad \text{ where }\medskip\\
	 \overline{p_{(T,x)}} &=& \begin{cases}
			p_{x}^T & \text{ if } x \in V^T\\
			0 & \text{ otherwise}
	 \end{cases}
         \end{array} \]
		This means that the probability for each node $x$ that is not a split-node in our tree $\F{T}_i$ is equal to the sum over all trees $T$ that contain $x$, where we multiply the probability of the tree $T$ by the probability of node $x$ in $T$.
	\end{itemize}

	In order to create the set $\text{Split}(\F{T}_i)$, we will iteratively remove the split-nodes and instead split the tree into multiple ones.
  For example, regard the following tree with a single split-node marked by $S$, where the
  probabilities for the
  path are depicted in red and the probabilities for the nodes are blue.
	\begin{center}
		\begin{tikzpicture}
			\tikzstyle{adam}=[circle,thick,draw=black!100,fill=white!100,minimum size=3mm]
			\tikzstyle{empty}=[shape=rectangle,draw=black!100,thick,minimum size=4mm]
			
			\node[empty,pin={[pin distance=0.1cm, pin edge={-}] 135:\tiny \textcolor{blue}{$1$}}] at (0, 0)  (a) {};

			\node[empty,pin={[pin distance=0.1cm, pin edge={-}] 135:\tiny \textcolor{blue}{$\tfrac{1}{2}$}}] at (-3, -1.5)  (aa) {$S$};
			\node[empty,pin={[pin distance=0.1cm, pin edge={-}] 45:\tiny \textcolor{blue}{$\tfrac{1}{2}$}}] at (3, -1.5)  (ab) {};

			\node[empty,pin={[pin distance=0.1cm, pin edge={-}] 135:\tiny \textcolor{blue}{$\tfrac{1}{4}$}}] at (-4.5, -3)  (aaa) {};
			\node[empty,pin={[pin distance=0.1cm, pin edge={-}] 45:\tiny \textcolor{blue}{$\tfrac{1}{4}$}}] at (-1.5, -3)  (aab) {};
		
			\draw (a) edge[->] node[above] {$\textcolor{red}{\tfrac{1}{2}}$} (aa);
			\draw (a) edge[->] node[above] {$\textcolor{red}{\tfrac{1}{2}}$} (ab);
			\draw (aa) edge[->] node[above] {$\textcolor{red}{\tfrac{1}{2}}$} (aaa);
			\draw (aa) edge[->] node[above] {$\textcolor{red}{\tfrac{1}{2}}$} (aab);
		\end{tikzpicture}
	\end{center}
  Then we would transform this tree into the following two trees together with the probability $\tfrac{1}{2}$.
	\begin{center}
    \begin{tikzpicture}
      \tikzstyle{adam}=[circle,thick,draw=black!100,fill=white!100,minimum size=3mm]
      \tikzstyle{empty}=[shape=rectangle,draw=black!100,thick,minimum size=4mm]
      \tikzstyle{realempty}=[shape=rectangle,draw=black!00,thick,minimum size=4mm]
      
      \node[empty,pin={[pin distance=0.1cm, pin edge={-}] 135:\tiny \textcolor{blue}{$1$}}] at (-3, 0)  (a) {};

      \node[empty] at (-4.5, -1.5)  (aa) {};
      \node[empty,pin={[pin distance=0.1cm, pin edge={-}] 45:\tiny \textcolor{blue}{$\tfrac{1}{2}$}}] at (-1.5, -1.5)  (ab) {};

      \node[realempty] at (-6, 0)  (prop1) {$\tfrac{1}{2}$:};

      \node[empty,pin={[pin distance=0.1cm, pin edge={-}] 135:\tiny \textcolor{blue}{$\tfrac{1}{2}$}}] at (-5, -3)  (aaa) {};
      \node[empty] at (-4, -3)  (aab) {};
    
      \draw (a) edge[->] (aaa);
      \draw (a) edge[->] (ab);

      \node[empty,pin={[pin distance=0.1cm, pin edge={-}] 135:\tiny \textcolor{blue}{$1$}}] at (3, 0)  (a) {};

      \node[empty] at (1.5, -1.5)  (aa) {};
      \node[empty,pin={[pin distance=0.1cm, pin edge={-}] 45:\tiny \textcolor{blue}{$\tfrac{1}{2}$}}] at (4.5, -1.5)  (ab) {};

      \node[realempty] at (0, -0)  (prop1) {$\tfrac{1}{2}$:};

      \node[empty] at (1, -3)  (aaa) {};
      \node[empty,pin={[pin distance=0.1cm, pin edge={-}] 45:\tiny \textcolor{blue}{$\tfrac{1}{2}$}}] at (2, -3)  (aab) {};
    
      \draw (a) edge[->] (aab);
      \draw (a) edge[->] (ab);
    \end{tikzpicture}
  \end{center}
  In this case, the set $\text{Split}(\F{T}_i)$ would contain the two pairs $(\tfrac{1}{2},T_1)$,$(\tfrac{1}{2},T_2)$, where $T_1$ and $T_2$ are the two trees depicted above.

  \medskip
			
  \noindent
  \underline{\textbf{Construction of $\text{Split}(\F{T}_i)$}}

  \noindent
	Now we present the formal construction of $\text{Split}(\F{T}_i)$.
	We recursively remove all split-nodes and construct a set $M$ that satisfies the following properties during the induction.
	\begin{itemize}
		\item[(a)] For all $(p_T,T) \in M$,  the tree $T$ is a finite $(\PP, \SSS)$-chain tree (possibly with split-nodes).
		\item[(b)] $\sum_{(p_T,T) \in M} p_T = 1$.
		\item[(c)] For every $x \in V_i$ that is not a split-node in $\F{T}_i$, we have $p_x^{\F{T}_i} = \sum_{(p_T,T) \in M} p_T \cdot \overline{p_{(T,x)}}$.
	\end{itemize}

	We start with $M \coloneqq \{(1,\F{T}_i)\}$.
	Here, clearly all of the three properties are satisfied.
	Now, assume that there is still a tree $\F{t}$ with $(p_{\F{t}}, \F{t}) \in M$ that contains a split-node $v \in V^{\F{t}}$.
  We will now split $\F{t}$ into multiple trees that do not contain $v$ anymore but move directly to one of its successors.

	First, assume that $v$ is not the root of $\F{t}$.
	Let $vE^\F{t} = \{w_1, \ldots, w_m\}$ be the direct successors of $v$ in $\F{t}$ and let $z$ be the predecessor of $v$ in $\F{t}$.
	Instead of one tree $\F{t}$ with the edges $(z,v), (v,w_1), \ldots, (v,w_m)$, we split the tree into $m$ different trees $\F{t}_1, \ldots, \F{t}_m$ such that for every $1 \leq h \leq m$, the tree $\F{t}_h$ contains a direct edge from $z$ to $w_h$.
	In addition to that, the unreachable nodes are removed, and we also have to adjust the probabilities of all (not necessarily direct) successors of $w_h$ (including $w_h$ itself) in $\F{t}_h$.
	\begin{figure}[H]
    \centering
		\begin{tikzpicture}
			\tikzstyle{adam}=[circle,thick,draw=black!100,fill=white!100,minimum size=3mm]
			\tikzstyle{empty}=[shape=rectangle,draw=black!100,thick,minimum size=4mm]
			\tikzstyle{realempty}=[shape=rectangle,draw=black!00,thick,minimum size=4mm]
			
			\node[realempty] at (-2, 0)  (a) {$z$};

			\node[realempty] at (-2, -1.5)  (aa) {$v$};

			\node[realempty] at (-3, -3)  (aaa) {$w_1$};
			\node[realempty] at (-2, -3)  (aab) {$\ldots$};
			\node[realempty] at (-1, -3)  (aac) {$w_m$};

			\node[realempty] at (0, -1.5)  (mid) {\LARGE $\leadsto$};
			
			\node[realempty] at (2, 0)  (x) {$z$};

			\node[realempty] at (2, -1.5)  (xx) {$v$};

			\node[realempty] at (1, -3)  (xxx) {$w_1$};
			\node[realempty] at (2, -3)  (xxy) {$\ldots$};
			\node[realempty] at (3, -3)  (xxz) {$w_m$};

			\node[realempty] at (4.5, -1.5)  (dots) {$\ldots$};
			
			\node[realempty] at (7, 0)  (tx) {$z$};

			\node[realempty] at (7, -1.5)  (txx) {$v$};

			\node[realempty] at (6, -3)  (txxx) {$w_1$};
			\node[realempty] at (7, -3)  (txxy) {$\ldots$};
			\node[realempty] at (8, -3)  (txxz) {$w_m$};

			\draw (a) edge[->] (aa);
			\draw (aa) edge[->] (aaa);
			\draw (aa) edge[->] (aab);
			\draw (aa) edge[->] (aac);

			\draw (x) edge[->] (xxx);

			\draw (tx) edge[->] (txxz);
		\end{tikzpicture}
		\caption{Splitting node $v$ to create $m$ different trees, where we directly move to one of its successors}
		\label{fig:split-construction-definition}
  \end{figure}
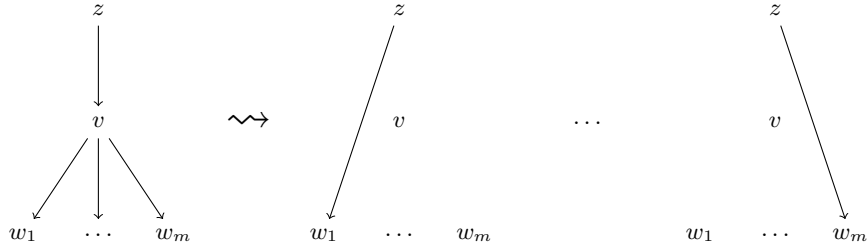
	To be precise, we set $\F{t}_h \coloneqq (V^{\F{t}_h}, E^{\F{t}_h}, L^{\F{t}_h},
        P^{\F{t}_h})$, with
	\begin{align*}
		V^{\F{t}_h} &\coloneqq (V^\F{t} \setminus v(E^\F{t})^*) \cup w_h(E^\F{t})^*\\
		E^{\F{t}_h} &\coloneqq (E^\F{t} \setminus (v(E^\F{t})^* \times v(E^\F{t})^*))
                \cup \{(z,w_h)\} \cup (E^\F{t} \cap (w_h(E^\F{t})^* \times w_h(E^\F{t})^*))\\
                 P^{\F{t}_h}&\coloneqq P^\F{t} \cap V^{\F{t}_h}
	\end{align*}
	Furthermore, let $p_h \coloneqq \tfrac{p_{w_h}^{\F{t}}}{p_{v}^{\F{t}}}$.
	Then, the labeling is defined by
	\[
		L^{\F{t}_h}(x) = \begin{cases}
			(\frac{1}{p_h} \cdot p_{x}^{\F{t}},t_x^{\F{t}}) & \text{ if } x \in w_h(E^\F{t})^*\\
			(p_x^{\F{t}},t_x^{\F{t}}) & \text{ otherwise}
		\end{cases}
	\]
	Note that 
	\begin{equation} \label{p_h sum is one}
		\sum_{1 \leq h \leq m} p_h = \sum_{1 \leq h \leq m} \tfrac{p_{w_h}^{\F{t}}}{p_v^{\F{t}}} = \tfrac{1}{p_v^{\F{t}}} \cdot \sum_{1 \leq h \leq m} p_{w_h}^{\F{t}} = \tfrac{1}{p_v^{\F{t}}} \cdot p_v^{\F{t}} = 1
	\end{equation}
	since $\F{t}$ is a $(\PP,\SSS)$-chain tree by the induction hypothesis.

	If $v$ is the root of $\F{t}$, then we use the same construction, but we have no predecessor $z$ of $v$ and directly start with the node $w_h$ as the new root.
	Hence, we have to use the edge relation 
	\[E^{\F{t}_h} \coloneqq (E \cap (w_h(E^\F{t})^* \times w_h(E^\F{t})^*))\]
	and the rest stays the same.
	In the end, we set 
	\[
		M' \coloneqq M \setminus \{(p_\F{t},\F{t})\} \cup \{(p_\F{t} \cdot p_1,\F{t}_1), \ldots, (p_\F{t} \cdot p_m,\F{t}_m)\}
	\]
	It remains to prove that our properties (a), (b), and (c) are still satisfied for $M'$.
	\begin{itemize}
    \item[(a)]
		Let $1 \leq h \leq m$.
		We have to prove that $\F{t}_h$ is a $(\PP,\SSS)$-chain tree.
		We constructed $\F{t}_h$ by removing the node $v$ in $\F{t}$ and directly moving from $z$ to $w_h$ (or starting with $w_h$ if $v$ was the root node).
		Hence, $(V^{\F{t}_h}, E^{\F{t}_h})$ is still a finitely branching, directed tree.
		Let $x \in V^{\F{t}_h}$ with $xE^{\F{t}_h} \neq \emptyset$.
		If $x \in w_h(E^\F{t})^*$, then $xE^{\F{t}_h} = xE^{\F{t}}$ and thus
		\[
			\sum_{y \in xE^{\F{t}_h}} p_y^{\F{t}_h}
			= \sum_{y \in xE^{\F{t}_h}} \frac{1}{p_h} \cdot p_{y}^{\F{t}}
			= \frac{1}{p_h} \cdot \sum_{y \in xE^{\F{t}_h}} p_{y}^{\F{t}}
			= \frac{1}{p_h} \cdot \sum_{y \in xE^{\F{t}}} p_{y}^{\F{t}}
			= \frac{1}{p_h} \cdot p_{x}^{\F{t}}
			= p_x^{\F{t}_h}
		\]
		If $v$ was not the root and $x = z$, then $zE^{\F{t}_h} = (zE^{\F{t}} \setminus \{v\}) \cup \{w_h\}$ and thus
		\[\begin{array}{rclcl}
			\sum\limits_{y \in zE^{\F{t}_h}} p_y^{\F{t}_h}
			&=& \sum\limits_{y \in (zE^{\F{t}} \setminus \{v\}) \cup \{w_h\}} p_{y}^{\F{t}_h}
			&=& \sum\limits_{y \in (zE^{\F{t}} \setminus \{v\})} p_{y}^{\F{t}_h} + p_{w_h}^{\F{t}_h}\\
			&=& \sum\limits_{y \in (zE^{\F{t}} \setminus \{v\})} p_{y}^{\F{t}} + \frac{1}{p_h} \cdot p_{w_h}^{\F{t}}
			&=& \sum\limits_{y \in (zE^{\F{t}} \setminus \{v\})} p_{y}^{\F{t}} + \tfrac{p_{v}^{\F{t}}}{p_{w_h}^{\F{t}}} \cdot p_{w_h}^{\F{t}}\\
			& =& \sum\limits_{y \in (zE^{\F{t}} \setminus \{v\})} p_{y}^{\F{t}} + p_{v}^{\F{t}}
			&=& \sum\limits_{y \in zE^{\F{t}}} p_{y}^{\F{t}}
			\; = \; p_z^{\F{t}}
		\end{array}\]
		Otherwise, we have $x \in V^{\F{t}} \setminus (v(E^\F{t})^* \cup \{z\})$.
		This means $p_y^{\F{t}_h} = p_y^{\F{t}}$ for all $y \in xE^{\F{t}_h}$ and $xE^{\F{t}_h} = xE^{\F{t}}$, and thus
		\[
			\sum_{y \in xE^{\F{t}_h}} p_y^{\F{t}_h} 
			= \sum_{y \in xE^{\F{t}_h}} p_{y}^{\F{t}} 
			= \sum_{y \in xE^{\F{t}}} p_{y}^{\F{t}} 
			= p_x^{\F{t}}
			= p_x^{\F{t}_h}
		\]
		For the last property, note that if $v$ is not the root in $\F{t}$, then the root and its labeling did not change, so that we have $p_{\ctroot^{\F{t}_h}}^{\F{t}_h} = p_{\ctroot^{\F{t}}}^{\F{t}} = 1$.
		If $v$ was the root, then $w_h$ is the new root with 
		\[
			p_{w_h}^{\F{t}_h} = \frac{1}{p_h} \cdot p_{w_h}^{\F{t}} = \tfrac{p_v^{\F{t}}}{p_{w_h}^{\F{t}}} \cdot p_{w_h}^{\F{t}} = \tfrac{1}{p_{w_h}^{\F{t}}} \cdot p_{w_h}^{\F{t}} = 1
		\]

		\item[(b)]
		We have
		\[
			\begin{array}{cl}
				  & \sum_{(p_{T},T) \in M'} p_T \\
				= & \sum_{(p_{T},T) \in M \setminus \{(p_\F{t},\F{t})\}} p_T + \sum_{(p_{T},T) \in \{(p_\F{t} \cdot p_1,\F{t}_1), \ldots, (p_\F{t} \cdot p_m,\F{t}_m)\}} p_T \\
				= & \sum_{(p_{T},T) \in M \setminus \{(p_\F{t},\F{t})\}} p_T + \sum_{1 \leq j \leq m} p_\F{t} \cdot p_j \\
				= & \sum_{(p_{T},T) \in M \setminus \{(p_\F{t},\F{t})\}} p_T + p_\F{t} \cdot \sum_{1 \leq j \leq m} p_j \\
				= & \sum_{(p_{T},T) \in M \setminus \{(p_\F{t},\F{t})\}} p_T + p_\F{t} \cdot 1 \hspace*{1cm} \text{(by \eqref{p_h sum is one})}\\ 
				= & \sum_{(p_{T},T) \in M \setminus \{(p_\F{t},\F{t})\}} p_T + p_\F{t}\\
				= & \sum_{(p_{T},T) \in M} p_T\\
				\stackrel{IH}{=} & 1
			\end{array}
		\]
		
		\item[(c)] 
		Let $x \in V_i$.
		If we have $x \not\in V^{\F{t}}$, then also $x \not\in V^{\F{t}_h}$ for all $1 \leq h \leq m$ and thus
		\[
			\begin{array}{cl}
				& \sum_{(p_T,T) \in M'} p_T \cdot \overline{p_{(T,x)}} \\
				= & \sum_{(p_T,T) \in M \setminus \{(p_\F{t},\F{t})\} \cup \{(p_\F{t} \cdot p_1,\F{t}_1), \ldots, (p_\F{t} \cdot p_m,\F{t}_m)\}} p_T \cdot \overline{p_{(T,x)}} \\
				= & \sum_{(p_{T},T) \in M \setminus \{(p_\F{t},\F{t})\}} p_T \cdot \overline{p_{(T,x)}} + \sum_{(p_{T},T) \in \{(p_\F{t} \cdot p_1,\F{t}_1), \ldots, (p_\F{t} \cdot p_m,\F{t}_m)\}} p_T \cdot \overline{p_{(T,x)}} \\
				= & \sum_{(p_{T},T) \in M \setminus \{(p_\F{t},\F{t})\}} p_T \cdot \overline{p_{(T,x)}} + \sum_{(p_{T},T) \in \{(p_\F{t} \cdot p_1,\F{t}_1), \ldots, (p_\F{t} \cdot p_m,\F{t}_m)\}} p_T \cdot 0 \\
				= & \sum_{(p_T,T) \in M \setminus \{(p_\F{t},\F{t})\}} p_T \cdot \overline{p_{(T,x)}} \\
				= & \sum_{(p_T,T) \in M \setminus \{(p_\F{t},\F{t})\}} p_T \cdot \overline{p_{(T,x)}} + p_{\F{t}} \cdot 0\\
				= & \sum_{(p_T,T) \in M \setminus \{(p_\F{t},\F{t})\}} p_T \cdot \overline{p_{(T,x)}} + p_{\F{t}} \cdot \overline{p_{(\F{t},x)}}\\
				= & \sum_{(p_T,T) \in M} p_T \cdot \overline{p_{(T,x)}}\\
				\stackrel{IH}{=} & p_x^{\F{T}_i}
			\end{array}
		\]
		If we have $x \in V^{\F{t}}$ and $x \not\in v(E^{\F{t}})^*$, then $x \in V^{\F{t}_h}$ for all $1 \leq h \leq m$ and $p_x^{\F{t}} = p_x^{\F{t}_h}$, and hence
		\[
			\begin{array}{cl}
				& \sum_{(p_T,T) \in M'} p_T \cdot \overline{p_{(T,x)}} \\
				= & \sum_{(p_T,T) \in M \setminus \{(p_\F{t},\F{t})\} \cup \{(p_\F{t} \cdot p_1,\F{t}_1), \ldots, (p_\F{t} \cdot p_m,\F{t}_m)\}} p_T \cdot \overline{p_{(T,x)}} \\
				= & \sum_{(p_T,T) \in M \setminus \{(p_\F{t},\F{t})\}} p_T \cdot \overline{p_{(T,x)}} + \sum_{(p_T,T) \in \{(p_{\F{t}} \cdot p_1,{\F{t}}_1), \ldots, (p_{\F{t}} \cdot p_m,{\F{t}}_m)\}} p_T \cdot \overline{p_{(T,x)}} \\
				= & \sum_{(p_T,T) \in M \setminus \{(p_\F{t},\F{t})\}} p_T \cdot \overline{p_{(T,x)}} + \sum_{1 \leq h \leq m} p_{\F{t}} \cdot p_h \cdot p_x^{\F{t}_h} \\
				= & \sum_{(p_T,T) \in M \setminus \{(p_\F{t},\F{t})\}} p_T \cdot \overline{p_{(T,x)}} + \sum_{1 \leq h \leq m} p_{\F{t}} \cdot p_h \cdot p_x^{\F{t}} \\
				= & \sum_{(p_T,T) \in M \setminus \{(p_\F{t},\F{t})\}} p_T \cdot \overline{p_{(T,x)}} + p_{\F{t}} \cdot p_x^{\F{t}} \cdot \sum_{1 \leq h \leq m} p_h\\
				= & \sum_{(p_T,T) \in M \setminus \{(p_\F{t},\F{t})\}} p_T \cdot \overline{p_{(T,x)}} + p_{\F{t}} \cdot p_x^{\F{t}} \cdot 1\\
				= & \sum_{(p_T,T) \in M} p_T \cdot \overline{p_{(T,x)}}\\
				\stackrel{IH}{=} & p_x^{\F{T}_i}
			\end{array}
		\]
		Otherwise we have $x \in V^{\F{t}}$ and $x \in v(E^{\F{t}})^*$ and thus,
$x \in v(E^{\F{t}})^+$.
		This means that we have $x \in V^{\F{t}_h}$ and $x \in w_h(E^{\F{t}})^*$ for some $1 \leq h \leq m$ and $x \not\in V^{\F{t}_g}$ for all $g \neq h$.
		Furthermore, we have $p_x^{\F{t}_h} = \frac{1}{p_h} \cdot p_x^{\F{t}}$, and hence
		\[
			\begin{array}{cl}
				& \sum_{(p_T,T) \in M'} p_T \cdot \overline{p_{(T,x)}} \\
				= & \sum_{(p_T,T) \in M \setminus \{(p_\F{t},\F{t})\} \cup \{(p_\F{t} \cdot p_1,\F{t}_1), \ldots, (p_\F{t} \cdot p_m,\F{t}_m)\}} p_T \cdot \overline{p_{(T,x)}} \\
				= & \sum_{(p_T,T) \in M \setminus \{(p_\F{t},\F{t})\}} p_T \cdot \overline{p_{(T,x)}} + \sum_{(p_T,T) \in \{(p_\F{t} \cdot p_1,\F{t}_1), \ldots, (p_\F{t} \cdot p_m,\F{t}_m)\}} p_T \cdot \overline{p_{(T,x)}} \\
				= & \sum_{(p_T,T) \in M \setminus \{(p_\F{t},\F{t})\}} p_T \cdot \overline{p_{(T,x)}} + p_\F{t} \cdot p_h \cdot p_x^{\F{t}_h} \\
				= & \sum_{(p_T,T) \in M \setminus \{(p_\F{t},\F{t})\}} p_T \cdot \overline{p_{(T,x)}} + p_\F{t} \cdot p_h \cdot \frac{1}{p_h} \cdot p_x^{\F{t}} \\
				= & \sum_{(p_T,T) \in M \setminus \{(p_\F{t},\F{t})\}} p_T \cdot \overline{p_{(T,x)}} + p_\F{t} \cdot p_x^{\F{t}}\\
				= & \sum_{(p_T,T) \in M} p_T \cdot \overline{p_{(T,x)}}\\
				\stackrel{IH}{=} & p_x^{\F{T}_i}
			\end{array}
		\]
	\end{itemize}

	In the end, we result in a set $M$ where all contained trees do not have any split-node anymore.
	Then $M$ is our desired set and we define $\text{Split}(\F{T}_i) \coloneqq M$.

  \medskip
			
  \noindent
  \underline{\textbf{Induction Step Continued for $\F{F}$}}

  \noindent
  Remember that in the tree $\F{F}$, we have a node in the $(i+1)$-th depth for every tree $T$ in $\text{Split}(\F{T}_i)$ such that $\sum_{x \in \ctleaf^{T}, x \in \ctleaf^{\F{T}}} p_x^{T} \leq c$.
In $\F{F}$, we draw an edge from a node $X$ in depth $i$ to the node $Y$ in depth $i+1$ if the corresponding trees are the same, if the tree for node $Y$ is the result of the tree for node $X$ if we add the successors of $x_i$, or if the tree for node $Y$ is the result of the tree for node $X$ if we split at node $x_i$, i.e., if we remove $x_i$ and directly move to one of its successors.
	A node $X$ can only have a finite number of direct successors in the tree $\F{F}$
        since the node $x_i$ can only have a finite number of direct successors in $\F{T}$ so that the split at node $x_i$ can only result in a finite number of possible trees.
	If $x_i$ is not a split-node, then there is a unique successor for each tree.
	Hence, $\F{F}$ is finitely branching.

	We now prove that there exists a node in every depth of the tree $\F{F}$.
	Let $i \in \IN$.
	We have to show that there exists a pair $(p_T,T) \in \text{Split}(\F{T}_i)$ such that $\sum_{x \in \ctleaf^{T}, x \in \ctleaf^{\F{T}}} p_x^{T} \leq c$.
	Assume for a contradiction that $\sum_{x \in \ctleaf^{T}, x \in \ctleaf^{\F{T}}} p_x^{T} > c$ holds for all $(p_T,T) \in \text{Split}(\F{T}_i)$.
	Then we would have
	\[
		\begin{array}{cl}
			  & \sum_{x \in \ctleaf^{\F{T}_i}, x \in \ctleaf^{\F{T}}} p_x^{\F{T}_i} \\
			= & \sum_{x \in \ctleaf^{\F{T}_i}, x \in \ctleaf^{\F{T}}} \sum_{(p_{T},T) \in \text{Split}(\F{T}_i)} p_T \cdot \overline{p_{(T,x)}}\\
        & \hspace*{1cm} \text{(by (C) and since a leaf in $\F{T}_i$ cannot be a split-node in $\F{T}_i$)}\\
			= & \sum_{(p_{T},T) \in \text{Split}(\F{T}_i)} \sum_{x \in \ctleaf^{\F{T}_i}, x \in \ctleaf^{\F{T}}} p_T \cdot \overline{p_{(T,x)}} \\
			= & \sum_{(p_{T},T) \in \text{Split}(\F{T}_i)} p_T \cdot \sum_{x \in \ctleaf^{\F{T}_i}, x \in \ctleaf^{\F{T}}} \overline{p_{(T,x)}} \\
			= & \sum_{(p_{T},T) \in \text{Split}(\F{T}_i)} p_T \cdot \sum_{x \in \ctleaf^{T}, x \in \ctleaf^{\F{T}}} p_x^T \\
			> & \sum_{(p_{T},T) \in \text{Split}(\F{T}_i)} p_T \cdot c \hspace*{1cm} \text{(as $\sum_{x \in \ctleaf^{T}, x \in \ctleaf^{\F{T}}} p_x^{T} > c$)} \\
			= & c \cdot \sum_{(p_{T},T) \in \text{Split}(\F{T}_i)} p_T \\
			= & c \cdot 1 \hspace*{1cm} \text{(by (B))} \\
			= & c
		\end{array}
	\]
	But this is a contradiction since this means 
	\[
		c < \sum_{x \in \ctleaf^{\F{T}_i}, x \in \ctleaf^{\F{T}}} p_x^{\F{T}_i} \leq \sum_{x \in \ctleaf^{\F{T}}} p_x^{\F{T}} = c
	\]
	Hence, there exists a node in every depth of the tree $\F{F}$.
	This means that $\F{F}$ is an infinite and finitely branching tree, so there must exist an infinite path by König's Lemma.

	Finally, we show that this infinite path corresponds to a $(\PP, \SSS)$-chain tree $\F{T}_{\lim}$ without splits such that $\ctleaf^{\F{T}_{\lim}} \subseteq \ctleaf^{\F{T}}$.
	Let $\F{T}_1, \F{T}_2, \ldots$ be the finite $(\PP, \SSS)$-computation trees without splits of the nodes in the infinite path in $\F{F}$.
	We define the tree $\F{T}_{\lim}$ by $\F{T}_{\lim} \coloneqq \lim_{i \to \infty} \F{T}_i$.
	Note that there is some natural number $N_j \in \IN$ for every $j \in \IN$, such
        that everything above
          the $j$-th depth is the same for all trees $\F{T}_a$ with $a > N_j$.
	Hence, the limit exists and all local properties for a $(\PP, \SSS)$-chain tree are satisfied for $\F{T}_{\lim}$.
	For the only global property regarding the infinite paths in $\F{T}_{\lim}$, note that every infinite path in $\F{T}_{\lim}$ corresponds to an infinite path in $\F{T}$ that is only missing split-nodes.
	Since split-nodes are contained in $\SSS$, we can be sure that every infinite path in $\F{T}_{\lim}$ must contain an infinite number of $P$ nodes.

Finally, we have to show that $\ctleaf^{\F{T}_{\lim}} \subseteq \ctleaf^{\F{T}}$.
	Let $x \in \ctleaf^{\F{T}_{\lim}}$.
	The depth of $x$ in
        $\F{F}$ is  some $j \in \IN$ and by our previous observation, there must be an $N_j \in \IN$ such that $x \in \ctleaf^{\F{T}_{a}}$ for all $a > N_j$.
	But this is only possible if $x$ is already a leaf in $\F{T}$ because otherwise, we would add one of its successors into the tree afterwards, or we would skip the node $x$ if it is a split-node.
	Thus we have $\ctleaf^{\F{T}_{\lim}} \subseteq \ctleaf^{\F{T}}$ and this ends the proof.
\end{myproof}

\begin{lemma}[Starting Lemma]\label{lemma:starting}
  If a DP problem $(\PP, \SSS)$ is not iAST, then there exists a $(\PP, \SSS)$-chain tree
  $\F{T}$ with $|\F{T}|_{\ctleaf} < 1$ that starts with $(1:\Com{1}(s^\#))$, where
  $s^\# = \ell^\# \sigma$ for some substitution $\sigma$ and some dependency tuple $\langle
  \ell^\#,\ell \rangle \to \{p_1: \langle d_1,r_1 \rangle, \ldots, p_k: \langle d_k,r_k \rangle\} \in \PP$, and every proper subterm of $\ell^\# \sigma$ is in normal form w.r.t.\ $\SSS$.
\end{lemma}

\begin{myproof}
	We prove the contraposition.
        Assume that every $(\PP, \SSS)$-chain tree $\F{T}$
converges with probability $1$ if it
 starts with $(1:\Com{1}(s^\#))$ and $s^\# = \ell^\# \sigma$ for some substitution
 $\sigma$, some dependency tuple $\langle \ell^\#,\ell \rangle \to \{p_1: \langle d_1,r_1
 \rangle, \ldots, p_k:\langle d_k,r_k \rangle\} \in \PP$, and every proper subterm of
 $\ell^\# \sigma$ is in normal form w.r.t.\ $\SSS$.  
	We now prove that then also every $(\PP, \SSS)$-chain tree $\F{T}$ 
	that starts with $(1:t)$ for some arbitrary term $t$ converges with probability $1$, 
	and thus $(\PP, \SSS)$ is iAST\@.
  Note that we can only apply rewrite rules with $\setitops$ to a term $t$ if after
  normalization,
  $t$ has the form $t = \Com{n}(s_1^\#,\ldots,s_n^\#)$.
	Hence, we prove the claim by structural induction on $|\cont(t)|$.

  For $|\cont(t)| = 0$, we  have $t = \Com{0}$, which is a normal form. So a chain tree
  starting with $(1:t)$ is trivially finite and hence, it converges with probability $1$.
    If $|\cont(t)| = 1$, then we have $t = \Com{1}(s_1^\#)$.
  If $s_1^\#$ is in normal form, then the claim is again trivial. 
  Otherwise, let $\F{T} = (V,E,L,P)$ be a chain tree that starts with $(1:t)$ and let $F$ be the set of all nodes $v \in P$ such that no proper ancestor of $v$ is in $P$.
  Before a node from $F$ is reached, only $\setitos$-steps are performed, so all these nodes are labeled with terms of the form $\Com{1}(\ldots)$.
  Hence, the nodes that do not have a proper ancestor in $F$ form a finitely branching tree without infinite paths (by Condition 6 of \cref{def:chain-tree}), which is finite by König's Lemma.
  Every $v \in F$ is labeled with $(p_v:\Com{1}(\ell^\# \sigma))$ for some dependency tuple $\langle
  \ell^\#,\ell \rangle \to \mu \in \PP$ and some substitution 
	$\sigma$ such that every proper subterm of $\ell^\# \sigma$ is in normal form w.r.t.\ $\SSS$.
  Then we know by our assumption that the sub chain tree $\F{T}[vE^*]$ that starts at $v$ converges with probability $1$.
  So $|\F{T}|_{\ctleaf}$ is the sum of the probabilities $p_v$ of the leaves $v$ of this finite tree (as $|\F{T}[vE^*]|_{\ctleaf} = 1$ for all $v \in F$), which is $1$.

	Now we regard the induction step, and assume that for $t =
        \Com{n}(s_1^\#,\ldots,s_n^\#)$ with $n > 1$, there is a chain tree $\F{T}$ that
        converges with probability $< 1$.
	Here, our induction hypothesis is that every $(\PP, \SSS)$-chain tree $\F{T}$ that starts with $(1:t')$ for some term $t'$ with $|\cont(t')| < n$ converges with probability $1$.
	Now let $1 \leq m < n$ and consider the two terms $\Com{m}(s_1^\#,\ldots,s_m^\#)$ and $\Com{n-m}(s_{m+1}^\#,\ldots,s_n^\#)$.
  By our induction hypothesis, we know that every $(\PP, \SSS)$-chain tree that starts with $(1:\Com{m}(s_1^\#,\ldots,s_m^\#))$ or  $(1:\Com{n-m}(s_{m+1}^\#,\ldots,s_n^\#))$ converges with probability $1$.
  We now use this to prove that every $(\PP, \SSS)$-chain tree $\F{T}$ that starts with $(1:\Com{n}(s_1^\#,\ldots,s_n^\#))$  converges with probability $1$ as well.
  Let $\F{T}_1$ be the tree that starts with $\Com{m}(s_1^\#,\ldots,s_m^\#)$ and uses the same rules as we did in $\F{T}$.
  If the rule that we used in $\F{T}$ is not applicable (because the terms originated from
  the second part $\Com{n-m}(s_{m+1}^\#,\ldots,s_n^\#)$), then we use a split-node
with the same probabilities as in the multi-distribution on the right-hand side of the
rule used in $\F{T}$ in order 
  to copy the tree structure.
  Analogously, let $\F{T}_2$ be the tree that starts with $\Com{n-m}(s_{m+1}^\#,\ldots,s_n^\#)$ and uses the same rules as we did in $\F{T}$.

  We now do the following:
	\begin{itemize}
    \item[1)] Partition the set $P$ into the sets $P_1$ and $P_2$ according to whether the
      rewrite step with the first component $\projOne(\PP)$ takes place in $\F{T}_1$ or $\F{T}_2$.
    \item[2)] Use the P-Partition Lemma (\Cref{lemma:p-partition}) to obtain a
      $(\PP,\SSS)$-chain tree $\F{T}'$ with $|\F{T}'|_{\ctleaf} < 1$ where every infinite
      path has an infinite number of $P_2$ nodes.
    \item[3)] Create a $(\PP, \SSS)$-chain tree $\F{T}_2'$ that starts with $(1:\Com{n-m}(s_{m+1}^\#,\ldots,s_n^\#))$ and $|\F{T}_2'|_{\ctleaf} < 1$.
	\end{itemize}
  The last step is a contradiction since we must have $|\F{T}_2'|_{\ctleaf} = 1$ by our induction hypothesis.
  
  \medskip
			
  \noindent
  \textbf{\underline{1) Partition the set $P$}}

  \noindent
  We can partition $P$ into the sets
  \[ 
    \begin{array}{lcrl}
      P_1 &:=& \{x \in P \mid &\text{ the rewrite step at node } x \text{ with the first
        component $\projOne(\PP)$}\\
      &&&\text{ takes place in } \F{T}_1 \}\\
      P_2 &:=& \{x \in P \mid &\text{ the rewrite step at node } x \text{ with the first
        component $\projOne(\PP)$}\\
      &&&\text{ takes place in } \F{T}_2 \}
    \end{array}
  \]

  First, note that the first component of a dependency tuple can only be used on a single argument of the compound symbol.
  Hence, this is really a partition.
  Second, note that every local property of a $(\PP, \SSS)$-chain tree (i.e., properties
  1-5 of \Cref{def:chain-tree}) is satisfied for $\F{T}_1$ and $\F{T}_2$.
  However, due to the replacement of $P$ steps by splits, there might exist infinite paths in $\F{T}_1$ (or $\F{T}_2$) that do not use an infinite number of $P_1$ (or $P_2$) nodes anymore.
  Hence, $\F{T}_1$ and $\F{T}_2$ might not be $(\PP, \SSS)$-chain trees anymore as the
  global property 6 of \Cref{def:chain-tree} might not be satisfied.
  However, every  sub chain tree of $\F{T}_1$ such that every infinite path has an infinite number of $P_1$ nodes is a $(\PP, \SSS)$-chain tree with splits.
  This also includes every finite  sub chain tree as there does not exist any infinite path in such trees.

  \medskip
    
  \noindent
  \textbf{\underline{2) Use the P-Partition Lemma}}

  \noindent
  In order to use the P-Partition Lemma for the tree $\F{T}$, we have to show that every  sub chain tree $\F{T}'_1$ of $\F{T}$ that only contains $P$ nodes from $P_1$ converges with probability $1$.
  Let $\F{T}'_1 = (V',E',L',P')$ be a sub chain tree of $\F{T}$ that does not contain nodes from $P_2$.
  There exists a set $W$ satisfying the conditions of \cref{def:chain-tree-induced-sub} such that $\F{T}'_1 = \F{T}[W]$.
      Let $w \in W$ be the root of $\F{T}'_1$.
  Since $\F{T}$ and $\F{T}_1$ have the same tree structure,  $\F{T}_1[W]$ is a sub chain tree of $\F{T}_1$ with $\F{T}_1[W] = \F{T}'_1$.
  Moreover, $\F{T}_1[W]$ is a $(\PP, \SSS)$-chain tree, since the set $W$ does not contain
  any inner nodes from $P_2$.
  We now show that $\F{T}_1[W]$ is a  sub chain tree of a $(\PP, \SSS)$-chain tree with
  splits that starts with
  $(1:\Com{m}(s_1^\#,\ldots,s_m^\#))$ and thus, we know that $\F{T}_1[W]$ must be
  converging with probability $1$ as well by our induction hypothesis.
(Due to \Cref{lemma:splitting}, we can also apply the induction hypothesis to chain trees
  with splits.)

  Let $(y_1, \ldots y_e)$ with $y_e = w$ be the path from the root to $w$ in
  $\F{T}_1$.
  The tree we are looking for is $T \coloneqq \F{T}_1[\bigcup_{1 \leq j \leq e} y_jE \cup W]$.
  This is a grounded sub chain tree.
  Every infinite path in $T$ must visit the node $w$ and corresponds to an infinite path in $\F{T}_1[W]$ so that it has an infinite number of $P_1$ nodes.
  Furthermore, $\bigcup_{1 \leq j \leq e} y_jE \cup W$ satisfies the conditions of
  \cref{def:chain-tree-induced-sub}, as the set is weakly connected and it always contains
  all successors or no successor at all.

  Now, we have shown that the conditions for the P-Partition Lemma (\cref{lemma:p-partition}) are satisfied.
  We can now apply the P-Partition Lemma to get a grounded sub chain tree
    $\F{T}'$ of $\F{T}$ with $|\F{T}'|_{\ctleaf} < 1$ such that on every infinite path, we have an infinite number of $P_2$ nodes.

  \medskip
    
  \noindent
  \textbf{\underline{3) Create a $(\PP, \SSS)$-chain tree $\F{T}_2'$}}

  \noindent
  Again, let $\F{T}_2'$ be the tree that starts with $\Com{n-m}(s_{m+1}^\#,\ldots,s_n^\#)$
  and uses the same rules as we did in $\F{T}'$.
If the rules in $\F{T}'$ are not applicable, then we again use a split-node  to copy the
tree structure.
  
  Again, all local properties for a $(\PP, \SSS)$-chain tree with splits are satisfied.
  Additionally, this time we know that every infinite path has an infinite number of $P_2$ nodes in $\F{T}'$, hence we also know that the global property for $\F{T}_2'$ is satisfied.
  This means that $\F{T}_2'$ is a $(\PP, \SSS)$-chain tree with splits that starts with
  $\Com{n-m}(s_{m+1}^\#,\ldots,s_n^\#)$ and with $|\F{T}_2'|_{\ctleaf} < 1$.
  By \Cref{lemma:splitting}, we can also obtain a corresponding $(\PP, \SSS)$-chain tree without
  splits that starts with
  $\Com{n-m}(s_{m+1}^\#,\ldots,s_n^\#)$
and converges with probability $< 1$.
This is our desired contradiction, which proves the induction step.
\end{myproof}

\ProbDepGraphProc*

\begin{myproof}
  \smallskip
 
         \noindent
         \underline{\emph{Completeness}}
 
         \noindent    
        Every $(\PP_i, \SSS)$-chain tree is also a $(\PP,\SSS)$-chain tree.
        Hence, if $(\PP_i, \SSS)$ is not iAST, then $(\PP,\SSS)$ is also not iAST\@.
   \medskip
 
         \noindent
         \underline{\emph{Soundness}}
 
         \noindent
  Let $\F{G}$ be the $(\PP, \SSS)$-dependency graph.
  Suppose that every $(\PP_i,\SSS)$-chain tree converges with probability $1$ for all $1 \leq i \leq n$.
  We prove that then also every $(\PP,\SSS)$-chain tree converges with probability 1.
  Let $W = \{\PP_1, \ldots, \PP_n\} \cup \{\{v\} \subseteq \PP \mid v$ is not in an SCC  of $\F{G}\}$ be the set of all SCCs and all singleton sets of nodes that do not belong to any SCC\@.
  The core steps of this proof are the following:
  \begin{enumerate}
    \item[1.] We show that every DP problem $(X, \SSS)$ with $X \in W$ is iAST\@.
    \item[2.] We show that composing SCCs maintains the iAST property.
    \item[3.] We show that for every $X \in W$, the DP problem $(\bigcup_{X >_{\F{G}}^* Y}
      Y, \SSS)$ is iAST by induction on $>_{\F{G}}$.
    \item[4.] We conclude that $(\PP,\SSS)$ must be iAST\@.
  \end{enumerate}
  Here, for two $X_1,X_2 \in W$ we say that $X_2$ is a direct successor of $X_1$ (denoted
  $X_1 >_{\F{G}} X_2$) if there exist nodes $v \in X_1$ and $w \in X_2$ such that there is
  an edge from $v$ to $w$ in $\F{G}$.

  \medskip

  \noindent
  \textbf{\underline{1. Every DP problem $(X, \SSS)$ with $X \in W$ is iAST\@.}}

  \noindent
  We start by proving the following:
  \begin{equation}
    \label{W is iAST}
    \mbox{Every DP problem $(X, \SSS)$ with $X \in W$ is iAST\@.}
  \end{equation}
  To prove~\eqref{W is iAST}, note that if $X$ is an SCC, then it follows from our assumption that $(X,\SSS)$ is iAST\@.
  If $X$ is a singleton set of a node that does not belong to any SCC,
                    then assume for a contradiction that $(X,\SSS)$ is not iAST\@.
  By \cref{lemma:starting} there exists an $(X,\SSS)$-chain tree $\F{T} = (V,E,L,P)$ that
  converges with a probability $<1$ and starts with $(1:\Com{1}(s^\#))$ such that $s^\# =
  \ell^\# \sigma$ for some substitution $\sigma$ and the only dependency tuple
  $\langle \ell^\#,\ell \rangle \to \{p_1:\langle d_1,r_1 \rangle, \ldots, p_k: \langle
  d_k,r_k \rangle\} \in X$, and every proper subterm of $\ell^\# \sigma$ is in normal form w.r.t.\ $\SSS$.
  Assume for a contradiction that there exists a node $x \in P$ in $\F{T}$ that is not the root.
  W.l.o.G., let $x$ be reachable from the root without traversing any other node from $P$.
  This means that for the corresponding term $t_x = \Com{n}(s_1'^\#, \ldots, s_n'^\#)$ for
  node $x$ there is an $1 \leq i' \leq n$ such that $s_{i'}'^\# = \ell^\# \sigma'$ for
  some substitution $\sigma'$ and the only dependency tuple $\langle \ell^\#,\ell \rangle
  \to \{p_1: \langle d_1,r_1 \rangle, \ldots, p_k: \langle d_k,r_k \rangle\} \in X$, and every proper subterm of $\ell^\# \sigma'$ is in normal form w.r.t.\ $\SSS$.
  Let $(z_0, \ldots, z_m)$ with $z_m = x$ be the path from the root to $x$ in $\F{T}$.
  The first rewrite step at the root must be $\Com{1}(s^\#) \setitops \{p_1:d_1 \sigma,
  \ldots, p_k:d_k \sigma\}$ so that $t_{z_1} = d_j \sigma =\Com{n}(s_1^\# \sigma, \ldots, s_n^\#
  \sigma)$ for some $1 \leq j
  \leq k$ where $d_j = \Com{n}(s_1^\#, \ldots, s_n^\#)$.
  After that, we only use $\setitos$ steps in the path since all of the nodes $z_{1}, \ldots, z_{m-1}$ are contained in $S$.
  Note that rewriting with $\setitos$ cannot add new arguments to the compound symbol but can only rewrite the existing ones.
  Therefore, we must have an $1 \leq i \leq n$ such that $s_i^\# \sigma \itononprobsstar
  s_{i'}'^\# = \ell^\# \sigma'$, which means that there must be a self-loop for the only
  dependency tuple in $X$, which is a contradiction to our assumption that $X$ is a
  singleton consisting of a dependency tuple that is not on any SCC of $\F{G}$.

  Now, we have proven that the $(X, \SSS)$-chain tree $\F{T}$ does not contain a node $x \in P$ that is not the root.
  By definition of a $(\PP, \SSS)$-chain tree, every infinite path must contain an infinite number of nodes in $P$.
  Thus, every path in $\F{T}$ must be finite, which means that $\F{T}$ is finite itself.
  But every finite chain tree converges with probability $1$, which is a contradiction to
  our assumption that $\F{T}$
  converges with probability $<1$. 

  \medskip

  \noindent
  \textbf{\underline{2. Composing SCCs maintains the iAST property.}}

  \noindent
  Next, we show that composing SCCs maintains the iAST property. More precisely, we prove
  the following:
  \begin{equation}
  \label{Composing iAST}
    \parbox{.8\textwidth}{Let $\overline{X} \subseteq W$ and $\overline{Y} \subseteq W$
      such that there are no $X_1,X_2 \in \overline{X}$ and $Y \in \overline{Y}$ which
      satisfy both $X_1 >_{\F{G}}^* Y >_{\F{G}}^* X_2$ and $Y \not\in \overline{X}$, and such that there are no $Y_1,Y_2 \in \overline{Y}$ and $X \in \overline{X}$ which
      satisfy both $Y_1 >_{\F{G}}^* X >_{\F{G}}^* Y_2$ and $X \not\in \overline{Y}$.
    If both $ (\bigcup_{X \in \overline{X}} X,\SSS) $ and $ (\bigcup_{Y \in \overline{Y}} Y, \SSS) $ are iAST, then $ (\bigcup_{X \in \overline{X}} X \cup \bigcup_{Y \in \overline{Y}} Y, \SSS) $ is iAST as well.}
  \end{equation}
  To show~\eqref{Composing iAST}, we assume that both $(\bigcup_{X \in \overline{X}} X,\SSS)$ and $(\bigcup_{Y \in \overline{Y}} Y, \SSS)$ are iAST\@.
  Let $Z = \bigcup_{X \in \overline{X}} X \cup \bigcup_{Y \in \overline{Y}} Y$.
  The property in~\eqref{Composing iAST} for $\overline{X}$ and $\overline{Y}$ says that a
  path between two nodes from $\bigcup_{X \in \overline{X}} X$ that only traverses nodes
  from $Z$ must also be a path that only traverses
  nodes from $\bigcup_{X \in \overline{X}} X$, so that $\bigcup_{Y \in \overline{Y}} Y$ cannot be used to ``create'' new paths between two nodes from $\bigcup_{X \in \overline{X}} X$, and vice versa.
  Assume for a contradiction that $(Z, \SSS)$ is not iAST\@.
  By \cref{lemma:starting} there exists a $(Z,\SSS)$-chain tree $\F{T} = (V,E,L,P)$ that
  converges with probability $<1$ and starts with $(1:\Com{1}(s^\#))$ such that $s^\# =
  \ell^\# \sigma$ for some substitution $\sigma$, a dependency tuple $\langle \ell^\#,\ell
  \rangle \to \{p_1: \langle d_1,r_1 \rangle, \ldots, p_k: \langle d_k,r_k \rangle\} \in Z$, and every proper subterm of $\ell^\# \sigma$ is in normal form w.r.t.\ $\SSS$.
  
  W.l.o.G., we may assume that the dependency tuple that is used for the rewrite step at
  the root is in
  $\bigcup_{X \in \overline{X}} X$.
  Otherwise, we simply swap $\bigcup_{X \in \overline{X}} X$ with $\bigcup_{Y \in \overline{Y}} Y$.

  We can partition the set $P$ of our $(Z,\SSS)$-chain tree $\F{T}$ into the sets
  \begin{itemize}
    \item[$\bullet$] $P_1 := \{x \in P \mid x$ together with the labeling and its successors represents a step with a dependency tuple from $\bigcup_{X \in \overline{X}} X\}$
    \item[$\bullet$] $P_2 := P \setminus P_1$
  \end{itemize}
  Note that in the case of $x \in P_2$, we know that $x$ together with its successors and the labeling represents a step with a dependency tuple from $\bigcup_{Y \in \overline{Y}} Y$ that is not in $\bigcup_{X \in \overline{X}} X$.
  We know that every $(\bigcup_{Y \in \overline{Y}} Y,\SSS)$-chain tree converges with probability $1$, since $(\bigcup_{Y \in \overline{Y}} Y, \SSS)$ is iAST\@.
  Thus, also every $(\bigcup_{Y \in \overline{Y}} Y \setminus \bigcup_{X \in \overline{X}} X,\SSS)$-chain tree converges with probability $1$.
  Furthermore, we have $|\F{T}|_{\ctleaf} < 1$ by our assumption.
  By the P-Partition Lemma (\cref{lemma:p-partition}) we can find a grounded  sub $(Z,\SSS)$-chain tree $\F{T}' = (V',E',L',P')$ with $|\F{T}'|_{\ctleaf} < 1$ such that every infinite path has an infinite number of $P_1$ edges.
  Since $\F{T}'$ is a grounded  sub chain tree of $\F{T}$ it must also start with $(1:\Com{1}(s^\#))$.

  We now construct a $(\bigcup_{X \in \overline{X}} X,\SSS)$-chain tree $\F{T}'' =
  (V',E',L'',P_1 \cap P')$ with splits
    that has the same underlying tree structure and adjusted labeling
  such that all nodes get the same probabilities
  as in $\F{T}'$.
  Since the tree structure and the probabilities are the same, we then get $|\F{T}'|_{\ctleaf} = |\F{T}''|_{\ctleaf}$.
  To be precise, the set of leaves in $\F{T}'$ is equal to the set of leaves in $\F{T}''$, and every leaf has the same probability.
  Since $|\F{T}'|_{\ctleaf} < 1$ we thus have $|\F{T}''|_{\ctleaf} < 1$, which (by
  \Cref{lemma:splitting})
  is a contradiction to our
  assumption that $(\bigcup_{X \in \overline{X}} X,\SSS)$ is iAST\@.

  \begin{figure}[H]
    \centering
    \small
    \begin{tikzpicture}
      \tikzstyle{adam}=[rectangle,thick,draw=black!100,fill=white!100,minimum size=4mm]
      \tikzstyle{empty}=[rectangle,thick,minimum size=4mm]
      
      \node[adam,pin={[pin distance=0.1cm, pin edge={,-}] 135:\tiny \textcolor{blue}{$P_1$}}] at (-3.5, 0)  (a) {$1:\Com{1}(s^\#)$};
      \node[adam] at (-5, -1)  (b) {$p_1:a_{1}'$};
      \node[adam,pin={[pin distance=0.1cm, pin edge={,-}] 45:\tiny \textcolor{blue}{$P_2$}}] at (-2, -1)  (c) {$p_2:a_{2}'$};
      \node[adam,pin={[pin distance=0.1cm, pin edge={,-}] 135:\tiny \textcolor{blue}{$P_1$}}] at (-6, -2)  (d) {$p_3:a_3'$};
      \node[adam,pin={[pin distance=0.1cm, pin edge={,-}] 45:\tiny \textcolor{blue}{$P_2$}}] at (-4, -2)  (e) {$p_4:a_4'$};
      \node[adam,pin={[pin distance=0.1cm, pin edge={,-}] 135:\tiny \textcolor{blue}{$P_1$}}] at (-2, -2)  (f) {$p_5:a_5'$};
      \node[empty] at (-6, -3)  (g) {$\ldots$};
      \node[empty] at (-4, -3)  (h) {$\ldots$};
      \node[empty] at (-2, -3)  (i) {$\ldots$};

      \node[empty] at (-0.5, -1)  (arrow) {\Huge $\leadsto$};
      
      \node[adam,pin={[pin distance=0.1cm, pin edge={,-}] 135:\tiny \textcolor{blue}{$P_1$}}] at (3.5, 0)  (a2) {$1:\Com{1}(s^\#)$};
      \node[adam] at (2, -1)  (b2) {$p_1:a''_{1}$};
      \node[adam] at (5, -1)  (c2) {$p_2:a''_{2}$};
      \node[adam,pin={[pin distance=0.1cm, pin edge={,-}] 135:\tiny \textcolor{blue}{$P_1$}}] at (1, -2)  (d2) {$p_3:a''_3$};
      \node[adam] at (3, -2)  (e2) {$p_4:a''_4$};
      \node[adam,pin={[pin distance=0.1cm, pin edge={,-}] 135:\tiny \textcolor{blue}{$P_1$}}] at (5, -2)  (f2) {$p_5:a''_5$};
      \node[empty] at (1, -3)  (g2) {$\ldots$};
      \node[empty] at (3, -3)  (h2) {$\ldots$};
      \node[empty] at (5, -3)  (i2) {$\ldots$};
    
      \draw (a) edge[->] (b);
      \draw (a) edge[->] (c);
      \draw (b) edge[->] (d);
      \draw (b) edge[->] (e);
      \draw (c) edge[->] (f);
      \draw (d) edge[->] (g);
      \draw (e) edge[->] (h);
      \draw (f) edge[->] (i);

      \draw (a2) edge[->] (b2);
      \draw (a2) edge[->] (c2);
      \draw (b2) edge[->] (d2);
      \draw (b2) edge[->] (e2);
      \draw (c2) edge[->] (f2);
      \draw (d2) edge[->] (g2);
      \draw (e2) edge[->] (h2);
      \draw (f2) edge[->] (i2);
    \end{tikzpicture}
    \caption{Construction for this proof. Every node $x \in P_2$ in $\F{T}'$
      is removed from $P$, which yields $\F{T}''$.}\label{fig:dep-graph-proof-construction}
  \end{figure}
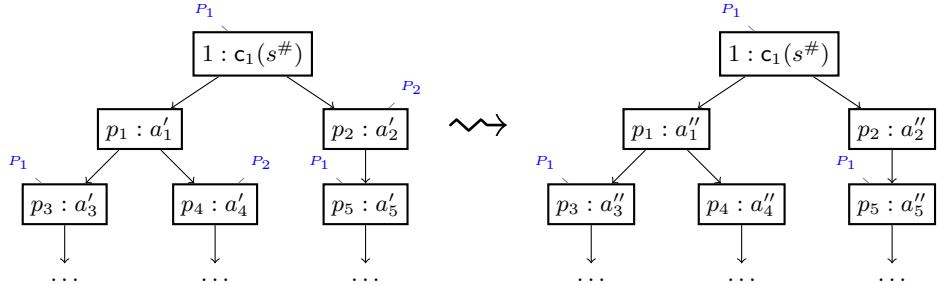

  The core idea of this construction is that terms introduced by rewrite steps at a node
  $x \in P_2$ are not
  important for our computation.
  The reason is that the rewrite step is done using a dependency tuple from $\bigcup_{Y \in \overline{Y}} Y$ that is not contained in $\bigcup_{X \in \overline{X}} X$.
  By the prerequisite of~\eqref{Composing iAST} we know that such a dependency tuple has no path in the dependency graph to a dependency tuple in $\bigcup_{X \in \overline{X}} X$.
  Hence, by definition of the dependency graph, we are never able to use these terms for a rewrite step with a dependency tuple from $\bigcup_{X \in \overline{X}} X$.
  We can therefore apply the PTRS $\SSS$ to perform the rewrite step with the second component of the used dependency tuple from $\bigcup_{Y \in \overline{Y}} Y$, and we completely ignore the first component.
  If the second component is not applicable (e.g., if the projection to the second component of the used dependency tuple is not contained in $\SSS$), then we use split-nodes which do nothing but mirror the tree structure. 
  This means that we remove every node $x \in P_2$ from $P$, but since every infinite path contains an infinite number of $P_1$ nodes, we can be sure that the resulting tree still has an infinite number of $P$ nodes on every infinite path and thus the resulting tree is a $(\bigcup_{X \in \overline{X}} X, \SSS)$-chain tree.

  For example, consider our DP problem $(\DTuple{\R_{\tpdiv}},\R_{\tpdiv})$ and
  regard the following $(\DTuple{\R_{\tpdiv}},\R_{\tpdiv})$-chain tree $\F{T}$, where we
  abbreviate $\ts^n(\O)$ by $n$ for all $n \in \IN$:
  \begin{center}
    \centering
    \footnotesize
    \begin{tikzpicture}[node distance=1cm,>=stealth',bend angle=20,auto]
      \tikzstyle{adam}=[rectangle,thick,draw=black!100,minimum size=4mm]
      \begin{scope}
        \node [adam,pin={[pin distance=0.1cm, pin edge={,-}] 135:\tiny \textcolor{blue}{$P$}}] at (0, 0) (x1) []{$1:\Com{1}(\tD(4,2))$};
          \node [adam,pin={[pin distance=0.1cm, pin edge={,-}] 135:\tiny \textcolor{blue}{$P$}}] at (-3, -1) (x2) []{$\tfrac{1}{2}:\Com{1}(\tD(4,2))$};
          \node [adam,pin={[pin distance=0.1cm, pin edge={,-}] 45:\tiny \textcolor{blue}{$P$}}] at (3, -1) (x4) []{$\tfrac{1}{2}:\Com{2}(\tD(\tminus(3,1),2), \tM(3,1))$};
          \node [adam] at (-4, -2) (x5) []{$\tfrac{1}{4}:\Com{1}(\tD(4,2))$};
          \node [adam] at (-2, -3) (x7) []{$\tfrac{1}{4}:\Com{2}(\tD(\tminus(3,1),2), \tM(3,1))$};
          \node [adam] at (3, -2) (x10) []{$\tfrac{1}{2}:\Com{2}(\tD(\tminus(2,0),2), \tM(2,0))$};

        \draw (x1) edge[->] (x2);
        \draw (x1) edge[->] (x4);
        \draw (x2) edge[->] (x5);
        \draw (x2) edge[->, in=30, out=340, looseness=2] (x7);
        \draw (x4) edge[->] (x10);
      \end{scope}
    \end{tikzpicture}
  \end{center}
  We can partition the set of all SCCs and all the singleton sets of the dependency graph into the sets $\overline{X} = \{\{\eqref{R-div-deptup-3}\}, \{\eqref{R-div-deptup-4}\}\}$ (the dependency tuples for $\tD$) and $\overline{Y} = \{\{\eqref{R-div-deptup-1}\}, \{\eqref{R-div-deptup-2}\}\}$ (the dependency tuples for $\tM$).
  Then $\overline{X}$ and $\overline{Y}$ satisfy the prerequisites of~\eqref{Composing iAST}.
  Instead of the dependency tuples from $\overline{Y}$, our construction now uses
  the PTRS rules $\eqref{eq:ptrs-div-1}$ and $\eqref{eq:ptrs-div-2}$ to evaluate the inner $\tminus$ subterm.
  This results in the following $(\DTuple{\R_{\tpdiv}},\R_{\tpdiv})$-chain tree that does not use any dependency tuples from $\overline{Y}$:
  \begin{center}
    \centering
    \footnotesize
    \begin{tikzpicture}[node distance=1cm,>=stealth',bend angle=20,auto]
      \tikzstyle{adam}=[rectangle,thick,draw=black!100,minimum size=4mm]
      \begin{scope}
        \node [adam,pin={[pin distance=0.1cm, pin edge={,-}] 135:\tiny \textcolor{blue}{$P$}}] at (0, 0) (x1) []{$1:\Com{1}(\tD(4,2))$};
          \node [adam,pin={[pin distance=0.1cm, pin edge={,-}] 135:\tiny \textcolor{blue}{$P$}}] at (-3, -1) (x2) []{$\tfrac{1}{2}:\Com{1}(\tD(4,2))$};
          \node [adam] at (3, -1) (x4) []{$\tfrac{1}{2}:\Com{2}(\tD(\tminus(3,1),2), \tM(3,1))$};
          \node [adam] at (-4, -2) (x5) []{$\tfrac{1}{4}:\Com{1}(\tD(4,2))$};
          \node [adam] at (-2, -3) (x7) []{$\tfrac{1}{4}:\Com{2}(\tD(\tminus(3,1),2), \tM(3,1))$};
          \node [adam] at (3, -2) (x10) []{$\tfrac{1}{2}:\Com{2}(\tD(\tminus(2,0),2), \tM(3,1))$};
  
        \draw (x1) edge[->] (x2);
        \draw (x1) edge[->] (x4);
        \draw (x2) edge[->] (x5);
        \draw (x2) edge[->, in=30, out=340, looseness=2] (x7);
        \draw (x4) edge[->] (x10);
      \end{scope}
    \end{tikzpicture}
  \end{center}
  For the right child of the root, we use the PTRS rule $\eqref{eq:ptrs-div-2}$ instead of the dependency tuple $\eqref{R-div-deptup-2}$.
  Thus, the term $\tM(3,1)$ remains the same.
  However, this term will never be used for a rewrite step with
$\mathrel{\oset{\scriptsize $\mathsf{i}$\qquad}{{\rightarrowtriangle}_{\bigcup_{X \in \overline{X}} X,\SSS}}}$,
so we can simply ignore this term.

  We now construct the new labeling $L''$ for the $(\bigcup_{X \in \overline{X}} X,\SSS)$-chain tree $\F{T}''$ recursively.
  Let $Q \subseteq V$ be the set of nodes where we have already defined the labeling $L''$.
  During our construction, we ensure that the following property holds:
  \begin{equation}\label{dep-graph-construction-induction-hypothesis}
    \parbox{.8\textwidth}{For every node $x \in Q$ we have $\cont(a'_x) \setminus \Junk(a'_x, \overline{X}) \subseteq \cont(a''_{x})$.}
  \end{equation}
  Here, for any normalized term $a_x' = \Com{n}(s_1^\#, \ldots, s_n^\#)$, let
  $\Junk(a'_x, \overline{X})$ denote the set of all terms $s_i^\#$ that can never be used
  for a rewrite step with a dependency tuple from $\overline{X}$, as indicated by the dependency graph.
  To be precise, we define $s_i^\# \in \Junk(a'_x, \overline{X})$:$\Leftrightarrow$ there
  is no $Z \in W$ with $Z >_{\F{G}}^* X$ for some $X \in \overline{X}$ such that there is
  a dependency tuple $\langle \ell^\#, \ell \rangle \to \{p_1: \langle d_1,r_1 \rangle,
  \ldots, p_k: \langle d_k,r_k \rangle\} \in Z$, and a substitution $\sigma$ with $s_i^\# \itononprobsstar \ell^\# \sigma$, and every proper subterm of $\ell^\# \sigma$ is in normal form w.r.t.\ $\SSS$.

  We start by setting $a''_{v} = a'_{v} = \Com{1}(s^\#)$ for the root $v$ of $\F{T}'$.
  Here, our property~\eqref{dep-graph-construction-induction-hypothesis} is clearly satisfied as we have 
  \[\cont(a'_{v}) \setminus \Junk(a'_{v}, \overline{X}) \subseteq \cont(a'_{v}) = \cont(a''_{v}).\]

  As long as there is still an inner node $x \in Q$ such that its successors are not contained in $Q$, we do the following.
  Let $xE = \{y_1, \ldots, y_k\}$ be the set of its successors.
  We need to define the corresponding terms for the nodes $y_1, \ldots, y_k$ in $\F{T}''$.

  Since $x$ is not a leaf and $\F{T}'$ is a $(Z, \SSS)$-chain tree, we have 
  \[a'_x \setitozs \{\tfrac{p_{y_1}}{p_x}:a'_{y_1}, \ldots, \tfrac{p_{y_k}}{p_x}:a'_{y_k}\} \text{ or } a'_x \setitos \{\tfrac{p_{y_1}}{p_x}:a'_{y_1}, \ldots, \tfrac{p_{y_k}}{p_x}:a'_{y_k}\}\]
  and hence, we have to deal with the following three cases:
  \begin{enumerate}
  \item If we used a step with $\setitozs$ and $x \in P_1$ in $\F{T}'$, then we perform
    the rewrite step with the same dependency tuple and the same substitution in $\F{T}''$.
    \item If we used a step with $\setitozs$ and $x \in P_2$ in $\F{T}'$, then we ignore
      the first component of the used dependency tuple and mimic all rewrite steps
on terms in $\cont(a'_{x}) \setminus \Junk(a'_{x}, \overline{X})$ by performing them with
$\setitos$ in  $\F{T}''$.
\item If we used a step with $\setitos$
on a term in $\cont(a'_{x}) \setminus \Junk(a'_{x}, \overline{X})$,
 then we perform the same rewrite steps in  $\F{T}''$.
  \end{enumerate}
  If in case (2) and (3) there is no rewrite step performed on a term in $\cont(a'_{x}) \setminus \Junk(a'_{x}, \overline{X})$, then we use a split-node to mirror the tree structure without rewriting anything.
 
  \medskip

  \noindent
  \textbf{(1) If we have} $a'_x \setitozs \{\tfrac{p_{y_1}}{p_x}:a'_{y_1}, \ldots,
  \tfrac{p_{y_k}}{p_x}:a'_{y_k}\}$ using a dependency tuple $\langle \ell^\#,\ell \rangle
  \to \{ p_1: \langle d_1,r_1 \rangle, \ldots, p_k: \langle d_k,r_k \rangle\} \in
  \bigcup_{X \in \overline{X}} X$ (i.e., $x \in P_1$) and a substitution $\sigma$, then we
  have $a'_x = \Com{n}(s_1^\#, \ldots, s_n^\#)$ and there is an $1 \leq i \leq n$ such that $s_i^\# = \ell^\# \sigma$ and all proper subterms of $\ell^\# \sigma$ are in normal form w.r.t.\ $\SSS$.
  Let $1 \leq j \leq k$.
  Then we have $a'_{y_j} = \Com{n}(s_1'^\#, \ldots, d_j \sigma, \ldots, s_n'^\#)$.
  Furthermore, let $\{i'_1, \ldots, i'_m\}$ be the set of indices such that for all $i'
  \in \{i'_1, \ldots, i'_m\}$ we have $s_{i'}'^\# = s_{i'}^\#[r_j \sigma]_{\tau_{i'}}$ for some
  position $\tau_{i'}$ such that $s_{i'}^\#|_{\tau_{i'}} = \ell \sigma$, and for all $i' \in
  \{1,\ldots,n\} \setminus \{i,i'_1, \ldots, i'_m\}$ we have $s_{i'}'^\#  = s_{i'}^\#$.

  Since $\langle \ell^\#,\ell \rangle \to \{ p_1: \langle d_1,r_1 \rangle, 
  \ldots, p_k:\langle d_k,r_k \rangle\} \in \bigcup_{X \in \overline{X}} X$ (i.e., we use a dependency tuple that is in $\bigcup_{X \in \overline{X}} X$),  the term $s_i^\#$ cannot be in $\Junk(a'_{x}, \overline{X})$.
  Hence, we must have $s_i^\# \in \cont(a'_{x}) \setminus \Junk(a'_{x}, \overline{X}) \subseteq_{(IH)} \cont(a''_{x})$.
  Let $a''_{x} = \Com{h}(t_1^\#, \ldots, t_h^\#)$.
  There exists an $1 \leq l \leq h$ such that $t_l^\# = s_i^\#$ and thus, we can rewrite
  the term $a''_{x} = \Com{h}(t_1^\#, \ldots, t_h^\#)$ using the dependency tuple $\langle
  \ell^\#,\ell \rangle \to \{ p_1:\langle d_1,r_1 \rangle, \ldots, p_k:\langle d_k,r_k
  \rangle\}
  \in \bigcup_{X \in \overline{X}} X$ and the substitution $\sigma$, as we have $t_l^\# = s_i^\# = \ell^\# \sigma$ and all proper subterms of $\ell^\# \sigma$ are in normal form w.r.t.\ $\SSS$.

  For each $i' \in \{i'_1, \ldots, i'_m\}$ such that $s_{i'}^\# \in \cont(a'_{x}) \setminus \Junk(a'_{x}, \overline{X})$, by the induction hypothesis we can find a (unique) $1 \leq l' \leq h$ with $t_{l'}^\# = s_{i'}^\#$.
  This $t_{l'}^\#$ will be rewritten so that we have $t_{l'}'^\# = t_{l'}^\#[r_j \sigma]_{\tau_{i'}}$ for the same position $\tau_{i'}$ as we used for $s_{i'}'^\#$.
  All other terms remain the same.
  This results in the term $a_{y_j}'' = \Com{h}(t_1'^\#, \ldots, d_j \sigma, \ldots, t_h'^\#)$.
        
  It remains to show that our property~\eqref{dep-graph-construction-induction-hypothesis}
  is still satisfied for this new labeling, i.e., we have $\cont(a'_{y_j}) \setminus \Junk(a'_{y_j}, \overline{X}) \subseteq \cont(a''_{y_j})$ for all $1 \leq j \leq k$.

  Let $1 \leq j \leq k$ and $b^\# \in \cont(a'_{y_j}) \setminus \Junk(a'_{y_j}, \overline{X})$.
  \begin{itemize}
    \item[$\bullet$] If $b^\#$ was introduced by $d_j \sigma$, then also $b^\# \in \cont(a''_{y_j})$ as we use the same dependency tuple with the same substitution for the rewrite step.
    
    \item[$\bullet$] If $b^\# = s_{i'}'^\#$ for some $i' \in \{i'_1, \ldots, i'_m\}$, then we have $b^\# = s_{i'}^\#[r_j \sigma]_{\tau_{i'}}$.

    If we have $s_{i'}^\# \in \Junk(a'_{x}, \overline{X})$, then we also have $b^\# \in \Junk(a'_{y_j}, \overline{X})$, since $s_{i'}^\# \itononprobs s_{i'}^\#[r_j \sigma]_{\tau_{i'}} = b^\#$, and this is a contradiction to our assumption that $b^\# \in \cont(a'_{y_j}) \setminus \Junk(a'_{y_j}, \overline{X})$.
    Thus, we have $s_{i'}^\# \in \cont(a'_{x}) \setminus \Junk(a'_{x}, \overline{X}) \subseteq_{(IH)} \cont(a''_{x})$, and we find a (unique) $1 \leq l' \leq h$ such that $t_{l'}^\# = s_{i'}^\#$.
    By construction, we performed the same rewrite step at the same position, such that we get $b^\# = s_{i'}^\#[r_j \sigma]_{\tau_{i'}} = t_{l'}^\#[r_j \sigma]_{\tau_{i'}} = t_{l'}'^\#$ so that $b^\# \in \cont(a''_{y_j})$.

    \item[$\bullet$] If $b^\# = s_{i'}'^\#$ for some $i' \not\in \{i, i'_1, \ldots, i'_m\}$, then we have $b^\# = s_{i'}^\#$.
    If we have $s_{i'}^\# \in \Junk(a'_{x}, \overline{X})$, then we also have $b^\# =
    s_{i'}'^\# \in \Junk(a'_{y_j}, \overline{X})$ since $s_{i'}^\# =
    s_{i'}'^\#$,
      and this is a contradiction to our assumption that $b^\# \in \cont(a'_{y_j}) \setminus \Junk(a'_{y_j}, \overline{X})$.
    Thus, we have $b^\# \in \cont(a'_{x}) \setminus \Junk(a'_{x}, \overline{X}) \subseteq_{(IH)} \cont(a''_{x})$, and we find a (unique) $1 \leq l' \leq h$ such that $t_{l'}^\# = s_{i'}^\#$.
    By construction, we did not change the term, such that we get $b^\# = s_{i'}^\# = t_{l'}'^\#$ so that $b^\# \in \cont(a''_{y_j})$.
  \end{itemize}

  \medskip

  \noindent
  \textbf{(2) If we have} $a'_x \setitozs \{\tfrac{p_{y_1}}{p_x}:a'_{y_1}, \ldots,
  \tfrac{p_{y_k}}{p_x}:a'_{y_k}\}$ using a dependency tuple $\langle \ell^\#,\ell \rangle
  \to \{ p_1:\langle d_1,r_1\rangle, \ldots, p_k:\langle d_k,r_k\rangle\} \in \bigcup_{Y
    \in \overline{Y}} Y \setminus \bigcup_{X \in \overline{X}} X$ (i.e., $x \in P_2$) and
  a substitution $\sigma$, then we have $a'_x = \Com{n}(s_1^\#, \ldots, s_n^\#)$
  and there is an $1 \leq i \leq n$ such that $s_i^\# = \ell^\# \sigma$ and all proper subterms of $\ell^\# \sigma$ are in normal form w.r.t.\ $\SSS$.
  Let $1 \leq j \leq k$.
  Then we have $a'_{y_j} = \Com{n}(s_1'^\#, \ldots, d_j \sigma, \ldots, s_n'^\#)$.
  Furthermore, let $\{i'_1, \ldots, i'_m\}$ be the set of indices such that for all $i'
  \in \{i'_1, \ldots, i'_m\}$ we have $s_{i'}'^\# = s_{i'}^\#[r_j \sigma]_{\tau_{i'}}$ for some
  position $\tau_{i'}$ such that $s_{i'}^\#|_{\tau_{i'}} = \ell \sigma$, and for all $i' \in
  \{1,\ldots,n\} \setminus \{i,i'_1, \ldots, i'_m\}$ we have $s_{i'}'^\# = s_{i'}^\#$.

  Since $\langle \ell^\#,\ell \rangle \to \{ p_1:\langle d_1,r_1\rangle, \ldots,
  p_k:\langle d_k,r_k\rangle\} \in \bigcup_{Y \in \overline{Y}} Y \setminus \bigcup_{X \in
    \overline{X}} X$ (i.e., $x \in P_2$),
   the term $s_i^\#$ is in $\Junk(a'_{x}, \overline{X})$.
  Here, we use $\setitos$ to mimic the rewrite steps with the second component but we ignore the first component.
  Let $a''_{x} = \Com{h}(t_1^\#, \ldots, t_h^\#)$.
  For each $i' \in \{i'_1, \ldots, i'_m\}$ such that $s_{i'}^\# \in \cont(a'_{x})
  \setminus \Junk(a'_{x}, \overline{X})$, by the induction hypothesis
  we can find a (unique) $1 \leq l' \leq h$ with $t_{l'}^\# = s_{i'}^\#$.
  Each of these $t_{l'}^\#$ will be rewritten so that we have $t_{l'}'^\# = t_{l'}^\#[r_j \sigma]_{\tau_{i'}}$ for the same position $\tau_{i'}$ as we used for $s_{i'}'^\#$.
  All other terms remain the same.
  This results in the term $a_{y_j}'' = \Com{h}(t_1'^\#, \ldots, 
  t_h'^\#)$.
  Note that $\{i'_1, \ldots, i'_m\}$ may also be empty. 
  In this case, we have to use a split-node, where we do not perform any rewrite step but only mirror the tree structure.
        
  It remains to show that our property~\eqref{dep-graph-construction-induction-hypothesis}
  is still satisfied for this new labeling, i.e., we have $\cont(a'_{y_j}) \setminus \Junk(a'_{y_j}, \overline{X}) \subseteq \cont(a''_{y_j})$ for all $1 \leq j \leq k$.

  Let $1 \leq j \leq k$ and $b^\# \in \cont(a'_{y_j}) \setminus \Junk(a'_{y_j}, \overline{X})$.
  \begin{itemize}
  \item[$\bullet$] First assume that
    $b^\#$ was introduced by $d_j \sigma$.
    Since our used dependency tuple is in $Y$ for some $Y \in \overline{Y}$ but not contained in $\bigcup_{X \in \overline{X}} X$, we know by~\eqref{Composing iAST} that $Y \not >_{\F{G}}^* X$ for all $X \in \overline{X}$.
    Intuitively, we started with a dependency tuple from $\overline{X}$ and then we reached $Y$, so we cannot reach any $X \in \overline{X}$ anymore.
    By definition of the dependency graph, this means that there is no $Z \in W$ with $Z
    >_{\F{G}}^* X$ for some $X \in \overline{X}$ such that there is a dependency tuple
    $\langle \ell'^\#, \ell'\rangle \to \ldots \in Z$ and a substitution $\sigma'$ with $b^\# \itononprobsstar \ell'^\# \sigma'$, where every proper subterm of $\ell'^\# \sigma'$ is in normal form w.r.t.\ $\SSS$.
    Hence, we have $b^\# \in \Junk(a'_{y_j}, \overline{X})$ and this is a contradiction to our assumption that $b^\# \in \cont(a'_{y_j}) \setminus \Junk(a'_{y_j}, \overline{X})$.
    Thus, this case is not possible.
    
    \item[$\bullet$] If $b^\# = s_{i'}'^\#$ for some $i' \in \{i'_1, \ldots, i'_m\}$, then we have $b^\# = s_{i'}^\#[r_j \sigma]_{\tau_{i'}}$.

    If we have $s_{i'}^\# \in \Junk(a'_{x}, \overline{X})$, then we also have $b^\# \in \Junk(a'_{y_j}, \overline{X})$, since $s_{i'}^\# \itononprobs s_{i'}^\#[r_j \sigma]_{\tau_{i'}} = b^\#$, and this is a contradiction to our assumption that $b^\# \in \cont(a'_{y_j}) \setminus \Junk(a'_{y_j}, \overline{X})$.
    Thus, we have $s_{i'}^\#  \in \cont(a'_{x}) \setminus \Junk(a'_{x}, \overline{X}) \subseteq_{(IH)} \cont(a''_{x})$, and we find a (unique) $1 \leq l' \leq h$ such that $t_{l'}^\# = s_{i'}^\#$.
    By construction, we performed the same rewrite step at the same position, such that we get $b^\# = s_{i'}^\#[r_j \sigma]_{\tau_{i'}} = t_{l'}^\#[r_j \sigma]_{\tau_{i'}} = t_{l'}'^\#$ so that $b^\# \in \cont(a''_{y_j})$.

    \item[$\bullet$] If $b^\# = s_{i'}'^\#$ for some $i' \not\in \{i, i'_1, \ldots, i'_m\}$, then we have $b^\# = s_{i'}^\#$.
    If we have $s_{i'}^\# \in \Junk(a'_{x}, \overline{X})$, then we also have $b^\# =
    s_{i'}'^\# \in \Junk(a'_{y_j}, \overline{X})$ since $s_{i'}^\# =  s_{i'}'^\#$,
    and this is a contradiction to our assumption that $b^\# \in \cont(a'_{y_j}) \setminus \Junk(a'_{y_j}, \overline{X})$.
    Thus, we have $b^\# \in \cont(a'_{x}) \setminus \Junk(a'_{x}, \overline{X}) \subseteq_{(IH)} \cont(a''_{x})$, and we find a (unique) $1 \leq l' \leq h$ such that $t_{l'}^\# = s_{i'}^\#$.
    By construction, we did not change the term, such that we get $b^\# = s_{i'}^\# = t_{l'}'^\#$ so that $b^\# \in \cont(a''_{y_j})$.
  \end{itemize}

  \medskip

  \noindent
  \textbf{(3) In the last case we have} $a'_x \setitos \{\tfrac{p_{y_1}}{p_x}:a'_{y_1}, \ldots, \tfrac{p_{y_k}}{p_x}:a'_{y_k}\}$ using a rewrite rule $\ell \to \{ p_1:r_1, \ldots, p_k:r_k\} \in \SSS$ (i.e., $x \in S$) and a substitution $\sigma$.
  Let $1 \leq j \leq k$, $a'_x = \Com{n}(s_1^\#, \ldots, s_n^\#)$, and $a'_{y_j} = \Com{n}(s_1'^\#, \ldots, s_n'^\#)$.
  Furthermore, let $\{i'_1, \ldots, i'_m\}$ be the set of indices such that for all $i'
  \in \{i'_1, \ldots, i'_m\}$ we have $s_{i'}'^\# = s_{i'}^\#[r_j \sigma]_{\tau_{i'}}$ for some
  position $\tau_{i'}$ such that $s_{i'}^\#|_{\tau_{i'}} = \ell \sigma$, and for all $i' \in
  \{1,\ldots,n\} \setminus \{i,i'_1, \ldots, i'_m\}$ we have $s_{i'}'^\# = s_{i'}^\#$.

  Let $a''_{x} = \Com{h}(t_1^\#, \ldots, t_h^\#)$.
  For each $i' \in \{i'_1, \ldots, i'_m\}$ such that $s_{i'}^\# \in \cont(a'_{x})
  \setminus \Junk(a'_{x}, \overline{X})$, by the induction hypothesis
  we can find a (unique) $1 \leq l' \leq h$ with $t_{l'}^\# = s_{i'}^\#$.
  Each of these $t_{l'}^\#$ will be rewritten so that we have $t_{l'}'^\# = t_{l'}^\#[r_j \sigma]_{\tau_{i'}}$ for the same position $\tau_{i'}$ as we used for $s_{i'}'^\#$.
  All other terms remain the same.
  This results in the term $a_{y_j}'' = \Com{h}(t_1'^\#, \ldots, t_h'^\#)$.
  Note that $\{i'_1, \ldots, i'_m\}$ may also be empty. 
  In this case, we have to use a split-node, where we do not perform any rewrite step but only mirror the tree structure.
        
  It remains to show that our property~\eqref{dep-graph-construction-induction-hypothesis}
 is still satisfied for this new labeling, i.e., we have $\cont(a'_{y_j}) \setminus \Junk(a'_{y_j}, \overline{X}) \subseteq \cont(a''_{y_j})$ for all $1 \leq j \leq k$.

  Let $1 \leq j \leq k$ and $b^\# \in \cont(a'_{y_j}) \setminus \Junk(a'_{y_j}, \overline{X})$.
  \begin{itemize}
      \item[$\bullet$] If $b^\# = s_{i'}'^\#$ for some $i' \in \{i'_1, \ldots, i'_m\}$, then we have $b^\# = s_{i'}^\#[r_j \sigma]_{\tau_{i'}}$.

      If we have $s_{i'}^\# \in \Junk(a'_{x}, \overline{X})$, then we also have $b^\# \in \Junk(a'_{y_j}, \overline{X})$, since $s_{i'}^\# \itononprobs s_{i'}^\#[r_j \sigma]_{\tau_{i'}} = b^\#$, and this is a contradiction to our assumption that $b^\# \in \cont(a'_{y_j}) \setminus \Junk(a'_{y_j}, \overline{X})$.
      Thus, we have $s_{i'}^\# \in \cont(a'_{x}) \setminus \Junk(a'_{x}, \overline{X}) \subseteq_{(IH)} \cont(a''_{x})$, and we find a (unique) $1 \leq l' \leq h$ such that $t_{l'}^\# = s_{i'}^\#$.
      By construction, we performed the same rewrite step at the same position, such that we get $b^\# = s_{i'}^\#[r_j \sigma]_{\tau_{i'}} = t_{l'}^\#[r_j \sigma]_{\tau_{i'}} = t_{l'}'^\#$ so that $b^\# \in \cont(a''_{y_j})$.

      \item[$\bullet$] If $b^\# = s_{i'}'^\#$ for some $i' \not\in \{i, i'_1, \ldots, i'_m\}$, then we have $b^\# = s_{i'}^\#$.
      If we have $s_{i'}^\# \in \Junk(a'_{x}, \overline{X})$, then we also have $b^\# =
      s_{i'}'^\# \in \Junk(a'_{y_j}, \overline{X})$ since $s_{i'}^\# =  s_{i'}'^\#$,
      and this is a contradiction to our assumption that $b^\# \in \cont(a'_{y_j}) \setminus \Junk(a'_{y_j}, \overline{X})$.
      Thus, we have $b^\# \in \cont(a'_{x}) \setminus \Junk(a'_{x}, \overline{X}) \subseteq_{(IH)} \cont(a''_{x})$, and we find a (unique) $1 \leq l' \leq h$ such that $t_{l'}^\# = s_{i'}^\#$.
      By construction, we did not change the term, such that we get $b^\# = s_{i'}^\# = t_{l'}'^\#$ so that $b^\# \in \cont(a''_{y_j})$.
  \end{itemize}

  This was the last case and ends the construction and this part of the proof.
  We have now shown that~\eqref{Composing iAST} holds.

  \medskip

  \noindent
  \textbf{\underline{3. For every $X \in W$, the DP problem $(\bigcup_{X >_{\F{G}}^* Y} Y, \SSS)$ is iAST\@.}}

  \noindent
  Using~\eqref{W is iAST} and~\eqref{Composing iAST}, by induction on $>_{\F{G}}$ we now prove that
  \begin{equation}
    \label{SCC induction} \mbox{for every $X \in W$, the DP problem $(\bigcup_{X >_{\F{G}}^* Y} Y, \SSS)$ is iAST\@.}
  \end{equation}
  Note that $>_{\F{G}}$ is well founded, \pagebreak[2] since $\F{G}$ is finite.
  
  For the base case, we consider an $X \in W$ that is minimal w.r.t.\ $>_{\F{G}}$.
  Hence, we have $\bigcup_{X >_{\F{G}}^* Y} Y = X$.
  By~\eqref{W is iAST}, $(X, \SSS)$ is iAST\@.

  For the induction step, we consider an $X \in W$ and assume that
  $(\bigcup_{Y >_{\F{G}}^* Z} Z, \SSS)$ is iAST for every $Y \in W$ with $X >_{\F{G}}^+ Y$.
  Let $\mathtt{Succ}(X) = \{Y \in W \mid X >_{\F{G}} Y\} = \{Y_1, \ldots Y_m\}$ be the set of all direct successors of $X$.
  The induction hypothesis states that $(\bigcup_{Y_u >_{\F{G}}^* Z} Z,\SSS)$ is iAST for all $1 \leq u \leq m$.
  We first prove by induction that for all $1 \leq u \leq m$,
  $(\bigcup_{1 \leq i \leq u} \bigcup_{Y_i >_{\F{G}}^* Z} Z,\SSS)$ is iAST\@.
  
  In the inner induction base, we have $u = 1$ and hence $(\bigcup_{1 \leq i \leq u} \bigcup_{Y_i >_{\F{G}}^* Z} Z,\SSS) = (\bigcup_{Y_1 >_{\F{G}}^* Z} Z,\SSS)$.
  By our outer induction hypothesis we know that $(\bigcup_{Y_1 >_{\F{G}}^* Z} Z,\SSS)$ is iAST\@.
  
  In the inner induction step, assume that the claim holds for some $1 \leq u < m$.
  Then $(\bigcup_{Y_{u+1} >_{\F{G}}^* Z} Z,\SSS)$ is iAST by our outer induction hypothesis and\linebreak $(\bigcup_{1 \leq i \leq u} \bigcup_{Y_{i} >_{\F{G}}^* Z} Z,\SSS)$ is iAST by our inner induction hypothesis.
  By~\eqref{Composing iAST}, we know that then $(\bigcup_{1 \leq i \leq u+1} \bigcup_{Y_{i} >_{\F{G}}^* Z} Z,\SSS)$ is iAST as well.
  The conditions for~\eqref{Composing iAST} are clearly satisfied, as we use the
  reflexive, transitive closure $>_{\F{G}}^*$
  in both\linebreak $\bigcup_{1 \leq i \leq u} \bigcup_{Y_{i} >_{\F{G}}^* Z} Z$ and $\bigcup_{Y_{u+1} >_{\F{G}}^* Z} Z$.
  
  Now we have shown that $(\bigcup_{1 \leq i \leq m} \bigcup_{Y_i >_{\F{G}}^* Z} Z,\SSS)$ is iAST\@.
  We know that $(X,\SSS)$ is iAST by our assumption and that $(\bigcup_{1 \leq i \leq m} \bigcup_{Y_i >_{\F{G}}^* Z} Z,\SSS)$ is iAST\@.
  Hence, by~\eqref{Composing iAST} we obtain that $(\bigcup_{X >_{\F{G}}^* Y} Y,\SSS)$ is iAST\@.
  Again, the conditions of~\eqref{Composing iAST} are satisfied, since $X$ is strictly
  greater w.r.t.\ $>_{\F{G}}^+$ than all $Z$ with $Y_i >_{\F{G}}^* Z$.

  \medskip

  \noindent
  \textbf{\underline{4. $(\PP,\SSS)$ is iAST\@.}}

  \noindent
 In~\eqref{SCC induction} we have shown that $(\bigcup_{X >_{\F{G}}^* Y} Y,\SSS)$ for every $X \in W$ is iAST\@.
  Let $X_1, \ldots, X_m\linebreak \in W$ be the maximal elements of $W$ w.r.t.\ $>_{\F{G}}$.
By induction, one can prove that $(\bigcup_{1 \leq i \leq u} \bigcup_{X_i >_{\F{G}}^* Y}
Y,\SSS)$ is iAST for all $1 \leq u \leq m$ by~\eqref{Composing iAST}, analogous to the previous induction.
Again, the conditions of~\eqref{Composing iAST} are satisfied as we use
the reflexive, transitive closure of $>_{\F{G}}$.
  In the end, we know that $(\bigcup_{1 \leq i \leq m} \bigcup_{X_i >_{\F{G}}^* Y} Y,\SSS) = (\PP, \SSS)$ is iAST and this ends the proof.
\end{myproof}

\UsableTermsProc*

\pagebreak[2]

\begin{myproof}
  \smallskip
 
         \noindent
         \underline{\emph{Completeness}}
 
         \noindent 
  Assume that $(\C{T}_\mathtt{UT}(\PP, \SSS), \SSS)$ is not iAST\@.
  By \cref{lemma:starting} there exists a $(\C{T}_\mathtt{UT}(\PP, \SSS), \SSS)$-chain tree $\F{T} = (V,E,L,P)$ that
  converges with probability $<1$ and starts with $(1:\Com{1}(s^\#))$ such that $s^\# =
  \ell^\# \sigma$ for some substitution $\sigma$ and a dependency tuple $\langle
  \ell^\#,\ell \rangle \to \{p_1:\langle \upairs(d_1)_{\ell^\#\!,\PP,\SSS},r_1\rangle, \ldots, p_k:\langle\upairs(d_k)_{\ell^\#\!,\PP,\SSS},r_k\rangle\} \in \C{T}_\mathtt{UT}(\PP, \SSS)$, and every proper subterm of $\ell^\# \sigma$ is in normal form w.r.t.\ $\SSS$.
  We will now construct a $(\PP, \SSS)$-chain tree $\F{T}' = (V,E,L',P)$ that also starts
  with $(1:\Com{1}(s^\#))$ with the same underlying tree structure and adjusted labeling
  such that all nodes get the same probabilities as in $\F{T}$.
  Since the tree structure and the probabilities are the same, we then obtain $|\F{T}|_{\ctleaf} = |\F{T}'|_{\ctleaf}$.
  To be precise, the set of leaves in $\F{T}$ is the same as the set of leaves in $\F{T}'$, and they have the same probabilities.
  Since $|\F{T}|_{\ctleaf} < 1$, we thus have $|\F{T}'|_{\ctleaf} < 1$.
  Therefore, there exists a $(\PP, \SSS)$-chain tree that converges with  probability $<1$ and thus $(\PP, \SSS)$ is not iAST\@.

  \begin{figure}[tbp]
    \centering
    \small
    \begin{tikzpicture}
      \tikzstyle{adam}=[rectangle,thick,draw=black!100,fill=white!100,minimum size=4mm]
      \tikzstyle{empty}=[rectangle,thick,minimum size=4mm]
      
      \node[adam,pin={[pin distance=0.1cm, pin edge={,-}] 135:\tiny \textcolor{blue}{$P$}}] at (-3.5, 0)  (a) {$1:\Com{1}(s^\#)$};
      \node[adam] at (-5, -1)  (b) {$p_1:a_{1}$};
      \node[adam] at (-2, -1)  (c) {$p_2:a_{2}$};
      \node[adam,pin={[pin distance=0.1cm, pin edge={,-}] 135:\tiny \textcolor{blue}{$P$}}] at (-6, -2)  (d) {$p_3:a_3$};
      \node[adam] at (-4, -2)  (e) {$p_4:a_4$};
      \node[adam,pin={[pin distance=0.1cm, pin edge={,-}] 135:\tiny \textcolor{blue}{$P$}}] at (-2, -2)  (f) {$p_5:a_5$};
      \node[empty] at (-6, -3)  (g) {$\ldots$};
      \node[empty] at (-4, -3)  (h) {$\ldots$};
      \node[empty] at (-2, -3)  (i) {$\ldots$};

      \node[empty] at (-0.5, -1)  (arrow) {\Huge $\leadsto$};
      
      \node[adam,pin={[pin distance=0.1cm, pin edge={,-}] 135:\tiny \textcolor{blue}{$P$}}] at (3.5, 0)  (a2) {$1:\Com{1}(s^\#)$};
      \node[adam] at (2, -1)  (b2) {$p_1:a'_{1}$};
      \node[adam] at (5, -1)  (c2) {$p_2:a'_{2}$};
      \node[adam,pin={[pin distance=0.1cm, pin edge={,-}] 135:\tiny \textcolor{blue}{$P$}}] at (1, -2)  (d2) {$p_3:a'_3$};
      \node[adam] at (3, -2)  (e2) {$p_4:a'_4$};
      \node[adam,pin={[pin distance=0.1cm, pin edge={,-}] 135:\tiny \textcolor{blue}{$P$}}] at (5, -2)  (f2) {$p_5:a'_5$};
      \node[empty] at (1, -3)  (g2) {$\ldots$};
      \node[empty] at (3, -3)  (h2) {$\ldots$};
      \node[empty] at (5, -3)  (i2) {$\ldots$};
    
      \draw (a) edge[->] (b);
      \draw (a) edge[->] (c);
      \draw (b) edge[->] (d);
      \draw (b) edge[->] (e);
      \draw (c) edge[->] (f);
      \draw (d) edge[->] (g);
      \draw (e) edge[->] (h);
      \draw (f) edge[->] (i);

      \draw (a2) edge[->] (b2);
      \draw (a2) edge[->] (c2);
      \draw (b2) edge[->] (d2);
      \draw (b2) edge[->] (e2);
      \draw (c2) edge[->] (f2);
      \draw (d2) edge[->] (g2);
      \draw (e2) edge[->] (h2);
      \draw (f2) edge[->] (i2);
    \end{tikzpicture}
    \caption{Construction in this proof direction.}\label{fig:usable-pairs-soundness-proof-construction}
   \vspace*{-.2cm}
  \end{figure}
  The core idea of this construction is that adding terms to the first components of the right-hand side of a dependency tuple does not interfere with the possibility of a rewrite step.
  Hence, we can add the terms and completely ignore them for our chain tree and all occurring rewrite steps.
  The general construction is similar to the construction used in the proof of the dependency graph processor (\cref{theorem:prob-DGP}).

  For example, consider the DP problem $(\{ \eqref{R-div-deptup-5}\},\R_{\tpdiv})$
  and regard the following $(\{ \eqref{R-div-deptup-5}\},\R_{\tpdiv})$-chain tree $\F{T}$,
  where we again abbreviate $\ts^n(\O)$ by $n$ for all $n \in \IN$: 
  \begin{center}
    \centering
    \footnotesize
    \begin{tikzpicture}[node distance=1cm,>=stealth',bend angle=20,auto]
        \tikzstyle{adam}=[rectangle,thick,draw=black!100,minimum size=4mm]
        \begin{scope}
            \node [adam,pin={[pin distance=0.1cm, pin edge={,-}] 135:\tiny \textcolor{blue}{$P$}}] at (0, 0) (x1) []{$1:\Com{1}(\tD(4,2))$};
                \node [adam,pin={[pin distance=0.1cm, pin edge={,-}] 135:\tiny \textcolor{blue}{$P$}}] at (-3, -1) (x2) []{$\tfrac{1}{2}:\Com{1}(\tD(4,2))$};
                \node [adam] at (3, -1) (x4) []{$\tfrac{1}{2}:\Com{1}(\tD(\tminus(3,1),2))$};
                \node [adam] at (-4, -2) (x5) []{$\tfrac{1}{4}:\Com{1}(\tD(4,2))$};
                \node [adam] at (-2, -3) (x7) []{$\tfrac{1}{4}:\Com{1}(\tD(\tminus(3,1),2))$};
                \node [adam] at (3, -2) (x10) []{$\tfrac{1}{2}:\Com{1}(\tD(\tminus(2,0),2))$};

            \draw (x1) edge[->] (x2);
            \draw (x1) edge[->] (x4);
            \draw (x2) edge[->] (x5);
            \draw (x2) edge[->, in=30, out=340, looseness=2] (x7);
            \draw (x4) edge[->] (x10);
        \end{scope}
    \end{tikzpicture}
  \end{center}
  Then we can also add the non-usable term $\tM(x,y)$, so that we are working with the dependency tuple $\eqref{R-div-deptup-4}$ instead of $\eqref{R-div-deptup-5}$.
  Now, we can add the terms corresponding to this added term $\tM(x,y)$ to our tree to get:
  \begin{center}
    \centering
    \footnotesize
    \begin{tikzpicture}[node distance=1cm,>=stealth',bend angle=20,auto]
        \tikzstyle{adam}=[rectangle,thick,draw=black!100,minimum size=4mm]
        \begin{scope}
            \node [adam,pin={[pin distance=0.1cm, pin edge={,-}] 135:\tiny \textcolor{blue}{$P$}}] at (0, 0) (x1) []{$1:\Com{1}(\tD(4,2))$};
                \node [adam,pin={[pin distance=0.1cm, pin edge={,-}] 135:\tiny \textcolor{blue}{$P$}}] at (-3, -1) (x2) []{$\tfrac{1}{2}:\Com{1}(\tD(4,2))$};
                \node [adam] at (3, -1) (x4) []{$\tfrac{1}{2}:\Com{2}(\tD(\tminus(3,1),2),\tM(3,1))$};
                \node [adam] at (-4, -2) (x5) []{$\tfrac{1}{4}:\Com{1}(\tD(4,2))$};
                \node [adam] at (-2, -3) (x7) []{$\tfrac{1}{4}:\Com{2}(\tD(\tminus(3,1),2),\tM(3,1))$};
                \node [adam] at (3, -2) (x10) []{$\tfrac{1}{2}:\Com{2}(\tD(\tminus(2,0),2),\tM(3,1))$};

            \draw (x1) edge[->] (x2);
            \draw (x1) edge[->] (x4);
            \draw (x2) edge[->] (x5);
            \draw (x2) edge[->, in=30, out=340, looseness=2] (x7);
            \draw (x4) edge[->] (x10);
        \end{scope}
    \end{tikzpicture}
  \end{center}
This is a $(\{\eqref{R-div-deptup-4}\},\R_{\tpdiv})$-chain tree, and as one can see,
adding the term $\tM(x,y)$ does not affect
the computation at all since we can ignore it completely.

  We now construct the new labeling $L'$ for the $(\PP, \SSS)$-chain tree $\F{T}'$ recursively.
  Let $X \subseteq V$ be the set of nodes where we have already defined the labeling $L'$.
  During our construction, we ensure that the following property holds for every node $x \in X$:
  \begin{equation}\label{usable-terms-completeness-induction-hypothesis}
      \parbox{.8\textwidth}{For every node $x \in X$ we have $\cont(a_x) \subseteq \cont(a'_{x})$.}
  \end{equation}

  We start by setting $a'_{v} = a_{v} = \Com{1}(s^\#)$ for the root $v$ of the tree.
  Here, our property~\eqref{usable-terms-completeness-induction-hypothesis} is clearly satisfied as we have $a'_{v} = a_{v}$.

  As long as there is still an inner node $x \in X$ such that its successors are not contained in $X$, we do the following.
  Let $xE = \{y_1, \ldots, y_k\}$ be the set of its successors.
  We need to define the corresponding terms for the nodes $y_1, \ldots, y_k$ in $\F{T}'$.

  Since $x$ is not a leaf and $\F{T}$ is a $(\C{T}_\mathtt{UT}(\PP, \SSS), \SSS)$-chain tree, we have 
  \[
  a_x \setitoUsablePairsps \{\tfrac{p_{y_1}}{p_x}:a_{y_1}, \ldots, \tfrac{p_{y_k}}{p_x}:a_{y_k}\} \text{ or } a_x \setitos \{\tfrac{p_{y_1}}{p_x}:a_{y_1}, \ldots, \tfrac{p_{y_k}}{p_x}:a_{y_k}\}
  \]

  \medskip

  \noindent
  \textbf{(1) If we have} $a_x \setitops \{\tfrac{p_{y_1}}{p_x}:a_{y_1}, \ldots,
  \tfrac{p_{y_k}}{p_x}:a_{y_k}\}$ using the dependency tuple $\langle \ell^\#,\ell\rangle
  \to \{ p_1:\langle\upairs(d_1)_{\ell^\#\!,\PP,\SSS},r_1\rangle, \ldots,
  p_k:\langle\upairs(d_k)_{\ell^\#\!,\PP,\SSS},r_k\rangle\}$ and a substitution $\sigma$,
  then we have $a_x = \Com{n}(s_1^\#, \ldots, s_n^\#)$ and there is an $1 \leq i \leq n$ such that $s_i^\# = \ell^\# \sigma$ and all proper subterms of $\ell^\# \sigma$ are in normal form w.r.t.\ $\SSS$.
  Let $1 \leq j \leq k$.
  Then we have $a_{y_j} = \Com{n}(s_1'^\#, \ldots, \upairs(d_j)_{\ell^\#\!,\PP,\SSS} \sigma, \ldots, s_n'^\#)$.
  Furthermore, let $\{i'_1, \ldots, i'_m\}$ be the set of indices such that for all $i'
  \in \{i'_1, \ldots, i'_m\}$ we have $s_{i'}'^\# = s_{i'}^\#[r_j \sigma]_{\tau_{i'}}$ for some
  position $\tau_{i'}$ such that $s_{i'}^\#|_{\tau_{i'}} = \ell \sigma$, and for all $i' \in
  \{1,\ldots,n\} \setminus \{i,i'_1, \ldots, i'_m\}$ we have $s_{i'}'^\# = s_{i'}^\#$.

  We have $s_i^\# \in \cont(a_{x}) \subseteq_{(IH)} \cont(a'_{x})$.
  Let $a'_{x} = \Com{h}(t_1^\#, \ldots, t_h^\#)$.

  There exists a (unique) $1 \leq l \leq h$ such that $t_l^\# = s_i^\#$ and thus, we can
  rewrite the term
  $a'_{x}$ using the dependency tuple $\langle\ell^\#,\ell\rangle \to \{ p_1:\langle
  d_1,r_1\rangle, \ldots, p_k:\langle d_k,r_k\rangle\} \in \PP$ and the substitution $\sigma$, as we have $t_l^\# = \ell^\# \sigma$ and all proper subterms of $\ell^\# \sigma$ are in normal form w.r.t.\ $\SSS$.

  For each $i' \in \{i'_1, \ldots, i'_m\}$, by the induction hypothesis we can find a (unique) $1 \leq l' \leq h$ with $t_{l'}^\# = s_{i'}^\#$.
  Each of these $t_{l'}^\#$ will be rewritten so that we have $t_{l'}'^\# = t_{l'}^\#[r_j \sigma]_{\tau_{i'}}$ for the same position $\tau_{i'}$ as we used for $s_{i'}'^\#$.
  All other terms remain the same.
  This results in the term $a_{y_j}' = \Com{h}(t_1'^\#, \ldots, d_j \sigma, \ldots, t_h'^\#)$.
        
  It remains to show that our property \eqref{usable-terms-completeness-induction-hypothesis}
is still satisfied for this new labeling, i.e., we have $\cont(a_{y_j}) \subseteq \cont(a'_{y_j})$ for all $1 \leq j \leq k$.

  Let $1 \leq j \leq k$ and $b^\# \in \cont(a_{y_j})$.
  \begin{itemize}
    \item[$\bullet$] If $b^\#$ was introduced by $\upairs(d_j)_{\ell^\#\!,\PP,\SSS} \sigma$, then $b^\# \in \cont(d_j\sigma)$.
    Hence, we also have $b^\# \in \cont(a'_{y_j})$.
    
    \item[$\bullet$] If $b^\# = s_{i'}'^\#$ for some $i' \in \{i'_1, \ldots, i'_m\}$, then we have $b^\# = s_{i'}^\#[r_j \sigma]_{\tau_{i'}}$.
    We have $s_{i'}^\# \in \cont(a_{x}) \subseteq_{(IH)} \cont(a'_{x})$, and hence we find a (unique) $1 \leq l' \leq h$ such that $t_{l'}^\# = s_{i'}^\#$.
    By construction, we performed the same rewrite step at the same position, such that we get $b^\# = s_{i'}^\#[r_j \sigma]_{\tau_{i'}} = t_{l'}^\#[r_j \sigma]_{\tau_{i'}} = t_{l'}'^\#$ so that $b^\# \in \cont(a'_{y_j})$.

    \item[$\bullet$] If $b^\# = s_{i'}'^\#$ for some $i' \not\in \{i, i'_1, \ldots, i'_m\}$, then we have $b^\# = s_{i'}^\#$.
    We have $b^\# \in \cont(a_{x}) \subseteq_{(IH)} \cont(a'_{x})$, and hence we find a (unique) $1 \leq l' \leq h$ such that $t_{l'}^\# = s_{i'}^\#$.
    By construction, we did not change the term, such that we get $b^\# = s_{i'}^\# = t_{l'}'^\#$ so that $b^\# \in \cont(a'_{y_j})$.
  \end{itemize}

  \medskip

  \noindent
  \textbf{(2) In the second case we have} $a_x \setitos \{\tfrac{p_{y_1}}{p_x}:a_{y_1}, \ldots, \tfrac{p_{y_k}}{p_x}:a_{y_k}\}$ using a rewrite rule $\ell \to \{ p_1:r_1, \ldots, p_k:r_k\} \in \SSS$ (i.e., $x \in S$) and a substitution $\sigma$.
  Let $1 \leq j \leq k$, $a_x = \Com{n}(s_1^\#, \ldots, s_n^\#)$, and $a_{y_j} = \Com{n}(s_1'^\#, \ldots, s_n'^\#)$.
  Furthermore, let $\{i'_1, \ldots, i'_m\}$ be the set of indices such that for all $i'
  \in \{i'_1, \ldots, i'_m\}$ we have $s_{i'}'^\# = s_{i'}^\#[r_j \sigma]_{\tau_{i'}}$ for some
  position $\tau_{i'}$ such that $s_{i'}^\#|_{\tau_{i'}} = \ell \sigma$, and for all $i' \in
  \{1,\ldots,n\} \setminus \{i,i'_1, \ldots, i'_m\}$ we have $s_{i'}'^\# = s_{i'}^\#$.

  Let $a'_{x} = \Com{h}(t_1^\#, \ldots, t_h^\#)$.
  For each $i' \in \{i'_1, \ldots, i'_m\}$, by the induction hypothesis we can find a (unique) $1 \leq l' \leq h$ with $t_{l'}^\# = s_{i'}^\#$.
  Each of these $t_{l'}^\#$ will be rewritten so that we have $t_{l'}'^\# = t_{l'}^\#[r_j \sigma]_{\tau_{i'}}$ for the same position $\tau_{i'}$ as we used for $s_{i'}'^\#$.
  All other terms remain the same.
  This results in the term $a_{y_j}' = \Com{h}(t_1'^\#, \ldots, t_h'^\#)$.
        
  It remains to show that our property
  \eqref{usable-terms-completeness-induction-hypothesis}
  is still satisfied for this new labeling, i.e., we have $\cont(a_{y_j}) \subseteq \cont(a'_{y_j})$ for all $1 \leq j \leq k$.

  Let $1 \leq j \leq k$ and $b^\# \in \cont(a_{y_j})$.
  \begin{itemize}
      \item[$\bullet$] If $b^\# = s_{i'}'^\#$ for some $i' \in \{i'_1, \ldots, i'_m\}$, then we have $b^\# = s_{i'}^\#[r_j \sigma]_{\tau_{i'}}$.
      We have $s_{i'}^\# \in \cont(a_{x}) \subseteq_{(IH)} \cont(a'_{x})$, and hence we find a (unique) $1 \leq l' \leq h$ such that $t_{l'}^\# = s_{i'}^\#$.
      By construction, we performed the same rewrite step at the same position, such that we get $b^\# = s_{i'}^\#[r_j \sigma]_{\tau_{i'}} = t_{l'}^\#[r_j \sigma]_{\tau_{i'}} = t_{l'}'^\#$ so that $b^\# \in \cont(a'_{y_j})$.

      \item[$\bullet$] If $b^\# = s_{i'}'^\#$ for some $i' \not\in \{i, i'_1, \ldots, i'_m\}$, then we have $b^\# = s_{i'}^\#$.
      We have $b^\# \in \cont(a_{x}) \subseteq_{(IH)} \cont(a'_{x})$, and hence we find a (unique) $1 \leq l' \leq h$ such that $t_{l'}^\# = s_{i'}^\#$.
      By construction, we did not change the term, such that we get $b^\# = s_{i'}^\# = t_{l'}'^\#$ so that $b^\# \in \cont(a'_{y_j})$.
  \end{itemize}
  This was the last case and ends the construction and the proof of completeness.
  \medskip
 
  \noindent
  \underline{\emph{Soundness}}

  \noindent
  Assume that $(\PP, \SSS)$ is not iAST\@.
  By \cref{lemma:starting} there exists a $(\PP, \SSS)$-chain tree $\F{T} = (V,E,L,P)$ that
  converges with probability $<1$ and starts with $(1:\Com{1}(s^\#))$ such that $s^\# =
  \ell^\# \sigma$ for some substitution $\sigma$ and a dependency tuple $\langle
  \ell^\#,\ell\rangle \to \{p_1:\langle d_1,r_1\rangle, \ldots, p_k:\langle d_k,r_k\rangle\} \in \PP$, and every proper subterm of $\ell^\# \sigma$ is in normal form w.r.t.\ $\SSS$.
  We will now construct a $(\C{T}_\mathtt{UT}(\PP, \SSS), \SSS)$-chain tree $\F{T}' =
  (V,E,L',P)$ with splits
  that also starts with $(1:\Com{1}(s^\#))$ with the same underlying tree structure and
  adjusted labeling such that all nodes get the same probabilities as in $\F{T}$.
  Since the tree structure and the probabilities are the same, we then obtain $|\F{T}|_{\ctleaf} = |\F{T}'|_{\ctleaf}$.
  To be precise, the set of leaves in $\F{T}$ is the same as the set of leaves in $\F{T}'$, and they have the same probabilities.
  Since $|\F{T}|_{\ctleaf} < 1$, we thus have $|\F{T}'|_{\ctleaf} < 1$.
  Therefore, there exists a $(\C{T}_\mathtt{UT}(\PP, \SSS), \SSS)$-chain tree that converges with probability $<1$ and thus  (by
  \Cref{lemma:splitting}) $(\C{T}_\mathtt{UT}(\PP, \SSS), \SSS)$ is not iAST\@.

  \begin{figure}[tbp]
    \centering
    \small
    \begin{tikzpicture}
      \tikzstyle{adam}=[rectangle,thick,draw=black!100,fill=white!100,minimum size=4mm]
      \tikzstyle{empty}=[rectangle,thick,minimum size=4mm]
      
      \node[adam,pin={[pin distance=0.1cm, pin edge={,-}] 135:\tiny \textcolor{blue}{$P$}}] at (-3.5, 0)  (a) {$1:\Com{1}(s^\#)$};
      \node[adam] at (-5, -1)  (b) {$p_1:a_{1}$};
      \node[adam] at (-2, -1)  (c) {$p_2:a_{2}$};
      \node[adam,pin={[pin distance=0.1cm, pin edge={,-}] 135:\tiny \textcolor{blue}{$P$}}] at (-6, -2)  (d) {$p_3:a_3$};
      \node[adam] at (-4, -2)  (e) {$p_4:a_4$};
      \node[adam,pin={[pin distance=0.1cm, pin edge={,-}] 135:\tiny \textcolor{blue}{$P$}}] at (-2, -2)  (f) {$p_5:a_5$};
      \node[empty] at (-6, -3)  (g) {$\ldots$};
      \node[empty] at (-4, -3)  (h) {$\ldots$};
      \node[empty] at (-2, -3)  (i) {$\ldots$};

      \node[empty] at (-0.5, -1)  (arrow) {\Huge $\leadsto$};
      
      \node[adam,pin={[pin distance=0.1cm, pin edge={,-}] 135:\tiny \textcolor{blue}{$P$}}] at (3.5, 0)  (a2) {$1:\Com{1}(s^\#)$};
      \node[adam] at (2, -1)  (b2) {$p_1:a'_{1}$};
      \node[adam] at (5, -1)  (c2) {$p_2:a'_{2}$};
      \node[adam,pin={[pin distance=0.1cm, pin edge={,-}] 135:\tiny \textcolor{blue}{$P$}}] at (1, -2)  (d2) {$p_3:a'_3$};
      \node[adam] at (3, -2)  (e2) {$p_4:a'_4$};
      \node[adam,pin={[pin distance=0.1cm, pin edge={,-}] 135:\tiny \textcolor{blue}{$P$}}] at (5, -2)  (f2) {$p_5:a'_5$};
      \node[empty] at (1, -3)  (g2) {$\ldots$};
      \node[empty] at (3, -3)  (h2) {$\ldots$};
      \node[empty] at (5, -3)  (i2) {$\ldots$};
    
      \draw (a) edge[->] (b);
      \draw (a) edge[->] (c);
      \draw (b) edge[->] (d);
      \draw (b) edge[->] (e);
      \draw (c) edge[->] (f);
      \draw (d) edge[->] (g);
      \draw (e) edge[->] (h);
      \draw (f) edge[->] (i);

      \draw (a2) edge[->] (b2);
      \draw (a2) edge[->] (c2);
      \draw (b2) edge[->] (d2);
      \draw (b2) edge[->] (e2);
      \draw (c2) edge[->] (f2);
      \draw (d2) edge[->] (g2);
      \draw (e2) edge[->] (h2);
      \draw (f2) edge[->] (i2);
    \end{tikzpicture}
    \caption{Construction in this proof direction.}\label{fig:usable-pairs-completeness-proof-construction}
  \end{figure}

  The core idea of this construction is that terms that cannot be rewritten to the left-hand side of a dependency tuple using $\setitos$ steps can never be used for a $\setitops$ step.
  This means that we can simply remove those terms from the
arguments of compound symbols in our rules and in 
 chain trees, and again result in valid chain trees.
  The general construction is similar to the construction used in the proof of the dependency graph processor (\cref{theorem:prob-DGP}).

  For example, consider the DP problem $(\{\eqref{R-div-deptup-4}\},\R_{\tpdiv})$
  and regard the following $(\{ \eqref{R-div-deptup-4}\},\R_{\tpdiv})$-chain tree $\F{T}$:
  \begin{center}
    \centering
    \footnotesize
    \begin{tikzpicture}[node distance=1cm,>=stealth',bend angle=20,auto]
        \tikzstyle{adam}=[rectangle,thick,draw=black!100,minimum size=4mm]
        \begin{scope}
            \node [adam,pin={[pin distance=0.1cm, pin edge={,-}] 135:\tiny \textcolor{blue}{$P$}}] at (0, 0) (x1) []{$1:\Com{1}(\tD(4,2))$};
                \node [adam,pin={[pin distance=0.1cm, pin edge={,-}] 135:\tiny \textcolor{blue}{$P$}}] at (-3, -1) (x2) []{$\tfrac{1}{2}:\Com{1}(\tD(4,2))$};
                \node [adam] at (3, -1) (x4) []{$\tfrac{1}{2}:\Com{2}(\tD(\tminus(3,1),2),\tM(3,1))$};
                \node [adam] at (-4, -2) (x5) []{$\tfrac{1}{4}:\Com{1}(\tD(4,2))$};
                \node [adam] at (-2, -3) (x7) []{$\tfrac{1}{4}:\Com{2}(\tD(\tminus(3,1),2),\tM(3,1))$};
                \node [adam] at (3, -2) (x10) []{$\tfrac{1}{2}:\Com{2}(\tD(\tminus(2,0),2),\tM(3,1))$};

            \draw (x1) edge[->] (x2);
            \draw (x1) edge[->] (x4);
            \draw (x2) edge[->] (x5);
            \draw (x2) edge[->, in=30, out=340, looseness=2] (x7);
            \draw (x4) edge[->] (x10);
        \end{scope}
    \end{tikzpicture}
  \end{center}
  Then we can also remove the non-usable term $\tM(x,y)$, so that we are working with the dependency tuple $\eqref{R-div-deptup-5}$ instead of $\eqref{R-div-deptup-4}$.
  Now, we can remove the terms corresponding to this removed term $\tM(x,y)$ from our tree to get:
  \begin{center}
      \centering
      \footnotesize
      \begin{tikzpicture}[node distance=1cm,>=stealth',bend angle=20,auto]
          \tikzstyle{adam}=[rectangle,thick,draw=black!100,minimum size=4mm]
          \begin{scope}
              \node [adam,pin={[pin distance=0.1cm, pin edge={,-}] 135:\tiny \textcolor{blue}{$P$}}] at (0, 0) (x1) []{$1:\Com{1}(\tD(4,2))$};
                  \node [adam,pin={[pin distance=0.1cm, pin edge={,-}] 135:\tiny \textcolor{blue}{$P$}}] at (-3, -1) (x2) []{$\tfrac{1}{2}:\Com{1}(\tD(4,2))$};
                  \node [adam] at (3, -1) (x4) []{$\tfrac{1}{2}:\Com{1}(\tD(\tminus(3,1),2))$};
                  \node [adam] at (-4, -2) (x5) []{$\tfrac{1}{4}:\Com{1}(\tD(4,2))$};
                  \node [adam] at (-2, -3) (x7) []{$\tfrac{1}{4}:\Com{1}(\tD(\tminus(3,1),2))$};
                  \node [adam] at (3, -2) (x10) []{$\tfrac{1}{2}:\Com{1}(\tD(\tminus(2,0),2))$};

              \draw (x1) edge[->] (x2);
              \draw (x1) edge[->] (x4);
              \draw (x2) edge[->] (x5);
              \draw (x2) edge[->, in=30, out=340, looseness=2] (x7);
              \draw (x4) edge[->] (x10);
          \end{scope}
      \end{tikzpicture}
  \end{center}
  This is a $(\{\eqref{R-div-deptup-5}\},\R_{\tpdiv})$-chain tree, and as one can see,
  removing the term does not affect
  the computation at all since we cannot rewrite the term with a dependency tuple anyway.
  
  We now construct the new labeling $L'$ for the $(\C{T}_\mathtt{UT}(\PP, \SSS),\SSS)$-chain tree $\F{T}'$ recursively.
  Let $X \subseteq V$ be the set of nodes where we have already defined the labeling $L'$.
  During our construction, we ensure that the following property holds for every node $x \in X$:
  \begin{equation}\label{usable-terms-soundness-induction-hypothesis}
      \parbox{.8\textwidth}{For every node $x \in X$ we have $\cont(a_x) \setminus \Junk(a_x) \subseteq \cont(a'_{x})$.}
  \end{equation}
  Here, for any normalized term $a_x = \Com{n}(s_1^\#, \ldots, s_n^\#)$, let $\Junk(a_x)$
  denote the set of all terms $s_i^\#$ that can never be used for a rewrite step with a
  dependency tuple from $\PP$, similar to the soundness proof of \cref{theorem:prob-DGP}.
  To be precise, we define $s_i^\# \in \Junk(a_x)$:$\Leftrightarrow$ there is no
  dependency tuple $\langle \ell^\#, \ell\rangle \to \{p_1: \langle d_1,r_1\rangle,
  \ldots, p_k: \langle d_k,r_k\rangle\} \in \PP$ and substitution $\sigma$, such that
  $s_i^\# \itononprobsstar \ell^\# \sigma$, and every proper subterm of $\ell^\# \sigma$
  is in normal form w.r.t.\ $\SSS$.

  We start by setting $a'_{v} = a_{v} = \Com{1}(s^\#)$ for the root $v$ of the tree.
  Here, our property~\eqref{usable-terms-soundness-induction-hypothesis} is clearly satisfied as we have $a'_{v} = a_{v}$.

  As long as there is still an inner node $x \in X$ such that its successors are not contained in $X$, we do the following.
  Let $xE = \{y_1, \ldots, y_k\}$ be the set of its successors.
  We need to define the corresponding terms for the nodes $y_1, \ldots, y_k$ in $\F{T}'$.

  Since $x$ is not a leaf and $\F{T}$ is a $(\PP, \SSS)$-chain tree, we have 
  \[
  a_x \setitops \{\tfrac{p_{y_1}}{p_x}:a_{y_1}, \ldots, \tfrac{p_{y_k}}{p_x}:a_{y_k}\} \text{ or } a_x \setitos \{\tfrac{p_{y_1}}{p_x}:a_{y_1}, \ldots, \tfrac{p_{y_k}}{p_x}:a_{y_k}\}
  \]

  \medskip

  \noindent
  \textbf{(1) If we have} $a_x \setitops \{\tfrac{p_{y_1}}{p_x}:a_{y_1}, \ldots,
  \tfrac{p_{y_k}}{p_x}:a_{y_k}\}$ using the dependency tuple $\langle \ell^\#,\ell \rangle
  \to \{ p_1:\langle d_1,r_1 \rangle, \ldots, p_k:\langle d_k,r_k\rangle\}$ and a substitution $\sigma$, then we have $a_x = \Com{n}(s_1^\#, \ldots, s_n^\#)$ and an $1 \leq i \leq n$ such that $s_i^\# = \ell^\# \sigma$ and all proper subterms of $\ell^\# \sigma$ are in normal form w.r.t.\ $\SSS$.
  Let $1 \leq j \leq k$.
  Then we have $a_{y_j} = \Com{n}(s_1'^\#, \ldots, d_j \sigma, \ldots, s_n'^\#)$.
  Furthermore, let $\{i'_1, \ldots, i'_m\}$ be the set of indices such that for all $i'
  \in \{i'_1, \ldots, i'_m\}$ we have $s_{i'}'^\# = s_{i'}^\#[r_j \sigma]_{\tau_{i'}}$ for some
  position $\tau_{i'}$ such that $s_{i'}^\#|_{\tau_{i'}} = \ell \sigma$, and for all $i' \in
  \{1,\ldots,n\} \setminus \{i,i'_1, \ldots, i'_m\}$ we have $s_{i'}'^\# = s_{i'}^\#$.

  The term $s_i^\#$ cannot be in $\Junk(a_{x})$.
  Hence, we must have $s_i^\# \in \cont(a_{x}) \setminus \Junk(a_{x}) \subseteq_{(IH)} \cont(a'_{x})$.
  Let $a'_{x} = \Com{h}(t_1^\#, \ldots, t_h^\#)$.

  There exists a (unique) $1 \leq l \leq h$ such that $t_l^\# = s_i^\#$ and thus, we can
  rewrite the term
  $a'_{x}$ using the dependency tuple $\langle \ell^\#,\ell \rangle \to \{ p_1:\langle\upairs(d_1)_{\ell^\#\!,\PP,\SSS},r_1\rangle, \ldots, p_k:\langle\upairs(d_k)_{\ell^\#\!,\PP,\SSS},r_k\rangle\} \in \C{T}_\mathtt{UT}(\PP, \SSS)$ and the substitution $\sigma$, as we have $t_l^\# = \ell^\# \sigma$ and all proper subterms of $\ell^\# \sigma$ are in normal form w.r.t.\ $\SSS$.

  For each $i' \in \{i'_1, \ldots, i'_m\}$ such that $s_{i'}^\# \in \cont(a_{x}) \setminus \Junk(a_{x})$, by the induction hypothesis we can find a (unique) $1 \leq l' \leq h$ with $t_{l'}^\# = s_{i'}^\#$.
  Each of these $t_{l'}^\#$ will be rewritten so that we have $t_{l'}'^\# = t_{l'}^\#[r_j \sigma]_{\tau_{i'}}$ for the same position $\tau_{i'}$ as we used for $s_{i'}'^\#$.
  All other terms remain the same.
  This results in the term $a_{y_j}' = \Com{h}(t_1'^\#, \ldots, \upairs(d_j)_{\ell^\#\!,\PP,\SSS} \sigma, \ldots, t_h'^\#)$.
        
  It remains to show that our property \eqref{usable-terms-soundness-induction-hypothesis}
is still satisfied for this new labeling, i.e., we have $\cont(a_{y_j}) \setminus \Junk(a_{y_j}) \subseteq \cont(a'_{y_j})$ for all $1 \leq j \leq k$.

  Let $1 \leq j \leq k$ and $b^\# \in \cont(a_{y_j}) \setminus \Junk(a_{y_j})$.
  \begin{itemize}
    \item[$\bullet$] If $b^\#$ was introduced by $d_j \sigma$, then $b^\# \in \cont(d_j\sigma)$.
    If $b^\# \in \cont(\upairs(d_j)_{\ell^\#\!,\PP,\SSS}\sigma)$, then we also have $b^\# \in \cont(a'_{y_j})$.
    If $b^\# \not\in \cont(\upairs(d_j)_{\ell^\#\!,\PP,\SSS}\sigma)$, then there is no
    dependency tuple $\langle\ell'^\#, \ell'\rangle \to \ldots \in \PP$ and substitution
    $\sigma'$ such that $b^\# \itononprobsstar \ell'^\# \sigma'$, and every proper subterm
    of $\ell'^\# \sigma'$ is in normal form w.r.t.\ $\SSS$, by the definition of usable terms.
    Hence, we have $b^\# \in \Junk(a_{y_j})$ and this is a contradiction to our assumption that $b^\# \in \cont(a_{y_j}) \setminus \Junk(a_{y_j})$, so this case is not possible.
    
    \item[$\bullet$] If $b^\# = s_{i'}'^\#$ for some $i' \in \{i'_1, \ldots, i'_m\}$, then we have $b^\# = s_{i'}^\#[r_j \sigma]_{\tau_{i'}}$.

    If we have $s_{i'}^\# \in \Junk(a_{x})$, then we also have $b^\# \in \Junk(a_{y_j})$, since $s_{i'}^\# \itononprobs s_{i'}^\#[r_j \sigma]_{\tau_{i'}} = b^\#$, and this is a contradiction to our assumption that $b^\# \in \cont(a_{y_j}) \setminus \Junk(a_{y_j})$.
    Thus, we have $s_{i'}^\# \in \cont(a_{x}) \setminus \Junk(a_{x}) \subseteq_{(IH)} \cont(a'_{x})$, and we find a (unique) $1 \leq l' \leq h$ such that $t_{l'}^\# = s_{i'}^\#$.
    By construction, we performed the same rewrite step at the same position, such that we get $b^\# = s_{i'}^\#[r_j \sigma]_{\tau_{i'}} = t_{l'}^\#[r_j \sigma]_{\tau_{i'}} = t_{l'}'^\#$ so that $b^\# \in \cont(a'_{y_j})$.

    \item[$\bullet$] If $b^\# = s_{i'}'^\#$ for some $i' \not\in \{i, i'_1, \ldots, i'_m\}$, then we have $b^\# = s_{i'}^\#$.
    If we have $s_{i'}^\# \in \Junk(a_{x})$, then we also have $b^\# = s_{i'}'^\# \in
    \Junk(a_{y_j})$
since $s_{i'}^\# = s_{i'}'^\#$, and this is a contradiction to our assumption that $b^\# \in \cont(a_{y_j}) \setminus \Junk(a_{y_j})$.
    Thus, we have $b^\# \in \cont(a_{x}) \setminus \Junk(a_{x}) \subseteq_{(IH)} \cont(a'_{x})$, and we find a (unique) $1 \leq l' \leq h$ such that $t_{l'}^\# = s_{i'}^\#$.
    By construction, we did not change the term, such that we get $b^\# = s_{i'}^\# = t_{l'}'^\#$ so that $b^\# \in \cont(a'_{y_j})$.
  \end{itemize}

  \medskip

  \noindent
  \textbf{(2) In the second case we have} $a_x \setitos \{\tfrac{p_{y_1}}{p_x}:a_{y_1}, \ldots, \tfrac{p_{y_k}}{p_x}:a_{y_k}\}$ using a rewrite rule $\ell \to \{ p_1:r_1, \ldots, p_k:r_k\} \in \SSS$ (i.e., $x \in S$) and a substitution $\sigma$.
  Let $1 \leq j \leq k$, $a_x = \Com{n}(s_1^\#, \ldots, s_n^\#)$, and $a_{y_j} = \Com{n}(s_1'^\#, \ldots, s_n'^\#)$.
  Furthermore, let $\{i'_1, \ldots, i'_m\}$ be the set of indices such that for all $i'
  \in \{i'_1, \ldots, i'_m\}$ we have $s_{i'}'^\# = s_{i'}^\#[r_j \sigma]_{\tau_{i'}}$ for some
  position $\tau_{i'}$ such that $s_{i'}^\#|_{\tau_{i'}} = \ell \sigma$, and for all $i' \in
  \{1,\ldots,n\} \setminus \{i,i'_1, \ldots, i'_m\}$ we have $s_{i'}'^\# = s_{i'}^\#$.

  Let $a'_{x} = \Com{h}(t_1^\#, \ldots, t_h^\#)$.
  For each $i' \in \{i'_1, \ldots, i'_m\}$ such that $s_{i'}^\# \in \cont(a_{x}) \setminus
  \Junk(a_{x})$, by the induction hypothesis we can find a (unique) $1 \leq l' \leq h$ with $t_{l'}^\# = s_{i'}^\#$.
  Each of these $t_{l'}^\#$ will be rewritten so that we have $t_{l'}'^\# = t_{l'}^\#[r_j \sigma]_{\tau_{i'}}$ for the same position $\tau_{i'}$ as we used for $s_{i'}'^\#$.
  All other terms remain the same.
  This results in the term $a_{y_j}' = \Com{h}(t_1'^\#, \ldots, t_h'^\#)$.
  Note that $\{i'_1, \ldots, i'_m\}$ may also be empty. 
  In this case, we have to use a split-node, where we do not perform any rewrite step but only mirror the tree structure.
        
  It remains to show that our property \eqref{usable-terms-soundness-induction-hypothesis}
is still satisfied for this new labeling, i.e., we have $\cont(a_{y_j}) \setminus \Junk(a_{y_j}) \subseteq \cont(a'_{y_j})$ for all $1 \leq j \leq k$.

  Let $1 \leq j \leq k$ and $b^\# \in \cont(a_{y_j}) \setminus \Junk(a_{y_j})$.
  \begin{itemize}
      \item[$\bullet$] If $b^\# = s_{i'}'^\#$ for some $i' \in \{i'_1, \ldots, i'_m\}$, then we have $b^\# = s_{i'}^\#[r_j \sigma]_{\tau_{i'}}$.

      If we have $s_{i'}^\# \in \Junk(a_{x})$, then we also have $b^\# \in \Junk(a_{y_j})$, since $s_{i'}^\# \itononprobs s_{i'}^\#[r_j \sigma]_{\tau_{i'}} = b^\#$, and this is a contradiction to our assumption that $b^\# \in \cont(a_{y_j}) \setminus \Junk(a_{y_j})$.
      Thus, we have $s_{i'}^\# \in \cont(a_{x}) \setminus \Junk(a_{x}) \subseteq_{(IH)} \cont(a'_{x})$, and we find a (unique) $1 \leq l' \leq h$ such that $t_{l'}^\# = s_{i'}^\#$.
      By construction, we performed the same rewrite step at the same position, such that we get $b^\# = s_{i'}^\#[r_j \sigma]_{\tau_{i'}} = t_{l'}^\#[r_j \sigma]_{\tau_{i'}} = t_{l'}'^\#$ so that $b^\# \in \cont(a'_{y_j})$.

      \item[$\bullet$] If $b^\# = s_{i'}'^\#$ for some $i' \not\in \{i, i'_1, \ldots, i'_m\}$, then we have $b^\# = s_{i'}^\#$.
      If we have $s_{i'}^\# \in \Junk(a_{x})$, then we also have $b^\# = s_{i'}'^\# \in
      \Junk(a_{y_j})$
since $s_{i'}^\# = s_{i'}'^\#$, and this is a contradiction to our assumption that $b^\#
\in \cont(a_{y_j}) \setminus \Junk(a_{y_j})$. 
      Thus, we have $b^\# \in \cont(a_{x}) \setminus \Junk(a_{x}) \subseteq_{(IH)} \cont(a'_{x})$, and we find a (unique) $1 \leq l' \leq h$ such that $t_{l'}^\# = s_{i'}^\#$.
      By construction, we did not change the term, such that we get $b^\# = s_{i'}^\# = t_{l'}'^\#$ so that $b^\# \in \cont(a'_{y_j})$.
  \end{itemize}

  This was the last case and ends the construction and the proof of soundness.
\end{myproof}

\ProbUsRulesProc*

\begin{myproof}
  \smallskip
 
  \noindent
  \underline{\emph{Soundness}}

  \noindent 
  Assume that $(\PP, \SSS)$ is not iAST\@.
  By \cref{lemma:starting}, there exists a $(\PP, \SSS)$-chain tree $\F{T} = (V,E,L,P)$
  that converges with probability $<1$ and starts with $\Com{1}(s^\#)$ such that $s^\# =
  \ell^\# \sigma$ for some dependency tuple $\langle \ell^\#,\ell \rangle \to \{p_1:
  \langle d_1,r_1\rangle, \ldots, p_k:\langle d_k,r_k\rangle\} \in \PP$ and some substitution $\sigma$, and every proper subterm of $\ell^\# \sigma$ is in normal form w.r.t.\ $\SSS$.

  By the definition of usable rules, as in the non-probabilistic case, 
  rules $\ell \to \mu \in \SSS$ that are not usable (i.e., $\ell \to \mu \not\in \urules(\PP,\SSS)$) will never be used in such a $(\PP, \SSS)$-chain tree.
  Hence, we can also view $\F{T}$ as a $(\PP, \urules(\PP,\SSS))$-chain tree that
  converges with probability $<1$ and thus $(\PP, \urules(\PP,\SSS))$ is not iAST\@.
\end{myproof}

As in the non-probabilistic case, the usable rules processor is not complete.

\begin{lemma}[Counterexample for Completeness of the Usable Rules Processor]\label{lemma:counterexample-prob-urp-completeness}
	\emergencystretch=1em
	There exists a DP problem $(\PP, \SSS)$ such that $(\PP, \SSS)$ is iAST, but $(\PP, \urules(\PP,\SSS))$ is not iAST\@.
\end{lemma}

\begin{myproof}
Let the PTRS $\SSS$ consist of the following rule:
  \begin{align}
		\label{R-prob-usable-rules-no-complete-1} & \ta \to \{1:\ta\}
	\end{align}
  and let $\PP$ consist of the following dependency tuple:
  \begin{align}
		\label{R-prob-usable-dep-pair-no-complete-1} & \langle
                \tF(\ta,x),\tf(\ta,x)\rangle \to
                \{1: \langle \Com{1}(\tF(x,x)), \tf(x,x) \rangle\}
	\end{align}

  \noindent
	\begin{minipage}{.70\textwidth}
		Then, the DP problem $(\PP,\SSS)$ is iAST, since the only dependency tuple
                \eqref{R-prob-usable-dep-pair-no-complete-1} cannot be used in innermost reductions.
		The reason is that the first component
$\tF(\ta,x)$
                in the left-hand side contains the proper subterm $\ta$
                that is not in normal form w.r.t.\ $\SSS$.
		However, the rule \eqref{R-prob-usable-rules-no-complete-1} is not usable since the right-hand side of the only dependency tuple does not contain $\ta$.
		We get $\Proc_{\mathtt{UR}}(\PP,\SSS) = \{(\PP,\urules(\PP,\SSS))\} = \{(\PP,\emptyset)\}$.
		The DP problem $(\PP,\emptyset)$ is not iAST anymore, since we
                have the $(\PP,\emptyset)$-chain tree depicted on the right that converges with probability $0$.
		The reason is that the proper subterm $\ta$ is now in normal form w.r.t.\ the empty PTRS.
	\end{minipage}%
	\begin{minipage}{.30\textwidth}
		\centering
		\small
		\begin{tikzpicture}
			\tikzstyle{adam}=[rectangle,thick,draw=black!100,fill=white!100,minimum size=4mm]
			\tikzstyle{empty}=[rectangle,thick,minimum size=4mm]
			
			\node[adam,pin={[pin distance=0.1cm, pin edge={,-}] 135:\tiny
                            \textcolor{blue}{$P$}}] at (0, 0)  (a) {$1:\Com{1}(\tF(\ta,\ta))$};
			\node[adam,pin={[pin distance=0.1cm, pin edge={,-}] 135:\tiny \textcolor{blue}{$P$}}] at (0, -1.5)  (b) {$1:\Com{1}(\tF(\ta,\ta))$};
			\node[empty] at (0, -3)  (c) {$\ldots$};
		
			\draw (a) edge[->] (b);
			\draw (b) edge[->] (c);
		\end{tikzpicture}
		\label{fig:counterexample-prob-urp-completeness}
			\end{minipage}
\end{myproof}

The proof of the reduction pair processor is split into two parts. 
First, we consider the technical part of the proof and then we conclude the soundness and completeness of the processor.

\begin{lemma}[Proving iAST on CTs with Reduction Pair Processor]\label{lemma:prob-RPP-CT-lemma}
	Let $(\PP, \SSS)$ be a DP problem and let $\Pol:\TSet{\SigmaDP}{\VSet} \to
        \IN[\VSet]$ be a weakly monotonic, multilinear, and $\Com{}$-additive polynomial interpretation. 

	Suppose that we have $\PP = \PP_{\geq} \uplus \PP_{>}$ with $\PP_>
        \neq \emptyset$ and the following conditions hold.
	\begin{itemize}
		\item[(1)] For every $\ell \to \{ p_1:r_1, \ldots,p_k:r_k \} \in \SSS$ we have 
		\[\Pol(\ell) \geq \sum_{1 \leq j \leq k} p_j \cdot \Pol(r_j).\]
		\item[(2)] For every  $\langle \ell^\#,\ell \rangle \to \{ p_1:\langle d_1,r_1\rangle,
                \ldots,p_k:\langle d_k,r_k\rangle \} \in \PP$, we have 
		\[\Pol(\ell^\#) \geq \sum_{1 \leq j \leq k} p_j \cdot \Pol(d_j).\]
		\item[(3)]
		For every $\langle \ell^\#,\ell \rangle \to \{ p_1:\langle d_1,r_1\rangle,
                \ldots,p_k:\langle d_k,r_k\rangle \} \in \PP_{>}$, there exists a $1 \leq j \leq k$ with 
		\[\Pol(\ell^\#) > \Pol(d_j).\]
		If $\ell \to \{ p_1:r_1, \ldots,p_k:r_k \} \in \SSS$, then we additionally require that 
		\[\Pol(\ell) \geq \Pol(r_j).\]
	\end{itemize}
	Let $\F{T} = (V,E,L,P)$ be a $(\PP, \SSS)$-chain tree.
	We can partition $P = P_{\geq} \uplus P_{>}$ according to $\PP = \PP_{\geq} \uplus \PP_{>}$.
	If $\F{T}$ satisfies
	\begin{itemize}
		\item[(+)] Every infinite path has an infinite number of $P_{>}$ nodes,
	\end{itemize}
	then $|\F{T}|_{\ctleaf} = 1$.
\end{lemma}

\begin{myproof}
 This proof proceeds similar to the proof
  of~\cref{theorem:ptrs-direct-application-poly-interpretations} and thus it also uses the
  proof idea for AST from~\cite{mciver2017new}.
	The difference to \cref{theorem:ptrs-direct-application-poly-interpretations} is
        that this time we consider
        chain trees and we have two different rewrite relations that we need to deal with.
For a  $(\PP, \SSS)$-chain tree $\F{T}$ that satisfies (+), the core steps of the proof are again the following:
	\begin{enumerate}
		\item[1.] We extend the conditions (1), (2), and (3) to rewrite steps instead of just rules (and thus, to edges of a chain tree).
		\item[2.] We create a chain tree $\F{T}^{\leq N}$ for any $N \in \IN$.
		\item[3.] We prove that $|\F{T}^{\leq N}|_{\ctleaf} \geq p_{min}^{N}$ for any $N \in \IN$.
		\item[4.] We prove that $|\F{T}^{\leq N}|_{\ctleaf}=1$ for any $N \in \IN$.
		\item[5.] Finally, we prove that $|\F{T}|_{\ctleaf}=1$.
	\end{enumerate}
	Parts (1.) and (3.) are much more involved in this proof compared to \cref{theorem:ptrs-direct-application-poly-interpretations} due to the more complex rewrite relation that we are dealing with.
	The other parts are nearly the same as before. 
	We only have to adjust everything to chain trees instead of rewrite sequences.
	Here, let $p_{min}>0$ be the minimal probability that occurs in the rules of $\PP$
        or $\SSS$.
	As $\PP$ and $\SSS$ both have only finitely many rules and all occurring multi-distributions are finite as well, this minimum is well defined.
    Again, we have $p_{min}^{N} = \underbrace{p_{min} \cdot \ldots \cdot p_{min}}_\text{$N$ times}$.

  \medskip
        
	\noindent 
  \textbf{\underline{1. We extend the conditions to rewrite steps instead of just rules}}

  \noindent
	We start by showing that the conditions (1), (2), and (3) of the lemma extend to rewrite steps instead of just rules.
	\begin{enumerate}
		\item[(a)] If $s \itos \{ p_1:t_1, \ldots, p_k:t_k \}$ using the rule $\ell \to \{ p_1:r_1, \ldots, p_k:r_k \} \in \SSS$ with $\Pol(\ell) \geq \Pol(r_j)$ for some $1 \leq j \leq k$, then we have $Pol(s) \geq Pol(t_j)$.
		\item[(b)] If $a \setitopsgt \{ p_1:b_1, \ldots, p_k:b_k \}$, then $\Pol(a) > \Pol(b_j)$ for some $1 \leq j \leq k$.
		\item[(c)] If $s \itos \{ p_1:t_1, \ldots, p_k:t_k \}$, then $Pol(s) \geq \sum_{1 \leq j \leq k} p_j \cdot \Pol(t_j)$.
		\item[(d)] If $a \setitops \{ p_1:b_1, \ldots, p_k:b_k\}$, then $\Pol(a) \geq \sum_{1 \leq j \leq k} p_j \cdot \Pol(b_j)$.
		\item[(e)] If $a \setitos \{p_1:b_1, \ldots, p_k:b_k\}$, then $\Pol(a) \geq \sum_{1 \leq j \leq k} p_j \cdot \Pol(b_j)$.
	\end{enumerate}

	\begin{itemize}
    	\item[(a)] In this case, there exist a rule $\ell \to \{ p_1:r_1, \ldots, p_k:r_k \} \in \SSS$ with $\Pol(\ell) \geq \Pol(r_j)$ for some $1 \leq j \leq k$, a substitution $\sigma$, and a position $\pi$ of $s$ such that $s|_\pi =\ell\sigma$, all proper subterms of $\ell\sigma$ are in normal form w.r.t.\ $\SSS$, and $t_h = s[r_h \sigma]_\pi$ for all $1 \leq h \leq k$.
        
      We perform structural induction on $\pi$.
      So in the induction base, let $\pi = \epsilon$.
      Hence, we have $s = \ell\sigma \itos \{ p_1:r_1 \sigma, \ldots, p_k:r_k \sigma \}$.
      By assumption we have $\Pol(\ell) \geq \Pol(r_j)$ for some $1 \leq j \leq k$.
      As these inequations hold for all instantiations of the occurring variables, for $t_j = r_j\sigma$ we have
      \[ \Pol(s) = \Pol(\ell\sigma) \geq \Pol(r_j\sigma) = \Pol(t_j). \]
      
      In the induction step, we have $\pi = i.\pi'$, $s = f(s_1,\ldots,s_i,\ldots,s_n)$,
      $s_i \itos \{ p_1:t_{i,1}, \ldots, p_k:t_{i,k} \}$, and $t_h =
      f(s_1,\ldots,t_{i,h},\ldots,s_n)$ with $t_{i,h} = s_i[r_h\sigma]_{\pi'}$ for all $1 \leq h \leq k$.
      Then by the induction hypothesis we have $Pol(s_i) \geq Pol(t_{i,j})$.
      For $t_j = f(s_1,\ldots,t_{i,j},\ldots,s_n)$ we obtain
      \[
      \begin{array}{lcl}
        \Pol(s) & = & \Pol(f(s_1,\ldots,s_i,\ldots,s_n)) \\
            & = & f_{\Pol}(\Pol(s_1),\ldots,\Pol(s_i),\ldots,\Pol(s_n)) \\
            & \geq & f_{\Pol}(\Pol(s_1),\ldots,\Pol(t_{i,j}),\ldots,\Pol(s_n)) \\
            &  & \hspace*{1cm} \text{ (by weak monotonicity of $f_{\Pol}$ and $Pol(s_i) \geq Pol(t_{i,j})$)} \\
        & = & \Pol(f(s_1,\ldots,t_{i,j},\ldots,s_n)) \\
            & = & \Pol(t_j).
      \end{array}
      \]

	    \item[(b)] In this case, there exist a dependency tuple $\langle \ell^\#,\ell
              \rangle \to \{ p_1:\langle d_1,r_1\rangle, \ldots, p_k:\langle d_k,r_k \rangle \} \in \PP_>$, and a substitution $\sigma$.
      Furthermore, let $a = \Com{n}(s_1^\#, \ldots,\linebreak s_n^\#)$ such that $s_i^\# = \ell^\#\sigma$ and all proper subterms of $\ell^\#\sigma$ are in normal form w.r.t.\ $\SSS$.
      Let $I_1$ be the set of indices such that for all $i' \in I_1$ we rewrite $s_{i'}^\#$ to $s_{i'}^\#[r_j \sigma]_{\tau_{i'}}$ for some position $\tau_{i'}$ such that $s_{i'}^\#|_{\tau_{i'}} = \ell \sigma$, and for all $i'' \in I_2 = \{1,\ldots,n\} \setminus (I_1 \cup \{i\}) $ the term $s_{i''}^\#$ remains the same.
      By Requirement (3), there exists a $1 \leq j \leq k$ with $\Pol(\ell^\#) >
      \Pol(d_j)$ and $\Pol(\ell) \geq \Pol(r_j)$ if $I_1 \neq \emptyset$.
      As these inequations hold for all instantiations of the occurring variables, we have
      \[
        \begin{array}{@{\hspace*{1cm}}lcl}
          \Pol(a) & = & \sum_{s^{\#} \in \cont(a)} \Pol(s^{\#})   \\
           && \hspace*{.5cm} \text{(as $\Pol$ is $\Com{}$-additive)}\\
              & = & \Pol(s_i^{\#}) + \sum_{i' \in I_1} \Pol(s_{i'}^{\#}) + \sum_{i'' \in I_2} \Pol(s_{i''}^{\#}) \\
              & = & \Pol(\ell^\#\sigma) + \sum_{i' \in I_1} \Pol(s_{i'}^{\#}) + \sum_{i'' \in I_2} \Pol(s_{i''}^{\#}) \\
                  && \hspace*{.5cm} \text{(as $s_i^{\#} = \ell^\#\sigma$)} \\
              & > & \Pol(d_j\sigma) + \sum_{i' \in I_1} \Pol(s_{i'}^{\#}) + \sum_{i'' \in I_2} \Pol(s_{i''}^{\#}) \\
                  && \hspace*{.5cm} \text{(as $\Pol(\ell^\#) > \Pol(d_j)$, and hence $\Pol(\ell^\#\sigma) > \Pol(d_j\sigma)$)}\\
              & \geq & \Pol(d_j\sigma) + \sum_{i' \in I_1} \Pol(s_{i'}^{\#}[r_j \sigma]_{\tau_{i'}}) + \sum_{i'' \in I_2} \Pol(s_{i''}^{\#}) \\
                            && \hspace*{.5cm} \text{(by $\Pol(\ell) \geq \Pol(r_j)$ and (a))}\\
              & = & \sum_{t^{\#} \in \cont(b_j)} \Pol(t^{\#}) \\
          & = & \Pol(b_j)\\
               && \hspace*{.5cm} \text{(as $\Pol$ is $\Com{}$-additive).}     
        \end{array}
      \]

	    \item[(c)] In this case, there exist a rule $\ell \to \{ p_1:r_1, \ldots, p_k:r_k \} \in \SSS$ with $\Pol(\ell) \geq \sum_{1 \leq j \leq k} p_j \cdot \Pol(r_j)$, a substitution $\sigma$, and a position $\pi$ of $s$ such that $s|_\pi =\ell\sigma$, all proper subterms of $\ell\sigma$ are in normal form w.r.t.\ $\SSS$, and $t_h = s[r_h \sigma]_\pi$ for all $1 \leq h \leq k$.
        
      We perform structural induction on $\pi$.
      So in the induction base $\pi = \epsilon$ we have $s = \ell\sigma \itos \{ p_1:r_1\sigma, \ldots, p_k:r_k\sigma \}$.
      As $\Pol(\ell) \geq \sum_{1 \leq j \leq k} p_j \cdot \Pol(r_j)$ holds for all instantiations of the occurring variables, for $t_j = r_j\sigma$ we obtain
      \[
        \Pol(s) \;=\; \Pol(\ell\sigma) \;\geq\;\sum_{1 \leq j \leq k} p_j \cdot \Pol(r_j\sigma) \;=\; \sum_{1 \leq j \leq k} p_j \cdot \Pol(t_j). 
      \]
      
      In the induction step, we have $\pi = i.\pi'$, $s = f(s_1,\ldots,s_i,\ldots,s_n)$,
      $s_i \itos \{ p_1:t_{i,1}, \ldots, p_k:t_{i,k} \}$, and $t_j =
      f(s_1,\ldots,t_{i,j},\ldots,s_n)$
      with $t_{i,j} = s_i[r_j\sigma]_{\pi'}$ for all $1 \leq j \leq k$.
      Then by the induction hypothesis we have $\Pol(s_i) \geq \sum_{1 \leq j \leq k} p_j \cdot \Pol(t_{i,j})$.
      Thus, we have
      \[
        \begin{array}{lcl}
          \Pol(s) & =    & \Pol(f(s_1,\ldots,s_i,\ldots,s_n)) \\
              & =    & f_{\Pol}(\Pol(s_1),\ldots,\Pol(s_i),\ldots,\Pol(s_n)) \\
              & \geq & f_{\Pol}(\Pol(s_1),\ldots,\sum_{1 \leq j \leq k} p_j \cdot \Pol(t_{i,j}),\ldots,\Pol(s_n)) \\
              &  & \; \text{(by weak monotonicity of $f_{\Pol}$ and $\Pol(s_i) \geq \sum_{1 \leq j \leq k} p_j \cdot \Pol(t_{i,j})$)} \\
              & = & \sum_{1 \leq j \leq k} p_j \cdot f_{\Pol}(\Pol(s_1),\ldots,\Pol(t_{i,j}),\ldots,\Pol(s_n)) \\
              &  & \; \text{(as $f_{\Pol}$ is multilinear)} \\
                & =    & \sum_{1 \leq j \leq k} p_j \cdot
          \Pol(f(s_1,\ldots,t_{i,j},\ldots,s_n))\\
          & = & \sum_{1 \leq j \leq k} p_j \cdot
          \Pol(t_j).
        \end{array}
      \]

	    \item[(d)] In this case, there exist a dependency tuple $\langle \ell^\#,\ell
              \rangle \to \{ p_1:\langle d_1,r_1\rangle, \ldots, p_k:\langle d_k,r_k \rangle \} \in \PP$ and a substitution $\sigma$.
      Furthermore, let $a = \Com{n}(s_1^\#, \ldots,\linebreak s_n^\#)$ such that $s_i^\# = \ell^\#\sigma$ and all proper subterms of $\ell^\#\sigma$ are in normal form w.r.t.\ $\SSS$.
      Let $I_1$ be the set of indices such that for all $i' \in I_1$ we rewrite $s_{i'}^\#$ to $s_{i'}^\#[r_j \sigma]_{\tau_{i'}}$ for some position $\tau_{i'}$ such that $s_{i'}^\#|_{\tau_{i'}} = \ell \sigma$, and for all $i'' \in I_2 = \{1,\ldots,n\} \setminus (I_1 \cup \{i\}) $ the term $s_{i''}^\#$ remains the same.
      By Requirement (2), we have $\Pol(\ell^\#) \geq \sum_{1 \leq j \leq k} p_j \cdot \Pol(d_j)$ and by (1) we have $\Pol(\ell) \geq \sum_{1 \leq j \leq k} p_j \cdot \Pol(r_j)$.
      As these inequations hold for all instantiations of the occurring variables, we have
        \[
        \begin{array}{lcl}
          \Pol(a) & = & \sum_{s^{\#} \in \cont(a)} \Pol(s^{\#}) \quad \text{(as $\Pol$ is $\Com{}$-additive)}\\
        & = & \Pol(s_i^{\#}) + \sum_{i' \in I_1} \Pol(s_{i'}^{\#}) + \sum_{i'' \in I_2} \Pol(s_{i''}^{\#}) \\
              & = & \Pol(\ell^\#\sigma) + \sum_{i' \in I_1} \Pol(s_{i'}^{\#}) + \sum_{i'' \in I_2} \Pol(s_{i''}^{\#}) \\
                  && \hspace*{1cm} \text{(as $s_i^{\#} = \ell^\#\sigma$)} \\
              & \geq & \sum_{1 \leq j \leq k} p_j \cdot \Pol(d_j\sigma) + \sum_{i' \in I_1} \Pol(s_{i'}^{\#}) + \sum_{i'' \in I_2} \Pol(s_{i''}^{\#}) \\
                  &  &  \hspace*{2cm} \text{(by $\Pol(\ell^\#) \geq \sum_{1 \leq j \leq k} p_j \cdot \Pol(d_j)$,}  \\
                  &  &  \hspace*{3cm} \text{and hence $\Pol(\ell^\#\sigma) \geq \sum_{1 \leq j \leq k} p_j \cdot \Pol(d_j\sigma)$)}  \\
              & \geq & \sum_{1 \leq j \leq k} p_j \cdot Pol(d_j\sigma) + \sum_{i' \in I_1} \sum_{1 \leq j \leq k} p_j \cdot \Pol(s_{i'}^{\#}[r_j \sigma]_{\tau_{i'}})\\
              &  & + \sum_{i'' \in I_2} \Pol(s_{i''}^{\#})\\
                  &  &  \hspace*{2cm} \text{(by $\Pol(\ell) \geq \sum_{1 \leq j \leq k} p_j \cdot \Pol(r_j)$ and (c))} \\
              & = & \sum_{1 \leq j \leq k} p_j \cdot Pol(d_j\sigma) + \sum_{1 \leq j \leq k} \sum_{i' \in I_1} p_j \cdot \Pol(s_{i'}^{\#}[r_j \sigma]_{\tau_{i'}})\\
              &  & + \sum_{i'' \in I_2} \Pol(s_{i''}^{\#})\\
              & = & \sum_{1 \leq j \leq k} p_j \cdot Pol(d_j\sigma) + \sum_{1 \leq j \leq k} p_j \cdot \sum_{i' \in I_1} \Pol(s_{i'}^{\#}[r_j \sigma]_{\tau_{i'}})\\
              &  & + \sum_{i'' \in I_2} \Pol(s_{i''}^{\#})\\
              & = & \sum_{1 \leq j \leq k} p_j \cdot Pol(d_j\sigma) + \sum_{1 \leq j \leq k} p_j \cdot \sum_{i' \in I_1} \Pol(s_{i'}^{\#}[r_j \sigma]_{\tau_{i'}})\\
              &  & + \sum_{1 \leq j \leq k} p_j \cdot \sum_{i'' \in I_2} \Pol(s_{i''}^{\#})\\
              & = & \sum_{1 \leq j \leq k} p_j \cdot (Pol(d_j\sigma) + \sum_{i' \in I_1} \Pol(s_{i'}^{\#}[r_j \sigma]_{\tau_{i'}}) + \sum_{i'' \in I_2} \Pol(s_{i''}^{\#})) \\
              & = & \sum_{1 \leq j \leq k} p_j \cdot \sum_{t^{\#} \in \cont(b_j)} \Pol(t^{\#}) \\
              & = & \sum_{1 \leq j \leq k} p_j \cdot \Pol(b_j) \quad \text{(as $\Pol$ is $\Com{}$-additive).}
        \end{array}
        \]
	    \item[(e)] In this case, there exist a rule $\ell \to \{ p_1:r_1, \ldots,
              p_k:r_k \} \in \SSS$ and a substitution $\sigma$.
      Furthermore, let $a = \Com{n}(s_1^\#, \ldots, s_n^\#)$ such that for some $1 \leq i \leq n$ and some position $\tau \in \IN^+$ we have $s_i^\#|_\tau =\ell\sigma$ and all proper subterms of $\ell\sigma$ are in normal form w.r.t.\ $\SSS$.
      Let $I_1$ be the set of indices such that for all $i' \in I_1$ we rewrite $s_{i'}^\#$ to $s_{i'}^\#[r_j \sigma]_{\tau_{i'}}$ for some position $\tau_{i'}$ such that $s_{i'}^\#|_{\tau_{i'}} = \ell \sigma$, and for all $i'' \in I_2 = \{1,\ldots,n\} \setminus (I_1 \cup \{i\}) $ the term $s_{i''}^\#$ remains the same.
      Now, it follows that
       \[
        \begin{array}{lcl}
          \Pol(a) & = & \sum_{s^{\#} \in \cont(a)} \Pol(s^{\#})  \quad        \text{(as $\Pol$ is $\Com{}$-additive)}\\
            & = & \sum_{i' \in I_1} \Pol(s_{i'}^{\#}) + \sum_{i'' \in I_2} \Pol(s_{i''}^{\#}) \\
              & \geq & \sum_{i' \in I_1} \sum_{1 \leq j \leq k} p_j \cdot \Pol(s_{i'}^{\#}[r_j \sigma]_{\tau_{i'}}) + \sum_{i'' \in I_2} \Pol(s_{i''}^{\#}) \\
                  &  &  \hspace*{2cm} \text{(by $\Pol(\ell) \geq \sum_{1 \leq j \leq k} p_j \cdot \Pol(r_j)$ and (c))} \\
              & = & \sum_{1 \leq j \leq k} \sum_{i' \in I_1} p_j \cdot \Pol(s_{i'}^{\#}[r_j \sigma]_{\tau_{i'}}) + \sum_{i'' \in I_2} \Pol(s_{i''}^{\#}) \\
              & = & \sum_{1 \leq j \leq k} p_j \cdot \sum_{i' \in I_1} \Pol(s_{i'}^{\#}[r_j \sigma]_{\tau_{i'}}) + \sum_{i'' \in I_2} \Pol(s_{i''}^{\#}) \\
              & = & \sum_{1 \leq j \leq k} p_j \cdot \sum_{i' \in I_1} \Pol(s_{i'}^{\#}[r_j \sigma]_{\tau_{i'}}) + \sum_{1 \leq j \leq k} p_j \cdot \sum_{i'' \in I_2} \Pol(s_{i''}^{\#}) \\
              & = & \sum_{1 \leq j \leq k} p_j \cdot (\sum_{i' \in I_1} \Pol(s_{i'}^{\#}[r_j \sigma]_{\tau_{i'}}) + \sum_{i'' \in I_2} \Pol(s_{i''}^{\#})) \\
              & = & \sum_{1 \leq j \leq k} p_j \cdot \sum_{t^{\#} \in \cont(b_j)} \Pol(t^{\#}) \\
              & = & \sum_{1 \leq j \leq k} p_j \cdot \Pol(b_j) \quad
  \text{(as $\Pol$ is $\Com{}$-additive).}
       \end{array}
      \]
	\end{itemize}

	\noindent 
  \textbf{\underline{2. We create a chain tree $\F{T}^{\leq N}$ for any $N \in \IN$}}
    
  \noindent
	Let $\F{T} = (V,E,L,P)$ be a $(\PP, \SSS)$-chain tree that satisfies (+).
	We define the \emph{value} of any node $x \in V$ in our chain tree.
	\[
		\val: V \to \IN, \qquad x \mapsto
		\begin{cases}
			0,           & \text{ if } x \in \ctleaf\\
			\Pol_{0}(t_x) + 1, & \text{ otherwise }
		\end{cases}
	\]
  Remember that $t_x$ denotes the term in the labeling of the node $x$ and $\Pol_{0}$ denotes the function $\TSet{\SigmaDP}{\VSet} \to \IN$, $t \mapsto \Pol(t)(0,\ldots,0)$, i.e., we use the polynomial interpretation of $t$ and instantiate all variables with $0$.
  Note that indeed $\val(x) \in \IN$ for any node
  $x \in V$ and that we have $\val(x) = 0$ if and only if $x$ is a leaf.

  For any $N \in \IN$, we create a modified tree $\F{T}^{\leq N}$, where we cut everything below a node $x$ of the tree with $\val(x) \geq N+1$.
	Let $C = \ctleaf^{\F{T}^{\leq N}} \setminus \ctleaf^{\F{T}}$ be the set of all new leaves in $\F{T}^{\leq N}$ due to the cut.
	So for all $x \in C$ we have $\val(x) \geq N+1$.

	Our goal is to prove that we have $|\F{T}|_{\ctleaf} = 1$.
	First of all, we prove that $|\F{T}^{\leq N}|_{\ctleaf} = 1$ for any $N \in \IN$.
	However, this does not yet prove that $|\F{T}|_{\ctleaf} = 1$, which we will show afterwards.

  \medskip

	\noindent 
  \textbf{\underline{3. We prove that $|\F{T}^{\leq N}|_{\ctleaf} \geq p_{min}^{N}$ for any $N \in \IN$}}

  \noindent
	This part of the proof drastically changes from the proof of \cref{theorem:ptrs-direct-application-poly-interpretations}, due to the fact that in a rewrite step $a \setitos \{p_1:b_1, \ldots, p_k:b_k\}$ (or a rewrite step $a \setitops \{p_1:b_1, \ldots, p_k:b_k\}$ with a dependency tuple from $\PP_{\geq}$), we cannot guarantee that there exists a $1 \leq j \leq k$ with $\Pol(a) > \Pol(b_j)$.
	Here, it is also possible to have $\Pol(a) = \Pol(b_j)$ for all $1 \leq j \leq k$.
	Thus, there does not have to be a single witness path of length $N$ in $\F{T}^{\leq N}$ that shows termination with a chance of at least $p_{min}^{N}$.
	Instead, we have to use multiple witness paths of finite length to ensure that we have $|\F{T}^{\leq N}|_{\ctleaf} \geq p_{min}^{N}$.
	We first explain how to find such a set of witness paths in a finite sub $(\PP, \SSS)$-chain tree that starts with a node from $\PP_{>}$ and then we prove by induction that we have $|\F{T}^{\leq N}|_{\ctleaf} \geq p_{min}^{N}$ in a second step.

  \medskip

	\noindent 
  \textbf{\underline{3.1. We find witnesses in  sub chain trees of $\F{T}^{\leq N}$}}

	\noindent 
	\begin{minipage}{.57\textwidth}
		In this part, we prove a first observation regarding the existence of
                certain witnesses that shows a guaranteed  decrease of values.
                Let $P$, $P_>$, and $P_\geq$ now refer to $\F{T}^{\leq N}$, i.e., $P$ only
                contains inner nodes of $\F{T}^{\leq N}$ and thus, $P \cap C = \emptyset$.
		For every $x \in P_{>}$, let $T_x$ be the  sub chain tree that starts at $x$ and where we cut every
		edge after the second node from $P_{>}$. This is illustrated in \Cref{Tx figure},
                where an empty node stands for a leaf that already existed in $\F{T}^{\leq N}$
                (i.e., it is in  $\ctleaf^{\F{T}}$ or in $C$).
		Since $\F{T}$ satisfies (+) and is finitely branching, we know that $T_x$ must be finite.
		We want to prove that for such a tree $T_x$, we have a set of leaves (a
                set of witnesses) that show a certain decrease of values compared to the root value $\val(x)$.
		To be precise, we want to prove that there exists a set $W^{x} \subseteq \ctleaf^{T_x}$ of leaves in $T_x$ with the following two properties:
	\end{minipage}%
	\begin{minipage}{.42\textwidth}
		\centering
		\begin{tikzpicture}[scale=0.9]
			\tikzstyle{adam}=[circle,thick,draw=black!100,fill=white!100,minimum size=5mm]
			\tikzstyle{empty}=[circle,thick,minimum size=5mm]
			
			\node[adam, label=center:{\tiny $P_{>}$}] at (0, 0)  (a) {};
			\node[adam, label=center:{\tiny $P_{\geq}$}] at (2, -2.5)  (b) {};
			\node[adam, label=center:{\tiny }] at (-2, -2.5)  (c) {};
			\node[adam, label=center:{\tiny $P_{>}$}] at (0.75, -2.5)  (d) {};
			\node[adam, label=center:{\tiny $P_{\geq}$}] at (-0.75, -2.5)  (e) {};
			\node[empty, label=center:{\small $S$}] at (0, -1)  (middleA) {};

			\node[adam, label=center:{\tiny }] at (-1.5, -4.5)  (nf1) {};
			\node[adam, label=center:{\tiny }] at (0, -4.5)  (nf2) {};
			
			\node[adam, label=center:{\tiny $P_{>}$}] at (2.75, -4.5)  (bb) {};
			\node[adam, label=center:{\tiny }] at (1.25, -4.5)  (bc) {};
			\node[adam, label=center:{\tiny $P_{>}$}] at (2, -4.5)  (bd) {};
			\node[empty, label=center:{\small $S$}] at (2, -3.5)  (middleA) {};

			\node[empty, label=center:{\small $S$}] at (-0.75, -3.5)  (middleA) {};
		
			\draw (a) edge[-] (b);
			\draw (b) edge[-] (d);
			\draw (d) edge[-] (e);
			\draw (e) edge[-] (c);
			\draw (a) edge[-] (c);
			
			\draw (b) edge[-] (bb);
			\draw (bb) edge[-] (bd);
			\draw (bd) edge[-] (bc);
			\draw (b) edge[-] (bc);
			
			\draw (e) -- (nf1) -- (nf2) -- (e);
			
			\begin{scope}[on background layer]
				\fill[green!20!white,on background layer] (0, 0) -- (-2, -2.5) -- (2, -2.5);
				\fill[green!20!white,on background layer] (2, -2.5) -- (1.25, -4.5) -- (2.75, -4.5);
				\fill[green!20!white,on background layer] (-0.75, -2.5) -- (-1.5, -4.5) -- (0, -4.5);
			\end{scope}
		\end{tikzpicture}
		\captionof{figure}{$T_x$}\label{Tx figure}
	\end{minipage}
	\begin{enumerate}
		\item[](W-1) For all $w \in W^{x}$ we have $\adval(w) \leq \adval(x) - 1$.
		\item[](W-2) $\sum_{w \in W^{x}} p_w^{T_x} \cdot p_{min}^{\adval(w)} \geq p_{min}^{\adval(x)}$.
	\end{enumerate}
	Here, we use an adjusted value function $\adval$, so that for every node $x \in V^{\F{T}^{\leq N}}$ we have
	\[
		\adval(x) = 
		\begin{cases}
			0, & \text{if } x \text{ is a leaf in } \F{T}^{\leq N}\\
			\val(x), & \text{otherwise} 
		\end{cases}
	\]             
	This is the same value function as before, except for the nodes in $C$.
	For all $x \in C$ we have $\val(x) \geq N+1$ and $\adval(x) = 0$.
	The first property (W-1) states that all of our witnesses have a strictly smaller value than the root.
	Furthermore, we have to be careful that the probabilities for our witnesses are not too low.
	The second property (W-2) states that the sum of all probabilities for the witnesses is still big enough.
	The additional $p_{min}^{\adval(w)}$ is used to allow smaller probabilities for our witnesses if the value decrease is high enough.

	In order to show the existence of such a set $W^x$, we prove by induction on the height $H$ of $T_x$ that there exists a set of nodes $W^{x}_i$ such that for all $1 \leq i \leq H$ we have
	\begin{enumerate}
		\item[] (W-1!) For all $w \in W^{x}_i$ we have $\adval(w) \leq \adval(x) - 1$.
		\item[] (W-2!) $\sum_{w \in W^{x}_i} p_w^{T_x} \cdot p_{min}^{\adval(w)} \geq p_{min}^{\adval(x)}$.
		\item[] (W-3!) Every $w \in W^{x}_i$ is a leaf of $T_x$ before the $i$-th depth or
                  at depth $i$ of $T_x$. 
	\end{enumerate}
	Then in the end if $i = H$ is the height of the tree $T_x$, we get a set $W^{x} = W_H^{x}$ such that every node in $W_H^{x}$ is a leaf in $T_x$ (i.e., $W_H^{x} \subseteq \ctleaf^{T_x}$) and both (W-1) and (W-2) are satisfied.
		  
	In the induction base, we consider depth $i = 1$ and look at the rewrite step at the root.
	The first edge represents a rewrite step with $\PP_{>}$.
	Let $xE = \{y_1, \ldots, y_k\}$ be the set of the successors of $x$.
	We have $t_x \setitops \{p_1:t_{y_1}, \ldots, p_k:t_{y_k}\}$ using a rule from $\PP_{>}$.
	\begin{figure}[H]
    \centering
		\begin{tikzpicture}
			\tikzstyle{adam}=[rectangle,thick,draw=black!100,fill=white!100,minimum size=4mm]
			\tikzstyle{empty}=[rectangle,thick,minimum size=4mm]
			
			\node[adam] at (0, 0)  (a) {$1:t_x$};
			\node[adam] at (-1.5, -1)  (b) {$p_{1}:t_{y_1}$};
			\node[adam] at (1.5, -1)  (c) {$p_{k}:t_{y_k}$};
			\node[empty] at (0, -1)  (d) {$\ldots$};
			\node[empty] at (4, -0.5)  (e) {$\setitops$ with $\PP_{>}$};
			\node[empty] at (-4, -0.5)  (f) {};
		
			\draw (a) edge[->] (b);
			\draw (a) edge[->] (c);
			\draw (a) edge[->] (d);
		\end{tikzpicture}
		\caption{Nodes in the induction base}
  \end{figure}
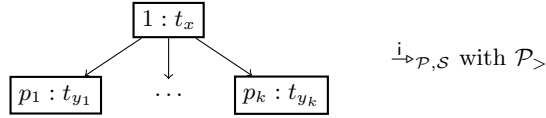
	Due to (b) there is a $1 \leq j \leq k$ with $\Pol(t_x) > \Pol(t_{y_j})$.
    This means $\Pol(t_x) \geq \Pol(t_{y_j}) + 1$, and hence $\val(x) \geq \val(y_j) + 1$, which also implies that $\adval(x) \geq \adval(y_j) + 1$. 
	Since $0 < p_{min} \leq 1$, we therefore have $p_{min}^{\adval(x)} \leq p_{min}^{\adval(y_j) + 1}$.
	Thus we can set $W_1^x = \{y_j\}$ and have (W-1!) satisfied, since $\adval(x) \geq \adval(y_j) + 1$ so $\adval(y_j) \leq \adval(x) - 1$.
	Property (W-3!) is clearly satisfied and (W-2!) holds as well since we have
	\[
		 \sum_{w \in W_1^x} p_{w}^{T_x} \cdot p_{min}^{\adval(w)} = p_j \cdot p_{min}^{\adval(y_j)} \stackrel{p_j \geq p_{min}}{\geq} p_{min} \cdot p_{min}^{\adval(y_j)} \geq p_{min} \cdot p_{min}^{\adval(x) - 1} = p_{min}^{\adval(x)}.
	\]
	
	In the induction step, we consider depth $i > 1$.
	Due to the induction hypothesis, there is a set $W_{i-1}^x$ that satisfies (W-1!), (W-2!),  and (W-3!).
	For every node $w \in W_{i-1}^x$ that is not a leaf of $T_x$, we proceed as
        follows:
        Let $wE = \{y_1, \ldots, y_k\}$ be the set of its successors.
	We rewrite $t_w$ either with $\setitops$ and a rule from $\PP_{\geq}$ or with $\setitos$ and a rule from $\SSS$,
	which rewrites $t_w$ to a multi-distribution $\{p_1:t_{y_1}, \ldots,p_k:t_{y_k}\}$.
	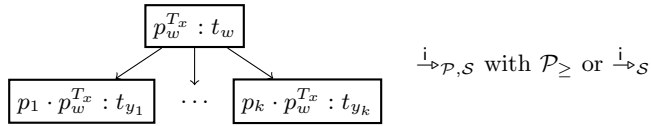
\begin{figure}[H]
	  \centering
          \vspace*{-.2cm}
		\begin{tikzpicture}
			\tikzstyle{adam}=[rectangle,thick,draw=black!100,fill=white!100,minimum size=4mm]
			\tikzstyle{empty}=[rectangle,thick,minimum size=4mm]
			
			\node[adam] at (0, 0)  (a) {$p_w^{T_x}:t_w$};
			\node[adam] at (-1.5, -1)  (b) {$p_{1} \cdot p_w^{T_x}:t_{y_1}$};
			\node[adam] at (1.5, -1)  (c) {$p_{k} \cdot p_w^{T_x}:t_{y_k}$};
			\node[empty] at (0, -1)  (d) {$\ldots$};
			\node[empty] at (4.5, -0.5)  (e) {$\setitops$ with $\PP_{\geq}$ or $\setitos$};
			\node[empty] at (-4.5, -0.5)  (f) {\textcolor{white}{$\setitops$ with $\PP_{\geq}$ or $\setitos$}};
		
			\draw (a) edge[->] (b);
			\draw (a) edge[->] (c);
			\draw (a) edge[->] (d);
		\end{tikzpicture}
		\caption{Induction Step}
                \vspace*{-.2cm}
	\end{figure}
	Due to (d) and (e), we have $\Pol_{0}(t_w) \geq \sum_{1 \leq j \leq k} p_{j} \cdot \Pol_{0}(t_{y_j})$.
	Hence, we either have $\Pol_{0}(t_w) = \Pol_{0}(t_{y_j})$ for all $1 \leq j \leq k$ or there exists at least one $1 \leq j \leq k$ with $\Pol_{0}(t_w) > \Pol_{0}(t_{y_j})$.
	We partition the set $W_{i-1}^x$ into the disjoint subsets $W_{i-1}^{x(1)}$, $W_{i-1}^{x(2)}$, and $W_{i-1}^{x(3)}$, where
	\begin{itemize}
		\item[$\bullet$] $w \in W_{i-1}^{x(1)}$:$\Leftrightarrow$ $w \in
                  W_{i-1}^{x}$ and $w$ is a leaf of $T_x$.
		\item[$\bullet$] $w \in W_{i-1}^{x(2)}$:$\Leftrightarrow$ $w \in W_{i-1}^{x}$ and
 				  $\Pol_{0}(t_w) = \Pol_{0}(t_{y_j})$ for all $1 \leq j \leq k$.
  		\item[$\bullet$] $w \in W_{i-1}^{x(3)}$:$\Leftrightarrow$ $w \in W_{i-1}^{x}$ and
		  		  there exists a $1 \leq j \leq k$ with $\Pol_{0}(t_w) > \Pol_{0}(t_{y_j})$. 
				  We denote this node $y_j$ by $w^+$.
	\end{itemize}
	In the first case, $w$ is already contained in  $W_{i-1}^{x}$.
	In the second case, we get $\adval(w) \geq \adval(y_j)$ for all $1 \leq j \leq k$.
	And in the third case, we have $\Pol_{0}(t_w) > \Pol_{0}(t_{w^+})$, hence $\adval(w^+) \leq \adval(w) - 1$.
	So for all of these nodes, we can be sure that the value does not increase.
	We can now define $W_{i}^x$ as:
	\[
		\begin{array}{lcl}
			W_{i}^x &=& W_{i-1}^{x(1)} \\
			& \cup & \bigcup_{w \in W_{i-1}^{x(2)}} wE\\
			& \cup & \{w^+ \mid w \in W_{i-1}^{x(3)}\}\\
		\end{array}
	\]
	Intuitively, this means that every leaf remains inside of the set of witnesses ($W_{i-1}^{x(1)}$) and for every inner node $w$ we have two cases.
	If there exists a successor $w^+$ with a strictly smaller value, then we replace
        the node $w$ by this successor $w^+$ in our set of witnesses ($\{w^+ \mid w \in W_{i-1}^{x(3)}\}$).
	Otherwise, all of the successors $y_1, \ldots, y_k$ of the node $w$ have the same
        value as the node $w$ itself, so that we have to replace $w$ with all of its
        successors in our set of witnesses as there is no single node with a guaranteed
        value decrease ($\bigcup_{w \in W_{i-1}^{x(2)}} wE$). 

	It remains to show that our properties (W-1!),  (W-2!),  and (W-3!) are still
        satisfied 
for $W_{i}^x$.
	In order to see that (W-1!) is satisfied, note that we have $\adval(w) \leq \adval(x) - 1$, for all $w \in W_{i-1}^x$ by our induction hypothesis.
        Thus, for all $w' \in W_{i-1}^{x(1)} \subseteq W_{i-1}^{x}$, we also obtain
        $\adval(w')\leq \adval(x) - 1$.
	For all $w' \in \bigcup_{w \in W_{i-1}^{x(2)}} wE$, we have $\adval(w') \leq \adval(w)$ for some $w \in W_{i-1}^x$ and thus $\adval(w') \leq \adval(w) \leq \adval(x) - 1$.
	Finally, for all $w' \in \{w^+ \mid w \in W_{i-1}^{x(3)}\}$, we have $\adval(w') \leq \adval(w) - 1$ for some $w \in W_{i-1}^x$ and thus $\adval(w') \leq \adval(w) - 1 \leq \adval(x) - 2 \leq \adval(x) - 1$.

	Property (W-3!) holds as well, since every node $w \in W_{i-1}$ that is not a leaf
        of $T_x$ is at depth $i-1$ of the tree $T_x$ by our induction hypothesis.
	We exchange each such node with one or all of its successors.
	The leaves in $W_{i-1}$ are at a depth of at most $i$ by induction hypothesis and remain in $W_{i}$.
	Hence, all of the nodes of $W_{i}$ are at a depth of at most $i$, and the nodes that are no leaves are at a depth of precisely $i$.
    
	Finally, we regard (W-2!).
	For $\bigcup_{w \in W_{i-1}^{x(2)}} wE$ we have:
	\[
		\begin{array}{lcll}
		& & \sum_{w' \in \bigcup_{w \in W_{i-1}^{x(2)}} wE} p_{w'}^{T_x} \cdot p_{min}^{\adval(w')}\\
		&=& \sum_{w \in W_{i-1}^{x(2)}} \sum_{w' \in wE}p_{w'}^{T_x} \cdot p_{min}^{\adval(w')}\\
		&=& \sum_{w \in W_{i-1}^{x(2)}, wE = \{y_1,\ldots,y_k\}} \sum_{1 \leq j \leq k} p_{y_j}^{T_x} \cdot p_{min}^{\adval(y_j)}\\
		&\geq& \sum_{w \in W_{i-1}^{x(2)}, wE = \{y_1,\ldots,y_k\}} \sum_{1 \leq j
                    \leq k} p_{y_j}^{T_x} \cdot p_{min}^{\adval(w)}\\
                  && \qquad \text{(as $\adval(y_j) \leq \adval(w)$ for all $1 \leq j \leq k$)} \\
		&=& \sum_{w \in W_{i-1}^{x(2)}, wE = \{y_1,\ldots,y_k\}} \sum_{1 \leq j
                    \leq k} p_j \cdot p_w^{T_x} \cdot p_{min}^{\adval(w)} \\
                  && \qquad \text{(as $p_{y_j}^{T_x} = p_j \cdot p_w^{T_x}$ for all $1 \leq j \leq k$)} \\
		&=& \sum_{w \in W_{i-1}^{x(2)}, wE = \{y_1,\ldots,y_k\}} p_w^{T_x} \cdot p_{min}^{\adval(w)} \cdot \sum_{1 \leq j \leq k} p_j\\
		&=& \sum_{w \in W_{i-1}^{x(2)}} p_w^{T_x} \cdot p_{min}^{\adval(w)} \cdot 1 & \text{(as $\sum_{1 \leq j \leq k} p_j = 1$)} \\
		&=& \sum_{w \in W_{i-1}^{x(2)}} p_w^{T_x} \cdot p_{min}^{\adval(w)}.
		\end{array}
	\]

	For $\{w^+ \mid w \in W_{i-1}^{x(3)}\}$ we have:
	\[
		\begin{array}{lcll}
			& & \sum_{w' \in \{w^+ \mid w \in W_{i-1}^{x(3)}\}} p_{w'}^{T_x} \cdot p_{min}^{\adval(w')}\\
			&=& \sum_{w \in W_{i-1}^{x(3)}} p_{w^+}^{T_x} \cdot p_{min}^{\adval(w^+)}\\
			&\geq& \sum_{w \in W_{i-1}^{x(3)}} p_{min} \cdot p_w^{T_x} \cdot p_{min}^{\adval(w^+)} &\text{(as $p_{w^+}^{T_x} \geq p_{min} \cdot p_w^{T_x}$)} \\
			&\geq& \sum_{w \in W_{i-1}^{x(3)}} p_{min} \cdot p_w^{T_x} \cdot p_{min}^{\adval(w)-1} &\text{(as $\adval(w^+) \leq \adval(w) - 1$)} \\
			&=& \sum_{w \in W_{i-1}^{x(3)}} p_w^{T_x} \cdot p_{min}^{\adval(w)}.
		\end{array}
	\]

	All in all, we have
	\begin{align*}
		&\phantom{{}={}}\textstyle\sum_{w' \in W_{i}^{x}} p_{w'}^{T_x} \cdot p_{min}^{\adval(w')}\\
		&=\textstyle\sum_{w' \in W_{i-1}^{x(1)}} p_{w'}^{T_x} \cdot p_{min}^{\adval(w')}\\
		&\phantom{{}={}}+ \textstyle\sum_{w' \in \bigcup_{w \in W_{i-1}^{x(2)}} wE} p_{w'}^{T_x} \cdot p_{min}^{\adval(w')}\\
		&\phantom{{}={}}+ \textstyle\sum_{w' \in \{w^+ \mid w \in W_{i-1}^{x(3)}\}} p_{w'}^{T_x} \cdot p_{min}^{\adval(w')}\\
		&\geq\textstyle\sum_{w \in W_{i-1}^{x(1)}} p_{w}^{T_x} \cdot p_{min}^{\adval(w)}\\
		&\phantom{{}={}}+ \textstyle\sum_{w \in W_{i-1}^{x(2)}} p_{w}^{T_x} \cdot p_{min}^{\adval(w)}\\
		&\phantom{{}={}}+ \textstyle\sum_{w \in W_{i-1}^{x(3)}} p_{w}^{T_x} \cdot p_{min}^{\adval(w)}\\
		&=\textstyle\sum_{w \in W_{i-1}^{x}} p_{w}^{T_x} \cdot p_{min}^{\adval(w)}\\
		&\stackrel{IH}{\geq} p_{min}^{\adval(x)}.
	\end{align*}

	\medskip

	\noindent 
  \textbf{\underline{3.2. We prove that $|\F{T}^{\leq N}|_{\ctleaf} \geq p_{min}^{N}$ for any $N \in \IN$ by induction}}

	\noindent 
	We now prove that $|\F{T}^{\leq N}|_{\ctleaf} \geq p_{min}^{N}$ holds for any $N \in \IN$.
	Let $Z_{k}$ denote the set of nodes in $\F{T}^{\leq N}$ from $P_{>}$ that have precisely $k-1$ nodes from $P_{>}$ above them (so they are themselves the $k$-th node from $P_{>}$).
        Here, $P$, $P_\geq$, and $P_>$ again refer to $\F{T}^{\leq N}$, i.e., we again
        have $P \cap C = \emptyset$.
	Let $\ctleaf_k$ denote the set of all leaves in $\F{T}^{\leq N}$ that are
        reachable by a path that uses less than $k$ nodes from $P_{>}$.
	We show by induction that for all $k \in [1,N+1]$, we have 
	\[
		\sum_{x \in Z_{k} \cup \ctleaf_{k}, 0 \leq \adval(x) \leq N+1-k} p_x^{\F{T}^{\leq N}} \cdot p_{min}^{\adval(x)} \geq p_{min}^{N}
	\]
	Then, for $k = N+1$ we finally have:
	\begin{align*}
		p_{min}^{N}&\leq\textstyle\sum_{x \in Z_{N+1} \cup \ctleaf_{N+1}, 0 \leq \adval(x) \leq N+1-(N+1)} p_x^{\F{T}^{\leq N}} \cdot p_{min}^{\adval(x)}\\
		&=\textstyle\sum_{x \in Z_{N+1} \cup \ctleaf_{N+1}, 0 \leq \adval(x) \leq 0} p_x^{\F{T}^{\leq N}} \cdot p_{min}^{\adval(x)}\\
		&=\textstyle\sum_{x \in Z_{N+1} \cup \ctleaf_{N+1}, \adval(x) = 0} p_x^{\F{T}^{\leq N}} \cdot p_{min}^{\adval(x)}\\
		&=\textstyle\sum_{x \in Z_{N+1} \cup \ctleaf_{N+1}, \adval(x) = 0} p_x^{\F{T}^{\leq N}} \cdot 1\\
		&=\textstyle\sum_{x \in Z_{N+1} \cup \ctleaf_{N+1}, \adval(x) = 0} p_x^{\F{T}^{\leq N}}\\
		&=\textstyle\sum_{x \in \ctleaf_{N+1}} p_x^{\F{T}^{\leq N}}\\
                &\phantom{=}\quad \text{(as $\adval(x) = 0$ iff $x$ is a leaf of $\F{T}^{\leq N}$)}\\
		&\leq\textstyle\sum_{x \in \ctleaf^{\F{T}^{\leq N}}} p_x^{\F{T}^{\leq N}}\\
		&=|\F{T}^{\leq N}|_{\ctleaf}.
	\end{align*}
	In the induction base, we have $k = 1$, and thus
	\[
		\begin{array}{lcl}
		&& \sum_{x \in Z_{1} \cup \ctleaf_{1}, 0 \leq \adval(x) \leq N+1-1} p_x^{\F{T}^{\leq N}} \cdot p_{min}^{\adval(x)}\\
		&=& \sum_{x \in Z_{1} \cup \ctleaf_{1}, 0 \leq \adval(x) \leq N} p_x^{\F{T}^{\leq N}} \cdot p_{min}^{\adval(x)}\\
		&\geq& \sum_{x \in Z_{1} \cup \ctleaf_{1}, 0 \leq \adval(x) \leq N} p_x^{\F{T}^{\leq N}}
				\cdot p_{min}^{N} \quad \text{(since $\adval(x) \leq N$)}\\
		&=& p_{min}^{N} \cdot \sum_{x \in Z_{1} \cup \ctleaf_{1}, 0 \leq \adval(x) \leq N} p_x^{\F{T}^{\leq N}}\\
		&=& p_{min}^{N} \cdot \sum_{x \in Z_{1} \cup \ctleaf_{1}} p_x^{\F{T}^{\leq N}}\\
		&=& p_{min}^{N} \cdot 1 \quad \text{(since $\sum_{x \in Z_{1} \cup \ctleaf_{1}} p_x^{\F{T}^{\leq N}} = 1$)}\\
		&=& p_{min}^{N}.
		\end{array}
	\]
	Here, we have $\sum_{x \in Z_{1} \cup \ctleaf_{1}} p_x^{\F{T}^{\leq N}} = 1$,
        since $Z_{1} \cup \ctleaf_{1}$ are the leaves of the finite sub chain tree where
        we cut everything below the first node of $\PP_{>}$ (i.e., we cut directly after
        the nodes in $Z_{1}$). This tree is finite, because by (+) there is no infinite
        path without $P_>$ nodes.
  All finite chain trees converge with probability $1$.

	In the induction step, we assume that the statement holds for some $k \in [1,N]$.
	Then we have 
	\[\mbox{\small $
		\begin{array}{cl}
		  &p_{min}^{N}\\
                  \stackrel{IH}{\leq}& \sum_{x \in Z_{k} \cup \ctleaf_{k}, 0 \leq \adval(x) \leq N+1-k} p_x^{\F{T}^{\leq N}} \cdot p_{min}^{\adval(x)}\\
		=& \sum_{x \in \ctleaf_{k}} p_x^{\F{T}^{\leq N}} \cdot p_{min}^{\adval(x)} \; + \; \sum_{x \in Z_{k}, 1 \leq \adval(x) \leq N+1-k} p_x^{\F{T}^{\leq N}} \cdot p_{min}^{\adval(x)}\\
		\leq& \sum_{x \in \ctleaf_{k}} p_x^{\F{T}^{\leq N}} \cdot p_{min}^{\adval(x)} \; + \; \sum_{x \in Z_{k}, 1 \leq \adval(x) \leq N+1-k} p_x^{\F{T}^{\leq N}} \cdot \sum_{w \in W^x} p_w^{T_x} \cdot p_{min}^{\adval(w)}\\
		  & \hspace*{2cm} \text{(existence of the set $W^x$ by the previous Step 3.1 and (W-2))} \\
		=& \sum_{x \in \ctleaf_{k}} p_x^{\F{T}^{\leq N}} \cdot p_{min}^{\adval(x)} \; + \; \sum_{x \in Z_{k}, 1 \leq \adval(x) \leq N+1-k} \sum_{w \in W^x} p_x^{\F{T}^{\leq N}} \cdot p_w^{T_x} \cdot p_{min}^{\adval(w)}\\
		=& \sum_{x \in \ctleaf_{k}} p_x^{\F{T}^{\leq N}} \cdot p_{min}^{\adval(x)} \; + \; \sum_{x \in Z_{k}, 1 \leq \adval(x) \leq N+1-k} \sum_{w \in W^x} p_w^{\F{T}^{\leq N}} \cdot p_{min}^{\adval(w)}\\
		  & \hspace*{2cm} \text{(as $p_x^{\F{T}^{\leq N}} \cdot p_w^{T_x} = p_w^{\F{T}^{\leq N}}$).}
	\end{array}$}
	\]
	Every node in $W^x$ is either contained in $\ctleaf_{k+1}$ or contained in $Z_{k+1}$.
	The reason for that is that $W^x$ only contains leaves of $T_x$ and a leaf in
        $T_x$ is either also a leaf in $\F{T}^{\leq N}$ so that it is contained in $\ctleaf_{k+1}$, or contained in $\PP_{>}$ and thus in $Z_{k+1}$, since $x$ is contained in $Z_{k}$ and there is no other inner node from $\PP_{>}$ in $T_x$.
	Furthermore, we know that $\adval(w) \leq \adval(x) - 1$ for all $w \in W^x$ by
        (W-1).
        Moreover, we have $\ctleaf_{k} \subseteq \ctleaf_{k+1}$.
	Thus, we get
	\[
		\begin{array}{lcl}
			&& \sum_{x \in \ctleaf_{k}} p_x^{\F{T}^{\leq N}} \cdot p_{min}^{\adval(x)} \; + \; \sum_{x \in Z_{k}, 1 \leq \adval(x) \leq N+1-k} \sum_{w \in W^x} p_w^{\F{T}^{\leq N}} \cdot p_{min}^{\adval(w)}\\
			&\leq& \sum_{w \in Z_{k+1} \cup \ctleaf_{k+1}, 0 \leq \adval(w) \leq N+1-(k+1)} p_w^{\F{T}^{\leq N}} \cdot p_{min}^{\adval(w)}.
		\end{array}
	\]
	Now, we have shown that $|\F{T}^{\leq N}|_{\ctleaf} \geq p_{min}^{N}$.

  \medskip

	\noindent 
  \textbf{\underline{4. We prove that $|\F{T}^{\leq N}|_{\ctleaf}=1$ for any $N \in \IN$}}
        
	\noindent 
	This part is completely analogous to the fourth part of the proof of \cref{theorem:ptrs-direct-application-poly-interpretations}.
	We only have to consider chain trees instead of rewrite sequences.
	We have proven that $|\F{T}^{\leq N}|_{\ctleaf} \ge p_{min}^N$ holds for all
        $(\PP, \SSS)$-chain trees $\F{T}$ that satisfy (+).
	Hence, for any $N \in \IN$, we have
	\begin{equation}
		\label{RPP-part-4-inf-definition}
	  	p^{\star}_{N}:=\inf_{\F{T} \text{ is a $(\PP, \SSS)$-chain tree satisfying (+)}}
            (|\F{T}^{\leq N}|_{\ctleaf}) \geq p_{min}^{N} > 0.
	\end{equation}
	We now prove by contradiction that this is enough to ensure $p^{\star}_N = 1$.
	So assume that $p^{\star}_N < 1$. 
	Then we define $\epsilon := \frac{p^{\star}_N \cdot (1-p^{\star}_N)}{2}>0$. 
	By definition of the infimum, $p^{\star}_N + \epsilon$ is not a lower bound of
        $|\F{T}^{\leq N}|_{\ctleaf}$ for all $(\PP, \SSS)$-chain trees $\F{T}$ that satisfy (+).
	Hence, there must exist a $(\PP, \SSS)$-chain tree $\F{T}$ satisfying (+) such that   
	\begin{equation}\label{RPP-part-4-pstarEpsilon}
		p^{\star}_N\leq |\F{T}^{\leq N}|_{\ctleaf} < p^{\star}_N + \epsilon.
    \end{equation}
	For readability, let $Z = \ctleaf^{\F{T}^{\leq N}}$ be the set of leaves of the
        tree $\F{T}^{\leq N}$ and let $\overline{Z} = V^{\F{T}^{\leq N}} \setminus
        \ctleaf^{\F{T}^{\leq N}}$ be the set of inner nodes of the tree $\F{T}^{\leq N}$.
Since $p_x^{\F{T}^{\leq N}} = p_x^{\F{T}}$ for all $x \in V^{\F{T}^{\leq N}}$, in the
following we just write $p_x$.        
	By the monotonicity of $|\cdot|_{\ctleaf}$ w.r.t.\ the depth of the tree $\F{T}^{\leq N}$, there must exist a natural number $m^{\star} \in \IN$ such that     
	\begin{equation}
		\label{RPP-part-4-p-star-half}
	 	\sum_{x \in Z, \ctdepth(x) \leq m^*} p_x > \tfrac{p^{\star}_N}{2}.
	\end{equation} 
  Here, $\ctdepth(x)$ denotes the depth of node $x$ in $\F{T}^{\leq N}$. 
	For every $x \in V^{\F{T}^{\leq N}}$ with $\ctdepth(x) = m^{\star}$ and $x \not\in
        Z$, we define the  sub $(\PP, \SSS)$-chain tree of $\F{T}^{\leq N}$ starting at node $x$ by $\F{T}^{\leq N}(x) = \F{T}^{\leq N}[x(E^{\F{T}^{\leq N}})^*]$.
	Then we have 
	\begin{align}\label{RPP-part-4-limit}
		|\F{T}^{\leq N}|_{\ctleaf} = \sum_{x \in Z, \ctdepth(x) \leq m^*} p_x + \sum_{x \in \overline{Z}, \ctdepth(x) = m^*} p_x \cdot |\F{T}^{\leq N}(x)|_{\ctleaf}.
	\end{align}
	Furthermore, we have
	\begin{equation}
		\label{RPP-part-4-first-observation}
		\sum_{x \in \overline{Z}, \ctdepth(x) = m^*} p_x = 1 - \sum_{x \in Z, \ctdepth(x) \leq m^*} p_x,
	\end{equation}
	since $\sum_{x \in \overline{Z}, \ctdepth(x) = m^*} p_x + \sum_{x \in Z,
          \ctdepth(x) \leq m^*} p_x = 1$, as
the nodes $x \in \overline{Z}$ with $d(x) = m^*$
are the leaves of the finite, grounded  sub chain tree of $\F{T}^{\leq N}$ where we cut every edge after the nodes of depth $m^*$.
	We obtain
	\begin{align*}
		& p^{\star}_N+\epsilon 
		\\ 
		{}>{}    & |\F{T}^{\leq N}|_{\ctleaf} \qquad \text{(by~\eqref{RPP-part-4-pstarEpsilon})}
		\\
		{}={}    & \sum_{x \in Z, \ctdepth(x) \leq m^*} p_x + 
		\sum_{x \in \overline{Z}, \ctdepth(x) = m^*} p_x \cdot              
		\underbrace{|\F{T}^{\leq N}(x)|_{\ctleaf}}_{\geq \,
                  p^{\star}_N} \qquad \text{(by~\eqref{RPP-part-4-limit} and~\eqref{RPP-part-4-inf-definition})}
		\\
		{}\geq{} &\sum_{x \in Z, \ctdepth(x) \leq m^*} p_x + \sum_{x \in \overline{Z}, \ctdepth(x) = m^*} p_x \cdot p^{\star}_N
		\\
		{}={} &\sum_{x \in Z, \ctdepth(x) \leq m^*} p_x + p^{\star}_N \cdot \sum_{x \in \overline{Z}, \ctdepth(x) = m^*} p_x
		\\ 
		{}={}    & \sum_{x \in Z, \ctdepth(x) \leq m^*} p_x + p^{\star}_N \cdot (1 - \sum_{x \in Z, \ctdepth(x) \leq m^*} p_x)  \qquad \text{(by~\eqref{RPP-part-4-first-observation})}
		\\   
		{}={}    & p^{\star}_N + \sum_{x \in Z, \ctdepth(x) \leq m^*} p_x - p^{\star}_N \cdot \sum_{x \in Z, \ctdepth(x) \leq m^*} p_x
		\\       
		{}={}    & p^{\star}_N + \sum_{x \in Z, \ctdepth(x) \leq m^*} p_x \cdot (1-p^{\star}_N)
		\\       
		{}>{}    & p^{\star}_N + \left(1-p^{\star}_N\right)\cdot
		\tfrac{p^{\star}_N}{2} \qquad
		\text{(by~\eqref{RPP-part-4-p-star-half})}\\
		{}={}    & p^{\star}_N + \epsilon, \quad \textcolor{red}{\lightning}
	\end{align*}
	a contradiction.
	So $p^{\star}_N=1$.
    In particular, this means that for every $N \in \IN$ and every $(\PP,\SSS)$-chain tree $\F{T}$ satisfying (+), we have
	\begin{equation}
		\label{RPP-part-4-limitMuLeqNIs1}
		|\F{T}^{\leq N}|_{\ctleaf} = 1.
	\end{equation}

  \pagebreak[3]

	\noindent 
  \textbf{\underline{5. Finally, we prove that $|\F{T}|_{\ctleaf}=1$}}
	
	\noindent 
	We adjust the value function for this new tree $\F{T}^{\leq N}$ once again and define:
	\[
		\advaltwo: V^{\F{T}^{\leq N}} \to \IN, \qquad x \mapsto
		\begin{cases}
			N+1, & \text{ if } x \in C \\
			0,   & \text{ if } x \in \ctleaf^{\F{T}}\\
			\Pol_{0}(t_x) + 1, & \text{ otherwise }
		\end{cases}
	\]

	Now for a node $x \in V^{\F{T}^{\leq N}}$ we have $0 \leq \advaltwo(x) \leq N+1$.
	Furthermore, we have $\val(x) \geq \advaltwo(x)$ for all
$x \in V^{\F{T}^{\leq N}}$.
	Let $|\F{T}^{\leq N}|_{\advaltwo} = \sum_{x \in \ctleaf^{\F{T}^{\leq N}}} p_x
        \cdot \advaltwo(x)$.
This (possibly infinite) sum is well defined (i.e., it is a convergent series), because
all addends are non-negative and the sum is  bounded from above by $N+1$, since 
	\[\mbox{\small $\sum\limits_{x \in \ctleaf^{\F{T}^{\leq N}}} \hspace*{-.3cm} p_x
          \cdot \advaltwo(x) \leq \hspace*{-.3cm} \sum\limits_{x \in \ctleaf^{\F{T}^{\leq
                N}}} \hspace*{-.3cm} p_x \cdot (N+1) = (N+1) \cdot \hspace*{-.3cm} \sum\limits_{x \in
            \ctleaf^{\F{T}^{\leq N}}} \hspace*{-.3cm} p_x \leq (N+1) \cdot 1 = N+1$.}\]

        For the root $\ctroot$ of $\F{T}$ and $\F{T}^{\leq N}$, we 
have $\advaltwo(\ctroot) \geq |\F{T}^{\leq N}|_{\advaltwo}$ because every edge is a $\setitops$ or $\setitos$ step
	and due to (d) and (e), we know that the (expected) value is non-increasing.
	
	Now we fix $N \in \IN$ and a chain tree $\F{T}$ satisfying (+), and obtain the corresponding transformed tree $\F{T}^{\leq N}$.
	Note that by~\eqref{RPP-part-4-limitMuLeqNIs1} we have $|\F{T}^{\leq N}|_{\ctleaf}
        = 1 = q_N + q_N'$, where $q_N := \sum_{x \in \ctleaf^{\F{T}^{\leq N}},
          \advaltwo(x) = 0} p_x$
and $q_N' = \sum_{x \in \ctleaf^{\F{T}^{\leq N}},
  \advaltwo(x) > 0} p_x = \sum_{x \in C} p_x$.
Now we can determine 
        $|\F{T}^{\leq N}|_{\advaltwo}$.
	The probabilities of zero-valued nodes (i.e., leaves in the original tree) add up
        to $q_N$, while the probabilities of new leaves due to a cut add up to probability
        $q_N' = 1-q_N$. 
	So $|\F{T}^{\leq N}|_{\advaltwo} = q_N \cdot 0 + (1-q_N) \cdot (N+1) = (1-q_N)\cdot (N+1)$.
	Thus,
	\[
		\val(\ctroot) \geq \advaltwo(\ctroot) \geq |\F{T}^{\leq N}|_{\advaltwo}
                \geq (1-q_N) \cdot (N+1),
	\]
	which implies $q_N \geq 1-\tfrac{\val(\ctroot)}{N+1}$.
	Note that $q_N$ is weakly monotonically increasing and bounded from above by 1 for $N \to \infty$. Hence, $q := \lim_{N \to \infty} q_N$ exists and $1 \geq q \geq \lim_{N \to \infty} (1-\tfrac{\val(\ctroot^{\F{T}})}{N+1}) = 1$, i.e., $q = 1$. 
	Hence, we obtain $|\F{T}|_{\ctleaf} = \lim_{N \to \infty} q_N = q = 1$.
\end{myproof}

By \Cref{lemma:prob-RPP-CT-lemma}, 
we have finally proven the main technical part of the reduction pair processor in the probabilistic setting.
Next, we want to prove the soundness and completeness of the actual processor, which is
fairly easy using \Cref{lemma:prob-RPP-CT-lemma} in addition to the P-Partition lemma.

\ProbRPP*

\begin{myproof}
 \smallskip

        \noindent
        \underline{\emph{Completeness}}

        \noindent    
		  Every $(\PP_{\geq}, \SSS)$-chain tree is also a $(\PP,\SSS)$-chain tree.
	    Hence, if $(\PP_{\geq}, \SSS)$ is not iAST, then $(\PP,\SSS)$ is also not iAST\@.
  \medskip

        \noindent
        \underline{\emph{Soundness}}

        \noindent
			Suppose that every $(\PP_{\geq}, \SSS)$-chain tree converges with probability $1$.
			Assume for a contradiction that there exists a $(\PP, \SSS)$-chain tree $\F{T} = (V,E,L,P)$ that converges with probability $<1$.
			We can partition $P = P_{\geq} \uplus P_{>}$.
			Then we apply the P-Partition Lemma (\cref{lemma:p-partition}),
                        which means that there must be a $(\PP, \SSS)$-chain tree $\F{T}'$
                        with $|\F{T}'|_{\ctleaf} < 1$ such that every infinite path has an infinite number of edges corresponding to $\PP_{>}$ steps.
			But this is a contradiction to \cref{lemma:prob-RPP-CT-lemma}.
\end{myproof}

\NPP*

\begin{myproof}
  Let $(\PP, \SSS)$ be a probabilistic DP problem such that every dependency tuple in
  $\PP$ has the form $\langle \ell^\#, \ell \rangle \to \{1:\langle d,r\rangle\}$ and every probabilistic rewrite rule in $\SSS$ has the form $\ell' \to \{1:r'\}$.
	Note that every $(\PP,\SSS)$-chain tree is a single (not necessarily finite) path.
	For such a chain tree $\F{T}$ that is only a single path, we have only two possibilities for $|\F{T}|_{\ctleaf}$.
	If the path is finite, then $|\F{T}|_{\ctleaf} = 1$, since we have a single leaf in this tree with probability $1$.
	Otherwise we have an infinite path, which means that there is no leaf at all and hence $|\F{T}|_{\ctleaf} = 0$.
        
        \medskip

        \noindent
        \underline{\emph{``only if''}}

        \noindent        
		Assume that $(\nonprob(\PP), \nonprob(\SSS))$ is not innermost terminating.
		Then there exists an infinite innermost $(\nonprob(\PP), \nonprob(\SSS))$-chain 
		\[t_0^\# \itononprobPS \circ \itononprobsstar t_1^\# \itononprobPS \circ \itononprobsstar t_2^\# \itononprobPS \circ \itononprobsstar ...\] 
		such that for all $i \in \IN$ we have $t_i^\# = \ell_i^\# \sigma_i$ for
                some dependency pair $\ell_i^\# \to r_i^\# \in \nonprob(\PP)$ that we use
                in the $i$-th rewrite step and some substitution $\sigma_i$, and every proper subterm of $\ell_i^\# \sigma_i$ is in normal form w.r.t.\ $\SSS$.
		Note that a term is in normal form w.r.t.\ $\SSS$ iff it is in normal form w.r.t.\ $\nonprob(\SSS)$ since they have the same left-hand sides.
		From this infinite innermost $(\nonprob(\PP), \nonprob(\SSS))$-chain, we will now construct an infinite $(\PP,\SSS)$-chain tree $\F{T} = (V,E,L,P)$.
		As explained above, we then know that this infinite innermost $(\PP,\SSS)$-chain tree must be an infinite path, and thus $|\F{T}|_{\ctleaf} = 0$, which means that $(\PP,\SSS)$ is not iAST\@.
		\begin{center}
			\small
			\begin{tikzpicture}
				\tikzstyle{adam}=[rectangle,thick,draw=black!100,fill=white!100,minimum size=4mm]
				\tikzstyle{empty}=[rectangle,thick,minimum size=4mm]
				
				\node[empty] at (-3, -1.5)  (a) {$t_0^\# \itononprobPS \circ \itononprobsstar t_1^\# \itononprobPS \circ \itononprobsstar \ldots$};

				\node[empty] at (1.5, -1.5)  (arrow) {\Huge $\leadsto$};
				
				\node[adam,pin={[pin distance=0.1cm, pin edge={,-}] 135:\tiny \textcolor{blue}{$P$}}] at (3, 0)  (a2) {$1:\Com{1}(t_0^\#)$};
				\node[adam] at (3, -1)  (b2) {$1:a_{1}$};
				\node[adam] at (3, -2)  (c2) {$1:a_{2}$};
				\node[empty] at (3, -3)  (d2) {$\ldots$};

				\draw (a2) edge[->] (b2);
				\draw (b2) edge[->] (c2);
				\draw (c2) edge[->] (d2);
			\end{tikzpicture}
    \end{center}
		We start our chain tree with $(1:\Com{1}(t_0^\#))$.
		In the non-probabilistic rewrite sequence, we have $t_0^\# \itononprobPS
                \circ \itononprobsstar t_1^\#$, so there exists a natural number $m_0 \geq
                1$ such that 
		\[\mbox{\small $t_0^\# = \ell_0^\# \sigma_0 \itononprobPS r_0^\# \sigma_0 = v_1^\#
                \itononprobs v_2^\# \itononprobs
                \ldots \itononprobs v_{m_0}^\# = t_1^\# = \ell_1^\# \sigma_1$} \]
		We now prove that we can mirror this rewrite sequence in $\F{T}$ such that the term $v_i^\#$ is contained in $\cont(a_i)$ for all $1 \leq i \leq m_0$. 
		Let $\langle \ell_0^\#, \ell_0 \rangle \to \{1: \langle d_0, r_0' \rangle\}
                \in \PP$ be the
                dependency tuple that was used to create the dependency pair $\ell_0^\# \to r_0^\#$ in  $\nonprob(\PP)$.
		This means that we have $r_0^\# \in \cont(d_0)$.
		Since we have $t_0^\# = \ell_0^\# \sigma_0$ such that every proper subterm
                of $\ell_0^\# \sigma_0$ is in normal form w.r.t.\ $\SSS$, we can also
                rewrite $\Com{1}(t_0^\#)$ with the dependency tuple $\langle\ell_0^\#,
                \ell_0\rangle  \to \{1: \langle d_0, r_0'\rangle\} \in \PP$ and the substitution $\sigma_0$.
		We result in $a_1 = d_0 \sigma_0$ and thus we have $v_1^\# = r_0^\#
                \sigma_0 \in \cont(d_0 \sigma_0) = \cont(a_1)$.

		In the induction step, we assume that we have $v_i^\# \in \cont(a_i)$ for some $1 \leq i < m_0$.
		In our non-probabilistic rewrite sequence we have
                $v_i^\# \itononprobs v_{i+1}^\#$ using a rule $\ell_i' \to r'_i \in \nonprob(\SSS)$, a substitution $\delta_i$, and a position $\tau \in \IN^+$ such that $v_i^\#|_{\tau} = \ell_i' \delta_i$, every proper subterm of $\ell_i' \delta_i$ is in normal form w.r.t.\ $\SSS$, and $v_{i+1}^\# = v_i^\#[r'_i \delta_i]_{\tau}$.
		We can mirror this rewrite step with the rule $\ell_i' \to \{1:r'_i\} \in \SSS$, since by construction we have $v_i^\# \in \cont(a_i)$.
    We can ignore every other term in $\cont(a_i)$ and simply obtain $a_i \setitos a_{i+1}$ by rewriting $v_i^\#$ to $v_{i+1}^\# = v_i^\#[r'_i \delta_i]_{\tau}$ and every other term in $\cont(a_i)$ remains the same.

    At the end of this induction, we result in $a_{m_0}$.
		Next, we can then mirror the step\linebreak $t_1^\# \itononprobPS \circ \itononprobsstar t_2^\#$ from our non-probabilistic rewrite sequence with the\linebreak same construction and so on.
    This results in an infinite $(\PP,\SSS)$-chain tree.
		To see that this is indeed a $(\PP,\SSS)$-chain tree, note that all of the
                local properties (1-5 of \Cref{def:chain-tree})
                are satisfied since every edge represents a rewrite step with $\setitops$ or $\setitos$.
	The global property (6 of \Cref{def:chain-tree}) is also satisfied since,
                in an infinite innermost $(\nonprob(\PP), \nonprob(\SSS))$-chain, we use
                an infinite number of steps with  $\itononprobPS$ 
so that our resulting chain tree has an infinite number of nodes in $P$.

         \medskip

        \noindent
        \underline{\emph{``if''}}

        \noindent	Assume that $(\PP, \SSS)$ is not iAST\@.
		By \cref{lemma:starting} there exists a $(\PP,\SSS)$-chain tree $\F{T} =
                (V,E,L,P)$ that converges with probability $<1$ and starts with
                $(1:\Com{1}(t^\#))$ such that $t^\# = \ell^\# \sigma$ for some
                substitution $\sigma$ and a dependency tuple $\langle \ell^\#,\ell \rangle
                \to \{1:\langle d,r \rangle\} \in \PP$, and every proper subterm of $\ell^\# \sigma$ is in normal form w.r.t.\ $\SSS$.
	As explained above, this tree must be an infinite path.
		From $\F{T}$, we will now construct an infinite innermost $(\nonprob(\PP), \nonprob(\SSS))$-chain, which shows that $(\nonprob(\PP), \nonprob(\SSS))$ is not innermost terminating.
    \begin{center}
			\centering
			\small
			\begin{tikzpicture}
				\tikzstyle{adam}=[rectangle,thick,draw=black!100,fill=white!100,minimum size=4mm]
				\tikzstyle{empty}=[rectangle,thick,minimum size=4mm]
				
				\node[empty] at (-3, -1.5)  (a) {$t_0^\# \itononprobPS \circ \itononprobsstar t_1^\# \itononprobPS \circ \itononprobsstar \ldots$};

				\node[empty] at (1.5, -1.5)  (arrow) {\Huge $\leftleadsto$};
				
				\node[adam,pin={[pin distance=0.1cm, pin edge={,-}] 135:\tiny \textcolor{blue}{$P$}}] at (3, 0)  (a2) {$1:\Com{1}(t^\#)$};
				\node[adam] at (3, -1)  (b2) {$1:a_{1}$};
				\node[adam] at (3, -2)  (c2) {$1:a_{2}$};
				\node[empty] at (3, -3)  (d2) {$\ldots$};

				\draw (a2) edge[->] (b2);
				\draw (b2) edge[->] (c2);
				\draw (c2) edge[->] (d2);
			\end{tikzpicture}
		\end{center}
    We start our infinite chain with the term $t_0^\# = t^\#$.
		We have $\Com{1}(t^\#) \setitops \{1:a_1\}$, so that there is a dependency
                tuple $\langle \ell^\#,\ell\rangle \to \{1:\langle d,r\rangle\} \in \PP$ and a substitution $\sigma_0$ such that $t^\# = \ell^\# \sigma_0$ and all proper subterms of $\ell^\# \sigma_0$ are in normal form w.r.t.\ $\SSS$.
		Then $a_{1} = d \sigma_0$.

                There must be a term  $r_0^\# \in d$ (i.e.,
                $r_0^\# \sigma_0 \in \cont(a_1)$) such that if we replace $a_{1} = d
    \sigma_0$ with
    $\Com{1}(r_0^\# \sigma_0)$ and obtain the same $(\PP, \SSS)$-chain tree except when we
    would rewrite terms that do not exist anymore (i.e., we ignore these rewrite steps),
    then we still end up with an infinite number of nodes in $P$ (otherwise, $\F{T}$
    would not have an infinite number of nodes in $P$).

		We can rewrite the term $t_0^\#$ with the dependency pair $\ell^\# \to r_0^\# \in \nonprob(\PP)$, using the substitution $\sigma_0$ since $t_0^\# = t^\# = \ell^\# \sigma_0$ and all proper subterms of $\ell^\# \sigma_0$ are in normal form w.r.t.\ $\SSS$.
		Hence, we result in $t_0^\# \itononprobPS r_0^\# \sigma_0$.

                Let us assume that we have already defined the
                $(\nonprob(\PP),\nonprob(\SSS))$-chain  up to the $i$-th step with a
                dependency pair $t_i^\# \itononprobPS r_i^\# \sigma_i$.
    We can apply the rewrite rules from $\nonprob(\SSS)$ to $r_i^\#\sigma_i$ that
    correspond to the rewrite rules from $\SSS$ applied to the term
    $\Com{1}(r_i^\#\sigma_i)$ in our chain tree until the tree reaches
 $\Com{1}(t_{i+1}^\#)$ where
    the next  step with $ \setitops$ is performed.
Thus, the     $(\nonprob(\PP),\nonprob(\SSS))$-chain reaches $t_{i+1}^\#$. 
    Now, we can proceed as before.

		This construction creates a sequence $t_0^\#, t_1^\#, \ldots$ of terms such that 
		\[t_0^\# \; \itononprobPS r_0^\#\sigma_0 \;  \itononprobsstar \; t_1^\# \;
\itononprobPS r_1^\#\sigma_1\;  \itononprobsstar\; \ldots \]
		Therefore, $(\nonprob(\PP), \nonprob(\SSS))$ is not innermost terminating.
\end{myproof}

}
\end{document}